\documentclass[<preprint>]{elsarticle}

\RequirePackage[a4paper,top=12mm,bottom=12mm,left=23mm,right=23mm,head=2mm,includeheadfoot]{geometry}
\RequirePackage{tocloft}

\RequirePackage[nottoc,notlot,notlof]{tocbibind}
\RequirePackage[utf8x]{inputenc}
\RequirePackage{doi}
\usepackage{graphicx}
\usepackage{setspace}
\usepackage{float}
\usepackage{color}
\definecolor{darkgreen}{rgb}{0,.7,0}
\definecolor{linkblue}{rgb}{0.,0.,0.9333}
\definecolor{linkgrey}{cmyk}{0.00, 0.00, 0.00, 0.82}
\hypersetup{
    pdffitwindow=true,      
    pdftitle={Understanding Heavy and Light Quark Brownian Motion from AdS/CFT},
    pdfauthor={A. K. Mes, R. W. Moerman, J. P.Shock, W. A. Horowitz},  
    pdfnewwindow=true,      
    bookmarksnumbered=true,
    colorlinks=true,       
    citecolor=darkgreen,        
    linkcolor=linkgrey,          
    filecolor=black,      
    urlcolor=black,           
    menucolor=black,
    baseurl=http://www.arxiv.org/abs/,
    }
\usepackage{multicol}
\usepackage{slashed}
\usepackage[justification=centering]{caption}

\usepackage[bottom,flushmargin,multiple]{footmisc}
\usepackage{tablefootnote}
\usepackage{booktabs}

\RequirePackage{amsmath,amssymb} 
\DeclareMathOperator\csch{csch}

\newcommand*\Eval[3]{\left.#1\right\rvert_{#2}^{#3}}
\newcommand{\normord}[1]{:\mathrel{#1}:}
\usepackage[font=small]{caption}
\usepackage{subcaption}
\usepackage{enumerate}
\usepackage{natbib} 
\usepackage{hyperref} 

\begin{document}

\numberwithin{equation}{section}

\begin{center}{\Large \textbf{
Strongly Coupled Heavy and Light Quark Thermal Motion from AdS/CFT}}\end{center}

\begin{center}
A. K. Mes\textsuperscript{1*},
R. W. Moerman\textsuperscript{2, 3},
J. P.Shock\textsuperscript{1},
W. A. Horowitz\textsuperscript{2}
\end{center}

\begin{center}
{\bf 1} Department of Mathematics and Applied Mathematics, University of Cape Town, Private Bag X3, Rondebosch 7701, South Africa
\\
{\bf 2} Department of Physics, University of Cape Town, Private Bag X3, Rondebosch 7701, South Africa
\\
{\bf 3} Department of Physics, Astronomy and Mathematics, University of Hertfordshire, Hatfield, Hertfordshire, AL10 9AB, United Kingdom
\\

* msxale002@myuct.ac.za
\end{center}

\begin{center}
\today
\end{center}


\section*{Abstract}
{\bf
We give a pedagogical presentation of heavy and light probe quarks in a thermal plasma using the AdS/CFT correspondence.  Three cases are considered: external heavy and light quarks undergoing Brownian motion in the plasma, and an external heavy quark moving through the plasma with a constant velocity.  For the first two cases we compute the mean-squared transverse displacement of the string's boundary endpoint.  At early times the behaviour is ballistic, while the late time dynamics are diffusive.  We extract the diffusion coefficient for both heavy and light quarks in an arbitrary number of dimensions and comment on the relevance for relativistic heavy ion phenomenology.
}

\vspace{12pt}
\noindent\rule{\textwidth}{1pt}
\tableofcontents
\noindent\rule{\textwidth}{1pt}

\listoffigures
\vspace{3mm}
\listoftables
\noindent\rule{\textwidth}{1pt}

\section*{Note on Notation}
Throughout this article, lower-case \textit{Latin alphabet} indices (e.g. $a, b, c, d$) index over the string's worldsheet parameter space, typically over the standard parameter variables $(\tau, \sigma)$ or the light-cone parameter variables $(\sigma^+, \sigma^-)$. Lower-case \textit{Greek alphabet} indices (e.g. $\mu, \nu, \lambda, \gamma$) are used to index over the spacetime, typically in anti-de Sitter spacetime over the temporal $t$, radial $r$, and transverse directions $X^I$. The subspace of anti-de Sitter spacetime spanned by the transverse directions are always indexed by $I,J$. In some instances the Greek alphabet indices are enumerated -- for example, the isothermal coordinate $X^\mu$, where $\mu \in \{0,1\}$ indexes over the first and second directions in a two-dimensional subspace of spacetime.

\section*{Note on Details}
Some of the details which may prove useful to students have been pruned from this paper. They can still be found at arXiv:2008.09196 [hep-th], if desired. 

\section*{Note on Coloured Links}
In this article, citations are declared in \textcolor{darkgreen}{dark green}; while links to page numbers, figures, tables, sections and equations are designed to be less noticeable and are given in \textcolor{linkgrey}{dark gray}.
\vspace{2mm}

\noindent\rule{\textwidth}{1pt}

\section{Introduction}\label{secIntroduction}

It is both experimentally pertinent and theoretically interesting to examine the movement of a quark undergoing Brownian motion in a hot plasma. The main aim of this work is to present an instructional or pedagogical approach to using the AdS/CFT correspondence to explore the dynamical behaviour of probe heavy and light quarks immersed in a thermal plasma.\\

The plasma in question, the quark-gluon plasma (QGP), is created by colliding lead and gold nuclei at high energies in the Large Hadron Collider (LHC) at CERN; and the Relativistic Heavy Ion Collider (RHIC) in Brookhaven National Laboratory (BNL). Before the first results from RHIC in 2000 it was expected that at high energy densities QCD asymptotic freedom would result in a weakly-coupled system exhibiting gas-like behaviour. However, the experimental results \cite{aronson2005hunting} indicated that the produced QGP (a deconfined state of mater consisting of quarks and gluons) does not expand isotropically and behaves, in-fact, as a strongly-coupled medium \cite{Heinz_2002, gyulassy2004qgp}. The thermal plasma expands anisotropically in its azimuthal direction. The momentum anisotropy of the measured particles is known as \textit{elliptic flow}\footnote{For two comprehensive reviews on discoveries relating to the hydrodynamic description of relativistic
heavy-ion collisions (specifically collective flow and viscosity), see \cite{Heinz_2013, GALE_2013}.}. This discovery did not definitively settle the question as to whether the QGP is weakly- or strongly-coupled. There is evidence to support both. Weak coupling techniques from perturbative quantum chromodynamics (pQCD) have been successful in predicting the distributions of high transverse momentum observables \cite{Majumder_2011, Horowitz_2013, Djordjevic_2014}; while low transverse momentum observables described by near-ideal relativistic hydrodynamics \cite{Hirano_2011, Shen_2011, Qiu_2012, Gale_event_2013} can be understood within a strong coupling paradigm \cite{Gubser_1996, casalderrey2014gauge}. Further, jet suppression \cite{Morad_2014, Casalderrey_Solana_hybrid_2014} and heavy quark energy loss studies \cite{Horowitz_2015} support the theory of a strongly-coupled thermal plasma.\\

At strong coupling, the usual perturbation techniques are no longer applicable and many quantities in gauge theories become difficult to compute. In 1974, 't Hooft \cite{Hooft_1974} postulated that the generalisation of the quantum chromodynamic $SU(3)$ gauge group (where $N_c=3$ is the number of quark colours in the theory) would be an $SU(N_c)$ gauge group, where the large $N_c$ limit is taken ($N_c \rightarrow \infty$ while $\lambda := g_{YM} N_c$ is kept fixed and large). The Yang-Mills coupling is denoted by $g_{YM}$, and the quantity $\lambda$ is known as the 't Hooft coupling. The large $N_c$ limit provides an approximation to compute the gauge theory at strong coupling\footnote{In the large $N_c$ limit, a topological factor $N_c^{\zeta}$ (where $\zeta$ is the Euler characteristic) is assigned to each Feynman graph. Summing over these graphs can be thought of as summing over the worldsheets of the supposed \textit{`QCD string'} (the QCD string dual is unknown and is approximated by the $\mathcal{N}=4$ SYM string dual) to yield the partition function for the large $N_c$ theory \cite{Klebanov_2001, natsuume2015ads}.}.
Inspired by 't Hooft's idea, a promising approach in studying the strong coupling limit of non-abelian quantum field theories (such as QCD) was formulated in 1999 by Juan Maldacena \cite{maldacena1999large}. Known as the anti–de Sitter/Conformal Field Theory (AdS/CFT) correspondence, it is a study of the duality between bulk gravitational physics of a given $d$-dimensional spacetime and a ($d-1$)-dimensional gauge theory on its boundary.
Considering the phenomenologically relevant case, at finite temperature the equivalence exists between type IIB string theory on ten-dimensional spacetime approximated by Einstein's general relativity on a five-dimensional non-compact anti-de Sitter spacetime and a five-dimensional sphere (AdS$_5 \times S^5$) (referred to as either the \textit{gravity} or \textit{bulk} theory); and $\mathcal{N}=4$ supersymmetric Yang-Mills theory (SYM) on four-dimensional Minkowski spacetime \cite{natsuume2015ads} (referred to as either the \textit{gauge} or \textit{boundary} theory). Quantities which are difficult to compute in a strongly-coupled gauge theory can be calculated in a weakly-coupled gravity theory, and translated back via the AdS/CFT dictionary (see table (\ref{tabletranslation}), page \pageref{tabletranslation})\footnote{There is a hitherto unaddressed assumption at play here -- the theory of quantum chromodynamics is not exactly equivalent to $\mathcal{N}=4$ supersymmetric Yang-Mills theory.
The theories are however analogous: while the matter fields and quantum dynamics of the theories differ, their gauge fields and tree-level interactions are common \cite{Gubser_2006}. Qualitatively the plasmas of QCD and $\mathcal{N}=4$ SYM share many properties, such as Debye screening and finite spatial correlation lengths \cite{Bak_2007}. Hence, for studies in the high temperature regime it is assumed that the plasma of $\mathcal{N}=4$ supersymmetric Yang-Mills is similar to the quark-gluon plasma of QCD.}.
After its initial discovery, generalisations of the AdS/CFT correspondence followed and the field is now grouped under \textit{gauge/string dualities}. In the succeeding two decades since the initial publications of \cite{maldacena1999large, witten1998anti, GubserPolyakov_1998, Aharony_1998, Aharony_2000} there has been much interest (and success) in using the gauge/string duality to determine properties of the given gauge theory's plasma state.\\

Propelled by experiments at RHIC and the LHC, theoretical developments in this regard have been inspired by an early calculation \cite{Policastro_2001} which found the ratio of shear viscosity to volume entropy density ($\eta/s$) in the strong coupling regime using the AdS/CFT correspondence. This ratio is experimentally measurable, and data provided by RHIC supported the theoretical result \cite{Adler_2003, Adams_2005}. Building on this, Kovtun \textit{et. al.} \cite{Kovtun_2003, Kovtun_2005} discovered the shear viscosity/entropy density ratio is universal for a large class of these QFTs and, further, that this ratio is a universal lower bound for the viscosity in general\footnote{In addition, see Buchel \textit{et al.}'s contributions \cite{Buchel_2004, Buchel_2005}.}.
This discovery was essential in understanding the elliptic flow that had been observed at RHIC, and sparked interest in the link between string theory and relativistic heavy-ion phenomenology\footnote{A worthwhile review on the topic is given by Casalderrey-Solana \textit{et al.} \cite{casalderrey2014gauge}.}. Exploring this link, two pertinent examples this article will focus on are (i) the dissipative and diffusive behaviour of a massive quark moving through a field theory plasma explained by studying trailing strings in the AdS spacetime
\cite{herzog2006energy, Liu_2006, Gubser_2006, Herzog_2006, Casalderrey_Solana_2006, Liu_2007, Casalderrey_Solana_2007, Gubser_2008};
and (ii) heavy and light quarks undergoing Brownian motion studied by examining transverse fluctuations on open strings in an AdS black hole background \cite{Boer_2009, Son_2009, Fischler_2012, Atmaja_2013, Atmaja_2014, Banerjee_2014, Chakrabortty_2014, Sadeghi_2014_godel, Sadeghi_2014, Fischler_2014, moerman2016semi}.\\ 

Motivated by these past studies, this article seeks to better understand the fluctuating energy loss of light and heavy quarks in a thermal medium. Specifically, heavy and light probe quarks in the gauge theory are considered to undergo Brownian motion\footnote{Brownian motion \cite{brown1828} concerns the ceaseless, random motion of a given particle undergoing microscopic collisions with the constituent particles of the fluid it is immersed in. The motion is responsible for the dissipative nature of a system and its approach to thermal equilibrium. Any particle suspended in a finite temperature fluid undergoes this motion, and as such a probe quark immersed in the QGP behaves the same way. The AdS/CFT correspondence can be used to study the behaviour a probe quark exhibits while interacting with the strongly-coupled thermal plasma.}, 
and the mean-squared transverse displacement of the string's boundary endpoint $s^2(t;d)$ (equivalent to the mean-squared displacement travelled by the external quark which is initially at rest in the thermal plasma) is computed.
Functioning as a pedagogical work for these types of gauge/string calculations, this article aims to present a consistent framework\footnote{A necessary pursuit considering how versatile the field is -- publications on gauge/string dualities span all physics arXivs \cite{natsuume2015ads}.} -- providing insights into published results and including proofs of some of the more vague statements in the literature. In terms of original advancement, previous generalisations of $s^2(t;3)$ in AdS$_3$-Schwarzschild to AdS$_d$-Schwarzschild for the light quark's case are challenged and the method in which to generalise to $s^2(t;d)$ correctly given for the first time. Notwithstanding this, the other main contribution of the article to this field is presenting a definitive, consolidated theoretical derivation -- from which further research, including analytical calculations with different test string configurations or numerical analysis confirming previous analytic results, can be undertaken.

\subsection{Structure of the Article} \label{subsecStructure}

This article is organised in two parts. The first part provides some useful theoretical background needed to understand the bulk calculations in Part II. The central results of this article are given in the second part.
\vspace{0.5cm}

\textbf{In Part I:}\\

First, the gauge/string duality is introduced in section (\ref{secGaugeString}). Particular attention is paid to the \textit{throat} construction of anti-de Sitter spacetime as a limit of D3-brane geometry, and the justification of the AdS/CFT conjecture. 
In section (\ref{secBrownMot}) the basic theory of Brownian motion and the non-retarded and retarded Langevin models are explored. The first and second fluctuation-dissipation theorems are derived. The mean-squared displacement $s^2(t)$ is calculated and its behaviour in the early and late time limits are discussed. At early times, the Brownian particle’s behaviour is proportional to time and the motion is expected to be ballistic $s(t) \sim t$; while, at late times, the Brownian particle's motion is diffusive, $s(t) \sim \sqrt{t}$.
\vspace{0.3cm}

\textbf{In Part II:}\\

To begin Part II, section (\ref{secHeavyQuark}) focuses on calculating an external heavy quark's diffusion coefficient in the bulk AdS$_3$-Schwarzschild spacetime. The dual description of the on-mass-shell heavy quark exhibiting Brownian motion in the thermal plasma -- an open string attached at the anti-de Sitter boundary\footnote{The boundary of a D7-brane stretching from $r=\infty$ to the string's boundary endpoint \cite{Karch_2002}.} and hanging towards the horizon undergoing transverse fluctuations -- is used to compute the mean-squared transverse displacement of the string's boundary endpoint $s^2(t;3)$. The diffusion coefficient is extracted at late times.
Similarly, in section (\ref{secLightQuark}) an off-mass-shell light quark exhibiting Brownian motion in the thermal plasma is modelled as an open string undergoing transverse fluctuations (initially stretched between the AdS boundary and just above the horizon) whose AdS boundary endpoint is released to fall at the local speed of light in the presence of a space filling D7-brane \cite{Karch_2002}. Initially this section focuses on calculating an external light quark's diffusion coefficient in the bulk AdS$_3$-Schwarzschild spacetime. Then, importantly, the results are generalised to AdS$_d$-Schwarzschild by expanding and truncating the near-horizon metric.
Further, in section (\ref{secDragForce}), an infinitely massive probe quark moving through the thermal plasma with a constant velocity is considered. An open string trailing out behind the quark (arcing down into AdS-Schwarzschild) is used to model the situation. From this, the drag force on the test string is calculated in the bulk and rewritten -- via the AdS/CFT correspondence -- in terms of relevant quantities in the gauge theory\footnote{Historically, the drag force calculation was the first application of AdS/CFT to understand the behaviour of a probe heavy quark in the quark-gluon plasma (see \cite{Gubser_2006, herzog2006energy}).}. The Langevin equation is then used to extricate the friction coefficient, and (by means of the Einstein-Sutherland relation) the diffusion coefficient for a heavy quark in AdS$_d$-Schwarzschild.
Section (\ref{secConclusion}) concludes the article and proposes avenues for future research.\\

There are a number of appendices to this article. Appendix (\ref{appPolyakov}) derives the string equations of motion by calculating the functional derivative of the Polyakov action with respect to the string worldsheet coordinates and setting this variation to zero. In an analogous fashion, the string equations of motion for the transverse fluctuations are found in appendix (\ref{appNambuGoto}) by varying the Nambu-Goto action. The Virasoro constraints and string equations of motion in isothermal
coordinates are given in appendix (\ref{appVirasoroEoMIso}); while a derivation of the energy of a test string in an AdS-Schwarzschild background can be found in appendix (\ref{appEnergyCurvedSpace}).
Appendix (\ref{appCommutationRels}) aims to fix the normalization constant $A_\omega$, thereby completely determining the general solution for the transverse equations of motion. As the penultimate, appendix (\ref{appTortoiseCoord}) calculates the leading order contributions of the tortoise coordinate in the near-horizon region. Appendix (\ref{appCode}) provides details for accessing the GitHub repository where the latest annotated Mathematica code notebooks supporting the analytical analysis of open string evolution presented here, can be found. The authors believe it is important to make the Mathematica notebooks available for reference purposes and in the spirit of open collaboration.

\part{\Large{Theory}} \label{partTheory}
\section{The Gauge/String Duality}  \label{secGaugeString}

As recognised by 't Hooft \cite{Hooft_1974}, understanding gauge theories with an $SU(N_c)$ gauge group in the large $N_c$ limit perhaps offers the best course to illuminating the strong coupling behaviour of QCD. In the late 1900's it was suspected that string theory could describe the large $N_c$ limit \cite{Polyakov_1998}. Maldacena \cite{maldacena1999large} first made the suggestion that a ($d-1$)-dimensional, conformally invariant field theory in the limit of large $N_c$ corresponds to string theory and supergravity on anti-de Sitter spacetime in $d$ dimensions times by a $d$-dimensional spherical compact manifold (AdS$ \times S$). This idea, based on the holographic principle\footnote{\textit{Holographic} as proposed by 't Hooft, Susskind and Thorn in \cite{hooft1993dimensional, Susskind_1995, thorn1992reformulating} respectively. In this sense, a holographic theory encodes a theory in $d$ dimensions by a theory in ($d-1$) dimensions.} and formally known as the AdS/CFT correspondence, was developed in the months that followed by \cite{witten1998anti, GubserPolyakov_1998, Aharony_1998, Aharony_2000}, to name a few. The AdS/CFT correspondence is an example of gauge/string dualities which have subsequently been developed into a powerful tool in understanding strongly-coupled systems\footnote{A duality between two theories describes a situation where one theory is in the strong coupling limit, while the other theory is in the weak coupling limit \cite{natsuume2015ads}. In order to study a strongly-coupled gauge theory, the equivalent weakly-coupled gravitational string theory (i.e. a theory where the curvature of the spacetime is small) can instead be used.}.

\subsection{Anti-de Sitter Spacetime as a limit of D3-brane Geometry} \label{secAdSSpaceTime}

The solution of low energy, type IIB string theory containing D3-branes prompted the formulation of the gauge/string dualities. Solving the supergravity equations of motion yields the spacetime metric sourced by $N_c$ D$p$-branes \cite{gibbons1988black, horowitz1991black, garfinkle1991charged}. Specifically, the spacetime for the extremal D3-brane\footnote{Only the ground state of the $N_c$ D3-branes is being considered.} is

\begin{equation}
    ds^2 = H^{-1/2} \left(-dt^2 +d\vec{x}^2\right) + H^{1/2} \left(dr^2+r^2 d \Omega_5^2 \right) \, , \label{D3braneMetric}
\end{equation}
\vspace{0.1cm}

\noindent where the D3-brane is extended along the spatial coordinates $\vec{x} = \left( x_1, x_2, x_3\right)$; $H(r) = 1 + l^4/r^4$ is known as the \textit{warping factor}; and the second term metric describes the $y-$directions transverse to the D3-brane written in spherical coordinates (where $r^2 = y_1^2+y_2^2+...+y_6^2$ is the radial coordinate). The parameter $l$ is interpreted as the characteristic length scale of the range of the $N_c$ D3-branes' gravitational effects \cite{casalderrey2014gauge}. In the limit $r \gg l$, $H \simeq 1$ and the metric Eq.(\ref{D3braneMetric}) reduces to Minkowski spacetime\footnote{A small correction proportional to $l^4/r^4$ is present.}. In the limit $r \ll l$, $H \simeq l^4/r^4$ (which corresponds to a series expansion around $r/l=0$ to leading order). Hence, in this limit the metric Eq.(\ref{D3braneMetric}) becomes

\begin{equation}
    ds^2 = ds_{\text{AdS$_5$}}^2 + l^2 d \Omega_5^2 \, , \,\,\,\,\,\,\,\,\,\, \text{where} \,\,\,\,\,\,\,\,\,\,  ds_{\text{AdS$_5$}}^2 = \frac{r^2}{l^2} \left(-dt^2 +d\vec{x}^2\right) + \frac{l^2}{r^2} dr^2 \, . \label{D3braneMetric_NH}
\end{equation}
\vspace{0.1cm}

\begin{figure}[!htb]
    \centering
    \includegraphics[width=0.6\textwidth]{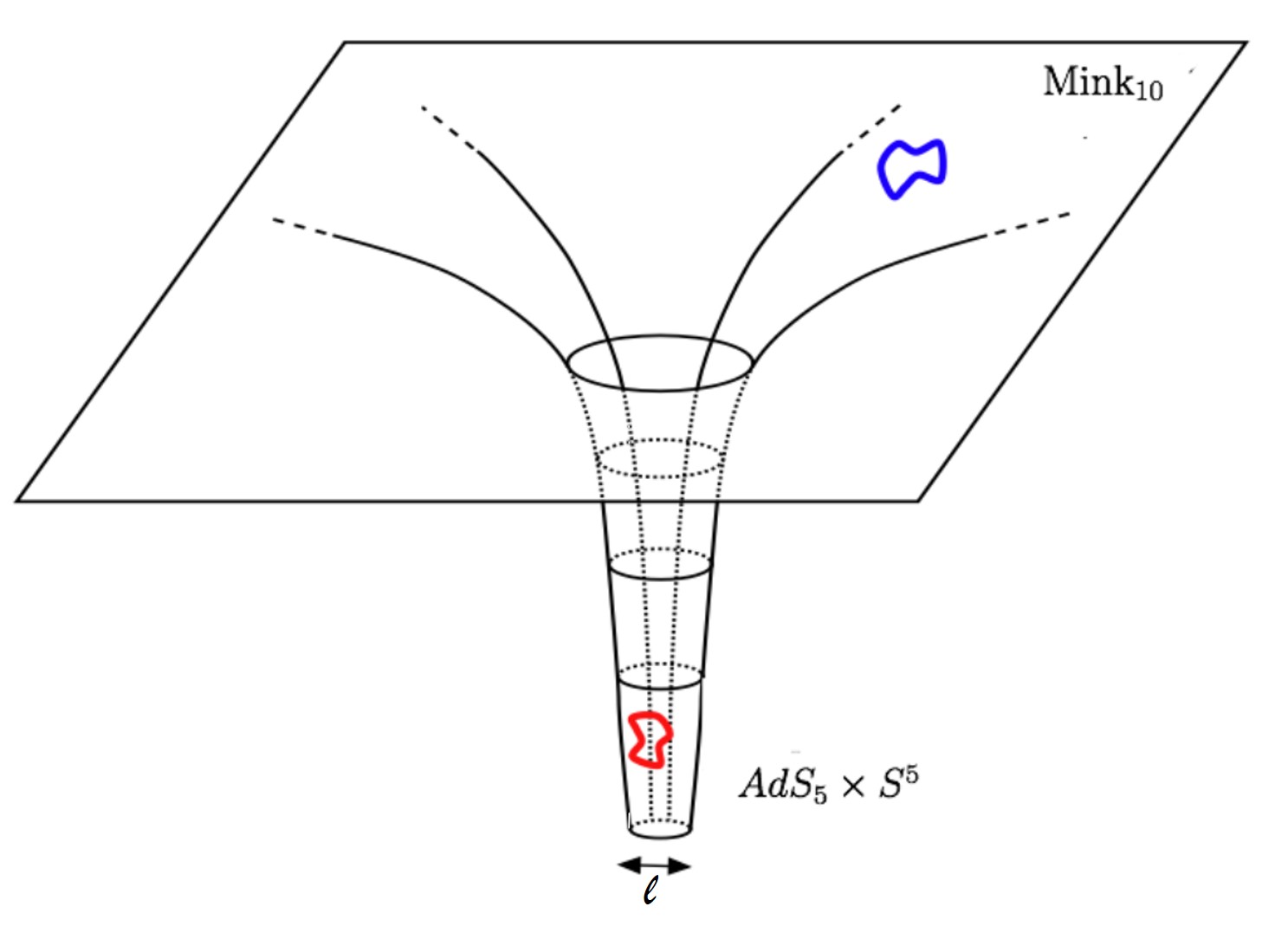}
    \caption[The spacetime around $N_c$ D3-branes in type IIB theory]{\label{fig:DeepThroat} The spacetime around $N_c$ D3-branes in type IIB theory, where $l$ is the radius of curvature of AdS$_d$ and $S^d$ (adapted from \cite{Mateos_2007, casalderrey2014gauge}). See page \pageref{DeepThroatExplanation} for details.}
    \end{figure}

\noindent Now $l$ is identified as the radius of curvature of AdS$_d$ and $S^d$ (see table (\ref{tabletranslation}), page \pageref{tabletranslation});
$ds_{\text{AdS$_5$}}^2$ is the metric of five-dimensional anti-de Sitter spacetime\footnote{The defining characteristic of AdS is a spacetime described by a constant negative curvature.}\footnote{The five-dimensional anti-de Sitter spacetime metric Eq.(\ref{D3braneMetric_NH}) represents a Poincar{\'e} chart of AdS$_5$ spacetime. There is an alternative formulation of anti-de Sitter spacetime in terms of global coordinates which provides a global coordinate chart (see, for example, \cite{Bayona_2007}).};
and $d \Omega_5^2$ is the metric on $S^5$ with unit radius.
Therefore, the spacetime sourced from a stack of D3-branes corresponds to ten-dimensional Minkowski spacetime far away from the branes (see blue patch, figure (\ref{fig:DeepThroat})),
while a $\textit{throat}$ geometry of the form AdS$_5 \times S^5$ becomes apparent close to the branes (see red patch, figure (\ref{fig:DeepThroat})). \label{DeepThroatExplanation}\\

Generalising to non-zero finite temperature $T$ systems by exciting the degrees of freedom on the D3-brane, adapts the anti-de Sitter spacetime part of the metric. Specifically, Eq.(\ref{D3braneMetric_NH}) becomes

\begin{equation}
     ds^2 = ds_{\text{AdS$_5$-Sch}}^2 + l^2 d \Omega_5^2 \, , \,\,\,\,\,\,\,\,\,\, \text{where} \,\,\,\,\,\,\,\,\,\, ds_{\text{AdS$_5$-Sch}}^2 = \frac{r^2}{l^2} \left(-h dt^2 +d\vec{x}^2\right) + \frac{l^2}{r^2\, h} dr^2 \, , \label{AdS-SchMetric}
\end{equation}

\noindent where $h(r)= 1 - r_H^4/r^4$ is known as the $\textit{blackening factor}$. The first term of the metric in Eq.(\ref{AdS-SchMetric}) describes an AdS$_5$ spacetime with a Schwarzschild black hole horizon at $r = r_H$\footnote{The Schwarzschild black hole is the solution to the Einstein equation with no matter fields or cosmological constant.}\footnote{The radial position of the black-brane horizon $r_H$ is proportional to the temperature $T$ (see table (\ref{tabletranslation}), page \pageref{tabletranslation}).}.\\

    \vspace{0.1cm}

A stack of $N_c$ D3-branes can be described by (i) closed strings propagating in a curved spacetime geometry, which -- in the low energy limit -- becomes closed IIB string theory in AdS$_5 \times S^5$ (as discussed above); but also (ii) open strings attached to a hyper-plane in  flat spacetime, where the low-energy limit is given by  $\mathcal{N}=4$, $SU(N_c)$ supersymmetric Yang-Mills theory \cite{Mateos_2007, casalderrey2014gauge}. The hypothesis that the two descriptions are equivalent was the first example of the AdS/CFT conjecture.

\subsection{The AdS/CFT Conjecture}

The best example of a gauge/string duality remains the equivalence between type IIB string theory in an AdS$_5 \times S^5$ background to $\mathcal{N}=4$, $SU(N_c)$ SYM theory on the four-dimensional Minkowski spacetime boundary of $AdS_5$. As has been seen in subsection (\ref{secAdSSpaceTime}), if the above duality is considered at finite temperature, the addition of a black hole in the anti-de Sitter spacetime is necessary. The spacetime is then referred to as the anti-de Sitter-Schwarzschild spacetime (or AdS-Schwarzschild).\\

The AdS/CFT correspondence claims that the partition functions of the gauge and gravitational theories are equivalent, i.e. $Z_{\text{gauge}} = Z_{\text{AdS}}$ \cite{natsuume2015ads}. In doing so the conjecture specifies a \textit{dictionary} between two physical theories which appear to be very different (these relations are summarised in the reference table (\ref{tabletranslation})). In practice, the strong/weak coupling nature of the duality means that it is difficult to explicitly test the validity of the conjecture\footnote{With contemporary knowledge, pertubative computations in $\lambda$ can mostly only be achieved in the field theory, while in the string theory only pertubative calculations in $1/\lambda$ are possible. Therefore, comparing the correlation functions of these theories is in principle not realisable \cite{Aharony_2000}.}. There are, however, several contiguous \textit{tests} -- coupling-independent properties of these theories that can be compared to investigate the duality. Specifically for AdS$_5 \times S^5$/$\mathcal{N}=4$ supersymmetric Yang-Mills, \textit{tests} include (i) comparing the global symmetries of the theories; (ii) matching the coupling-independent correlation functions (these are normally protected
from quantum corrections and related to anomalies); (iii) comparing the spectrum of chiral operators; and (iv) examining the qualitative behaviour of the theory deformed by chiral operators\footnote{Qualitative tests include the existence of confinement in finite temperature gauge theories \cite{witten1998anti_gauge}, and examining how the theory behaves on its moduli space \cite{douglas1998branes, Das_1999, Gonzalez_Rey_1999}.} \cite{Aharony_2000}.\\

\begin{table}[!ht]
\centering
 \begin{tabular}{|c|c|c|}
\hline
      \,\,\,\textbf{AdS$_d$}  \,\,\,& \,\,\,\textbf{$(d-1)$-dimensional Gauge Theory}  \,\,\, & \textbf{Description} \\
      \hline
      $l$  & $\sqrt{\alpha^{\prime}} \, \lambda^{1/4}$  & Radius of curvature of AdS$_d$ and $S^d$  \\
      $\ell_s$  & $\sqrt{\alpha^{\prime}} \, \equiv \, \lambda^{-1/4} l$  & Fundamental string length scale   \\
      $T_0$  & $1/2 \,\pi\, \alpha^{\prime}$  & String tension\\
      $(l/\ell_s)^4$  & $\lambda$  & 't Hooft coupling\footnotemark \\
      $r_H$  & $4 \pi\, l^2 \,T/(d-1)$  & Radial position of the black hole horizon  \\
     $ (d-1)\, r_H/4 \pi\, l^2$  & $T \, \equiv \, 1/\beta$  & Temperature of the gauge theory\footnotemark  \\
     $r_c \, \equiv \, (r_s + \ell_0)$  & $2 \,\pi\, \alpha^{\prime} (M_{\text{rest}}+ \Delta m)$  & Minimal radius of D7-brane\footnotemark \\
      $T_0 \, r_H$  & $\Delta m(T)$  & Thermal rest mass shift \\
     $T_0 \, (r_c - r_H)$  & $M_{\text{rest}}(T)$  & Static thermal mass of external particle\footnotemark \\
      \hline
    \end{tabular} \caption[The AdS/CFT dictionary - translating between quantities in the bulk and boundary theories]{The AdS/CFT dictionary - translating between quantities in the bulk and boundary theories (adapted from Herzog \textit{et al.} \cite{herzog2006energy})} \label{tabletranslation}
    \end{table}
    \vspace{0.1cm}

\footnotetext[20]{The 't Hooft coupling \cite{Hooft_1974} is specifically defined in the phenomenologically relevant AdS$_5$/$\mathcal{N}=4$ SYM case. In this case: $\lambda \, \equiv \, g_{\text{YM}}^2\,N_c$, where $N_c$ is the number of colours and the Yang-Mills coupling is related to the string coupling by:  $g_{\text{YM}}\, = \, 2 \sqrt{\pi \,g_s}$. However, this article follows \cite{Boer_2009} in using the same terminology to refer to $(l/\ell_s)^4$ for general $d$ dimensions.}
\footnotetext[21]{This is equivalent to the Hawking temperature of the black hole.}
\footnotetext[22]{A \textit{UV} cut-off surface imposed near the boundary in order to consider external particles of finite mass in the gauge theory.}
\footnotetext[23]{In the AdS$_5$/$\mathcal{N}=4$ SYM case, this is the free energy of a quark at rest in $\mathcal{N}=4$ SYM plasma. In the limit of zero temperature it is equal to the QCD Lagrangian quark mass, $m_q$ \cite{tanabashi2018m}.} 

The main advantage of the AdS/CFT correspondence is that otherwise intractable problems are able to be solved by their mapping onto the equivalent dual. Applications of the correspondence include gluon scattering amplitudes calculations in strongly-coupled $\mathcal{N}=4$ supersymmetric Yang-Mills \cite{alday2007gluon}; and the study of holographic superconductivity and critical phenomena \cite{Hartnoll_2008, Gubser_Breaking_2008, Maeda_2009}\footnote{See \cite{natsuume2015ads} for a extensive review on systems with phase transitions in AdS/CFT, and \cite{hartnoll2009lectures} for an introduction to holographic methods for condensed matter physics.}.

\section{Brownian Motion: the Dynamics of Langevin's Model}\label{secBrownMot}

Discovered by Robert Brown in 1827 \cite{brown1828} the ceaseless and irregular motion which small pollen particles undergo when suspended in water, became known as Brownian motion. Brown postulated that this movement had a physical rather than biological origin. In 1905 Albert Einstein \cite{einstein1905motion}\footnote{Original publication: \textit{{\"U}ber die von der molekularkinetischen Theorie der W{\"a}rme geforderte Bewegung von in ruhenden Fl{\"u}ssigkeiten suspendierten Teilchen}, Annalen der physik 322.8 (1905): 549-560. Translated from German.} combined statistical mechanics and the diffusion equation, to arrive at a theoretical explanation of the phenomenon.
His famous formula for the particle's mean-squared displacement, $\langle s^2(t) \rangle$, provided an observable for experimental physicists to measure.
Independently, Marian Smoluchowski \cite{smoluchowski1906kinetic}\footnote{Original publication: \textit{Zarys kinetycznej teoriji ruchów browna i roztworów metnych}, Rozprawy i Sprawozdania z Posiedzen Wydzialu Matematyczno-Przyrodniczego Akademii Umiejetnosci 3 (1906): 257-282. Reprinted in German: \textit{Zur kinetischen theorie der brownschen molekularbewegung und der suspensionen}, Annalen der physik 326.14 (1906): 756-780. Translated from German.} derived a similar result based on combinatorics and kinetic theory's mean-free-path approximation \cite{brush1968history}.\\

In a mathematical context, Louis Bachelier wrote his 1900 doctoral thesis on modelling prices on the Paris stock exchange as the limit of a random walk \cite{bachelier1900theory}\footnote{Original publication: \textit{Théorie de la spéculation}, Annales scientifiques de l'École normale supérieure. Vol. 17. (1900).
Translated from French.}. Neither Einstein nor Bachelier rigorously classified Brownian motion as a stochastic process. This was taken up by the American mathematician, Norbert Wiener, who ultimately proved the existence of Brownian motion by defining it as a stochastic process in 1923 \cite{wiener1923differential}.
Wiener's output includes a series of papers starting in 1918 giving the mathematical definition and properties of the physical process of Brownian motion abstracted as a stochastic process. Hence, Wiener processes (essentially the same concept as Brownian motion but emphasising the mathematical aspects) developed separately to the school of thought emphasising the physical aspects.\\

Turning back towards the explanations of physicists, Einstein and Smoluchowski failed to take into account the inertia of the particle undergoing Brownian motion. Applying Newtonian dynamics to a Brownian particle, Paul Langevin \cite{langevin1908}\footnote{Original publication: \textit{Sur la théorie du mouvement brownien}, Compt. Rendus 146 (1908): 530-533. Translated from French.} arrived at a useful phenomenological model in 1908.

\subsection{The Non-retarded Langevin Model}
Consider the following: a non-relativistic particle of mass $m$, undergoing Brownian motion in one spatial dimension, can be described by the Langevin equation

\begin{flalign}
& \begin{aligned}
    & \Dot{p}(t) \, = \, - \gamma_0 \, p(t) \, + \, F(t)\, , \\[.2cm]
    & \Rightarrow m\, \Dot{v}(t) \, = \, - m \, \gamma_0 \, v(t) \, + \, F(t)\, , \label{LangEq}
\end{aligned} &
\end{flalign}
\vspace{0.1cm}

\noindent where $\Dot{p} \equiv d p/dt $, $\gamma_0$ is a constant known as the friction coefficient,  and $v$ is the velocity of the particle. Eq.(\ref{LangEq}) is a stochastic differential equation and can be thought of as consisting of two parts: i). a systematic term $- m  \gamma_0  v$ corresponding to the friction force, and ii). a fluctuating part $F(t)$ corresponding to a random force \cite{uhlenbeck1930theory}.
It is worthwhile to note that although it is simpler to understand the Langevin equation in terms of a friction force and a random force -- on a microscopic scale both forces are caused by the particle's collision with the constituent modules of the immersive  fluid \cite{Boer_2009}. \\

Concerning the statistical properties of $F(t)$, three assumptions are made.
First, the average value of the random force at time $t$, over a particle ensemble with the same initial velocity, vanishes

\begin{equation}
    \langle F(t) \rangle \, = \, 0\, , \label{LangAss1}
\end{equation}
\vspace{0.1cm}

\noindent where an ensemble is defined as a large number of similar, but independent particles. Second, the autocorrelation of the random force is related to a constant, $\kappa_0$, which measures the fluctuation strength

\begin{equation}
    \langle  F(t) \, F(t')  \rangle \, = \, \kappa_0 \, \delta(t-t')\, , \label{LangAss2}
\end{equation}
\vspace{0.1cm}

\noindent where $\delta(x)$ is the Dirac delta function. This assumption implies

\begin{equation}
  \int_{-\infty}^{\infty} \langle  F(t) \, F(t')  \rangle \, dt'\, = \, \kappa_0
  \, . \label{DiracProp}
\end{equation}
\vspace{0.1cm}

\noindent Lastly, $F(t)$ is assumed to be a Gaussian process\footnote{The central limit theorem can be employed to justify this assumption.
The random force $F(t)$ is thought of as resulting from the superposition of many identically distributed random functions, since the Brownian particle has undergone multiple collisions \cite{pottier2010nonequilibrium}.}.\\

To find the solutions of Eq.(\ref{LangEq}) -- a first order linearly separable differential equation -- the variables are separated, each side is multiplied with respect to $e^{\gamma_0\,t}$, and integrated between initial time $t_0$ and $t$

\begin{flalign}
& \begin{aligned}
    & m\, \Dot{v}(t) \,e^{\gamma_0\,t} \, + \, m \, \gamma_0 \, v(t)\, e^{\gamma_0\,t} \, = \, F(t)\, e^{\gamma_0\,t}\,  \\[.2cm]
    & \Rightarrow \int_{t_0}^t \Dot{v}(t') \,e^{\gamma_0\,t^{'}}\, dt' \, + \,  \left[ \Eval{v(t')\, e^{\gamma_0\,t'}}{t_0}{t} \, - \, \int_{t_0}^t \Dot{v}(t') \,e^{\gamma_0\,t^{'}}\, dt'\right]\, = \, \frac{1}{m}\, \int_{t_0}^t\, F(t')\, e^{\gamma_0\,t'} \, dt'\,  \\[.2cm]
    & \Rightarrow  v(t)\, = \, v_0 \, e^{-\gamma_0\,t} \, + \,\frac{1}{m}\, \int_{t_0}^t\, F(t')\, e^{- \gamma_0\,(t -t')} \, dt' \, , \label{LangVelocity}
\end{aligned} &
\end{flalign}
\vspace{0.1cm}

\noindent where the initial velocity of the Brownian particle is denoted by $v(t_0)=v_0$, $\gamma_0 \in \mathbb{R}^+$ and it is assumed that $t \geq 0$. Unless otherwise stated, the system is thought of as having an initial time $t_0=0$, and being in equilibrium at $t = \infty$. As time progresses the velocity of the Brownian particle depends on the exponential decay of the initial velocity (the first term in Eq.(\ref{LangVelocity})) and the extra velocity caused by the random force (the second term in Eq.(\ref{LangVelocity})).\\

Since, on average, the random force vanishes (Eq.(\ref{LangAss1})) the average of the velocity (Eq.(\ref{LangVelocity})) is

\begin{equation}
    \langle v(t) \rangle \, = \,v_0 \, e^{-\gamma_0\,t}\, , \,\,\,\,\,\,\,\,\, t \geq 0 \,. \label{LangVelocityAv}
\end{equation}
\vspace{0.1cm}

\noindent Due to friction, the mean velocity decreases exponentially. Using Eq.(\ref{LangVelocity}), the autocorrelation function of the velocity is given by

\begin{equation}
    \langle v(t) v(t') \rangle \, = \,v_0^2 \, e^{-\,\gamma_0\,(t+t')}\,+ \,\frac{1}{m^2}\, \int_{t_0}^t\, dt'' \, \int_{t_0}^{t'} \, dt''' \,
    \langle F(t'')\,F(t''') \rangle \, e^{- \gamma_0\,(t' -t''')}\, e^{- \gamma_0\,(t -t'')}, \label{LangVelocityAutocor}
\end{equation}
\vspace{0.1cm}

\noindent where the two cross-terms which appeared when multiplying $v(t)v(t')$ are both first order in the noise and, as such, disappear when averaging over the noise. In the limit $t=t'$, Eq.(\ref{LangVelocityAutocor}) reduces to

\begin{equation}
    \langle v^2(t) \rangle \, = \,v_0^2 \, e^{-2\,\gamma_0\,t}\, + \,\frac{1}{m^2}\, \int_{t_0}^t\, dt' \, \int_{t_0}^t \, dt'' \,
    \langle F(t')\,F(t'') \rangle \, e^{- \gamma_0\,(t -t')}\, e^{- \gamma_0\,(t -t'')}\, . \label{LangVelocitySquaredAv}
\end{equation}
\vspace{0.1cm}

\noindent Using the assumption Eq.(\ref{LangAss2}), the mean-squared velocity becomes


\begin{equation}
    \langle v^2(t) \rangle \, = \,v_0^2 \, e^{-2\,\gamma_0\,t}\, + \,\frac{\kappa_0}{2\,m^2\, \gamma_0}\, \big( 1 \, - \,e^{- 2 \gamma_0 \, t} \big) \, \label{LangVelocitySquaredAv2}
\end{equation}
\vspace{0.1cm}

\noindent Taking the limit as $t \rightarrow \infty$, the value of the mean-squared velocity at equilibrium is
\begin{equation}
   \displaystyle{\lim_{t \rightarrow \infty}} \, \langle v^2(t) \rangle \, = \frac{\kappa_0}{2\,m^2\, \gamma_0}\, . \label{LangVelocityLimit}
\end{equation}
\vspace{0.1cm}

Assuming the fluid is in thermodynamic equilibrium at temperature $T$, the average energy of the particle takes on its equipartion value\footnote{The equipartition theorem of classical statistical mechanics relates the average energy of a system to its temperature.}\footnote{The Boltzmann constant is taken to be $k_B \, = \,1$.} $\langle E \rangle = T/2$ \cite{pottier2010nonequilibrium}.
Since $\langle E(t)\rangle := m \langle v^2(t) \rangle/2$, Eq.(\ref{LangVelocityLimit}) becomes

\begin{equation}
   \kappa_0 \, =\, 2\, \gamma_0 \, m \, T\, , \label{kappa0}
\end{equation}
\vspace{0.01cm}

\noindent which describes the relationship between the magnitude of the random force $\kappa_0$ and the friction coefficient $\gamma_0$. Using assumption Eq.(\ref{DiracProp}), Eq.(\ref{kappa0}) can be rewritten into the form

\begin{equation}
   \gamma_0 \, =\, \frac{1}{2 \, m\, T}\, \int_{-\infty}^{\infty} \langle  F(t) \, F(t')  \rangle \, dt'\, . \label{FD2}
\end{equation}
\vspace{0.1cm}

\noindent This equation is recognised as the second fluctuation-dissipation theorem \cite{kubo1966fluctuation}. The theorem states that the random fluctuations of the Brownian particle have the same origin as the dissipative frictional force acting on the Brownian particle as it moves in the medium. In order to calculate the position of the Brownian particle, $x(t):= v(t) t$, Eq.(\ref{LangVelocity}) can be integrated between the initial time $t_0$ and $t$

\begin{flalign}
& \begin{aligned}
    & \int_{t_0}^t \, v(t')\, dt'\, = \, \int_{t_0}^t \, v_0 \, e^{-\gamma_0\,t'}\,dt' \, + \,\frac{1}{m}\, \int_{t_0}^t\, \int_{t_0}^{t''}\, F(t')\, e^{- \gamma_0\,(t'' -t')} \, dt'\,dt'' \,  \\[.35cm]
    & \Rightarrow x(t)\, = \, x_0 \, + \,\frac{v_0}{\gamma_0}\,(1 \,- \, e^{ - \gamma_0 \, t}) \, + \,\frac{1}{m}\, \int_{t_0}^t\, F(t')\, \frac{1\, - \, e^{\gamma_0 (t-t')}}{\gamma_0} \, dt' \, , \label{LangPosition}
\end{aligned} &
\end{flalign}
\vspace{0.1cm}


The average of the displacement (remembering Eq.(\ref{LangAss1})) is

\begin{equation}
    \langle x(t) \rangle \, = \, x_0 \, + \,\frac{v_0}{\gamma_0}\,(1 \,- \, e^{ - \gamma_0 \, t})\, . \label{LangMeanPosition}
\end{equation}
\vspace{0.1cm}

\noindent The displacement $s(t):= x(t) -  x_0$, is easily read off from Eq.(\ref{LangPosition})

\begin{equation}
   s(t)\, = \,\frac{v_0}{\gamma_0}\,(1 \,- \, e^{ - \gamma_0 \, t}) \, + \,\frac{1}{m}\, \int_{t_0}^t\, F(t')\,
   \frac{1\, - \, e^{\gamma_0 (t-t')}}{\gamma_0} \, dt'\, . \label{LangDisplacement}
\end{equation}
\vspace{0.1cm}

\noindent In order to calculate the particle's mean-squared displacement, $\langle s^2(t) \rangle$, Eq.(\ref{LangDisplacement}) can be squared and averaged,
\begin{flalign}
& \begin{aligned}
    &  \langle s^2(t) \rangle \, = \,\frac{v_0^2}{\gamma_0^2}\,(1 \,- \, e^{ - \gamma_0 \, t})^2 \, + \,\frac{1}{m^2\,\gamma_0^2}\, \int_{t_0}^t\,dt' \int_{t_0}^t \, dt'' \,\langle F(t')F(t'') \rangle\, \big(1 - e^{\gamma_0 (t-t')}\big) \, \big(1 -  e^{\gamma_0 (t-t'')}\big) \,  \\[.35cm]
    & \,\,\,\,\,\,\,\,\,\,\,\,\,\,\,\,\,\,\, = \,\frac{v_0^2}{\gamma_0^2}\,(1 \,- \, 2\, e^{ - \gamma_0 \, t}\, +\, e^{-2\gamma_0 \, t}) \, + \,\frac{\kappa_0}{m^2\,\gamma_0^2}\, \int_{t_0}^t\,dt'  \, \big(1 - e^{\gamma_0 (t-t')}\big)^2  \\[.35cm]
    & \,\,\,\,\,\,\,\,\,\,\,\,\,\,\,\,\,\,\, = \,\frac{v_0^2}{\gamma_0^2}\,(1 \,- \, 2\, e^{ - \gamma_0 \, t}\, +\, e^{-2\gamma_0 \, t}) \, + \,\frac{\kappa_0}{2\,m^2\,\gamma_0^3}\, \Big(2\gamma_0\, t -\,3\, +  \,4\,e^{-\gamma_0\,t}\, - \,e^{-2\gamma_0\,t} \Big)\, . \label{LangDisplacementSquaredAv}
\end{aligned} &
\end{flalign}
\vspace{0.1cm}

\noindent Note that, in the second line, the two cross-terms vanish and the assumption Eq.(\ref{LangAss2}) has been used. Further, a second average of $\langle s^2(t) \rangle$ over all possible initial velocities needs to be taken. This yields,

\begin{equation}
   \langle s^2(t) \rangle \, = \, 2\,\frac{T}{m}\, \frac{1}{\gamma_0^2}\,\Big(\, \gamma_0\,t \,-\,1\, + \, \,e^{-\gamma_0\,t} \Big) \, , \label{LangDisplacementSquaredAv2}
\end{equation}
\vspace{0.1cm}

where the relation for $\kappa_0$ (Eq.(\ref{kappa0})), and the equipartition of energy ($\langle v_0^2 \rangle \, = \, T/m$) is used.\\

The Einstein-Sutherland relation defines the diffusion coefficient

\begin{equation}
    D \, := \, \frac{T}{\gamma_0\,m}\, . \label{DiffusionC}
\end{equation}
\vspace{0.1cm}

\noindent Hence, Eq.(\ref{LangDisplacementSquaredAv2}) becomes

\begin{equation}
     \langle s^2(t) \rangle \, = \, \frac{2\,D}{\gamma_0}\,
     \Big(\, \gamma_0\,t \,-\,1\, + \, \,e^{-\gamma_0\,t} \Big)\, . \label{LangDisplacementSquaredAv3}
\end{equation}
\vspace{0.1cm}

\noindent This result was first derived in 1917 by Dutch physicist Leonard Ornstein \cite{ornstein1917brownian}. The early and late time limits of the mean-squared displacement $\langle s^2(t) \rangle$ should be examined.

\begin{enumerate}[(i)]
  \item In the limit of early times $\big(t \ll \frac{1}{\gamma_0}\big)$: a Taylor expansion of $e^{-\gamma_0\,t}$ can be performed.
\begin{equation}
    \therefore \langle s^2(t) \rangle \, \approx \, \frac{2\,D}{\gamma_0}\, \left(\gamma_0 \, t \, - \, 1 \, + \,
    \left(1 \, - \, \gamma_0\,t \, + \, \frac{\gamma_0^2\,t^2}{2} \right) \right)\, = \,D\,\gamma_0 \, t^2 \, = \, \frac{T}{m}\, t^2 \, , \label{DisplacementSqLimit1}
    \end{equation}
     where the last equality comes from using the Einstein-Sutherland relation Eq.(\ref{DiffusionC}). In this limit, known as the ballistic regime, the Brownian particle's behaviour is proportional to time, $s(t) \sim t$. This behaviour is expected since the initial behaviour of the Brownian particle -- before it is bombarded by a substantial number of fluid particles -- is inertial (with a velocity  determined by the equipartition of energy $ v(t) \, = \, \sqrt{T/m}$ ) \cite{Boer_2009}.
  \item In the limit of late times $\big(t \gg \frac{1}{\gamma_0}\big)$: $e^{-\gamma_0\,t} \rightarrow 0 \,\,\,\text{as}\,\,\, t \rightarrow \infty, \,\,\,\text{and}\,\,\, \gamma_0\,t$ becomes the dominant term.
\begin{equation}
    \therefore \langle s^2(t) \rangle \, \approx \, 2\,D\,t\, . \label{DisplacementSqLimit2}
\end{equation}
Eq.(\ref{DisplacementSqLimit2}) is the famous equation found by Einstein in his 1905 paper \cite{einstein1905motion}. In this limit, referred to as the diffusive regime, $ s(t)  \sim \sqrt{t}$ and the particle experiences a random walk. The Brownian particle deviates from its initial course due to numerous collisions with the constituent fluid particles that cause the Brownian particle to `forget' its early time behaviour \cite{Boer_2009}.
\end{enumerate}
\vspace{0.1cm}

\noindent The cross-over time\footnote{In this article, the cross-over time is interchangeably referred to as the relaxation time.} between the diffusive regime and the ballistic regime is denoted by

\begin{equation}
    t_{\text{relax}} \, \sim \, \frac{1}{\gamma_0} \, , \label{trelax}
    \end{equation}

\noindent which represents the time it takes for a Brownian particle that had some initial velocity at $t \, = \, t_0$ to thermalize in the medium.\\

\subsection{The Generalised Langevin Model} \label{subsecGenLangevin}

The Langevin model can be generalised to include retardation effects. The retarded Langevin equation\footnote{The retarded Langevin Equation is also known as the generalised Langevin Equation.} is given by \cite{kubo1966fluctuation, mori1965transport}

\begin{equation}
    \Dot{p}(t) \, = \, - \, \int_{-\infty}^t dt' \,  \gamma(t-t') \, p(t') \, + \, F(t) \, + \, K(t) \, , \label{GenLangEq}
\end{equation}
\vspace{0.1cm}

\noindent where the memory kernel $\gamma(t-t')$ allows the friction term to depend on the past trajectory of the Brownian particle, and $K(t)$ is an external force which acts on the system. The system is taken to be in equilibrium in the limit $t_0 \rightarrow - \infty$. The non-retarded Langevin Equation Eq.(\ref{LangEq}) only holds if the Brownian particle is taken to have infinite mass with respect to the constituent fluid particles. The generalisation of the Langevin equation to Eq.(\ref{GenLangEq}) fixes two physical problems, (i) the friction is no longer considered to be instantaneous, (ii) a correlation can exist between random forces at different times. The random force in Eq.(\ref{GenLangEq}) is taken to satisfy

\begin{equation}
    \langle F(t) \rangle \, = \, 0  \, \,\,\,\, \text{and} \, \, \,\,\,\langle  F(t) \, F(t')  \rangle \, = \, \kappa(t-t')\, ,\label{GenLangAss}
\end{equation}
\vspace{0.1cm}

\noindent where $\kappa(t)$ is an unspecified function. As in the non-retarded case, $F(t)$ is assumed to be a Gaussian process.\\

The retarded Langevin equation is Fourier transformed\footnote{The Fourier and Inverse Fourier Transform are defined here as
\begin{flalign*}
& \begin{aligned}
    & G(\omega) \, := \, \mathcal{F}[\, g(t)\, ] \, = \, \int_{-\infty}^\infty \, g(t) \, e^{i  \omega  t }\, dt \,\,\,\,\,\, \text{and} \,\,\,\,\,\, g(t) \, := \, \mathcal{F}^{-1}[\, G(\omega)\,] \, = \, \int_{-\infty}^\infty \, G(\omega) \, e^{- i \omega  t } \, dt
\end{aligned} &
\end{flalign*}
where $w \,= \,- 2  \pi  f$.} in order to analyse its behaviour. 

Fourier transforming both sides of Eq.(\ref{GenLangEq}), and using the linearity and derivative properties of the Fourier Transform gives

\begin{equation}
    - i \, \omega \, p(\omega) \, + \,\mathcal{F}\Big[\, \int_{-\infty}^t dt' \,  \gamma(t-t') \, p(t') \, \Big] \, = \, F(\omega) \, + \, K(\omega)     \, . \label{FTGenLang}
\end{equation}
\vspace{0.1cm}

\noindent The convolution property would be useful in calculating the second term in Eq.(\ref{FTGenLang}). However, the asymmetric bounds of this integral ($t'\in [-\infty, t]$) are inconvenient. Hence, the memory kernel's bounds are redefined by introducing the causal memory kernel

\begin{equation}
    \Tilde{\gamma}(t) \, = \, \Theta(t)\, \gamma(t)\, ,\label{HeavySideMemory}
\end{equation}
\vspace{0.1cm}

\noindent where $\Theta(t)$ is the Heavy-side function. Notice that while $\gamma(t-t')$ is defined only for $t > 0$, $\Tilde{\gamma}(t-t')$ is defined for all $t$.\\

Rewriting Eq.(\ref{FTGenLang}) in terms of the causal memory kernel yields

\begin{flalign}
& \begin{aligned}
& \Rightarrow  - i \, \omega \, p(\omega) \, + \,\mathcal{F}\Big[\, \int_{-\infty}^\infty dt' \,  \Tilde{\gamma}(t-t') \, p(t') \, \Big] \, = \, F(\omega) \, + \, K(\omega)\\[0.25cm]
& \Rightarrow - i \, \omega \, p(\omega) \, + \, p(\omega) \, \Tilde{\gamma}(\omega) \, = \, F(\omega) \, + \, K(\omega) \, ,\label{FTGenLang2}
\end{aligned} &
\end{flalign}

\noindent where, in the second line, the symmetric bounds of the integral ensured that the convolution property can be applied. Note that in Eq.(\ref{FTGenLang2})

\begin{equation}
    \Tilde{\gamma}(\omega) \,=\, \int_{-\infty}^\infty dt \, \Tilde{\gamma}(t) \, e^{i\omega t} \, = \,  \int_{-\infty}^\infty dt \, \Theta(t) \, \gamma(t) \, e^{i\omega t} \, = \, \int_0^{\infty} dt\, \gamma(t) \, e^{i\omega t} := \, \gamma[\omega] \, ,\label{FTLapaceDef}
\end{equation}
\vspace{0.1cm}

\noindent is actually the Fourier-Laplace transform of $\gamma(t)$, while

\begin{equation}
    p(\omega) \, := \, \int_{-\infty}^\infty dt \, \, p(t) \, e^{i  \omega  t }\,  ,\label{FTDEf}
\end{equation}

\noindent $F(\omega)$ and $K(\omega)$ are the standard Fourier transforms of $p(t)$, $F(t)$ and $K(t)$ respectively. By rearranging Eq.(\ref{FTGenLang2}) into a simpler form

\begin{equation}
    p(\omega)\, = \, \frac{F(\omega) \, + \, K(\omega)}{ \gamma[\omega] \, - i \, \omega } \, , \label{FTGenLang3}
\end{equation}

\noindent it is now easy to take the statistical average. Remembering the assumptions regarding the random force (Eq.(\ref{GenLangAss})), yields

\begin{equation}
    \langle p(\omega) \rangle \, = \, \mu(\omega)\, K(\omega),  \, \,\,\,\, \text{where} \, \, \,\,\,\mu(\omega) \,
    := \, \frac{1}{\gamma[\omega] \, - \, i\, \omega}\, .\label{GenLangAvp}
\end{equation}
\vspace{0.1cm}

\noindent The quantity $\mu(\omega)$ is known as the admittance, and describes the system's response to an external perturbation. Measuring the response $\langle p(\omega) \rangle $ to an external force on the system $K(\omega)$, the admittance $\mu(\omega)$ can be calculated and hence $\gamma[w]$ determined.\\

The first fluctuation-dissipation theorem relates the admittance to the autocorrelation of the equilibrium velocity \cite{kubo1966fluctuation, pottier2010nonequilibrium}. Following Eq.(\ref{LangVelocityAutocor}) the autocorrelation function of the equilibrium velocity for the retarded Langevin equation Eq.(\ref{GenLangEq}) is

\begin{flalign}
& \begin{aligned}
& \langle v(t) v(0) \rangle \, = \,\frac{1}{m^2}\, \int_{-\infty}^t\, dt' \, \int_{-\infty}^{0} \, dt'' \, \langle F(t')\,F(t'') \rangle \, e^{\gamma\,t''}\, e^{- \gamma\,(t -t')} \\[0.2cm]
& \,\,\,\,\,\,\,\,\,\,\,\,\,\,\,\,\,\,\,\,\,\,\,\,\,\,= \,\frac{\kappa_0}{m^2}\, \int_{-\infty}^t\, dt' \, \int_{-\infty}^{0} \, dt'' \, \delta(t' - t'')\, e^{\gamma\,t''}\, e^{- \gamma\,(t -t')} \\[0.2cm]
&  \,\,\,\,\,\,\,\,\,\,\,\,\,\,\,\,\,\,\,\,\,\,\,\,\,\, =\,\frac{T}{m}\, e^{- \gamma\, |t|} \, , \label{LangVelocityAutocortp0}
\end{aligned} &
\end{flalign}
\vspace{0.1cm}

\noindent where $|t|$ is the absolute value of $t$; the random force correlation function is chosen, for simplicity, to have the form $\kappa(t-t') = \kappa_0 \, \delta(t - t')$; and Eq.(\ref{kappa0}) has been used in the third line. The Fourier-Laplace transform of Eq.(\ref{LangVelocityAutocortp0}) is

\begin{flalign}
& \begin{aligned}
& \int_0^{\infty} \langle v(t) v(0) \rangle\, e^{i \omega t}\, dt \, = \, \frac{T}{m}\, \frac{1}{\gamma[\omega]\,-\, i \omega} \\[0.2cm]
& \Rightarrow \mu(\omega) \,  =\, \frac{m}{T} \,\int_0^{\infty} \langle v(t) v(0) \rangle\, e^{i \omega t}\, dt   \, , \label{FD1}
\end{aligned} &
\end{flalign}
\vspace{0.1cm}

\noindent where the definition of the admittance is recognised. Eq.(\ref{FD1}) is known as the first fluctuation-dissipation theorem \cite{kubo1966fluctuation}.\\

The power spectrum

\begin{equation}
    I_{\mathcal{O}}(\omega) \, = \, \int_{-\infty}^\infty dt \, \langle \mathcal{O}(t_0) \, \mathcal{O}(t_0 + t) \rangle \, e^{i \omega t}  \, ,\label{PowerSp}
\end{equation}
\vspace{0.1cm}

\noindent is defined for a quantity $\mathcal{O}(t)$. The Wiener-Khintchine theorem\footnote{The Wiener-Khintchine theorem was proven for a deterministic function by Norbert Wiener in 1930 \cite{wiener1930generalised}.} states that a stationary random process' autocorrelation function has a spectral decomposition described by the process' power spectrum. The theorem allows one to compare the autocorrelation function and the power spectrum

\begin{equation}
    \langle \mathcal{O}(\omega)\, \mathcal{O}(\omega') \rangle \, = \, 2\, \pi \, \delta(\omega \, + \, \omega') \, I_{\mathcal{O}}(\omega)  \, ,\label{WKTheorem}
\end{equation}

of a quantity $\mathcal{O}(t)$. \\

As previously mentioned, $\gamma(t)$ and $\kappa(t)$ are related to each other by the second fluctuation-dissipation theorem. However they can be individually determined by considering two cases where different external forces are applied on the system:
\begin{enumerate}[(i)]
\item the external force is periodic with frequency $\omega$ (i.e. $K(t) \, = \, K_0 \, e^{-i\omega t}$), and Eq.(\ref{GenLangAvp}) becomes

\begin{equation}
   \langle p(t) \rangle \, = \, \mu(\omega) \, K_0 \, e^{-i\omega t}\, .\label{GenLangAvp2}
\end{equation}

The memory kernel of the system $\gamma(t)$ can be determined by measuring the response $\langle p(t) \rangle$ to $K(t)$.

\item the external force $K(t)$ is absent, and Eq.(\ref{FTGenLang3}) becomes

\begin{equation}
    p(\omega)\, = \, \frac{F(\omega)}{ \gamma[\omega] \, - i \, \omega}  \, .\label{FTGenLangNoExtForce}
\end{equation}
\vspace{0.1cm}

\noindent In order to determine $\gamma(t)$ and $\kappa(t)$ individually, $p(\omega)$ should first be squared and averaged to find the relation between the power spectrum of $p$ ($I_p(\omega)$) and the power spectrum of $F$ ($I_F(\omega)$). This yields, 


\begin{equation}
    I_p(\omega) \, = \, \frac{ I_F(\omega)}{ | \, \gamma[\omega] \, - i \, \omega \, |^2}\, , \label{GenLangAvpNoExtForce}
\end{equation}
\vspace{0.1cm}

\noindent where, the Wiener-Khintchine theorem has been applied to both sides. The measured response's autocorrelation $\langle p(\omega)p(\omega') \rangle$ defines the power spectrum $I_p(\omega)$ through Eq.(\ref{WKTheorem}). Since $\gamma(t)$ and $I_p(\omega)$ are known, $I_F(\omega)$ can be determined from Eq.(\ref{GenLangAvpNoExtForce}).
Using $I_F(t)$ the function $\kappa(t)$ is then calculated from Eq.(\ref{GenLangAss}) and Eq.(\ref{WKTheorem}), i.e.

\begin{equation}
    I_F(t) \, = \,\frac{\kappa(t-t')}{2\,\pi \, \delta(t+t')}  \, .\label{PowerSpectrumF}
\end{equation}
\vspace{0.1cm}

\end{enumerate}
\vspace{0.1cm}

\noindent For the generalised Langevin model, the relaxation time Eq.(\ref{trelax}) becomes

\begin{equation}
    t_{\text{relax}} \, = \, \frac{1}{\int_0^\infty dt\, \gamma(t)} \, = \, \frac{1}{\gamma[\omega = 0]} \, = \, \mu(\omega = 0)  \, ,\label{Gentrelax}
\end{equation}
\vspace{0.1cm}

\noindent where Eq.(\ref{FTLapaceDef}) and Eq.(\ref{GenLangAvp}) have been used in the above simplification.\\

In order to gain an intuition regarding the relaxation time for the generalised Langevin model, consider an example where the memory kernel $\gamma(t)$ is sharply peaked around $t=0$, i.e. the friction is approximated as being instantaneous. The retarded effect of the friction term in Eq.(\ref{GenLangEq}) is ignored, and the integral becomes

\begin{equation}
    \int_0^\infty dt' \, \gamma(t-t') \, p(t') \, \approx \, \int_0^\infty dt' \, \gamma(t') \, p(t) \, = \, \frac{1}{t_{\text{relax}}}\, p(t)  \, ,\label{GenLangEqPeak0}
\end{equation}
\vspace{0.1cm}

\noindent where Eq.(\ref{Gentrelax}) is used in the final equality. Inserting Eq.(\ref{GenLangEqPeak0}) into the generalised Langevin equation Eq.(\ref{GenLangEq}), returns the non-retarded Langevin equation Eq.(\ref{LangEq}). Hence the interpretation of $t_{\text{relax}}$ remains the same as for the non-retarded Langevin model: it is the time it takes for the Brownian particle to thermalize in the medium \cite{Boer_2009}.\\

Another relevant time scale, the correlation (or microscopic) time, $t_{\text{c}}$, is defined. The correlation time is considered to be the width of the random force correlator function. Specifically,

\begin{equation}
    t_{\text{c}} \, = \, \int_0^\infty dt \, \frac{\kappa(t)}{\kappa(0)}  \, .\label{tc}
\end{equation}
\vspace{0.1cm}

\noindent The quantity $t_{\text{c}}$ measures the time duration of a single scattering process, i.e. it indicates how long the random force is correlated for. In most cases $t_{\text{relax}} \gg t_{\text{c}}$, however this does not necessarily hold for Brownian motion dual to AdS black holes.\\

On a final note for this subsection, the Langevin Model can also be generalised to $d$ spatial dimensions. In Eq.(\ref{LangEq}), the momentum $p(t)$ and random force $F(t)$ become $d$-component vectors, while the assumptions Eq.(\ref{LangAss1}) and Eq.(\ref{LangAss2}) generalise to

\begin{equation}
    \langle F_i(t) \rangle \, = \, 0  \, \,\,\,\, \text{and} \, \, \,\,\,\langle  F_i(t) \, F_j(t')  \rangle \, = \, \kappa_0 \, \delta_{i\,j} \, \delta(t-t')\, , \label{GendLangAss}
\end{equation}
\vspace{0.1cm}

\noindent where $i,j = (1, \,2,\, ... \, ,\, n)$. The fluctuation-dissipation theorem is independent of $d$; as such $\kappa_0$ is still given by Eq.(\ref{kappa0}). The mean-squared displacement behaves like $\langle s^2(t) \rangle \, \approx \, 2\,d\, D\,t$ in the late time limit $\big(t \gg \frac{1}{\gamma_0}\big)$, where the diffusion coefficient $D$ is still given by the Einstein-Sutherland relation Eq.(\ref{DiffusionC}). \\

This concludes a brief foray into some of the salient theory about Brownian motion\footnote{For more detailed reading on the topic see \cite{uhlenbeck1930theory, wang1945theory} who provide insightful early reviews,
while \cite{Dunkel_2009, pottier2010nonequilibrium, kubo2012statistical} can be useful in understanding some of the more modern developments in the field.}. It is this motion which is responsible for the dissipative nature of a system and its approach to thermal equilibrium. Any particle suspended in a finite temperature fluid undergoes Brownian motion and as such a probe quark immersed in a thermal plasma behaves the same way. The AdS/CFT correspondence can be used to study the Brownian behaviour a probe quark exhibits while interacting with the strongly-coupled thermal plasma by modelling the external quark as a test string in the bulk theory. This is done in sections (\ref{secHeavyQuark}) and (\ref{secDragForce}) for the heavy quark case, and section (\ref{secLightQuark}) for the light quark case.

\part{\Large{Calculations in the Bulk}}\label{partBulkCal}

\section{Analysing Heavy Quark Brownian Motion}\label{secHeavyQuark}

\begin{figure}[!htb]
\centering
\includegraphics[width=0.65\textwidth]{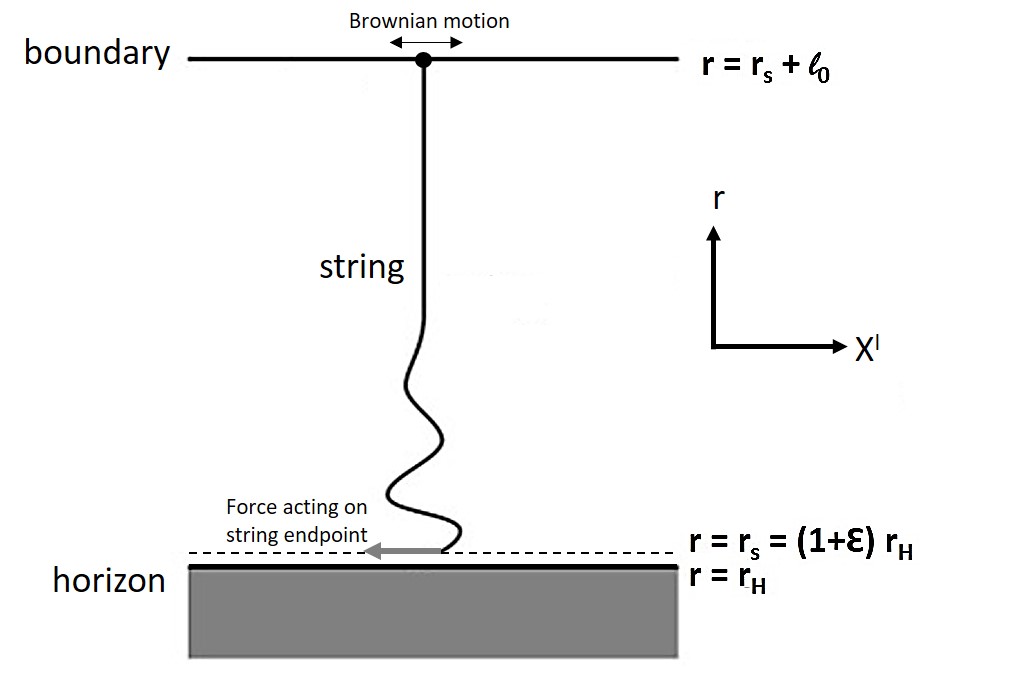}
\caption[A fundamental open string of length $\ell_0$ used as a probe in an AdS black hole background]{\label{fig:stringdiagram} A fundamental open string of length $\ell_0$ used as a probe in an AdS black hole background. The string starts at the boundary of the anti-de Sitter spacetime and hangs down to a \textit{stretched horizon} ($r_s \, = \, (1+ \epsilon)\,r_H$ where $0 < \epsilon \ll 1$) placed just above the Schwarzschild black hole horizon. This figure is adapted from \cite{Boer_2009}.}
\end{figure}

In the previous section the theory of Brownian motion (specifically the Langevin model) was reviewed. An external test quark in a thermal plasma is governed by these equations. In the AdS/CFT context, this is the gauge theory -- or boundary -- side. The following three sections focus on the gravitational -- or bulk -- side, in which a probe string is set up in an anti-de Sitter black hole background\footnote{As discussed in section (\ref{secGaugeString}), the bulk theory is in AdS$_d \times S^d$. However, the physics on the compact $S^d$ space corresponds to a set of scalar and fermion fields rotated among each other in $\mathcal{N}=4$ SYM. In this article, the rotational supersymmetric Yang-Mills charges are ignored. The calculations can be considered to take place in AdS$_d$ and at a point on $S^d$.} and the transverse fluctuations on this string (resulting from its proximity to the Schwarzschild black hole) are examined. The AdS/CFT correspondence is used to equate the two descriptions. Section (\ref{secHeavyQuark}) begins this incursion into the bulk theory by modelling an on-mass-shell quark of large finite mass, identified by particle physicists as an on-mass-shell heavy quark, as an open probe string stretched between the boundary and a Schwarzschild black hole in AdS spacetime\footnote{Throughout this article the test strings considered are in the probe approximation, i.e. the string's backreaction on the background is taken to be negligible. Further, it is assumed that no $B$-field is present in the background.}.
The presence of the black hole results in thermal fluctuations in the transverse $X^I$ directions on the string. The resultant random movement of the string's endpoint on the boundary corresponds to the heavy quark undergoing Brownian motion. This set-up is depicted in figure (\ref{fig:stringdiagram}). 

The stretched horizon depicted in figure (\ref{fig:stringdiagram}) is introduced to regulate an infrared divergence. Similarly, a UV cut-off is imposed near the boundary to ensure the mass of the probe particle is finite\footnote{Following how this terminology used in \cite{Boer_2009}, the terms \textit{infrared} (IR) and \textit{ultraviolet} (UV) are understood to be with respect to the boundary energy. In the bulk theory infrared means near the horizon, while ultraviolet means near the boundary.}. Dirichlet boundary conditions will be imposed on the fixed string endpoint attached to the stretched horizon, while Neumann boundary conditions will be imposed at the boundary. Using this setup, we will explicitly compute the mean distance squared $s^2(t)$ travelled by the endpoint of the string, which represents the position of the heavy quark in the thermal medium.  We will see that at late times $s^2(t)$ is diffusive, growing linearly with time,
\begin{align}
    s^2(t) & = 2\,D\,t,
\end{align}
with a diffusion coefficient that exactly respects Einstein's fluctuation-dissipation theorem for non-relativistic motion \cite{Boer_2009},

\begin{align}
    D = \frac{T}{\gamma m},
\end{align}
where the friction coefficient $\gamma$ will be determined in section (\ref{secDragForce}).\\

\subsection{Polyakov String Equations of Motion} \label{subsecPolyakovEoM} 

The Brink-Di Vecchia-Howe-Deser-Zumino action \cite{brink1976locally, deser1976complete} -- or Polyakov Action \cite{polyakov1989quantum} for short -- describes the leading order dynamics of the fundamental probe string and is given by

\begin{equation}
    S_P \, := \, \frac{1}{4 \pi \alpha'} \int_{\mathcal{M}} \, d^2 \sigma \, \mathcal{L}_P \,
    = \, - \frac{1}{4 \pi \alpha'} \, \int d^2 \sigma \, \sqrt{-\text{det}\, (\gamma_{a b})} \, \gamma^{a b}\, g_{ab} \, , \label{PolyakovAct}
\end{equation}

\noindent where the slope parameter $\alpha'$ is related to the string tension $T_0$ \cite{zwiebach2004first}; $\mathcal{L}_P$ is the Lagrangian density; the worldsheet parameter space is denoted by $\mathcal{M}$ (with coordinates $(t, \sigma) \in \, [0,t_f] \times [0, \sigma_f] \, = \, \mathcal{M}$) and the spacetime by $\mathcal{N}$. The induced worldsheet metric is given by

\begin{equation}
   g_{ab} \, := \, {\partial_a} X^\mu \,{\partial_b} X^\nu \, G_{\mu \nu}  \, , \label{InducedMetric}
\end{equation}
\vspace{0.1cm}

\noindent $G_{\mu \nu}$ is the spacetime metric, and $\gamma_{ab}$ is an auxiliary worldsheet metric. From here onwards, the determinant of $\gamma_{ab}$ will simply be denoted by $\gamma$. Notice that a gauge choice has been made: the static gauge is used. The reparameterization offered by the static gauge separates the time and space coordinates \cite{zwiebach2004first}, identifying the $\tau$ parameter with the time coordinate, $\tau = t$.\\

The string worldsheet is embedded into the target spacetime by the mapping functions $X^\mu: \mathcal{M} \rightarrow \mathcal{N}; (t, \sigma) \rightarrow X^\mu(t,\sigma)$. The canonical conjugate momentum densities $\Pi_\mu^a(t,\sigma)$ are easily determined once the functional derivative of the Polyakov Action with respect to the derivatives of these embedding functions has been calculated\footnote{The functional derivative (sometimes referred to as the variational or Fr\'ecet derivative) compares the change in a functional to the change in a function that the functional depends on. The functional derivative of functional $J$ with respect to function $f$ (evaluated at point $x$) is defined as
\begin{equation*}
\delta_{f} J\, = \, \int_{a}^b\, \frac{\delta J}{\delta f(x)} \, \delta f(x)\, d x \,.
\end{equation*}}.
Specifically,

\begin{equation}
   \delta_{(\partial_a X^\mu)}  S_P \, := \, \int_\mathcal{M}\, d^2 \sigma \,  \frac{\delta S_P}{\delta (\partial_a X^\mu(t, \sigma))}\,\delta (\partial_a X^\mu(t, \sigma))  \, . \label{Polyakov_dFunc}
\end{equation}
\vspace{0.1cm}

\noindent Using the definition of the Polyakov Action Eq.(\ref{PolyakovAct}),


\begin{equation}
   \delta_{(\partial_a X^\mu)}  S_P \,  = \, \int_\mathcal{M} d^2 \sigma \, \delta ({\partial_a} X^\mu) \, \Big(- \frac{1}{2 \pi \alpha'} \, \sqrt{-\gamma} \, \gamma^{ab}\, G_{\mu \nu}\, {\partial_b} X^\nu \Big)\, , \label{VaryPolyakovAct}
\end{equation}
\vspace{0.1cm}

\noindent where the factor of two arises from symmetry (the variation of ${\partial_a} X^\mu$ and ${\partial_b} X^\nu$ gives the same the result). Hence the canonical momentum densities, defined as the variation of the action with respect to the derivatives of the embedding functions, is given by

\begin{equation}
   \Pi_\mu^a(t,\sigma) \, := \, \frac{\delta \,\mathcal{S}_P}{\delta (\partial_a \, X^\mu(t, \sigma))} \,
   = \, - \frac{1}{2 \pi \alpha'} \, \sqrt{-\gamma} \, \gamma^{ab}\, G_{\mu \nu}\, {\partial_b} X^\nu  \, . \label{MomDensity}
\end{equation}
\vspace{0.1cm}

The energy-momentum tensor $T_{ab}$ can be defined by the variation of the Polyakov Action with respect to the auxiliary worldsheet metric \cite{polchinski1998string},

\begin{equation}
 T_{ab} \, :=\, - 4\,\pi \, \frac{1}{\sqrt{-\gamma}} \, \frac{\delta  S_P}{\delta \gamma^{ab}}\, . \label{EMdef}
\end{equation}
\vspace{0.1cm}

\noindent Before calculating $T_{ab}$, note that determining the functional derivative of the auxiliary worldsheet metric gives the two useful relations\footnote{To prove the first relation make use of the matrix property $\delta \big(\text{det}(A)\big) = \text{det}(A) \, \text{Tr}\big(A^{-1} \delta A \big)$, where $A$ is an $n \times n$ matrix.}

\begin{equation*}
\delta \big(\text{det}(\gamma_{ab}) \big) \, \equiv \, \delta \gamma\, = \, \gamma \big( \gamma^{ab} \, \delta \gamma_{ab} \big) \, = \, \gamma \big( \gamma_{ab} \, \delta \gamma^{ab} \big) \,,
\end{equation*}

\noindent and

\begin{equation*}
\delta \big(\sqrt{-\gamma} \big) \, = \, -
\frac{1}{2\, \sqrt{-\gamma}}\, \delta \gamma \, = \, -
\frac{\gamma}{2\, \sqrt{-\gamma}}\, \big(\gamma_{ab} \, \delta \gamma^{ab} \big) \, = \, -
\frac{1}{2}\, \sqrt{-\gamma}\,  \gamma_{ab} \, \delta \gamma^{ab} \,.
\end{equation*}
\vspace{0.1cm}

From the definition of the Polyakov Action Eq.(\ref{PolyakovAct}) and the relation for $\delta \big(\sqrt{-\gamma} \big)$, the functional derivative of the action with respect to the auxiliary worldsheet metric can be simplified to

\begin{flalign}
& \begin{aligned}
     \delta_\gamma S_P \, &= \, - \frac{1}{4 \pi \alpha'} \,\int_\mathcal{M} d^2 \sigma\, \Big( \delta(\sqrt{-\gamma}) \, \gamma^{ab}\, G_{\mu \nu}\, {\partial_a} X^\mu\, {\partial_b} X^\nu \, + \, \sqrt{-\gamma} \, \delta \gamma^{ab} \, G_{\mu \nu}\, {\partial_a} X^\mu\, {\partial_b} X^\nu \Big) \\[.2cm]
    & = \, - \frac{1}{4 \pi \alpha'} \,\int_\mathcal{M} d^2 \sigma\, \sqrt{-\gamma}\,  \delta \gamma^{ab} \,G_{\mu \nu} \left[ -\frac{1}{2}\, \gamma_{ab}\, \Big(\gamma^{cd}\, {\partial_c} X^\mu\, {\partial_d} X^\nu \Big)\, + \, {\partial_a} X^\mu\, {\partial_b} X^\nu \right]\\[.2cm]
    & \equiv \, \int_\mathcal{M} d^2 \sigma\, \delta \gamma^{ab} \, \frac{\delta  S_P}{\delta \gamma^{ab}} \, , \label{VaryMetric}
\end{aligned} &
\end{flalign}
\vspace{0.1cm}

\noindent where the final line is simply the definition of the functional derivative. Hence, the energy-momentum tensor $T_{ab}$ (Eq.(\ref{EMdef})), becomes

\begin{flalign}
& \begin{aligned}
     T_{ab} \, & =\, - 4\,\pi \, \frac{1}{\sqrt{-\gamma}} \, \Big[ - \frac{1}{4 \pi \alpha'} \, \sqrt{-\gamma} \,G_{\mu \nu} \Big(-\frac{1}{2}\, \gamma_{ab}\,\gamma^{cd}\, {\partial_c} X^\mu\, {\partial_d} X^\nu \, + \, {\partial_a} X^\mu\, {\partial_b} X^\nu \Big)\Big] \\[.2cm]
    & = \, \frac{1}{\alpha'}\Big( -\frac{1}{2}\, \gamma_{ab}\, \gamma^{cd}\, g_{cd} \, + \, g_{ab} \Big)  \, , \label{TabDef}
\end{aligned} &
\end{flalign}
\vspace{0.1cm}

\noindent where the definition of the induced worldsheet metric (Eq.(\ref{InducedMetric})) is used in the last line. Requiring by the principle of least action that the energy-momentum tensor vanishes, implies

\begin{equation}
g_{ab}\, = \, \frac{1}{2}\, \gamma_{ab}\, \gamma^{cd}\, g_{cd} \, . \label{Constraint1}
\end{equation}
\vspace{0.1cm}

Eq.(\ref{Constraint1}) sets the auxiliary metric $\gamma_{ab}$ proportional to the induced metric $g_{ab}$ at every point of the worldsheet. If the proportionality constant is defined to be positive (the notions of \textit{timelike} and \textit{spacelike} vectors defined by $\gamma_{ab}$ and $g_{ab}$ should agree); the metrics are said to be conformal to each other. This proportionality constant is denoted by $f^2$. Writing,

\begin{flalign}
& \begin{aligned}
    & \gamma_{ab} \, =\, f^2 \, g_{ab} \\[.2cm]
    & \Rightarrow \, \text{det}(\gamma_{ab}) \, =\, f^4 \, \text{det}(g_{ab}) \\[.2cm]
    & \Rightarrow \, (-\gamma)^{-\frac{1}{2}} \, =\, \frac{1}{f^2} \, (-g)^{-\frac{1}{2}} \\[.2cm]
    & \Rightarrow \, \frac{\gamma_{ab}}{\sqrt{-\gamma}} \, =\, \frac{g_{ab}}{\sqrt{-g}} \, , \label{Constraint2}
\end{aligned} &
\end{flalign}
\vspace{0.1cm}

\noindent a constraint equation is obtained. From Eq.(\ref{Constraint2}) it is evident that the Polyakov action has an additional symmetry to the Nambu-Goto string action (an action defined in terms of the induced metric $g_{ab}$) \cite{nambu1995duality, goto1971relativistic}, since the auxiliary worldsheet metric can be re-scaled arbitrarily. Indeed, the Polyakov action is locally scale invariant on the string worldsheet. This is known as conformal/Weyl invariance: under a local change of scale all angles are kept fixed on the worldsheet, while the length of the lines may change. \\

Due to this additional symmetry, another gauge choice presents itself. Choosing the conformal gauge $\gamma^{ab} = \eta^{ab}$ (which restricts the choice of worldsheet to one with vanishing Euler characteristic), the constraint Eq.(\ref{Constraint2}) simplifies to

\begin{equation}
\eta^{ab} \, =\, \frac{g_{ab}}{\sqrt{-g}} \, , \label{Constraint3}
\end{equation}

\noindent since $\eta^{ab} = \eta_{ab}$, and $\eta=\text{det}(\eta_{ab})=-1$. Hence, with this choice of gauge restricting the auxiliary metric $\gamma_{ab}$, the induced worlsheet metric $g_{ab}$ becomes conformally flat. In the conformal gauge, the energy-momentum tensor Eq.(\ref{TabDef}) becomes
\begin{equation}
T_{ab} \, = \, \frac{1}{\alpha'}\,G_{\mu \nu}\,\Big(- \frac{1}{2}\, \eta_{ab}\,\eta^{cd}  {\partial_c} X^\mu\, {\partial_d} X^\nu \,
+ \, {\partial_a} X^\mu\, {\partial_b} X^\nu \Big) \, . \label{TabDefConformal}
\end{equation}

\noindent The worldsheet parameter space coordinate system can be changed to light-cone coordinates, where the light-cone coordinates are defined as
\begin{equation}
\sigma^\pm \,= \,\frac{1}{\sqrt{2}}\big( \tau \, \pm \,  \sigma \big) \,\,\,\, \text{and} \,\,\,\, \partial_\pm \,
= \, \frac{1}{2} \big(\partial_\tau \, \pm \, \partial_\sigma \big)  \,. \label{LCDef}
\end{equation}

\noindent The energy-momentum tensor can be written in terms of light-cone coordinates. This is easily done after explicitly writing out the components of $T_{ab}$  (Eq.(\ref{TabDefConformal})),
\begin{flalign}
& \begin{aligned}
   T_{\tau \tau} \, =\, T_{\sigma \sigma} \, & = \, \frac{1}{2\, \alpha'}\,G_{\mu \nu}\,\big(\dot{X}^\mu\,\dot{X}^\nu\, + \, X^{'\mu}\, X^{'\nu}  \big)\, ,  \,\,\,\,\,\,\, \text{and} \\[.2cm]
    T_{\tau \sigma}\,=\,T_{\sigma \tau} \, & = \, \frac{1}{\alpha'}\,G_{\mu \nu}\,\dot{X}^\mu\, X^{'\nu} \, ,
   \label{TComponents}
\end{aligned} &
\end{flalign}
\vspace{0.1cm}

\noindent where $\dot{X}^\mu = \partial_{\tau} X^\mu$ and $X^{'\mu}= \partial_{\sigma} X^\mu$. Further, $X^{'\mu}\,\dot{X}^\nu = \dot{X}^\mu\,X^{'\nu}$
since $G_{\mu \nu}$ is a diagonal matrix. In light-cone coordinates the components of the energy-momentum tensor are

\begin{equation}
T_{++} \, = \, \frac{1}{2}(T_{\tau \tau}\,+\,T_{\tau \sigma}) \,\,\,\, \text{and} \,\,\,\, T_{--} \,
= \, \frac{1}{2}(T_{\tau \tau}\,-\,T_{\tau \sigma})  \,,\label{LCTab}
\end{equation}
\vspace{0.1cm}

\noindent which is easily proven from Eq.(\ref{TabDefConformal}) using the definition Eq.(\ref{LCDef}) and the light-cone metric (where $\eta_{+-}=\eta_{-+}=-1$, $\eta_{++}=\eta_{--}=0$). Inserting Eq.(\ref{TComponents}) into Eq.(\ref{LCTab}), yields

\begin{flalign}
& \begin{aligned}
   T_{\pm \pm} \, &= \,\frac{1}{2}\, \frac{1}{\alpha'}\,G_{\mu \nu}\,\Big[\frac{1}{2}\,\big( \partial_{\tau}X^\mu\,\partial_{\tau}X^\nu\, + \, \partial_{\sigma}X^\mu\, \partial_{\sigma}X^\nu  \big) \, \pm \, \partial_{\tau}X^\mu\,\partial_{\sigma}X^\nu \Big] \\[.2cm]
    & = \,\frac{1}{\alpha'}\,G_{\mu \nu}\,\partial_{\pm}\,X^\mu \, \partial_{\pm}\,X^\nu \, , \label{Tpmcomponent}
\end{aligned} &
\end{flalign}
\vspace{0.1cm}

As previously mentioned, due to energy conservation, the energy-momentum tensor vanishes: $T_{ab}\,=\,0$. In light-cone coordinates $T_{++}\,=\,T_{--}\,=\,0$. Hence, Eq.(\ref{Tpmcomponent}) becomes

\begin{equation}
G_{\mu \nu}\,\partial_{\pm}\,X^\mu \, \partial_{\pm}\,X^\nu \,=\, 0\, , \label{Virasoro}
\end{equation}

which are known as the Virasoro constraint equations.\\

Finally, varying the Polyakov Action with respect to the string worldsheet coordinates and setting this functional variation to zero, yields the string equations of motion in the static gauge

\begin{equation}
0 \, = \, \partial_a \,\Pi_\mu^a \, - \, \Gamma^\alpha_{\mu \nu}\, \partial_a \,X^\nu\,\Pi^a_\alpha \, =:\, \nabla_a\,\Pi^a_\mu\, , \label{StringEoM}
\end{equation}

where the Christoffel symbols are defined as

\begin{equation}
 \Gamma^\alpha_{\mu \nu}\, := \, \frac{1}{2}\, G^{\alpha \gamma} \big(\partial_\mu \, G_{\nu \gamma}  + \partial_\nu\,G_{\mu \gamma}
 - \partial_\gamma\,G_{\mu \nu}\big)  \, . \label{ChristoffelDef}
\end{equation}
\vspace{0.1cm}

\noindent The details of this derivation are given in appendix (\ref{appPolyakov}). The boundary conditions chosen in order to satisfy Eq.(\ref{StringEoM}) are

\begin{equation}
 \delta X^\mu\, \Pi_\mu^a \big\rvert^{\sigma = \sigma_f}_{\sigma = 0} \, = \, 0 \, . \label{BC}
\end{equation}
\vspace{0.1cm}

\noindent This becomes a Dirichlet boundary condition at the stretched horizon ($r=r_s$)\footnote{The stretched horizon is defined by $r_s \, = \, (1+ \epsilon)\,r_H$ where $0 < \epsilon \ll 1$. In the limit $\epsilon \rightarrow 0$ implementing a Neumann boundary condition here instead of a Dirichlet condition would be equivalent. This is done by de Boer \textit{et al.} \cite{Boer_2009}.}

\begin{equation}
 \delta X^\mu(t,0) \, = \, 0 \, , \label{BC_HQ_horizon}
\end{equation}
\vspace{0.1cm}

\noindent and a Neumann boundary condition at the boundary endpoint\footnote{ From the bulk perspective, this endpoint terminates on a space filling flavour D7-brane which is introduced in order to ensure the test quarks in the field theory have a finite mass \cite{Karch_2002}.}

\begin{equation}
 \Pi_\mu^a(t,\sigma_f) \, = \, 0 \, .\label{BC_HQ_boundary}
\end{equation}
\vspace{0.1cm}

\noindent Eq.(\ref{StringEoM}) is the equations of motion describing the leading order string behaviour. Considering additional behaviour, small transverse fluctuations for example, would results in supplementary equations of motion for $X^I$ (where $I$ denote the transverse directions).
This is precisely the focus of subsection (\ref{subsecHQ_TSDynamics}).\\

The general, leading order solution to the string equations of motion is found by solving Eqs.(\ref{Virasoro}, \ref{StringEoM}) with respect to the boundary conditions Eqs.(\ref{BC_HQ_horizon}, \ref{BC_HQ_boundary}) and the relevant initial conditions. The metric $G_{\mu \nu}$, the initial conditions, and hence the string solution will depend on the geometry of the spacetime the test string is set up in.
For example, in $\mathbb{R}^{1,1}$ the string's worldsheet parameter space $\mathcal{M}$ (with coordinates $(t, \sigma) \in \, [0,t_f] \times [0, \sigma_f] \, = \, \mathcal{M}$) is embedded into the target spacetime $\mathcal{N} = \mathbb{R}^{1,1}$ by the function

\begin{equation}
 X_{\text{Mink}}^\mu(t,\sigma) \, = \, \big(t,\, x(t, \sigma) \big)^{\mu} \, ,\label{HQ_GenSol_Flat}
\end{equation}
\vspace{0.1cm}

\noindent where $x(t, \sigma)$ still needs to be determined. This is the aim of the following two subsections, where the general solutions for the test string in an $\mathbb{R}^{1,1}$ and an AdS$_3$-Schwarzschild background respectively are found\footnote{A note to the reader: a more interesting embedding occurs when studying the off-mass-shell light quark, since the parameter space is divided into two separate regions. This will be explored in detail in subsection (\ref{subsecLQ_LOBehaviour}).}.

\subsection{Leading Order String Behaviour} \label{subsecHQ_LOBehaviour}

\subsubsection{Test Strings in \texorpdfstring{$\mathbb{R}^{1,1}$}{R-2d-flatspace}} \label{subsubsecHQinFlatSpace}
\vspace{0.1cm}

The embedding functions\footnote{Referred to interchangeably throughout this article as the leading order solution to the string equations of motion.}, $X^\mu_{\text{Mink}}$, given in Eq.(\ref{HQ_GenSol_Flat}) are found by solving the Virasoro constraints and string equations of motion in $\mathbb{R}^{1,1}$. For simplicity, a square parameter space is chosen i.e. $t_f=\sigma_f$. Working in the static gauge, a Dirichlet Boundary condition at $\sigma = 0$ is imposed (corresponding to the fixed initial position of the string endpoint)

\begin{equation}
x(t,0) \, = \, x_0 \, \in \, \mathbb{R} \,, \label{Dirichlet_BC_R11}
\end{equation}

and a Neumann Boundary condition at $\sigma = \sigma_f$ is imposed (enforcing zero flux through the other string endpoint)

\begin{equation}
\partial_\sigma\, x(t,\sigma)\big\rvert_{\sigma = \sigma_f} \, = \, 0 \,. \label{Neumann_BC_R11}
\end{equation}
\vspace{0.1cm}

For initial conditions, take the string to be static at $t=0$,

\begin{equation}
\partial_t \, x(t,\sigma)\big\rvert_{t= 0} \, = \, 0 \,, \label{Static_t0_IC_R11}
\end{equation}
\vspace{0.1cm}

and stretched between $x_0$ and $x_0 \, + \, \ell_0$ in the $x$ spatial direction,

\begin{equation}
x(0, \sigma) \, = \, x_0 \, + \, \frac{\sigma}{\sigma_f}\, \ell_0 \, , \, \,\,\,\, \ell_0 \, \in \, \mathbb{R}^+ \,. \label{Spatial_x0_IC_R11}
\end{equation}
\vspace{0.1cm}

The total energy and momentum of the string are given by

\begin{equation}
\text{E}\, = \,- \int \, d\sigma \, \Pi^{\tau}_{\,t} \,\,\,\,\,\,\,
\text{and} \,\,\,\,\,\,\, \text{p} \, = \, \int \, d\sigma \, \Pi^{\tau}_{\,x} \,, \label{EandMomDef}
\end{equation}
\vspace{0.1cm}

\noindent where $\Pi^a_\mu$ are the canonical momentum densities defined in Eq.(\ref{MomDensity}) \cite{zwiebach2004first}. In flat space the total energy of the static string\footnote{\textit{Static} in terms of the string's configuration, not in terms of gauge choice.} -- which would be equal to the mass of a heavy test quark in the boundary theory\footnote{Supposing for a moment that the gauge/string duality postulated the existence of such a boundary theory to $\mathbb{R}^{1,1}$ spacetime.} -- is given by


\begin{equation}
\text{E}\, = \, - \frac{1}{2 \pi \alpha'}\, \int_0^{\sigma_f}\, d \sigma \, \eta^{\tau b} \, \eta_{t \nu} \, \partial_b  X^\nu \, = \, - \frac{\ell_0}{2 \pi \alpha'} \, . \label{FlatSpaceEnergy}
\end{equation}
\vspace{0.1cm}

\noindent where it has been recognised that $\sigma_f = \ell_0$ (which follows from the Virasoro constraints in flat space \cite{moerman2016semi}). The static gauge choice ($\tau = t$), and $X^0 = t$ (Eq.(\ref{HQ_GenSol_Flat})) are also used. Therefore,

\begin{equation}
\text{E}^2\, \equiv \, m_q^2 \, = \, \frac{\ell_0^2}{4 \pi^2 \alpha'^2} \,, \label{FlatSpaceMass}
\end{equation}
\vspace{0.1cm}

\noindent where $E^2 = m^2$ since the string is initially static -- i.e. the total momentum vanishes. The length of the string and the magnitude of the quark's mass are directly proportional. This same relationship holds for a test string in an AdS-Schwarzschild background, and indeed Eq.(\ref{FlatSpaceMass}) remains true. For proof of this see appendix (\ref{appEnergyCurvedSpace}).\\

Since the leading order dynamics are that of a static string, the initial condition Eq.(\ref{Spatial_x0_IC_R11}) holds for all $t \in \mathbb{R}^+$. Hence, the embedding functions $ X_{\text{Mink}}^\mu$ are somewhat trivially\footnote{As $(t, \sigma)$ sweep out a square region of parameter space
$(t, \sigma) \in \, [0,\sigma_f] \times [0, \sigma_f] \, = \, \mathcal{M}$, the embedding functions map out a rectangular region (square if $x_0=0$) of target spacetime.} given by

\begin{equation}
 X_{\text{Mink}}^\mu(t,\sigma) \, = \, \big(t,\, x_0 + \sigma \big)^{\mu} \, ,\label{HQ_Sol_Flat}
\end{equation}
\vspace{0.1cm}

with coordinates $(t, \sigma) \in \, [0,\sigma_f] \times [0, \sigma_f] \, = \, \mathcal{M}$.

\subsubsection{Test Strings in \texorpdfstring{AdS$_3$-Schwarzschild}{AdS3-Schwarzschild}} \label{subsubsecHQinAdS3}

Consider a test string in AdS$_3$-Schwarzschild set up as in figure (\ref{fig:stringdiagram}), page \pageref{fig:stringdiagram}. Examining the leading order behaviour of the string (no transverse fluctuations are present), the equations of motion with respect to the boundary and initial conditions can be solved in order to find the embedding functions $X^{\mu}_\text{\text{AdS$_3$-Sch}} (t, \sigma) $. The AdS$_d$-Schwarzschild metric in $d$ dimensions is given by\footnote{
As mentioned in section (\ref{secGaugeString}), the anti-de Sitter-Schwarzschild spacetime metric Eq.(\ref{AdSd_Metric}) represents a Poincar{\'e} chart of AdS$_d$ spacetime. The global AdS-Schwarzschild spacetime yields two black hole solutions at the same temperature: a black hole with a specific heat which is negative, and a black hole in thermal equilibrium exhibiting Hawking radiation. In order to access both solutions, global AdS-Schwarzschild geometry with a compact spatial boundary would need to be considered \cite{Boer_2009} -- which is beyond the current scope of this work.}

\begin{equation}
ds_d^2 = \frac{r^2}{l^2} \, \left(-h(r;d) \, dt^2 \, + \, d\Vec{X}_{I}^2 \,\right) \,+\, \frac{l^2}{r^2}\frac{dr^2}{h(r;d)} \, , \label{AdSd_Metric}
\end{equation}
\vspace{0.1cm}

where $t \in [0, \infty)$ is the temporal coordinate, $r \in [0, \infty)$ is the radial coordinate, and the transverse spatial directions are denoted by $\Vec{X}_{I} = \left(X^2, X^3, ...., X^{(d-1)}\right) \in \mathbb{R}^{d-2}$. Further, $l \in \mathbb{R}^+$ is the curvature radius of AdS$_d$ and $S^d$, and the blackening factor of the Schwarzschild black hole situated at the horizon, $h(r;d)$, is given by
\begin{equation}
 h(r;d) \,:=\, 1 - \bigg( \frac{r_H}{r} \bigg)^{d-1} \, . \label{BlackFactors}
\end{equation}

The function $h(r;d) \in [0,1]$, where $h(r;d) = 0$ at the stretched horizon and $h(r;d) = 1$ at the AdS-Schwarzschild boundary. The radial position of the black-brane horizon is denoted by $r_H \in \mathbb{R}^+$. The Hawking temperature of the black-brane in AdS$_d$-Schwarzschild -- corresponding to the temperature of the thermal plasma in the boundary theory -- is

\begin{equation}
    T \, \equiv \, \frac{1}{\beta} \, = \, \frac{(d-1)\,r_H}{4\,\pi\,l^2} \, . \label{HawkingTemp}
\end{equation}
\vspace{0.1cm}

In order to consider whether $r$ is a good radial coordinate to use to find the embedding functions, the causal structure of the spacetime is briefly explored using probe light rays approaching the horizon ($r = r_H$). Considering a null geodesic ($ds_d^2=0$) along the radial direction ($\vec{X}_I =0$), the AdS$_d$-Schwarzschild metric Eq.(\ref{AdSd_Metric}) becomes
\begin{equation}
\frac{dt}{dr} = \pm \, \left( \frac{r^2}{l^2} \, h(r;d) \right)^{-1} \, , \label{AdSd_Metric_NullGeo}
\end{equation}
\vspace{0.1cm}

i.e. as a probe light ray approaches the black hole event horizon ($r = r_H$), $dt/dr \rightarrow \pm \infty$\footnote{Think of a series of light-cones drawn from each point along the trajectory of the light ray approaching the horizon -- as the light ray approaches $r=r_H$, the light-cones \textit{close up} \cite{carroll2004spacetime}.}. There appears to be singular behaviour at $r = r_H$ (as the event horizon is approached, movement in the radial direction with respect to the coordinate time $t$ becomes less and less successful), but it's actually highly dependent on the chosen coordinate system.
For instance, the singular behaviour can be assuaged if the time coordinate is replaced with a coordinate which moves \textit{appropriately slowly} along the null geodesic. Defining $t :=\pm r_*$ along this null geodesic, Eq.(\ref{AdSd_Metric_NullGeo}) can be integrated with respect to $r$ to yield

\begin{equation}
    r_*(d) \, = \, l^2 \, \int dr \frac{1}{r^2 \, h(r;d)} \, , \label{TortoiseCoord}
\end{equation}
\vspace{0.1cm}

where $r_*$ is known as the tortoise coordinate \cite{carroll2004spacetime}. Using Eq.(\ref{BlackFactors}) to solve the integral\footnote{The integral Eq.(\ref{TortoiseCoord}) was solved using Mathematica \cite{Mathematica},
and specifically an integration package called Rubi \cite{Rich2018}.} yields

\begin{equation}
    r_*(d) \, = \,- \frac{l^2}{r} \,{}_{2}F_{1}\,\left(1, \, \frac{1}{d-1}; \, \frac{d}{d-1}; \,
    \left( \frac{r_H}{r}\right)^{d-1} \right)\, , \label{gen_d_TortoiseCoord}
\end{equation}
\vspace{0.1cm}

where ${}_{2}F_{1}$ is the Gaussian hypergeometric function\footnote{See footnote \ref{footnotehypergeometric}, page \pageref{footnotehypergeometric}, for the definition of the Gaussian hypergeometric function.}. Using the tortoise coordinate $r_*$ as the new radial coordinate presents an advantage -- the horizon can be approached at the relevant rate ($dt/dr_*$ remains finite)\footnote{The cost of this coordinate replacement is that the horizon surface is pushed to infinity, i.e. at $r_* = \infty$ \cite{carroll2004spacetime}.}.
Hence, to proceed in solving the string equations of motion, the coordinate set $(t, r_*)$ is chosen since it allows the correct boundary conditions for the fluctuations at the black hole horizon to be specified. The tortoise coordinate, however, can only be inverted for $d=3$.
Therefore it is only trivial to solve the string equations of motion to find the embedding functions of the test string in AdS$_3$-Schwarzschild. For $d=3$, Eq.(\ref{gen_d_TortoiseCoord}) becomes

\begin{equation}
    r_*(3) \, = \, \frac{1}{2} \, \frac{l^2}{r_H}\, \ln \left(\frac{r - r_H}{r + r_H} \right) \, = \, \frac{l^2}{r_H}\,
    \coth^{-1} \left(- \frac{r}{r_H} \right)\, , \label{d=3_TortoiseCoord}
\end{equation}
\vspace{0.1cm}

which can be easily inverted to find $r$ in terms of $r_*(3)$:

\begin{equation}
     r \, = \, - r_H\, \coth \left(\frac{r_H\, r_*}{l^2} \right) \,,  \label{d=3_TortoiseCoordINVERTED}
\end{equation}
\vspace{0.1cm}
where $r_*(3)$ is given by $r_*$ for concision. In $d=3$ dimensions, the metric Eq.(\ref{AdSd_Metric}) is given by\footnote{In AdS$_3$-Schwarzschild there is only one transverse direction, refered to as $x$.}


\begin{equation}
    ds_3^2 \,=\, - \frac{r^2 -r_H^2}{l^2}\, dt^2 \, + \, \frac{l^2}{r^2 -r_H^2}\, dr^2 \, + \,  \frac{r^2}{l^2} \,dx^2  \, . \label{AdS3_Metric}
\end{equation}
\vspace{0.1cm}

This, incidentally, is also the metric for the non-rotating BTZ black hole \cite{Banados_1992}. The AdS$_3$-Schwarzschild metric Eq.(\ref{AdS3_Metric}) can be written in terms of the coordinate set $(t, r_*)$ using Eq.(\ref{d=3_TortoiseCoordINVERTED}) and its differential,

\begin{equation}
     d r \, = \, \frac{r_H^2}{l^2} \, \csch^2 \left(\frac{r_H\, r_*}{l^2} \right) \, d r_* \,.  \label{d=3_TortoiseCoordINVERTED_diff}
\end{equation}
\vspace{0.1cm}

Inserting Eqs.(\ref{d=3_TortoiseCoordINVERTED}, \ref{d=3_TortoiseCoordINVERTED_diff}) into the metric Eq.(\ref{AdS3_Metric})\footnote{Mathematica is used to help simplify the trigonometry. This computation (and others in this section) are explicitly shown in Mathematica Notebook [b] (\texttt{BrownianMotion.nb}). For access, see appendix (\ref{appCode}).}, yields


\begin{equation}
ds_3^2 \, = \,\frac{r_H^2}{l^2}\, \csch^2 \left(\frac{r_H r_*}{l^2} \right) \, \left(-dt^2 \, +\, dr_*^2 \right) \, +\, \frac{r_H^2}{l^2}\, \coth^2 \left(\frac{r_H r_*}{l^2} \right) \,dx^2 \, . \label{AdS3_Metric2}
\end{equation}
\vspace{0.1cm}

From Eq.(\ref{AdS3_Metric2}), notice that the AdS$_3$-Schwarzschild metric in the $(t, r_*)$ coordinate system is conformally flat, i.e. the chosen coordinates $(t, r_*)$ are a set of isothermal coordinates. \\

As discussed previously, to regulate an infrared divergence at the horizon, the stretched horizon $r_s =  (1+ \epsilon)\,r_H$ (where $0 < \epsilon \ll 1$) is introduced. At the stretched horizon, a Dirichlet boundary condition (comparable to Eq.(\ref{BC_HQ_horizon})) for the test string in AdS$_3$-Schwarzschild is implemented

\begin{equation}
X^r_{\text{AdS$_3$-Sch}}(t,0) \, = \, r_s \,,  \label{Dirichlet_BC_AdS3}
\end{equation}
\vspace{0.1cm}
while a Neumann boundary condition (comparable to Eq.(\ref{BC_HQ_boundary})) is implemented at the boundary endpoint

\begin{equation}
\partial_\sigma\, X^r_{\text{AdS$_3$-Sch}}(t,\sigma)\big\rvert_{\sigma = \sigma_f} \, = \, 0 \,. \label{Neumann_BC_AdS3}
\end{equation}
\vspace{0.1cm}

The boundary conditions Eq.(\ref{Dirichlet_BC_AdS3}) and Eq.(\ref{Neumann_BC_AdS3}) are rewritten in terms of the coordinate set $(t, r_*)$. The Dirichlet condition Eq.(\ref{Dirichlet_BC_AdS3}) becomes

\begin{equation}
X^{r_*}_{\text{AdS$_3$-Sch}}(t,0) \, = \, r_{s*} \,,  \label{Dirichlet_BC_AdS3_TortoiseCoord}
\end{equation}
\vspace{0.1cm}
where $r_{s*}$ is defined using Eq.(\ref{d=3_TortoiseCoord}),
\vspace{0.1cm}
\begin{equation}
r_{s*} \,= \, \frac{l^2}{r_H}\, \coth^{-1} \left(- \frac{r_s}{r_H} \right) \,,  \label{StrHor_TortoiseCoord}
\end{equation}
\vspace{0.1cm}

and the Neumann condition Eq.(\ref{Neumann_BC_AdS3}) becomes
\begin{flalign}
& \begin{aligned}
0 \, & = \, \frac{r_H^2}{l^2}\, \csch^2 \left(\frac{r_H}{l^2} \, X^{r_*}_{\text{AdS$_3$-Sch}}(t,\sigma) \right) \, \partial_\sigma\, X^{r_*}_{\text{AdS$_3$-Sch}}(t,\sigma)\big\rvert_{\sigma = \sigma_f}\\[0.2cm]
 &= \, \partial_\sigma\, X^{r_*}_{\text{AdS$_3$-Sch}}(t,\sigma)\big\rvert_{\sigma = \sigma_f} \,, \label{Neumann_BC_AdS3_TortoiseCoord}
\end{aligned} &
\end{flalign}
\vspace{0.1cm}

where Eq.(\ref{d=3_TortoiseCoordINVERTED_diff}) is used in the first line, and the last line follows since $\csch (x) \,\neq \,0\,, \, \forall \, x \in \mathbb{R}$.
The boundary conditions in the conformally flat description of AdS$_3$-Schwarzschild (Eqs.(\ref{Dirichlet_BC_AdS3_TortoiseCoord}, \ref{Neumann_BC_AdS3_TortoiseCoord})) are respectively analogous to the boundary conditions for the test string in $\mathbb{R}^{1+1}$,
 Eqs.(\ref{Dirichlet_BC_R11}, \ref{Neumann_BC_R11}).
 Hence, the embedding functions for the test string in the conformally flat description of AdS$_3$-Schwarzschild are the direct analogue of Eq.(\ref{HQ_Sol_Flat}). In $(t, r_*)$ coordinates, the embedding functions are

\begin{equation}
X^{\mu}_\text{\text{AdS$_3$-Sch}} (t, \sigma) \, = \, \big(t,\,  r_{s*} + \sigma \, , \, 0\big)^{\mu} \, ,\label{HQ_Sol_AdS3_TortoiseCoord}
\end{equation}
\vspace{0.1cm}

with coordinates $(t, \sigma) \in \, [0,\sigma_f] \times [0, \sigma_f] \, = \, \mathcal{M}$.
Using the inverse tortoise transformation Eq.(\ref{d=3_TortoiseCoordINVERTED}), the embedding functions Eq.(\ref{HQ_Sol_AdS3_TortoiseCoord}) can be written in terms of the original $(t,r)$ coordinates

\begin{equation}
X^{\mu}_\text{\text{AdS$_3$-Sch}} (t, \sigma) \, = \, \left(t,\,  - r_H\, \coth \left(\frac{r_H}{l^2}\,\left(r_{s*} \,+\,\sigma\right) \right)
+ \sigma \, , \, 0\right)^{\mu} \, ,\label{HQ_Sol_AdS3}
\end{equation}
\vspace{0.1cm}

where the position of the fixed string endpoint attached to the stretched horizon $r_{s*}$ is given by Eq.(\ref{StrHor_TortoiseCoord}) and the length of the string in tortoise coordinates is given by

\begin{equation}
\sigma_f \,= \, \frac{l^2}{r_H}\, \coth^{-1} \left(- \frac{r_s\, + \, \ell_0}{r_H} \right) \, - \, r_{s*} \,.  \label{simga_f_TortoiseCoor}
\end{equation}
\vspace{0.1cm}

Making the identification

\begin{equation}
r \,= \, -r_H\, \coth \left( \frac{r_H\, \left(r_{s*}\, + \, \sigma \right)}{l^2} \right) \,,  \label{InverseTortoiseCoord_Iden}
\end{equation}
\vspace{0.1cm}

leads to agreement between Eq.(\ref{HQ_Sol_AdS3}), and the stretched string of de Boer \textit{et al.}'s \cite{Boer_2009, Atmaja_2014} calculations.
Finally, notice that comparing Eq.(\ref{d=3_TortoiseCoordINVERTED}) with Eq.(\ref{InverseTortoiseCoord_Iden}), yields $r_* \, \equiv \,r_{s*}\, + \, \sigma $.\\

\subsection{Fluctuating Test String Dynamics} \label{subsecHQ_TSDynamics}

In this subsection transverse fluctuations of the test string in AdS$_3$-Schwarzschild are studied\footnote{Later, in subsection (\ref{subsubsecHQ_GenAdSd}), this will be generalised to  AdS$_d$-Schwarzschild.}, in order to explore (using the $AdS/CFT$ correspondence) the heavy quark's Brownian motion induced by the thermal noise in the plasma. The fluctuations are chosen to be small so as to not affect the leading order radial solution (discovered in subsection (\ref{subsubsecHQinAdS3})).
de Boer \textit{et al.} \cite{Boer_2009} outlays a semiclassical treatment of the test string's transverse motion in AdS$_3$-Schwarzschild by quantizing these transverse fluctuations and relating the behaviour of the modes on the string to the dynamics of the boundary endpoint.\\

The following subsection (\ref{subsubsecHQ_NambuTransverse}) derives the equations of motion for small transverse fluctuations on the test string in AdS$_d$-Schwarzschild. Following the method first laid out in \cite{Boer_2009}, subsection (\ref{subsubsecHQ_Displacement}) solves these equations of motion to find the general solution, and from there the mean-squared displacement of the test string's boundary endpoint, $s^2(t)$.

\subsubsection{Nambu-Goto String Equations of Motion for Transverse Fluctuations} \label{subsubsecHQ_NambuTransverse}

The standard Nambu-Goto action\footnote{Using Eq.(\ref{Constraint2}) the equivalence between the Polyakov action Eq.(\ref{PolyakovAct}) and the Nambu-Goto action Eq.(\ref{NGAction}) is apparent.} can be used to describe the dynamics of the transverse fluctuations on the test string

\begin{equation}
    S_{NG} \, := \, \frac{1}{2 \pi \alpha'}  \int_{\mathcal{M}} \, d^2 \sigma \, \mathcal{L}_{NG} \,
    = \,- \frac{1}{2 \pi \alpha'} \, \int_{\mathcal{M}} d^2 \sigma \, \sqrt{-g}  \, , \label{NGAction}
\end{equation}
\vspace{0.1cm}

where $\mathcal{L}_{NG}$ is the Lagrangian density, $g \, := \,\text{det}\,( g_{a b})$ and the induced worldsheet metric $g_{a b}$ is given by Eq.(\ref{InducedMetric}).\\

Take the embedding functions (given in AdS$_3$-Schwarzschild by Eq.(\ref{HQ_Sol_AdS3})) to be the leading order static string solution in a $(1+1)$-dimensional subspace of spacetime spanned by the temporal and radial directions $(t,r)$, denoted by $X^\mu_0 (t, \sigma)$ (where $X^\mu_0 : [0, t_f]  \times [0, \sigma_f] \rightarrow  \mathbb{R}^{d-1, 1}$; $(t, \sigma)  \rightarrow  X^\mu_0(t, \sigma)$).
Consider the addition of transverse fluctuations $X^I$ (where $I \in \left(2,\, 3, \, ...\, ,\, d-1 \right)$) to $X^\mu_0(t, \sigma)$. The effective action for the transverse fluctuations appears as a correction to the leading order Nambu-Goto action. It follows that to find the effective action for the transverse fluctuations (and from this the equations of motion) the Nambu-Goto action needs to be expanded about $X^\mu_0(t, \sigma)$, in terms of $X^I$,

\begin{equation}
    S_{NG} \, = \, \frac{1}{2 \pi \alpha'}  \int_{\mathcal{M}} \, d^2 \sigma \, \mathcal{L}_{NG}\, \Big\rvert_{X^\mu_0} \,
    + \, S_{NG}^{(2)}\, +\, S_{NG}^{(4)} \,+\, \mathcal{O}\left(S_{NG}^{(6)}\right) \, , \label{NGActionExpand}
\end{equation}

where

\begin{equation}
    S_{NG}^{(2)} := \frac{1}{2 \pi \alpha'} \int_{\mathcal{M}} \, d^2 \sigma \, \mathcal{L}_{NG}^{(2)} \,\,\,\,\,\,\,\, \text{and} \,\,\,\,\,\,\,\, S_{NG}^{(4)}
    := \frac{1}{2 \pi \alpha'} \int_{\mathcal{M}} \, d^2 \sigma \, \mathcal{L}_{NG}^{(4)} \, . \label{NGAction_SNG2_SNG4}
\end{equation}
\vspace{0.1cm}

In order to explicitly calculate the expansion in Eq.(\ref{NGActionExpand}), the determinant of the induced worldsheet metric is needed. This is derived in appendix (\ref{appEnergyCurvedSpace}) and is given below

\begin{equation}
  g \, := \, \text{det}\, (g_{a b }) \, = \,  G_{rr}\,G_{tt}\, \left( r'^2 \, + \, \frac{r'^2 }{G_{tt}}\,G_{I I}\,\Dot{X}_I^2 \,
  +\, \frac{1}{G_{rr}}\,G_{I I}\, X_I'^{\,2} \right) \,, \label{InducedDet}
\end{equation}
\vspace{0.1cm}

where $\dot{X}= \partial_{\tau} X$ and $X^{'}= \partial_{\sigma} X$; $I$ indexes over the transverse directions $X_I=\left(X^2,X^3, ...., X^{(d-1)}\right)$; and the AdS$_d$-Schwarzschild spacetime metric $G_{\mu \nu}$ is the given explicitly by Eq.(\ref{appTargetMatrix}).
Using Eq.(\ref{InducedDet}) the Nambu-Goto action Eq.(\ref{NGAction}) becomes

\begin{flalign}
& \begin{aligned}
  S_{NG}  & =  - \frac{1}{2 \pi \alpha'} \int_{\mathcal{M}} d^2 \sigma  \left(- r'^2 \, G_{rr} G_{tt} \right)^\frac{1}{2}\, \sqrt{ 1  \, + \, \frac{1 }{G_{tt}}\,G_{II}\,\Dot{X}_I^2 \, +\, \frac{1}{r'^2 \,G_{rr}}\,G_{I I}\, X_I'^{\,2} } \\[.2cm]
  & =  - \frac{1}{2 \pi \alpha'}  \int_{\mathcal{M}}  d^2 \sigma  \sqrt{-g}\big\rvert_{X^\mu_0} \ -  \frac{1}{4 \pi \alpha'}  \int_{\mathcal{M}}  d^2 \sigma  \sqrt{-g}\big\rvert_{X^\mu_0}   \left(  \frac{G_{II}}{G_{tt}} \Dot{X}_I^2 + \frac{G_{II}}{r'^2 \,G_{rr}} X_I'^{\,2} \right)   \Bigg\rvert_{X^\mu_0}\\[.2cm]
  &  \,\,\,\,\,\,+  \frac{1}{16 \pi \alpha'}  \int_{\mathcal{M}}  d^2 \sigma  \sqrt{-g}\big\rvert_{X^\mu_0}  \left(  \frac{G_{II}}{G_{tt}}\Dot{X}_I^2  + \frac{G_{II}}{r'^2 \,G_{rr}} X_I'^{\,2} \right)^2   \Bigg\rvert_{X^\mu_0} \,, \label{NGActionExpand2}
\end{aligned} &
\end{flalign}

where a Taylor expansion up to quadratic order (i.e. $(1 + x)^{1/2}  \approx 1 + \frac{x}{2} - \frac{x^2}{8}\, $) was first performed. Then it is noticed from Eq.(\ref{InducedDet}) that to leading order the determinant of the induced worldsheet metric is simply given by

\begin{equation}
  g\big\rvert_{X^\mu_0} \, = \,  r'^2 \,G_{rr}\,G_{tt}\,. \label{InducedDet_LO}
\end{equation}
\vspace{0.1cm}

The quadratic correction in the expansion of the Nambu-Goto action (defined in Eq.(\ref{NGAction_SNG2_SNG4})) is the effective action for the transverse fluctuations. Matching terms between Eq.(\ref{NGActionExpand}) and Eq.(\ref{NGActionExpand2}), the quadratic correction is given by

\begin{flalign}
& \begin{aligned}
  S_{NG}^{(2)} \, & =  - \frac{1}{4 \pi \alpha'} \int_{\mathcal{M}} d^2 \sigma  \sqrt{-g}\big\rvert_{X^\mu_0}   \left(  \frac{G_{II}}{G_{tt}}\Dot{X}_I^2  + \frac{G_{II}}{r'^2 \,G_{rr}} X_I'^{\,2} \right)   \bigg\rvert_{X^\mu_0} \\[.2cm]
  & = - \frac{1}{4 \pi \alpha'}  \int_{\mathcal{M}}  d^2 \sigma  \left( \frac{\sqrt{-g}}{g_{a b}}\, G_{IJ} \right)\bigg\rvert_{X^\mu_0}  {\partial_a X^I} {\partial_a X^J} \,, \label{NG_Quadratic}
\end{aligned} &
\end{flalign}
\vspace{0.05cm}

where $a,b$ indexes over the static gauge worldsheet coordinates $(t, \sigma)$, the spacetime metric (Eq.(\ref{appTargetMatrix})) is recognised as a diagonal matrix, and to leading order the induced worldsheet metric (see appendix (\ref{appEnergyCurvedSpace})) is given by
\begin{equation}
g_{a b}\big\rvert_{X^\mu_0}\, =\,
  \begin{bmatrix}
    G_{tt}\,+ \, G_{I I}\, \Dot{X}^2 & G_{I I}\,\Dot{X}_I\,X_I' \\
  G_{I I}\,X_I'\,\Dot{X}_I & r'^2\,G_{rr} \, + \,G_{I I} \, X_I'^{\,2}
  \end{bmatrix} \Bigg\rvert_{X^\mu_0} \, = \,
  \begin{bmatrix}
    G_{tt} & 0 \\
  0 & r'^2\,G_{rr}
  \end{bmatrix}  \,. \label{InducedMatrix_LO}
\end{equation}
\vspace{0.3cm}

Hence, the effective action for the transverse fluctuations $S_{NG}^{(2)}$ is given by

\begin{flalign}
& \begin{aligned}
S_{NG}^{(2)}   & =  - \frac{1}{4 \pi \alpha'}\, \int_{\mathcal{M}} \, d^2 \sigma \, \big(\sqrt{-g}\, g^{ab} \, G_{IJ} \big) \big\rvert_{X^\mu_0} \,\, \partial_a  X^I \partial_b  X^J \\[0.2cm]
 & \equiv \frac{1}{4 \pi \alpha'}\, \int_{\mathcal{M}} \, d^2 \sigma \, \dfrac{\partial^2 \, \mathcal{L}_{NG}}{\partial \big( \partial_a \, X^I\big) \, \partial\big( \partial_b \, X^J\big)} \bigg\rvert_{X^\mu_0} \,\, \partial_a  X^I \partial_b  X^J  \,, \label{NG_Quadratic2}
\end{aligned} &
\end{flalign}
\vspace{0.1cm}

where $I, J$ index over the transverse directions $\left(2, 3, ...\, , d-1 \right)$. Comparing Eq.(\ref{NG_Quadratic2}) with Eq.(\ref{PolyakovAct}), notice that if the auxiliary worldsheet metric is chosen to be the leading order induced worldsheet metric ($\gamma_{ab} = g_{a b}\rvert_{X^\mu_0}$) the action for transverse fluctuations Eq.(\ref{NG_Quadratic2}) can be understood as an effective Polyakov Action. It remains to be proven whether or not the quartic term $S_{NG}^{(4)}$ is significant and contributes to the effective action of the transverse fluctuations.
The quartic term $S_{NG}^{(4)}$ (defined in Eq.(\ref{NGAction_SNG2_SNG4})) is given by

\begin{equation}
S_{NG}^{(4)}   =  \frac{1}{16 \pi \alpha'} \int_{\mathcal{M}}  d^2 \sigma
\left( \big(\sqrt{-g} g^{ab}  G_{IJ} \big) \big\rvert_{X^\mu_0} \, \partial_a  X^I \,\partial_b  X^J\right)^2  \,. \label{NG_Quartic}
\end{equation}
\vspace{0.1cm}

Comparing the quadratic action term Eq.(\ref{NG_Quadratic2}) with the quartic action term Eq.(\ref{NG_Quartic}), yields

\begin{equation}
S_{NG}^{(2)} \, = \,- \frac{1}{\sqrt{\pi \alpha'}} \, \left(S_{NG}^{(4)}\right)^{\frac{1}{2}}\,. \label{NG_Quadratic_vs_Quartic}
\end{equation}
\vspace{0.1cm}

Hence the quartic correction term $S_{NG}^{(4)}$ will significantly contribute to the effective action for the transverse fluctuations when the test string is within a distance of $\sqrt{\alpha'}$ from the black-brane horizon.
Therefore the quadratic action $S_{NG}^{(2)}$ (Eq.(\ref{NG_Quadratic2})) can be considered to solely contribute to the effective action for the transverse fluctuations, as long as the fluctuations are small and the test string is in the region further than $\sqrt{\alpha'}$ away from the black-brane horizon, where\vspace{0.05cm}

\begin{equation}
  \sqrt{\alpha'} \, = \,\lambda^{-1/4} \, l  \,, \label{alpha_def}
\end{equation}
\vspace{0.05cm}

in AdS$_d$-Schwarzschild. Eq.(\ref{alpha_def}) defines the fundamental string length scale $\sqrt{\alpha'}$ is terms of the radius of curvature of AdS$_d$ spacetime $l$ and the 't Hooft coupling $\lambda$.\\

The canonical momentum densities conjugate to the transverse coordinates $X^I$ are defined as

\begin{equation}
  \Pi_{\,I}^a \, := \, \frac{\delta \,\mathcal{S}^{(2)}_{NG}}{\delta (\partial_a X^I)} \, =  \, - \frac{1}{2 \pi \alpha'} \,
  \big(\sqrt{-g}\, g^{ab} \, G_{IJ} \big) \big\rvert_{X^\mu_0}\,\, {\partial_b} X^J  \, , \label{MomDen_XI}
\end{equation}
\vspace{0.1cm}

where the factor of two arises from symmetry (the variation of ${\partial_a} X^I$ and ${\partial_b} X^J$ gives the same the result).\\

Varying the effective action for the transverse fluctuations $S_{NG}^{(2)}$ with respect to the transverse string worldsheet coordinates $X^I$ and setting this functional variation to zero, yields the equations of motion

\begin{equation}
0 \, = \, \partial_a \,\Pi_{\, I}^a \, =\, \nabla_a\,\Pi^a_{\,I}\, . \label{StringEoM_XI}
\end{equation}
\vspace{0.1cm}

These equations of motion are analogous to Eq.(\ref{StringEoM}); but in Eq.(\ref{StringEoM_XI}) the Christoffel symbols (defined in Eq.(\ref{ChristoffelDef})) are identically zero ($\Gamma_{I J}^{\alpha} \, = \, 0$) due to the symmetry between each of the transverse directions. The details of this derivation are given in appendix (\ref{appNambuGoto}). The chosen boundary conditions (analogous to Eq.(\ref{BC})) are

\begin{equation}
 \Pi_{\, I}^a \, \delta X^I\,  \big\rvert^{\sigma = \sigma_f}_{\sigma = 0} \, = \, 0 \, , \label{BC_XI}
\end{equation}
\vspace{0.1cm}

which ensures the correct boundary terms vanish in the derivation of the string equations of motion for the transverse fluctuations (Eq.(\ref{StringEoM_XI})).

\subsubsection{Calculating the Heavy Quark's Mean-Squared Displacement \texorpdfstring{$s^2(t)$}{s2t} in \texorpdfstring{AdS$_3$-Schwarzschild}{AdS3}} \label{subsubsecHQ_Displacement}

In AdS$_3$-Schwarzschild, the string boundary endpoint's mean-squared transverse displacement is given by 

\begin{equation*}
 s^2(t) \, = \frac{\beta^2}{4  \pi^2  \sqrt{\lambda}}  \int_0^\infty \frac{d \omega}{\omega}  \frac{1}{e^{\beta \omega} - 1}  \left[f_\omega (\sigma_f) f^{\,*}_{\omega} (\sigma_f) \left(2 -e^{- i \omega t}- e^{i \omega t} \right)\right] \, . 
\end{equation*}
\vspace{0.1cm}

A detailed derivation of this equation is given in this subsection.\\

The leading order static solution to the string equations of motion is given by Eq.(\ref{HQ_Sol_AdS3}) in AdS$_3$-Schwarzschild. Following subsection (\ref{subsubsecHQ_NambuTransverse}), this solution is renamed $X^\mu_0 (t, \sigma)$. Adding non-zero fluctuations in the transverse $x$-direction, the test string solution becomes

\begin{equation}
X^{\mu}_\text{\text{AdS$_3$-Sch}} (t, \sigma) \, = \, X^\mu_0(t, \sigma) \, + \, \big(0, \,0,\, X(t, \sigma) \big)^\mu \,
= \,\left( t, \, - r_H \coth \left(\frac{r_H}{l^2}\left(r_{s*} +\sigma\right) \right) , \, X(t, \sigma) \right)^{\mu}\,. \label{HQ_GenSolAdS3_XI}
\end{equation}
\vspace{0.1cm}

In AdS$_3$-Schwarzschild, the equations of motion for the transverse fluctuations, Eq.(\ref{StringEoM_XI}), become


\begin{equation}
0 \,= \, \partial_a  \left(  \left(\sqrt{-g}\, g^{ab} \, \frac{r^2}{l^2} \right) \bigg\rvert_{X^\mu_0} {\partial_b} X(t, \sigma)\right) \,, \label{StringEoM_XI_d=3}
\end{equation}
\vspace{0.1cm}

where the definition transverse momentum densities Eq.(\ref{MomDen_XI}) are used in the first line, and the spacetime metric Eq.(\ref{appTargetMatrix}) in the second line. To leading order the induced worldsheet metric is given by Eq.(\ref{InducedMatrix_LO}). For $d=3$, this metric simplifies to

\begin{equation}
g_{a b}\big\rvert_{X^\mu_0}\, := \,
  \begin{bmatrix}
   g_{tt} & g_{t \sigma} \\
   g_{\sigma t} & g_{\sigma \sigma}
  \end{bmatrix} \Bigg\rvert_{X^\mu_0} \, = \,
  \begin{bmatrix}
    G_{tt} & 0 \\
   0 & r'^2\, G_{rr}
  \end{bmatrix} \, = \,
  \begin{bmatrix}
    \left( -\frac{r^2-r_H^2}{l^2} \right) & 0 \\
   0 & r'^2  \left(\frac{l^2}{r^2-r_H^2} \right)
  \end{bmatrix} \,, \label{InducedMatrix_LO_d=3}
\end{equation}
\vspace{0.1cm}
where Eq.(\ref{AdS3_Metric}) is used, and $r'= \partial_{\sigma} r$. Hence, its inverse is given by

\begin{equation}
g^{a b}\big\rvert_{X^\mu_0}\, := \,
  \begin{bmatrix}
   g^{tt} & g^{t \sigma} \\
   g^{\sigma t} & g^{\sigma \sigma}
  \end{bmatrix} \Bigg\rvert_{X^\mu_0} \, = \,
  \frac{1}{\det \big(g_{a b}\rvert_{X^\mu_0} \big) }\,\begin{bmatrix}
    r'^2 \, G_{rr} & 0 \\
   0 & G_{tt}
  \end{bmatrix} \, = \,
  \begin{bmatrix}
    \left(- \frac{l^2}{r^2-r_H^2} \right)  & 0 \\
   0 & \frac{1}{r'^2} \left( \frac{r^2-r_H^2}{l^2} \right)
  \end{bmatrix} \,, \label{InducedMatrix_LO_d=3_INVERSE}
\end{equation}
\vspace{0.3cm}

where $\det \big(g_{a b}\rvert_{X^\mu_0} \big)$ is given in Eq.(\ref{InducedDet_LO}). The indices in Eq.(\ref{StringEoM_XI_d=3}) can be expanded
\begin{flalign}
& \begin{aligned}
0 \, & = \, \partial_\sigma  \left(  \left(\sqrt{-g}\, g^{\sigma \sigma} \, \frac{r^2}{l^2} \right) \bigg\rvert_{X^\mu_0}\,\, {\partial_\sigma} X(t,\sigma)\right) \,+\, \partial_t  \left(  \left(\sqrt{-g}\, g^{tt} \, \frac{r^2}{l^2} \right) \bigg\rvert_{X^\mu_0}\,\, {\partial_t} X(t,\sigma)\right)\\[0.2cm]
 & = \, - \, \partial_t^2 \, X(t,\sigma)\,+\, \frac{r^2-r_H^2}{l^4\,r^2}\,\frac{1}{\partial_\sigma r} \, \partial_\sigma \left( \frac{1}{\partial_\sigma r}\, r^2\, \left( r^2-r_H^2 \right) \, {\partial_\sigma} X(t,\sigma)\right) \,, \label{StringEoM_XI_d=3_tr1}
\end{aligned} &
\end{flalign}
\vspace{0.1cm}

where Eqs.(\ref{InducedDet_LO}, \ref{InducedMatrix_LO_d=3_INVERSE}) are used in the second line. The transverse equations of motion can be fully converted into spacetime variables $(t, r)$. The derivative operator $\partial_\sigma$ first needs to be converted into $\partial_r$ using Eq.(\ref{InverseTortoiseCoord_Iden}). The differential of Eq.(\ref{InverseTortoiseCoord_Iden}) is
\begin{flalign}
& \begin{aligned}
 \partial_\sigma &\, = \, \frac{r_H^2}{l^2} \, \csch^2 \left(\frac{r_H\, (r_{s*}\, + \, \sigma )}{l^2} \right) \, \partial_r\\[0.2cm]
 &\, = \, \frac{r^2 - r_H^2}{l^2}\, \partial_r  \,. \label{InverseTortoiseCoord_Iden_diff}
\end{aligned} &
\end{flalign}
\vspace{0.1cm}

Using Eq.(\ref{InverseTortoiseCoord_Iden_diff}), the transverse equations of motion Eq.(\ref{StringEoM_XI_d=3_tr1}) become

\begin{equation}
0\, = \, - \, \partial_t^2 \, X(t,r)\,+\, \frac{r^2-r_H^2}{l^4\,r^2}\, \partial_r \left(  r^2\, \left( r^2-r_H^2 \right) \, {\partial_r} X(t,r)\right) \,. \label{StringEoM_XI_d=3_tr}
\end{equation}
\vspace{0.1cm}

Further, the transverse equations of motion can be completely written in terms of worldsheet parameter coordinates $(t, \sigma)$ using Eqs.(\ref{InverseTortoiseCoord_Iden}, \ref{InverseTortoiseCoord_Iden_diff}), which yields

\begin{equation}
0\, = \,  - \, \partial_t^2 \, X(t,\sigma)\,+\, \frac{1}{\coth^2 \left(\frac{r_H}{l^2}\,\left(r_{s*} +\sigma\right) \right)}\, \partial_\sigma \left( \coth^2 \left(\frac{r_H}{l^2}\,\left(r_{s*} +\sigma\right) \right) \, {\partial_\sigma} X(t,\sigma)\right) \,. \label{StringEoM_XI_d=3_tsig}
\end{equation}
\vspace{0.1cm}

To proceed in solving the linear and homogeneous partial differential equation  Eq.(\ref{StringEoM_XI_d=3_tr}) -- or Eq.(\ref{StringEoM_XI_d=3_tsig}) in ($t, \sigma$) coordinates -- the method laid out in de Boer \textit{et al.} \cite{Boer_2009} is followed. Explicitly, consider a mode expansion in $X$,
\begin{equation}
X(t,r) \, = \, f_\omega (r) \, e^{- i \omega t} \,, \label{ModeExpansion_X}
\end{equation}
\vspace{0.1cm}

where $X(t,r)$ is a separable eigenmode solution to Eq.(\ref{StringEoM_XI_d=3_tr}) which oscillates in time with a well-defined angular frequency $\omega \in \mathbb{R}^+$. Inserting Eq.(\ref{ModeExpansion_X}) into Eq.(\ref{StringEoM_XI_d=3_tr}), it can been seen that $f_\omega (r)$ satisfies the ordinary differential equation

\begin{equation}
0  \, =\, \omega^2 \, f_\omega (r) \,+\, \frac{r^2-r_H^2}{l^4\,r^2}\, \partial_r  \left(  r^2\, \left(r^2-r_H^2 \right) \, {\partial_r}\, f_\omega (r) \right)  \,. \label{fw_r_ODE}
\end{equation}
\vspace{0.1cm}

Defining the dimensionless quantity

\begin{equation}
\nu \, := \, \frac{l^2 \, \omega}{r_H} \,, \label{dim_angular_f_def}
\end{equation}

the ordinary differential equation (Eq.(\ref{fw_r_ODE})) can be rewritten as

\begin{equation}
\left( \nu^2 \,  \,+\, \frac{r_H^2 \, (r^2-r_H^2)}{l^8\,r^2}\, \partial_r  \left(  r^2\, \big( r^2-r_H^2 \big) \, {\partial_r}\, \right) \right) f_\omega (r) \, = \, 0 \,. \label{nu_fw_r_ODE}
\end{equation}
\vspace{0.1cm}

This second order ODE has two linearly independent solutions\footnote{This is consistent with Eq.(2.36) in de Boer \textit{et al.}'s paper \cite{Boer_2009}.}

\begin{equation}
f^{(\pm)}_\omega (r) \, = \, \frac{1}{1 \,\pm \, i \nu}\, \frac{r \,\pm \, i r_H \nu}{r} \,
\left(\frac{r \,- \,  r_H }{r\,+\, r_H} \right)^{\pm i \nu /2 }\,. \label{nu_fw_r_ODE_Solution}
\end{equation}
\vspace{0.1cm}

In solving Eq.(\ref{nu_fw_r_ODE}) the initial condition near the horizon is chosen to be

\begin{equation}
    f_\omega^{(\pm)}(r) \, \rightarrow \, \left(\frac{r \,- \,  r_H }{r\,+\, r_H} \right)^{\pm i \nu /2 }  \, \, \, \, \, \,\, \, \, \, \, \,
    \text{as}\, \, \, \, \, \,\, \, \, \, \, \,  r \, \rightarrow \, r_H \, , \label{IC_horizon_fw}
\end{equation}
\vspace{0.1cm}

i.e. the normalization at the black-brane horizon ($r= r_H$) is taken to be $\frac{1}{1 \,\pm \, i \nu}\, \frac{r \,\pm \, i r_H \nu}{r} \, = \, 1$.
For convenience, the solutions Eq.(\ref{nu_fw_r_ODE_Solution}) can be partially rewritten in terms of the tortoise coordinate $r_*$. Rearranging Eq.(\ref{InverseTortoiseCoord_Iden}) yields

\begin{flalign}
& \begin{aligned}
&  r_{s*} +  \sigma \, = \, \frac{l^2}{r_H}\, \coth^{-1} \left(- \frac{r}{r_H} \right)\\[0.2cm]
 &\Rightarrow  \, e^{\pm i \omega \, ( r_{s*} +  \sigma)} \, = \,  \left( \frac{r-r_H}{r+r_H} \right)^{\pm i \nu /2} \,, \label{TC_def_for_nu_fw_r_ODE_Solution}
\end{aligned} &
\end{flalign}
\vspace{0.1cm}

where the definition for $\nu$, Eq.(\ref{dim_angular_f_def}), is used in the second line. Inputting Eq.(\ref{TC_def_for_nu_fw_r_ODE_Solution}) into Eq.(\ref{nu_fw_r_ODE_Solution}) gives

\begin{equation}
f^{(\pm)}_\omega (r) \, = \, \frac{1}{1 \,\pm \, i \nu}\,
\frac{r \,\pm \, i r_H \nu}{r} \, e^{\pm i \omega \, ( r_{s*}\, + \, \sigma)}\,, \label{TC_nu_fw_r_ODE_Solution}
\end{equation}
\vspace{0.1cm}

where the initial condition near the horizon Eq.(\ref{IC_horizon_fw}) becomes

\begin{equation}
    f^{(\pm)}_\omega (r) \, \rightarrow \, e^{\pm i \omega \, ( r_{s*}\, + \, \sigma)}  \, \, \, \, \, \,\, \, \, \, \, \,
    \text{as}\, \, \, \, \, \,\, \, \, \, \, \,  r \, \rightarrow \, r_H \, . \label{IC_horizon_fw_TC}
\end{equation}
\vspace{0.1cm}

From Eq.(\ref{IC_horizon_fw_TC}) it is apparent that $f^{(\pm)}_\omega (r)$ denotes out-going $(+)$ and in-falling $(-)$ basis modes\footnote{The mode $f^{+}_\omega (r)$ is outgoing at the horizon; while $f^{-}_\omega (r)$ is a mode which is reflected at the boundary and -- after experiencing a phase shift -- falls back towards the horizon.}. The general solution to the ODE, $f_\omega (r)$, is a linear combination of these modes

\begin{equation}
    f_\omega(r) \, = \,  f_\omega^{(+)}(r) \, + \, B_\omega \, f_\omega^{(-)}(r) \, , \label{gen_solution_sum_modes}
\end{equation}
\vspace{0.1cm}

where the constant $B_\omega$ measures the difference in phase between the $(+)$ and $(-)$ modes, and is yet to be determined. Imposing a Neumann boundary condition in the radial direction at the boundary endpoint,

\begin{flalign}
& \begin{aligned}
0\, &  = \, \partial_r \, f_\omega(r)\big\rvert_{r = r_s + \ell_o}\\[0.2cm]
 &= \, \partial_r  \left(\frac{1}{1 +  i \nu}\frac{r +  i r_H \nu}{r}  \left(\frac{r -   r_H }{r+ r_H} \right)^{ i \nu /2 }  + \, B_\omega \, \frac{1}{1 - i \nu} \frac{r - i r_H \nu}{r}  \left(\frac{r -  r_H }{r+ r_H} \right)^{- i \nu /2 } \right)\Bigg\rvert_{r = r_s + \ell_o}  \,, \label{NeumannBC_fixBw}
\end{aligned} &
\end{flalign}
\vspace{0.1cm}

enables $B_\omega$ to be fixed. Note that Eqs.(\ref{nu_fw_r_ODE_Solution}, \ref{gen_solution_sum_modes}) have been used in the second line.
Hence, Eq.(\ref{NeumannBC_fixBw}) is solved\footnote{Mathematica is used (see Mathematica Notebook [b]: \texttt{BrownianMotion.nb}).} to find $B_\omega$:

\begin{flalign}
& \begin{aligned}
B_\omega \, & = \,  \left(\frac{i +   \nu}{-i+ \nu} \right) \left(\frac{r \,\nu -   i r_H}{r\, \nu+ i r_H} \right) \left(\frac{r -   r_H }{r+ r_H} \right)^{ i \nu }\Bigg\rvert_{r = r_s + \ell_o} \\[0.2cm]
 & = \, \left(\frac{1 -   i\nu}{1+ i\nu} \right) \left(\frac{1+ i \tilde{r}_0 \,\nu}{1- i \tilde{r}_0 \,\nu} \right) \left(\frac{\tilde{r}_0 -   1}{\tilde{r}_0+ 1} \right)^{ i \nu }  \,  \,, \label{B_w}
\end{aligned} &
\end{flalign}
\vspace{0.1cm}

where the dimensionless quantity

\begin{equation}
    \tilde{r}_0 \, := \, \frac{r_s \, + \, \ell_0}{r_H} \, , \label{dim_r0}
\end{equation}
\vspace{0.1cm}

has been defined in the second line. The coefficient $B_\omega$ can be partially rewritten in terms of the tortoise coordinate $r_*$. To this end Eq.(\ref{simga_f_TortoiseCoor}) is rearranged,

\begin{flalign}
& \begin{aligned}
&  r_{s*} +  \sigma_f \, = \, \frac{l^2}{r_H}\, \coth^{-1} \left(- \frac{r_s \, + \, \ell_0}{r_H} \right)\\[0.2cm]
 &\Rightarrow  \, e^{ i 2\, \omega \, ( r_{s*} +  \sigma_f)} \, = \,  \left(\frac{\tilde{r}_0 -   1}{\tilde{r}_0+ 1} \right)^{ i \nu } \,, \label{part_B_w_TC}
\end{aligned} &
\end{flalign}
\vspace{0.1cm}

where the definition for $\nu$, Eq.(\ref{dim_angular_f_def}), and Eq.(\ref{dim_r0}) is used in the final line. Inputting Eq.(\ref{part_B_w_TC}) into Eq.(\ref{B_w}), yields

\begin{equation}
    B_\omega \, =\, \left(\frac{1 -   i\nu}{1+ i\nu} \right)
    \left(\frac{1+ i \tilde{r}_0 \,\nu}{1- i \tilde{r}_0 \,\nu} \right) e^{ i 2\, \omega \, ( r_{s*}\, + \, \sigma_f)}  \, . \label{B_w_TC}
\end{equation}
\vspace{0.1cm}

The Eqs.(\ref{TC_nu_fw_r_ODE_Solution}, \ref{gen_solution_sum_modes}, \ref{B_w_TC}) determine $f_\omega (r)$ completely. The linear superposition of $f_\omega(r)$ for all frequencies $\omega \in \mathbb{R}^+$, yields the general solution for $X(t,r)$,

\begin{equation}
    X(t,r) \, := \, \int_0^\infty \frac{d \omega}{2 \pi} \, A_\omega \, \left[f_\omega (r) \, e^{- i \omega t} \, a_\omega \,
    + \, f^{\,*}_\omega (r) \, e^{i \omega t} \, a^{\,*}_\omega \right]  \, , \label{gen_solution_linear_sup_X}
\end{equation}
\vspace{0.1cm}

where the mode expansion Eq.(\ref{ModeExpansion_X}) is used, and ($a_\omega, \, a^{\,*}_\omega$) are Fourier coefficients. The constant $A_\omega$ is fixed by demanding the normalization of the appropriate basis.\\

To quantize the theory, the scalar field $X(t, \sigma)$ and its canonically conjugate momentum $P^t(t,\sigma)$ are promoted to operators and suitable commutation relations are imposed. From Eq.(\ref{gen_solution_linear_sup_X}), the position operator is given by

\begin{equation}
    \hat{X}(t,\sigma) \, := \, \int_0^\infty \frac{d \omega}{2 \pi} \, A_\omega \,
    \left[f_\omega (\sigma) \, e^{- i \omega t} \, \hat{a}_\omega \,
    + \, f^{\,*}_\omega (\sigma) \, e^{i \omega t} \, \hat{a}^{\dagger}_\omega \right]  \, , \label{X_operator}
\end{equation}
\vspace{0.1cm}

and from Eq.(\ref{MomDen_XI}) the conjugate momentum operator is given by

\begin{flalign}
& \begin{aligned}
\hat{P}^t(t,\sigma)  & :=  - \frac{1}{2 \pi \alpha'} \, \big(\sqrt{-g}\, g^{tt} \, G_{II} \big) \big\rvert_{X^\mu_0}\,\, {\partial_t} \hat{X}(t,\sigma) \\[0.2cm]
 &  =  - \frac{1}{2 \pi \alpha'} \, \big(\sqrt{-g}\, g^{tt} \, G_{II} \big) \big\rvert_{X^\mu_0}\,\, \int_0^\infty \frac{d \omega}{2 \pi} \, (-i\, \omega) \, A_\omega \, \left[f_\omega (\sigma) \, e^{- i \omega t} \, \hat{a}_\omega \, - \, f^{\,*}_\omega (\sigma) \, e^{i \omega t} \, \hat{a}^{\,*}_\omega \right]  \\[0.2cm]
 & = - \frac{i}{2 \pi \alpha'} \, \frac{r_H^2}{l^2}\, \coth^2 \left( \frac{r_H \, (r_{s*} + \sigma)}{l^2}\right)\, \int_0^\infty \frac{d \omega}{2 \pi} \, \omega \, A_\omega \, \left[f_\omega (\sigma) \, e^{- i \omega t} \, \hat{a}_\omega \, - \, f^{\,*}_\omega (\sigma) \, e^{i \omega t} \, \hat{a}^{\,\dagger}_\omega \right]\\[0.2cm]
 & = - \frac{i}{2 \pi \alpha'} \, \frac{r^2}{l^2}\, \int_0^\infty \frac{d \omega}{2 \pi} \, \omega \, A_\omega \, \left[f_\omega (\sigma) \, e^{- i \omega t} \, \hat{a}_\omega \, - \, f^{\,*}_\omega (\sigma) \, e^{i \omega t} \, \hat{a}^{\,\dagger}_\omega \right]   \,  \,, \label{P_operator}
\end{aligned} &
\end{flalign}

where Eq.(\ref{X_operator}) is used in the second line, $r_*  \equiv r_{s*}\, + \, \sigma $ is used in the fourth line, and the identification Eq.(\ref{InverseTortoiseCoord_Iden}) in the final line. The metric entries in the fourth line follow from using the identification Eq.(\ref{InverseTortoiseCoord_Iden}) to convert the spacetime metric as well as the induced metric, its determinant and its inverse into ($t, r_*$) coordinates. From the spacetime metric Eq.(\ref{appTargetMatrix}),

\begin{equation}
    G_{II} \, = \, G_{II} \big\rvert_{X^\mu_0} \, = \, \frac{r_H^2}{l^2}\, \coth^2 \left( \frac{r_H\, \left(r_{s*}\, + \, \sigma \right)}{l^2} \right) \, , \label{GII_conformal}
\end{equation}
\vspace{0.1cm}

while the induced metric, its determinant and its inverse (Eqs.(\ref{InducedMatrix_LO_d=3}, \ref{InducedDet_LO}, \ref{InducedMatrix_LO_d=3_INVERSE})) are given by

\begin{equation}
g_{a b} \, \equiv \, g_{a b}\big\rvert_{X^\mu_0}\, = \,
\begin{bmatrix}
    G_{tt} & 0 \\
   0 & r'^2 \, G_{rr}
  \end{bmatrix} \, = \,
  \begin{bmatrix}
   - \frac{r_H^2}{l^2}\, \csch^2 \left( \frac{r_H\, \left(r_{s*}\, + \, \sigma \right)}{l^2} \right) & 0 \\
   0 & \frac{r_H^2}{l^2}\, \csch^2 \left( \frac{r_H\, \left(r_{s*}\, + \, \sigma \right)}{l^2} \right)
  \end{bmatrix}  \,, \label{g_ab_conformal}
\end{equation}
\vspace{0.1cm}

\begin{equation}
  g \, := \, \text{det}\, \big(g_{a b } \big) \, \equiv \,  \text{det}\, \big(g_{a b}\rvert_{X^\mu_0} \big) \,
  = \, - \frac{r_H^4}{l^4}\, \csch^4 \left( \frac{r_H\, \left(r_{s*}\, + \, \sigma \right)}{l^2} \right) \,, \label{det_g_ab_conformal}
\end{equation}
\vspace{0.1cm}

\begin{equation}
g^{a b} \, \equiv g^{a b}\big\rvert_{X^\mu_0}\, = \,
  \frac{1}{\det \big(g_{a b}\rvert_{X^\mu_0} \big) }\,\begin{bmatrix}
    r'^2\, G_{rr} & 0 \\
   0 & G_{tt}
  \end{bmatrix} \, = \,
  \begin{bmatrix}
    - \frac{l^2}{r_H^2}\, \sinh^2 \left( \frac{r_H\, \left(r_{s*}\, + \, \sigma \right)}{l^2} \right)  & 0 \\
   0 &  \frac{l^2}{r_H^2}\, \sinh^2 \left( \frac{r_H\, \left(r_{s*}\, + \, \sigma \right)}{l^2} \right)
  \end{bmatrix} \,, \label{g^ab_conformal}
\end{equation}
\vspace{0.1cm}

in the preferred isothermal set of coordinates ($t, r_*$). In order to fix the normalization constant $A_\omega$ canonical commutation relations between $\hat{X}(t,\sigma)$ and $\hat{P}^t(t,\sigma)$ are enforced:

\begin{flalign}
& \begin{aligned}
    &  \left[\hat{X}(t,\sigma), \, n_t \,\hat{P}^t(t,\sigma^\prime)\right]_\Sigma \, = \, i \, \delta(\sigma, \sigma^\prime) \, = \, i  \frac{\delta(\sigma - \sigma^\prime)}{\sqrt{\Tilde{g}}\rvert_\Sigma} \, ,  \\[0.2cm]
    & \left[\hat{X}(t,\sigma), \hat{X}(t,\sigma^\prime) \right]_\Sigma \, = \, 0 \, = \, \left[n_t \,\hat{P}^t(t,\sigma), \,  n_t \,\hat{P}^t(t,\sigma^\prime) \right]_\Sigma  \, , \label{commutation_relations}
\end{aligned} &
\end{flalign}
\vspace{0.1cm}

where $\Sigma$ is a Cauchy hypersurface in the $x^\mu = (t,r)^\mu$ part of spacetime that is chosen to be a constant time surface\footnote{Giving initial conditions on this hypersurface determines the future (and past) evolution uniquely.},
$\Tilde{g}$ is the induced metric on $\Sigma$, and $n_\mu$ is the future pointing normal to $\Sigma$ (where $n_\mu = \delta_{\mu t}/ \sqrt{-\Tilde{g}_{tt}}$).\\

Additionally, canonical creation and annihilation commutation relations on the Fourier coefficient operators ($\hat{a}_\omega, \hat{a}^{\dagger}_\omega$) are enforced:

\begin{flalign}
& \begin{aligned}
    &  \Big[\hat{a}_\omega, \, \hat{a}_{\omega^\prime}^{\dagger} \Big]_\Sigma \, = \, 2 \pi \delta(\omega - \omega^\prime) \, ,  \\[0.2cm]
    & \Big[\hat{a}_\omega, \, \hat{a}_{\omega^\prime} \Big]_\Sigma \, = \, 0 \, = \, \Big[\hat{a}_\omega^{\dagger} , \, \hat{a}_{\omega^\prime}^{\dagger} \Big]_\Sigma  \, . \label{FC_commutation_relations}
\end{aligned} &
\end{flalign}
\vspace{0.1cm}

Demanding consistency between the sets of commutation relations Eq.(\ref{commutation_relations}) and Eq.(\ref{FC_commutation_relations}), requires the normalization constant to be defined as

\begin{equation}
    A_\omega \, := \, \frac{l}{r_H} \, \sqrt{\frac{\pi \alpha'}{\omega}} \,
    = \, \frac{\beta}{2 \, \sqrt{\pi \omega} \, \lambda^{1/4}} \, , \label{Aw_normalization}
\end{equation}
\vspace{0.1cm}

where the second equality follows from using the definition of the AdS radius of curvature $l$ (Eq.(\ref{alpha_def})), and the relation $l  = \beta r_H^2 / 2 \pi$ from Eq.(\ref{HawkingTemp}). The derivation of Eq.(\ref{Aw_normalization}) is the subject of appendix (\ref{appCommutationRels}).\\

With the constants $B_\omega$ and $A_\omega$ (Eqs.(\ref{B_w_TC}, \ref{Aw_normalization}) respectively) found, the general solution for transverse equations of motion (Eq.(\ref{gen_solution_linear_sup_X})) is completely determined.
The displacement of the boundary endpoint as the test string undergoes these transverse fluctuations can now be calculated. The position of the endpoint of the test string on the boundary is defined by the operator

\begin{equation}
    \hat{X}_{\text{end}}(t) \, := \, \hat{X}(t, \, \sigma_f) \, . \label{Boundary_Endpoint_Operator}
\end{equation}
\vspace{0.1cm}

At the Hawking temperature, the transverse fluctuations on the string are excited. Assume -- as was done by de Boer \textit{et al.} \cite{Boer_2009} -- that these excitations are purely thermal and are therefore described by the Bose-Einstein distribution
\begin{equation}
\langle \hat{a}_{\omega}^{\dagger} \hat{a}_{\omega^\prime} \rangle \,
= \, \frac{2 \pi \, \delta(\omega - \omega^\prime)}{e^{\beta \omega} - 1} \, .\label{Bose_Einstein_Distribution}
\end{equation}
\vspace{0.1cm}

The Bose-Einstein distribution applies when (i) quantum effects are important, (ii) particles are indistinguishable, and (iii) are bosons (particles which do not obey the Pauli exclusion principle). The string boundary endpoint's mean-squared transverse displacement, $s^2(t)$, is defined as\footnote{The mean-squared transverse displacement, $s^2(t)$, is a gauge and coordinate independent quantity, i.e. it doesn't matter which coordinate system $s^2(t)$ is calculated in. Note that the position of the horizon is still dependent on the choice of coordinates (as discussed in subsection (\ref{subsubsecLQinAdS3}) when motivating for the use of ($t, r_*$) coordinates).}

\begin{equation}
s^2(t) \, := \, \langle  \normord{\big(\hat{X}_{\text{end}}(t) - \hat{X}_{\text{end}}(0) \big)^2}  \rangle \, = \,\langle \normord{\hat{X}^2_{\text{end}}(t)} \rangle \, + \,
\langle \normord{\hat{X}^2_{\text{end}}(0)} \rangle \, - \, 2 \langle \normord{\hat{X}_{\text{end}}(t) \hat{X}_{\text{end}}(0)}  \rangle \, . \label{displacement_def}
\end{equation}
\vspace{0.1cm}

In order to determine $s^2(t)$, begin by calculating the expectation value of the position of the boundary endpoint at two different times. Using Eq.(\ref{X_operator}),

\begin{flalign}
& \begin{aligned}
     & \langle  \normord{\hat{X}_{\text{end}}(t_1) \hat{X}_{\text{end}}(t_2)}  \rangle \\[0.2cm]
     &  =  \frac{\beta^2}{4  \pi   \sqrt{\lambda}}   \int_0^\infty \frac{d \omega \, d \omega^\prime}{(2 \pi)^2}  \frac{1}{\sqrt{\omega  \omega^\prime}} \Big[f_\omega (\sigma_f) f^{\,*}_{\omega^\prime} (\sigma_f)  e^{- i \omega t_1 + i {\omega^\prime} t_2} \langle \normord{ \hat{a}_\omega \hat{a}^{\,\dagger}_{\omega^\prime}} \rangle + f^{\,*}_\omega (\sigma_f) f_{\omega^\prime} (\sigma_f) e^{i \omega t_1 - i {\omega^\prime} t_2} \langle \normord{ \hat{a}^{\,\dagger}_\omega \hat{a}_{\omega^\prime}} \rangle \Big]\\[0.2cm]
     &  =  \frac{\beta^2}{4  \pi   \sqrt{\lambda}}  \int_0^\infty \frac{d \omega \, d \omega^\prime}{2 \pi} \frac{1}{\sqrt{\omega \, \omega^\prime}} \Big[\delta( \omega^\prime - \omega) f_\omega (\sigma_f) f^{\,*}_{\omega^\prime} (\sigma_f)  \frac{e^{- i \omega t_1 + i {\omega^\prime} t_2}}{e^{\beta \omega^\prime} - 1}  + \delta(\omega - \omega^\prime) f^{\,*}_\omega (\sigma_f) f_{\omega^\prime} (\sigma_f) \frac{ e^{i \omega t_1 - i {\omega^\prime} t_2}}{e^{\beta \omega} - 1} \Big]\\[0.2cm]
     &  =  \frac{\beta^2}{4  \pi   \sqrt{\lambda}}  \int_0^\infty \frac{d \omega}{2 \pi}  \frac{1}{\omega } \frac{1}{e^{\beta \omega} - 1} \Big[f_\omega (\sigma_f) f^{\,*}_{\omega} (\sigma_f)  e^{- i \omega (t_1 - t_2)} + f^{\,*}_\omega (\sigma_f) f_{\omega} (\sigma_f) e^{ i \omega (t_1 - t_2)}  \Big]
      \, , \label{Xt1_Xt2_expectation}
\end{aligned} &
\end{flalign}

where $\normord{ \hat{a}^{\,\dagger}_\omega \hat{a}_{\omega^\prime}} \, = \, \normord{ \hat{a}_{\omega^\prime} \hat{a}^{\,\dagger}_\omega} \,
= \, \hat{a}^{\,\dagger}_\omega \hat{a}_{\omega^\prime}$ is the normal ordering operator, and the definition of the Bose-Einstein distribution (Eq.(\ref{Bose_Einstein_Distribution})) is used in the third equality.
The final line follows  from the property of the Dirac delta function $\int_{0}^\infty dk  \, e^{- k x} \delta(k - a)  =  e^{- a x}$. Using the complex conjugate property $\text{Re}(z) \, =\, (z + \bar{z})/2$, Eq.(\ref{Xt1_Xt2_expectation}) simplifies to
\begin{equation}
   \langle  \normord{\hat{X}_{\text{end}}(t_1) \hat{X}_{\text{end}}(t_2)}  \rangle  =
   \frac{\beta^2}{4 \pi^2  \sqrt{\lambda}} \int_0^\infty \frac{d \omega}{\omega}  \frac{1}{e^{\beta \omega} - 1}  \text{Re}\left(f_\omega (\sigma_f) f^{\,*}_{\omega} (\sigma_f)  e^{- i \omega (t_1 - t_2)}\right) \, .
\end{equation}
\vspace{0.1cm}

The normal ordering operator is necessary to remove the logarithmic ultraviolet divergences. Had it not been used, extra terms would arise in the calculation of Eq.(\ref{Xt1_Xt2_expectation}). Specifically,

\begin{flalign}
& \begin{aligned}
     & \langle  \hat{X}_{\text{end}}(t_1) \hat{X}_{\text{end}}(t_2) \rangle \\[0.2cm]
      &  =  \frac{\beta^2}{4  \pi   \sqrt{\lambda}}   \int_0^\infty \frac{d \omega \, d \omega^\prime}{(2 \pi)^2}  \frac{1}{\sqrt{\omega  \omega^\prime}} \Big[f_\omega (\sigma_f) f^{\,*}_{\omega^\prime} (\sigma_f)  e^{- i \omega t_1 + i {\omega^\prime} t_2} \left(\langle  \hat{a}^{\,\dagger}_{\omega^\prime} \hat{a}_\omega \rangle + 2 \pi \, \delta(\omega - \omega^\prime) \right)\\[0.2cm]
      & \,\,\,\,\,\,+ f^{\,*}_\omega (\sigma_f) f_{\omega^\prime} (\sigma_f) e^{i \omega t_1 - i {\omega^\prime} t_2} \langle \hat{a}^{\,\dagger}_\omega \hat{a}_{\omega^\prime} \rangle \Big]\\[0.2cm]
     &  =  \frac{\beta^2}{4  \pi   \sqrt{\lambda}}  \int_0^\infty \frac{d \omega \, d \omega^\prime}{2 \pi} \frac{1}{\sqrt{\omega \, \omega^\prime}} \bigg[\delta( \omega^\prime - \omega) f_\omega (\sigma_f) f^{\,*}_{\omega^\prime} (\sigma_f)  \frac{e^{- i \omega t_1 + i {\omega^\prime} t_2}}{e^{\beta \omega^\prime} - 1} +   \delta(\omega - \omega^\prime)f_\omega (\sigma_f) f^{\,*}_{\omega^\prime} (\sigma_f)  e^{- i \omega t_1 + i {\omega^\prime} t_2}\\[0.2cm]
     & \,\,\,\,\,\,+ \delta(\omega - \omega^\prime) f^{\,*}_\omega (\sigma_f) f_{\omega^\prime} (\sigma_f) \frac{ e^{i \omega t_1 - i {\omega^\prime} t_2}}{e^{\beta \omega} - 1} \bigg] \\[0.2cm]
     &= \frac{\beta^2}{4 \pi^2  \sqrt{\lambda}} \int_0^\infty \frac{d \omega}{\omega}  \left[\frac{1}{e^{\beta \omega} - 1}  \text{Re}\left(f_\omega (\sigma_f) f^{\,*}_{\omega} (\sigma_f)  e^{- i \omega (t_1 - t_2)}\right) +  \frac{1}{2} f_\omega (\sigma_f) f^{\,*}_{\omega} (\sigma_f)  e^{- i \omega (t_1 - t_2)} \right]
      \, , \label{Xt1_Xt2_expectation_no}
\end{aligned} &
\end{flalign}

where the relation $\langle \hat{a}_\omega \hat{a}^{\,\dagger}_{\omega^\prime} \rangle
= \langle  \hat{a}^{\,\dagger}_{\omega^\prime} \hat{a}_\omega \rangle + 2 \pi \, \delta(\omega - \omega^\prime)$ results from the commutation relation Eq.(\ref{FC_commutation_relations}).
Following \cite{Boer_2009}, the term $ \frac{1}{2} f_\omega (\sigma_f) f^{\,*}_{\omega} (\sigma_f)  e^{- i \omega (t_1 - t_2)}$ is interpreted as a logarithmic UV divergence. Since this divergence stems from the zero-point energy and exists at zero temperature, it can be regularised by instituting normal ordering of the oscillators $a_\omega, \hat{a}^{\,\dagger}_\omega$ (as was done in Eq.(\ref{Xt1_Xt2_expectation})).
At specific times, the regularized correlator (Eq.(\ref{Xt1_Xt2_expectation})) becomes

\begin{equation}
\langle  \normord{\hat{X}_{\text{end}}(t) \hat{X}_{\text{end}}(0)}  \rangle  \,
= \,\frac{\beta^2}{4 \, \pi^2  \, \sqrt{\lambda}} \, \int_0^\infty \frac{d \omega}{\omega} \, \frac{1}{e^{\beta \omega} - 1} \,
\text{Re}\Big(f_\omega (\sigma_f) f^{\,*}_{\omega} (\sigma_f)  e^{- i \omega t}\Big) \, , \label{Xt_X0_expectation}
\end{equation}

\begin{equation}
\langle  \normord{\hat{X}_{\text{end}}(0) \hat{X}_{\text{end}}(0)}  \rangle \,=\,\langle  \normord{\hat{X}_{\text{end}}(t) \hat{X}_{\text{end}}(t)}  \rangle \,
= \,\frac{\beta^2}{4 \, \pi^2  \, \sqrt{\lambda}} \, \int_0^\infty \frac{d \omega}{\omega} \, \frac{1}{e^{\beta \omega} - 1} \,
\text{Re}\Big(f_\omega (\sigma_f) f^{\,*}_{\omega} (\sigma_f)\Big) \, . \label{X0_X0_Xt_Xt_expectation}
\end{equation}
\vspace{0.1cm}

Inputting Eqs.(\ref{Xt_X0_expectation}, \ref{X0_X0_Xt_Xt_expectation}) into Eq.(\ref{displacement_def}), the string boundary endpoint's mean-squared transverse displacement can be calculated
\begin{flalign}
& \begin{aligned}
     s^2(t) \, &  =  \frac{\beta^2}{4  \pi^2   \sqrt{\lambda}}  \int_0^\infty \frac{d \omega}{\omega}  \frac{1}{e^{\beta \omega} - 1}  \text{Re}\Big(2 f_\omega (\sigma_f) f^{\,*}_{\omega} (\sigma_f) - 2 f_\omega (\sigma_f) f^{\,*}_{\omega} (\sigma_f)  e^{- i \omega t} \Big)\\[0.2cm]
     & = \frac{\beta^2}{4  \pi^2  \sqrt{\lambda}}  \int_0^\infty \frac{d \omega}{\omega}  \frac{1}{e^{\beta \omega} - 1}  \left[f_\omega (\sigma_f) f^{\,*}_{\omega} (\sigma_f) \left(2 -e^{- i \omega t}- e^{i \omega t} \right)\right] \, , \label{st_squared}
\end{aligned} &
\end{flalign}

where in the second line the complex conjugate property $\text{Re}(z) \, =\, (z + \bar{z})/2$ is again used.

\subsubsection{The Limiting Cases of \texorpdfstring{$s^2(t)$}{s2t} } \label{subsubsecHQ_LimitCases}

In this subsection the behaviour of $s^2(t)$, Eq.(\ref{st_squared}), in the asymptotic early and late time limits is considered. Inputting\footnote{Mathematica is used to simplify the algebra (see Mathematica Notebook [b]: \texttt{BrownianMotion.nb}).} Eqs.(\ref{TC_nu_fw_r_ODE_Solution}, \ref{gen_solution_sum_modes}, \ref{B_w_TC}) to calculate $f_\omega (\sigma_f) f_\omega^{\,*} (\sigma_f)$, yields


\begin{equation}
   s^2(t) \,= \, \frac{4 \beta^2}{\pi^2 \sqrt{\lambda}} \, \int_0^\infty \,  \frac{d \nu}{\nu} \,\frac{1+\nu^2}{1+\Tilde{r}_0^2 \nu^2} \, \frac{\sin^2{\frac{\pi t \nu}{\beta}}}{ e^{2 \pi \nu} -1} \, , \label{st_squared2}
\end{equation}
\vspace{0.1cm}

where the change of variables was performed using Eqs.(\ref{HawkingTemp}, \ref{dim_angular_f_def}). The integral in Eq.(\ref{st_squared2}) is tricky to analytically evaluate. Following the insights of de Boer \textit{et al.} \cite{Boer_2009}, the integral can be broken into two parts,

\begin{flalign}
& \begin{aligned}
    s^2(t) \, &= \frac{4 \beta^2}{\pi^2 \sqrt{\lambda}} \left[4 \frac{\Tilde{r}_0^2 -1 }{\Tilde{r}_0^2} \left( \int_0^\infty \,  \frac{d \nu}{\nu}\, \frac{1}{1+\Tilde{r}_0^2 \nu^2} \, \frac{\sin^2{\frac{\pi \nu t}{\beta}}}{ e^{2 \pi \nu} -1} \right)\, + \, 4 \frac{1 }{\Tilde{r}_0^2} \left(\int_0^\infty \,  \frac{d \nu}{\nu}\, \frac{\sin^2{\frac{\pi \nu t}{\beta}}}{ e^{2 \pi \nu} -1}\right) \right]\\[0.25cm]
     & = \frac{\beta^2}{\pi^2 \sqrt{\lambda}} \left[\frac{\Tilde{r}_0^2 -1 }{\Tilde{r}_0^2}\, I_1 \, + \, \frac{1 }{\Tilde{r}_0^2}\, I_2\right]\,, \label{st_squared_I1I2}
\end{aligned} &
\end{flalign}
\vspace{0.1cm}

where the integrals $I_1$ and $I_2$ have been defined as,

\begin{equation}
    I_1 \,:= \, 4 \int_0^\infty \,  \frac{d \nu}{\nu}\, \frac{1}{1+\Tilde{r}_0^2 \nu^2} \, \frac{\sin^2{\frac{\pi \nu t}{\beta}}}{ e^{2 \pi \nu} -1} \, = \,  4  \int_0^\infty \,  \frac{d x}{x}\, \frac{1}{1+ a^2 x^2} \, \frac{\sin^2{\frac{k x}{2}}}{ e^{x} -1} \, , \label{I1}
\end{equation}

and

\begin{equation}
    I_2 \,:= \,4  \int_0^\infty \,  \frac{d \nu}{\nu}\, \frac{\sin^2{\frac{\pi \nu t}{\beta}}}{ e^{2 \pi \nu} -1} \, = \,  4 \int_0^\infty \,  \frac{d x}{x}\, \frac{\sin^2{\frac{k x}{2}}}{ e^{x} -1} \, . \label{I2}
\end{equation}
\vspace{0.1cm}

The change of variables in Eqs.(\ref{I1}, \ref{I2}) is made by defining the new variables

\begin{equation}
x := 2 \pi \nu \, , \,\,\,\,\,\,\,\,\, a:=\frac{\Tilde{r}_0}{2 \pi} \, , \,\,\,\,\,\,\,\,\, k:=\frac{t}{\beta} \, .\label{Var_axk}
\end{equation}
\vspace{0.1cm}
The integral $I_1$ is solved by deforming the contour on the complex $x$ plane. The solution is given by \cite{Boer_2009},

\begin{flalign}
& \begin{aligned}
    I_1  =  & \, \frac{1}{2} \left(\psi \left(1+\frac{1}{2 \pi  a}\right)+\psi \left(1-\frac{1}{2 \pi  a}\right)\right) +\frac{1}{2} \left(e^{\frac{k}{a}} \text{Ei}\left(-\frac{k}{a}\right)+e^{-\frac{k}{a}} \text{Ei}\left(\frac{k}{a}\right)\right) -\frac{\pi}{2}\left(1-e^{-\frac{\left| k\right| }{a}}\right) \cot \left(\frac{1}{2 a}\right)\\[0.2cm]
    &+\frac{e^{-2 \pi  \left| k\right| }}{2} \left(\frac{\, _2F_1\left(1,1+\frac{1}{2 \pi  a};2+\frac{1}{2 \pi  a};e^{-2 \pi  \left| k\right| }\right)}{\frac{1}{2 \pi  a}+1}+\frac{\, _2F_1\left(1,1-\frac{1}{2 \pi  a};2-\frac{1}{2 \pi  a};e^{-2 \pi  \left| k\right| }\right)}{1-\frac{1}{2 \pi  a}}\right)\\[0.2cm]
    &+\ln\left(\frac{2 a \sinh (\pi  k)}{k}\right) \,, \label{I1Sol}
\end{aligned} &
\end{flalign}
where $\text{Ei}(z)$ is the exponential integral\footnote{For real, non-zero values of $z$, the exponential integral is defined as $\text{Ei}(z) = - \int^{\infty}_{-z} dt \, (e^{-t}/t)$.}; $_2F_1(a,\, b;\,c;\, z)$ is the Gaussian hypergeometric function\footnote{\label{footnotehypergeometric}The hypergeometric function is a special function defined for $|z| <1$ by the hypergeometric series, $_2F_1(a,\, b;\,c;\, z) = \sum^{\infty}_{k=0} \frac{(a)_k (b)_k}{(c)_k}\frac{z^k}{k!}$.}; and $\psi(z)$ is the digamma function\footnote{The digamma function is defined in terms of derivatives of the gamma function: $\psi(z) := \frac{1}{\Gamma(z)}\frac{d \Gamma(z) }{dz}$. The function is meromorphic, and defined on the complex numbers $\mathbb{C}$.}. The integral $I_2$ can be performed analytically,

\begin{flalign}
& \begin{aligned}
    I_2 \, &=\,4 \, \sum_{n=1}^{\infty} \, \int_0^\infty \,  \frac{d x}{x} \, \sin^2{\left(\frac{kx}{2}\right)}\, e^{-n x}\\[0.2cm]
    &= \, \ln{\left(\prod_{n=1}^{\infty} \, \bigg(1 + \frac{k^2}{n^2} \bigg)\right)}\\[0.2cm]
    &= \, \ln{\left(\frac{\sinh\left(\pi k \right)}{\pi k}  \right)}\,, \label{I2Sol}
\end{aligned} &
\end{flalign}
\vspace{0.1cm}

where, in the first line, the geometric series $1/(1-x) \, = \, \sum_{n=0}^\infty x^n$ is used; and in the second line the integral was performed using Mathematica \cite{Mathematica}.\\

From Eq.(\ref{I2Sol}) it appears that $\beta = t/k$ naturally defines a cross-over time scale between the dynamics of early and late times\footnote{\label{footnoteCrossOverTime}More specifically, the cross-over time (defined in Eq.(\ref{tc})) is given by $t_c \, \sim \, \beta\,\Tilde{r}_0 \, \sim \, (\alpha' \,m_q)/(l^2\,T^2)$,
where Eqs.(\ref{alpha_def}, \ref{appCurveVirtualFINAL2}) have been used in the second approximation.}.
The integrals $I_1$ and $I_2$ can be examined in the limits $(t \ll \beta)$ and $(t \gg \beta)$ to yield the early and late time behaviour respectively. Since the probe particle in the boundary theory is an on-mass-shell heavy quark, the assumption $\Tilde{r}_0 \gg 1$ (equivalently $a \gg 1$) can be made. This is essentially the statement that the large mass of the external particle translates to the distance in the radial direction between the test string's boundary endpoint and the stretched horizon being large\footnote{The assumption $\Tilde{r}_0 \rightarrow \infty$ would correspond to an infinitely massive quark, such as the probe particle considered in section (\ref{secDragForce}).}.\\

\textbf{(I) Asymptotic Early Time Dynamics}\\

In the early time limit $t\ll\beta$ (equivalently $k \ll 1 \ll a$), the integrals $I_1$ and $I_2$ can be simplified. Defining a temporary variable $f:=1/a$, and series expanding the solution to $I_1$ (Eq.(\ref{I1Sol})) around $f=0$ yields

\begin{flalign}
& \begin{aligned}
    I_1 \, &= \, \left(-\ln\left(1-e^{-2 k \pi }\right)\,+\,\ln(f)+\ln\left(\frac{2 \sinh (k \pi )}{f}\right)-k \pi \right)\,+\,\frac{1}{2} k^2 \pi  f\,+ \,\mathcal{O}\left(f^2\right) \\[0.2cm]
    &= \, \frac{k^2 \pi}{2 a} \, + \, \mathcal{O}\left(\frac{1}{a^2}\right)\, ,\label{I1_EarlyTime}
\end{aligned} &
\end{flalign}
\vspace{0.1cm}

where, in the second line, the expression is rewritten in terms of the variable $a$ and appropriately truncated at leading order.  For the $I_2$ integral, notice there is no $a$-dependence -- hence a series expansion in $1/a$ yields $I_2 \, = \, \mathcal{O}\left(a^0\right)$.  Therefore, in the early time limit, $s^2(t)$ (Eq.(\ref{st_squared_I1I2})) becomes

\begin{equation}
s^2(t) \,\rvert_{(t \ll \beta)} \, \frac{\beta^2}{\pi^2 \sqrt{\lambda}} \left[\frac{\Tilde{r}_0^2 -1 }{\Tilde{r}_0^2}\, \left(\frac{\pi^2 t^2}{\beta^2 \Tilde{r}_0} \, + \, \mathcal{O}\left(\frac{1}{\Tilde{r}_0^2}\right) \right)\, + \, \frac{1 }{\Tilde{r}_0^2}\,  \mathcal{O}\left(\Tilde{r}_0\right)^0\right] \, = \,  \frac{t^2}{ \Tilde{r}_0  \sqrt{\lambda}} \,+ \, \mathcal{O}\left(\frac{1}{\Tilde{r}_0^2}\right)  \, , \label{st_squaredHQ_Early}
\end{equation}
\vspace{0.1cm}

where the variables have been changed using Eq.(\ref{Var_axk}). In terms of the relevant field theoretic quantities, Eq.(\ref{st_squaredHQ_Early}) can be converted using Eq.(\ref{appCurveVirtualFINAL2}). Specifically, $s^2(t) = (T \, t^2)/ m_q$, where $m_q$ is the mass of the probe quark and $T$ is the temperature of the thermal plasma. Since $ s^2(t) \sim t^2$, the early time dynamics exhibit ballistic behaviour (see Eq.(\ref{DisplacementSqLimit1}))\footnote{The result for $s^2(t)$ in the early time limit given in Eq.(\ref{st_squaredHQ_Early}) agrees exactly with de Boer \textit{et al.} \cite{Boer_2009}, Eq.(3.6) (after recognising $\Tilde{r}_0 = \rho_c$, and making use of Eq.(\ref{alpha_def})).}.\\

\textbf{(II) Asymptotic Late Time Dynamics}\\

At asymptotically late times, $t \gg \beta$ (equivalently $k \gg a \gg 1$), the integrals $I_1$ and $I_2$ have the same behaviour. To prove this, take the integral $I_1$ (Eq.(\ref{I1})) and make the change of variables $x := \frac{x'}{k}$,

\begin{equation}
    I_1 \,= \,4 \, \int_0^\infty \,  \frac{d x'}{x'}\, \frac{1}{1+\frac{a^2 x'}{k^2}} \, \frac{\sin^2{\frac{x'}{2}}}{ e^{\frac{x'}{k}} -1}\,\, \stackrel{(\beta \ll t)}{\longrightarrow}\,\, 4 \, \int_0^\infty \,  \frac{d x'}{x'} \, \frac{\sin^2{\frac{x'}{2}}}{ e^{\frac{x'}{k}} -1} \, \equiv \, I_2 \, ,\label{I1_Late_Times}
\end{equation}
\vspace{0.1cm}

where it is noticed that the term $\frac{a^2 x'}{k^2} \ll 1$ can be neglected in the late time limit. Expanding the solution to $I_2$ (Eq.(\ref{I2Sol})), yields

\begin{flalign}
& \begin{aligned}
    I_1\rvert_{\beta \ll t} \, \equiv \, I_2 \, & =\, \ln{\big(\sinh{\pi k}\big)} \, - \, \ln{\big(\pi k \big)}\\[0.2cm]
    &= \, \ln{\big(e^{\pi k}\big)} \, - \, \ln{\big(2\big)}\, - \, \ln{\big(\pi k\big)}\, \\[0.2cm]
    &= \pi k \, + \, \mathcal{O}(\ln{k}) \,, \label{I2_Late_Times}
\end{aligned} &
\end{flalign}
\vspace{0.1cm}

where the second line follows from trigonometric identity $\sinh(x)=(e^x - e^{-x})/2$, and in the third line the term $e^{-\pi k}/2 \rightarrow 0$ since $k \gg 1$ at late times.\\

Therefore, in the late time limit, $s^2(t)$ (Eq.(\ref{st_squared_I1I2})) becomes

\begin{equation}
s^2(t) \,\rvert_{(t \ll \beta)} \, \frac{\beta^2}{\pi^2 \sqrt{\lambda}} \left[\frac{\Tilde{r}_0^2 -1 }{\Tilde{r}_0^2}\, \left(\frac{\pi t}{\beta} \, + \, \mathcal{O}\left(\ln{\frac{t}{\beta}}\right) \right)\, + \, \frac{1 }{\Tilde{r}_0^2}\,  \left(\frac{\pi t}{\beta} \, + \, \mathcal{O}\left(\ln{\frac{t}{\beta}}\right) \right)\right] \, = \,  \frac{\beta \,t}{ \pi  \sqrt{\lambda}} \,+ \, \mathcal{O}\left(\ln{\frac{t}{\beta}}\right)  \, , \label{st_squaredHQ_Late}
\end{equation}
\vspace{0.1cm}

where the variables have been changed using Eq.(\ref{Var_axk}). Since $s^2(t) \sim t$, the late time dynamics exhibit diffusive behaviour\footnote{The result for $s^2(t)$ in the late time limit given in Eq.(\ref{st_squaredHQ_Late}) agrees exactly with de Boer \textit{et al.} \cite{Boer_2009}, Eq.(3.6) (where Eqs.(\ref{HawkingTemp}, \ref{alpha_def}) have been used to show the equivalence).}. Concretely, at late times $s^2(t) \, =\, 2 \,D \, t$, where $D$ is the diffusion coefficient (Eq.(\ref{DisplacementSqLimit2})). Comparing this to Eq.(\ref{st_squaredHQ_Late}), the diffusion coefficient in AdS$_3$-Schwarzschild is extricated,

\begin{equation}
D \, = \, \frac{\beta}{2 \, \pi  \sqrt{\lambda}}\, . \label{DifCoeff_HQ}
\end{equation}

As will be shown in section (\ref{secDragForce}), the drag, or friction, experienced by a slowly moving heavy quark in AdS$_3$ --- which is identical to the dissipation of momentum down the string into the black hole horizon --- is given by
\begin{equation}
\frac{dp}{dt} \, = \, -\gamma \, p \, , \,\,\,\,\, \text{and}  \,\,\,\,\,\gamma = \frac{2\,\pi \sqrt{\lambda} T^2}{m}
\end{equation}

One can see that the fluctuation-dissipation theorem, which connects the drag and diffusion of a probe particle in a thermal medium \cite{Boer_2009}, is exactly respected:
\begin{align}
    D = \frac{T}{\gamma m} \, .
\end{align}

\subsection{Generalising to \texorpdfstring{AdS$_d$-Schwarzschild}{AdSd-Schwarzschild}} \label{subsubsecHQ_GenAdSd}

For the off-mass-shell light quark case -- as will be seen in subsection (\ref{subsubsecLQ_GenAdSd}) -- generalising the diffusion coefficient to AdS$_d$-Schwarzschild\footnote{Remember that a general $d$ dimensions in the bulk theory corresponds to a ($d-1$)-dimensional thermal plasma in the boundary theory.} relies on the observation that the behaviour of an arbitrary virtuality light quark at asymptotically late times is encoded in the small virtuality light quark case. Since $s^2(t)$ in the latter case can easily be calculated for any dimensions $d\geq 3$ in AdS$_d$-Schwarzschild, the diffusion coefficient $D_{\text{LQ}}^{\text{AdS$_d$}}$ can be determined. Unfortunately, this method can not be applied to the on-mass-shell heavy quark case\footnote{To talk about the small virtuality/small mass limit of a massive, heavy probe quark, would be nonsensical.}. The avenue left is to determine, in AdS$_d$-Schwarzschild, the friction coefficient of the medium plasma and relate this -- via the Einstein-Sutherland equation (Eq.(\ref{DiffusionC})) -- to the diffusion coefficient. This is precisely what is done in section (\ref{secDragForce}), where the friction coefficient in AdS$_d$-Schwarzschild is found by first calculating the drag force on an infinitely massive, heavy quark moving through the thermal plasma with a constant velocity\footnote{Implies that the probe quark is under the influence of an external force.} in the bulk theory. As will be seen, the diffusion coefficient extracted from this method\footnote{These calculations predominantly follow the work of \cite{Gubser_2006, herzog2006energy}.}, agrees exactly with the the diffusion coefficient given in Eq.(\ref{DifCoeff_HQ}) for $d=3$.

\section{Analysing Light Quark Brownian Motion}\label{secLightQuark}

A light quark probe in a thermal medium on the boundary starts at $t=0$ as an off-mass-shell particle, radiates energy as it travels through the thermal medium and finally stops radiating as it becomes an on-mass-shell particle.
Dual to this light quark in the boundary theory, a test string in the bulk can be thought of as half the \textit{yo-yo string} in $\mathbb{R}^{1,1}$. The latter is a string which, at a time $t=0$, stretches a length of $2L$ along the x-axis; as time progresses the string shrinks till, at a time $t= L$, it is a point; then it begins to expand again and the process repeats. The yo-yo solution is a well-studied classical solution of the open string \cite{bardeen1976study}.
There is plenty of recent work \cite{Gubser2_2008, Chesler1_2009, Chesler_2009, Ficnar_2012, Ficnar_2013, Morad_2014} modelling gluons and light quarks as half the initial yo-yo string contracting to a point after a time $t$, in order to calculate energy loss
(see also \cite{Ficnar_2014, moerman2016semi} for other works using this string set-up). This idea of a \textit{falling string} is precisely what is used in order to study the off-mass-shell light quark in this article.
Such a string set-up is refered to, following the convention of \cite{moerman2016semi}, as the \textit{Limp Noodle}.\\

\begin{figure}[!htb]
\centering
\includegraphics[width=1\textwidth]{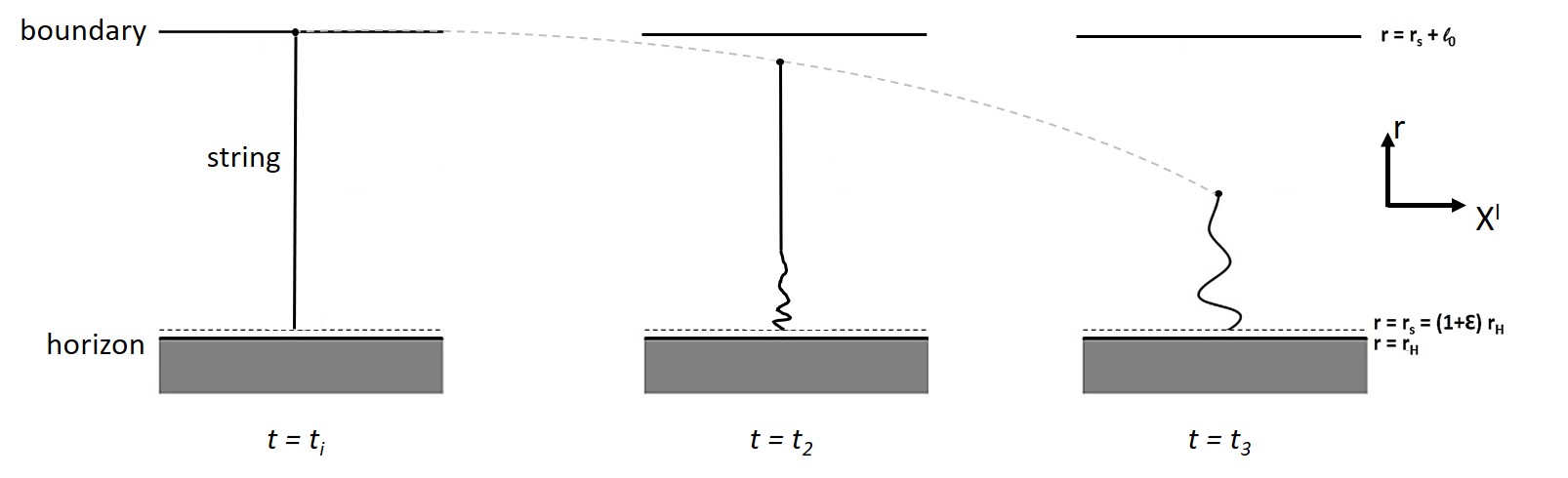}
\caption[Snapshots of a fundamental open string of length $\ell_0$ used as a probe in an AdS black hole background, falling as time increases from left to right]{\label{fig:stringdiagramLQ} A fundamental falling open string of length $\ell_0$ used as a probe in an AdS black hole background.
Initially, at $t=t_i$, the string starts at the boundary of the anti-de Sitter spacetime and hangs down to a \textit{stretched horizon} ($r_s \, = \, (1+ \epsilon)\,r_H$ where $0 < \epsilon \ll 1$) placed just above the Schwarzschild black hole horizon. As time evolves $t_i < t_2 < t_3$, the string endpoint at the boundary is released and the string collapses.
Its trajectory as it falls (grey, dashed line) is predicted by \cite{Gubser2_2008, Chesler1_2009, Chesler_2009, Ficnar_2012, Ficnar_2013, Morad_2014}.}
\end{figure}

If the test string is placed in an AdS-Schwarzschild background, the presence of the black hole results in thermal fluctuations in the transverse $X^I$ directions on the string. The resultant random movement of the string endpoint on the boundary corresponds to the light quark undergoing Brownian motion. This set-up is depicted in figure (\ref{fig:stringdiagramLQ}).\\

\vspace{0.1cm}

An initial static set-up of the Limp Noodle such that one of its endpoints is held fixed at the boundary for an asymptotically long time, allowing the string to be fully populated by thermal modes emanating from the black hole horizon, is then equivalent to the leading order static stretched string for the heavy quark described in subsection (\ref{subsecHQ_LOBehaviour}). The two situations diverge once the boundary endpoint of the string is released, and the endpoint starts to fall at the local speed of light. Information, however, is also restricted to travel at the local speed of light. Hence, all parts of the string below the falling boundary endpoint can not \textit{know} whether the boundary endpoint has remained fixed (as in the case of the heavy quark's test string) or been released to fall until the endpoint crashes through it. Intuitively, it is therefore expected that all parts of the string except the falling boundary endpoint are described by a solution (embedding functions) identical to Eq.(\ref{HQ_Sol_Flat}) in $\mathbb{R}^{1,1}$ and Eq.(\ref{HQ_Sol_AdS3}) in AdS$_3$-Schwarzschild, while the stretched endpoint falls at the local speed of light. This will be confirmed in the following subsection.

\subsection{Leading Order String Behaviour} \label{subsecLQ_LOBehaviour}

The basics of string theory discussed in subsection (\ref{subsecPolyakovEoM}) -- leading to the Virasoro constraints (Eq.(\ref{Virasoro})) and the string equations of motion (Eq.(\ref{StringEoM})) -- hold for the light quark's test string (Limp Noodle).
In order to solve Eqs.(\ref{Virasoro}, \ref{StringEoM}) (with respect to the boundary conditions Eqs.(\ref{BC_HQ_horizon}, \ref{BC_HQ_boundary}) and the relevant initial conditions) to yield the embedding functions requires the correct partitioning of the worldsheet parameter space. This partitioning was trivial in the heavy quark's test string case, since its leading order behaviour is static for all $t$.
It is however more complicated in the Limp Noodle case, considering that the boundary string endpoint is allowed to fall at the local speed of light for $t >0$. A method in which to partition the worldsheet in order to find the solution to the string equations of motion was given by Itzhak Bars in 1994 \cite{Bars_1994}, and expounded upon a year later, in 1995, by himself and a collaborator Jürgen Schulze \cite{Bars_1995}. The Bars \textit{et al.} method is summarised in the following subsection.

\subsubsection{Parameterising the Limp Noodle: The Bars \textit{et al.} method} \label{subsubsecBarsMethod}

The Bars \textit{et al.} method can be used to solve the classical motion of a relativistic string in any $(1+1)$-dimensional spacetime. Originally, the method was used by Bars \textit{et al.} to study the motion of folded strings in curved spacetime. However the Limp Noodle can be interpreted as half a folded (or yo-yo) string.
The application of the Bars \textit{et al.} method to parameterise the Limp Noodle and find the general solution to the string equations of motion was first proposed by Moerman \textit{et al.} in \cite{moerman2016semi}. The Bars \textit{et al.} method can be broken into the following steps:
\begin{enumerate}[(i)]
\item Define lattice-like patches on the worldsheet using the light-cone coordinates.

    \begin{figure}[!htb]
    \centering
    \includegraphics[width=0.4\textwidth]{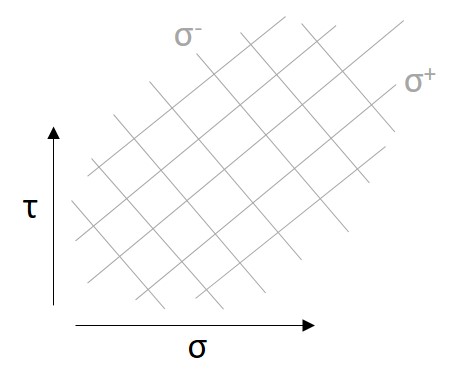}
    \caption[Light-cone coordinate lattice structure imposed on the worldsheet parameter space in order to determine the general string solution using the Bars \textit{et al.} method]{\label{fig:BarsLattice} Lattice structure imposed on the worldsheet parameter space using the light-cone coordinates $\sigma^+$ and $\sigma^-$. This structure is used in the Bars \textit{et al.} method \cite{Bars_1994, Bars_1995} in order to determine the general solution for a moving open string on a manifold.}
    \end{figure}

    \item Naive solutions to the string equations of motion come in four classes, referred to as: \textit{A, B, C,} and \textit{D}. They can be defined only on patches of the worldsheet, since the full solution needs to satisfy two conditions: (a) string solutions must be periodic in $\sigma$, and (b) the global time coordinate must be an increasing function of $\tau$ and $\sigma$ (in order to exclude anti-strings from appearing in the same solutions as strings). In the static gauge the latter condition is trivially satisfied. Hence to build the general solution from the solution classes \textit{A, B, C, D}, assign the pattern of solutions to ensure forward propagation and impose periodicity at a fixed value of $\tau$.

    \item Complete solutions are obtained by matching boundary conditions at the boundary of the patches.

    \item Define a transfer operation matrix to determine future evolution. The propagation of the solution into the future is performed by using this matrix to matching the boundary conditions while increasing $\tau$.

\end{enumerate}
\vspace{0.1cm}

The leading order dynamics of the Limp Noodle are constrained to the $(1+1)$-dimensional subspace of AdS$_d$-Schwarzschild described by the set of coordinates $(x^0,x^1)$, where $x^0$ denotes the temporal direction and $x^1$ denotes the radial direction. The metric of any $(1+1)$-dimensional subspace (where one of these dimensions is the temporal dimension) can be transformed into a conformally flat metric \cite{Bars_1994, Bars_1995}.
In order to perform this transformation the string field in the target spacetime is redefined by the coordinate reparametrization  $x^\mu(\tau, \sigma) = x^\mu(y^{\mu'}(\tau, \sigma))$. This yields the new spacetime metric in terms of isothermal coordinates $y^{\mu'}(\tau, \sigma)$, which can be related to the old spacetime metric by

\begin{equation}
G_{\mu \nu}(x) \, = \, \dfrac{\partial y^{\mu'}}{\partial x^\mu} \, \dfrac{\partial y^{\nu'}}{\partial x^\nu} \, G_{\mu'  \nu'}(y) \,, \label{STMetricTrans}
\end{equation}
\vspace{0.1cm}

where the isothermal coordinates in the static gauge $y^{\mu'}(t, \sigma)$, are given by

\begin{equation}
y^{0'} \, = \,\frac{1}{\sqrt{2}}\Big( x^0 \,+\, x^1 \Big) \,\,\,\,
\text{and} \,\,\,\, y^{1'} \, = \, \frac{1}{\sqrt{2}}\Big(x^0 \,-\, x^1 \Big)  \,. \label{IsothemalCo-ord}
\end{equation}
\vspace{0.1cm}

In this isothermal coordinate frame the metric is conformally flat

\begin{equation}
ds^2 \, = \, G_{\mu'  \nu'}(y)  \, dy^{\mu'} \, dy^{\nu'} \,, \label{ConMetric}
\end{equation}
\vspace{0.1cm}

and the new spacetime metric is found by inverting Eq.(\ref{STMetricTrans})

\begin{equation}
G_{\mu' \nu'}(y) \, = \, \dfrac{\partial x^\mu}{\partial y^{\mu'}} \, \dfrac{\partial x^\nu}{\partial y^{\nu'}} \, G_{\mu \nu}(x) \,
= \, \dfrac{\partial x^\mu}{\partial y^{\mu'}} \, \dfrac{\partial x^\nu}{\partial y^{\nu'}} \, G \, \eta_{\mu \nu} \,, \label{STMetricTransInverse}
\end{equation}
\vspace{0.1cm}

where the second equality follows from choosing the conformal gauge, and $G = G(y^0,y^1) \in \mathbb{R}^+$ is some non-zero coordinate dependent scalar function. The metric can be calculated component-wise. For example,

\begin{flalign}
& \begin{aligned}
  G_{0'0'}(y) \, &= \,G \left[-\dfrac{\partial x^0}{\partial y^{0'}}\,\dfrac{\partial x^0}{\partial y^{0'}} \, +\, \dfrac{\partial x^1}{\partial y^{0'}}\,\dfrac{\partial x^1}{\partial y^{0'}} \right] \, = \,G \left[-\Big(\frac{1}{\sqrt{2}}\Big)\,\Big(\frac{1}{\sqrt{2}}\Big) \, +\, \Big(\frac{1}{\sqrt{2}}\Big)\,\Big(\frac{1}{\sqrt{2}}\Big) \right] \, = \,0  \, , \\[0.2cm]
\end{aligned} &
\end{flalign}
\vspace{0.1cm}

where the definition of the isothermal coordinates Eq.(\ref{IsothemalCo-ord}) have been used. Similarly $G_{0'1'}(y) = -G$, $G_{1'0'}(y) = -G$, and $G_{1'1'}(y) = 0$. Hence $G_{\mu' \nu'}(y)$ has the explicit form

\begin{equation}
G_{\mu' \nu'}(y) \, =\, G\,
  \begin{bmatrix}
   0 & -1 \\
   -1 & 0
  \end{bmatrix} \,=\,  G \, \tilde{\eta}_{\mu' \nu'}\, , \label{NewSTMetric}
\end{equation}
\vspace{0.1cm}

where $\tilde{\eta}_{\mu \nu}$ is the light-cone metric, and $G := G(y^{0'},y^{1'}) \in \mathbb{R}^+$ is defined as some non-zero coordinate dependent scalar function. Using Eq.(\ref{NewSTMetric}), the metric Eq.(\ref{ConMetric}) becomes

\begin{equation}
ds^2 \, = \,-  G \big( dy^{0'} \, dy^{1'} \, + \, dy^{1'} \, dy^{0'}\big)\,. \label{ConMetric2}
\end{equation}
\vspace{0.1cm}

The string mapping functions $Y^{\mu'}(\sigma^+, \sigma^-)$ are given by

\begin{equation}
Y^{0'} \, = \,\frac{1}{\sqrt{2}}\Big( X^0 \,+\, X^1 \Big) \,\,\,\,
\text{and} \,\,\,\, Y^{1'} \, = \, \frac{1}{\sqrt{2}}\Big(X^0 \,-\, X^1 \Big)  \,. \label{IsoMapFunc}
\end{equation}
\vspace{0.1cm}

Note that since the coordinates are isothermal, $Y^0 \, = \, -Y_1$ and $Y^1 \, = \, - Y_0$. The canonically conjugate momentum densities Eq.(\ref{MomDensity}) can be rewritten in terms of the isothermal embedding functions, $Y^{\mu'}$. In the conformal gauge, the canonical momentum densities are given by

\begin{flalign}
& \begin{aligned}
   & \Pi_\mu^a(t,\sigma) \, = \, - \frac{1}{2 \pi \alpha'} \, \eta^{ab}\, G_{\mu \nu} \, {\partial_b} X^\nu  \\[.2cm]
    & \Rightarrow \Pi_\mu^{\pm}(t,\sigma) \, = \, \frac{1}{2 \pi \alpha'} \,  G \, \eta_{\mu \nu} \, {\partial_\mp} X^\nu  \, , \label{MomDensityLC}
\end{aligned} &
\end{flalign}
\vspace{0.1cm}

if the light-cone coordinate frame is chosen (where $\partial_{\mp} := \frac{\partial}{\partial \sigma^{\mp}}$). Specifically,

\begin{equation}
\Pi_0^{\pm}(t,\sigma) \,  = \, - \frac{1}{2 \pi \alpha'} \, G\, {\partial_\mp} X^0 \,\,\,\, \text{and} \,\,\,\, \Pi_1^{\pm}(t,\sigma) \,
= \,\frac{1}{2 \pi \alpha'} \, G\, {\partial_\mp} X^1  \,. \label{MomDensityLC_Components}
\end{equation}
\vspace{0.1cm}

Hence the conjugate momentum densities can be defined in terms of the string mapping functions $Y^{\mu'}(\sigma^+, \sigma^-)$

\begin{flalign}
& \begin{aligned}
   &\Pi^\pm_{0'}(t,\sigma) \, := \, \frac{1}{\sqrt{2}}\Big( \Pi^\pm_{0} \,+\, \Pi^\pm_{1} \Big)\, 
   \equiv \,- \frac{1}{2\, \pi \alpha'} \, G\, {\partial_\mp} Y^{1'}  \, , \,\,\,\,\,\,\,\, \text{and}\\[.2cm]
   &\Pi^\pm_{1'}(t,\sigma) \, := \, \frac{1}{\sqrt{2}}\Big( \Pi^\pm_{0} \,-\, \Pi^\pm_{1} \Big)\, 
   \equiv \,-  \frac{1}{2\, \pi \alpha'} \, G\, {\partial_\mp} Y^{0'} \, .
   \label{MomDensityLCIso_Components}
\end{aligned} &
\end{flalign}
\vspace{0.1cm}

Combining the components in Eq.(\ref{MomDensityLCIso_Components}), renaming the indices $\mu' = \mu$, and recognising the raising and lowering property of the isothermal coordinates, leaves

\begin{equation}
\Pi^\pm_{\mu}(t,\sigma) \, = \, \frac{1}{2\, \pi \alpha'} \, G\, {\partial_\mp} Y_{(1-\mu)}  \,. \label{MomDensityLCIso}
\end{equation}
\vspace{0.1cm}

The Virasoro constraint equations and the string equations of motion, Eqs.(\ref{Virasoro}, \ref{StringEoM}), can also be rewritten in terms the new string mapping functions $Y^{\mu'}(\sigma^+, \sigma^-)$:

\begin{flalign}
& \begin{aligned}
   &\big( \partial_{\pm}\,Y^0 \big)\, \big( \partial_{\pm}\,Y^1 \big)\,=\, 0 \, , \\[.2cm]
   & \partial_+ \left(G \partial_- Y^1\right) \,+\, \partial_- \left(G \partial_+ Y^1\right) \, = \, \left(\partial_0  G \right) \left[ \left(\partial_+ Y^0 \right)\left(\partial_- Y^1 \right) \,+\, \left(\partial_+ Y^1 \right)\left(\partial_- Y^0 \right)\right] \, , \\[.2cm]
   & \partial_+ \left(G \partial_- Y^0\right) \,+\, \partial_- \left(G \partial_+ Y^0\right) \, = \, \left(\partial_1  G \right)\left[ \left(\partial_+ Y^0 \right)\left(\partial_- Y^1 \right) \,+\, \left(\partial_+ Y^1 \right)\left(\partial_- Y^0 \right)\right] \, ,
   \label{Virasoro_EoM_Iso}
\end{aligned} &
\end{flalign}
\vspace{0.1cm}

where $\partial_{\mu} := \frac{\partial}{\partial Y^{\mu}}$ for $\mu = (0,1)$. Since, the derivation of Eq.(\ref{Virasoro_EoM_Iso}) is somewhat tedious, it has been relegated to appendix (\ref{appVirasoroEoMIso}).\\

There are four classes of solutions to the set of equations Eqs.(\ref{Virasoro_EoM_Iso}). Namely, 
\begin{flalign}
& \begin{aligned}
  &  \textit{A}: \,\,\,\,\,\, Y^0(\sigma^+,\sigma^-) \, = \, Y^0_{\textit{A}}(\sigma^+) \, , \,\,\,\,\,\,\,\,\,\,\,\,\,\,\,\, Y^1(\sigma^+,\sigma^-) \, = \, Y^1_{\textit{A}}(\sigma^-) \\[.2cm]
    & \textit{B}: \,\,\,\,\,\, Y^0(\sigma^+,\sigma^-) \, = \, Y^0_{\textit{B}}(\sigma^-) \, , \,\,\,\,\,\,\,\,\,\,\,\,\,\,\,\, Y^1(\sigma^+,\sigma^-) \, = \, Y^1_{\textit{B}}(\sigma^+) \\[.2cm]
    & \text{C}: \,\,\,\,\,\, Y^0(\sigma^+,\sigma^-) \, = \, Y^0_{\textit{C}}\, = \, c_1 \, , \,\,\,\,\,\,\,\,\,\,\,\,\, Y^1(\sigma^+,\sigma^-) \, = \, f_\textit{C} \big(\, \alpha_\textit{C}(\sigma^+) \, + \, \beta_\textit{C}(\sigma^-); \, Y^0_\textit{C} \, \big)\\[.2cm]
    & \text{D}: \,\,\,\,\,\, Y^0(\sigma^+,\sigma^-) \, = \, f_\textit{D} \big(\, \alpha_\textit{D}(\sigma^+) \, + \, \beta_\textit{D}(\sigma^-); \, Y^0_\textit{D} \, \big)\, , \,\,\,\,\,\,\,\,\,\,\,\,\, Y^1(\sigma^+,\sigma^-) \, = \, Y^1_{\textit{D}}\, = \, c_2  \, , \label{ABCDsolutions}
\end{aligned} &
\end{flalign}
\vspace{0.1cm}

where $c_{1,2}$ are constants, and $Y^0_{\textit{A}}(\sigma^+)$, $Y^1_{\textit{A}}(\sigma^-)$, $Y^0_{\textit{B}}(\sigma^-)$, $ Y^1_{\textit{B}}(\sigma^+)$,
 $\alpha_\textit{C}(\sigma^+)$, $\beta_\textit{C}(\sigma^-)$, $\alpha_\textit{D}(\sigma^+)$, and $\beta_\textit{D}(\sigma^-)$ are arbitrary functions \cite{Bars_1994, Bars_1995}.
 The string solutions $A$, $B$, $C$ and $D$ given in Eq.(\ref{ABCDsolutions}) may be verified by direct substitution into Eqs.(\ref{Virasoro_EoM_Iso}). This is easy for classes $A$ and $B$, since these classes exist for any metric. Taking the derivative with respect to the light-cone coordinates $\sigma_\pm$, yields

\begin{flalign}
& \begin{aligned}
  &  \textit{A}: \,\,\,\,\partial_- \, \big(Y^0_{\textit{A}}(\sigma^+)\big) \, = \, 0 \, , \,\,\,\,\,\,\,\,\,\,\,\,\,\,\partial_+\, \big(Y^1_{\textit{A}}(\sigma^-) \big) \, = \, 0 \\[.2cm]
    & \textit{B}:  \,\,\,\,\partial_+ \, \big(Y^0_{\textit{B}}(\sigma^+)\big) \, = \, 0 \, , \,\,\,\,\,\,\,\,\,\,\,\,\,\,\partial_- \, \big(Y^1_{\textit{B}}(\sigma^-) \big) \, = \, 0 \, , \label{ABcheck}
\end{aligned} &
\end{flalign}
\vspace{0.1cm}

which is used, together with the chain rule, to easily show that the $A$ and $B$ classes of string solutions satisfy the Virasoro constraints and the string equations of motion Eqs.(\ref{Virasoro_EoM_Iso}). This is harder to prove for classes $C$ and $D$. These classes are metric dependent solutions since the functions $f_\text{C}$ and $f_\text{D}$ are metric dependent. These functions are calculated by inverting the following relations that depend on the target spacetime metric,
\begin{flalign}
& \begin{aligned}
  &  \textit{C}: \,\,\,\,\,\, Y^0(\sigma^+,\sigma^-) \, = \, Y^0_{\text{C}} , \,\,\,\,\,\,\,\,\,\,\,\,\,\, \int_{f_\text{C}} \, ds\, G\big( Y^0_\text{C}, s\big) \, = \, \alpha_\text{C}(\sigma^+) \, + \, \beta_\text{C}(\sigma^-)\\[.2cm]
    & \textit{D}: \,\,\,\,\,\, Y^1(\sigma^+,\sigma^-) \, = \, Y^1_{\text{D}} , \,\,\,\,\,\,\,\,\,\,\,\,\,\, \int_{f_\text{D}} \, ds\, G\big(s, Y^1_\text{D}\big) \, = \, \alpha_\text{D}(\sigma^+) \, + \, \beta_\text{D}(\sigma^-) \, . \label{CDmetricdef}
\end{aligned} &
\end{flalign}
\vspace{0.1cm}

Taking the derivatives of Eq.(\ref{CDmetricdef}) with respect to the light-cone coordinates $\sigma_\pm$ and making use of Eq.(\ref{ABCDsolutions}),


\begin{flalign}
& \begin{aligned}
  &  \textit{C}: \,\,\,\,\,\, \partial_\pm Y^0(\sigma^+,\sigma^-) \, = \, 0 , \,\,\,\,\,\,\,\,\,\,\,\,\,\, G \, \partial_\pm Y^1(\sigma^+,\sigma^-)  \, = \, \partial_\pm \big( \alpha_\text{C}(\sigma^+) \, + \, \beta_\text{C}(\sigma^-) \big) \\[.3cm]
    & \textit{D}: \,\,\,\,\,\, \partial_\pm Y^1(\sigma^+,\sigma^-) \, = \, 0 , \,\,\,\,\,\,\,\,\,\,\,\,\,\, G \, \partial_\pm Y^0(\sigma^+,\sigma^-)  \, = \, \partial_\pm \big( \alpha_\text{D}(\sigma^+) \, + \, \beta_\text{D}(\sigma^-) \big) \, . \label{CDcheck}
\end{aligned} &
\end{flalign}
\vspace{0.1cm}

Using Eq.(\ref{CDcheck}), the $C$ and $D$ classes of string solutions can be easily proven to satisfy the Virasoro constraints and string equations of motion Eqs.(\ref{Virasoro_EoM_Iso}) by direct substitution. For the $C$ class, the Virasoro constraints and second equation of motion are trivially satisfied. It remains to be checked whether the first equation of motion is satisfied. To this end, Eq.(\ref{CDcheck}) can be substituted into the first equation of motion in Eq.(\ref{Virasoro_EoM_Iso})


\begin{equation}
 0 \,  = \, \partial_+ \left(G \partial_- Y^1\right) \,+\, \partial_- \left(G \partial_+ Y^1\right)= \, \partial_+\left( \partial_- \,\beta_\text{C}(\sigma^-)  \right) \,+\, \, \partial_-\left(\partial_+ \,\alpha_\text{C}(\sigma^+)  \right) \, = \, 0 \, . \label{Cproof}
\end{equation}
\vspace{0.1cm}

Similarly for class $D$, the Virasoro constraints and the first equation of motion are trivially satisfied, while the second equation of motion is verified by substituting Eq.(\ref{CDcheck}) into Eq.(\ref{Virasoro_EoM_Iso}).\\

It is not immediately obvious that the classes of solutions presented in Eq.(\ref{ABCDsolutions}) form a complete set. Completeness was shown in Bars \textit{et al.} \cite{Bars_1995} by using the curved spacetime approach based on \textit{G/H} gauged WZW models. Note that the pattern of solutions defined on the patches of the worldsheet is metric independent and spacetime dimension independent \cite{Bars_1994}.
Hence, in order to find the general string solution for the Limp Noodle in AdS$_d$-Schwarzschild, start by finding the pattern of solutions defined on the patches of the worldsheet for the Limp Noodle in flat space. This is done in the next subsection.\\

Prior to this, some intuition regarding the patch-parametrization of the Limp Noodle can be developed. At a fold in a string or a string endpoint, the determinant of the induced worldsheet metric $g_{ab}$ (Eq.(\ref{InducedMetric})) vanishes. Hence the string endpoint's trajectory in the target spacetime is a null geodesic (the point travels at the local speed of light). For classes $C$ and $D$ -- by virtue of either $Y^0$ or $Y^1$ reducing to a constant over the patch of parameter space where the class is a solution -- the induced worldsheet metric reduces to zero, $g_{ab} =0$.
From this it can be deduced that all points in $C$ or $D$ parameter space patches map to the trajectory of the string endpoint (or fold) in target spacetime. This is a many-to-one mapping: the string's endpoint in target spacetime is represented several times on the worldsheet \cite{Bars_1994}. The string (sans the endpoints) behaves like a massive state, hence it is expected that motion of the string is described by classes $A$ and $B$ -- ever-present, metric independent solutions.

\subsubsection{Test Strings in \texorpdfstring{$\mathbb{R}^{1,1}$}{R11}} \label{subsubsecLQinFlatSpace}

In order to study the Limp Noodle in $\mathbb{R}^{1,1}$ the static gauge ansatz is chosen

\begin{equation}
X^\mu \, : \, [0, t_f] \times [0, \sigma_f] \, 	\rightarrow \, \mathbb{R}^{1,1}  \,, \label{StaticGuageAnsatz1}
\end{equation}

\begin{equation}
(t, \sigma) \,  	\rightarrow \, x^\mu (t, \sigma) \, = \, (t, x(t, \sigma))^\mu  \,. \label{StaticGuageAnsatz2}
\end{equation}
\vspace{0.1cm}

The initial test string set-up is the same as in the heavy quark case. Eqs.(\ref{Dirichlet_BC_R11}, \ref{Neumann_BC_R11}) are chosen as the boundary conditions for the position of the fixed endpoint at the horizon and the boundary endpoint falling in the $x(t,\sigma)$ direction respectively.
For initial conditions Eqs.(\ref{Static_t0_IC_R11}, \ref{Spatial_x0_IC_R11}) are chosen i.e. at $t=0$ the string is static and stretches between $x_0$ and $x_0 \, + \, \ell_0$. Set $t_f$ to be the minimum time it takes for a string to shrink from its initial length $\ell_0$ to a point $x_0$.\\

In flat space the total energy of the string, which represents the mass of an initially off-mass-shell light quark, is given by Eq.(\ref{FlatSpaceMass})\footnote{The derivation of the total energy of the test string in an AdS-Schwarzschild background is given in appendix (\ref{appEnergyCurvedSpace}).}.
The equation of motion for $x(t, \sigma)$ can be easily adapted from Eqs.(\ref{StringEoM_mu=0_final}, \ref{StringEoM_mu=1_final}), since the spacetime is $\mathbb{R}^{1+1}$ (therefore $G=1$ and $(\partial_0 G) \, = \, (\partial_1 G) \, = \, 0$). Hence, the non-trivial string equation of motion reduces to

\begin{equation}
    \partial_+ \, \partial_- \, X^{1} \,  = \, 0 \,\,\, \longrightarrow \,\,\, \partial_+ \, \partial_- \, x(t, \sigma) \,  = \, 0  \,\,\,\,\,\,\, \text{in}\, \,\,\,\,\,\,\, \mathbb{R}^{1+1}\, , \label{EoMFlatSpace}
\end{equation}
\vspace{0.1cm}

which is simply the wave equation. The general solution to Eq.(\ref{EoMFlatSpace}) is a sum of continuous analytic functions of $\sigma^+$ and $\sigma^-$ (i.e. \textit{left and right movers})

\begin{equation}
x(t, \sigma) \, = \, \frac{1}{2}\, \left( f_1(\sigma^+) \, + \, f_2(\sigma^-)\right) \,. \label{GenEoMFlatSpace}
\end{equation}
\vspace{0.1cm}

Using the boundary condition Eq.(\ref{Dirichlet_BC_R11}) to redefine  $f(z) \, := \, f_1(z) \, = \, 2\, x_0 \, - \, f_2(z)$, Eq.(\ref{GenEoMFlatSpace}) becomes


\begin{equation}
x(t, \sigma) \, = x_0 \, +\, \frac{1}{2}\, \left( f(\sigma^+) \, - \,  f(\sigma^-)\right)\, . \label{GenEoMFlatSpace2}
\end{equation}
\vspace{0.1cm}

The Neumann boundary condition Eq.(\ref{Neumann_BC_R11}) becomes


\begin{equation}
0 \, = \frac{1}{2} \,  \left[ \partial_+ \, f(\sigma^+) \, + \,  \partial_- \, f(\sigma^-)\right] \big\rvert_{\sigma = \sigma_f} \, , \label{Neumann_BC_FlatSpace}
\end{equation}
\vspace{0.1cm}

which follows from the fact that $x_0 \in  \mathbb{R}$ and the definition Eq.(\ref{LCDef}),
since $\partial_{\sigma} f(\sigma^+) = \left(\partial_{+} - \partial_{-} \right) f(\sigma^+) = \partial_{+} f(\sigma^+)$. To simplify Eq.(\ref{Neumann_BC_FlatSpace}) let $z\, := \, t \, - \, \sigma_f$, then
\vspace{0.2cm}

\begin{equation}
\sigma^- \rvert_{\sigma = \sigma_f} \, = \, t \, - \, \sigma_f  \, = \, z \,\,\,\,\,\,\,\,\,\,
\text{and} \,\,\,\,\,\,\,\,\,\, \sigma^+ \rvert_{\sigma = \sigma_f} \, = \, t \, + \, \sigma_f  \, = \, z \,+\, 2\,\sigma_f\,. \label{LC_at_sigmaf}
\end{equation}
\vspace{0.1cm}

Using Eq.(\ref{LC_at_sigmaf}) and defining $F(z)\, := \, \dfrac{d}{d z} f(z)$, Eq.(\ref{Neumann_BC_FlatSpace}) becomes\\


\begin{equation}
F(z \,+\, 2\sigma_f) \, = \,  - F(z)\, , \label{Neumann_BC_FlatSpace2}
\end{equation}
\vspace{0.1cm}

which shows that $F(z)$ is anti-periodic with an interval of $2\, \sigma_f$. The Virasoro constraint equations Eq.(\ref{Virasoro}) can also be rewritten in terms of $F(z)$

\begin{flalign}
& \begin{aligned}
    0 \, & = \,  -\,\big( \partial_{\pm}\,t \big) \, \big( \partial_{\pm}\,t \big)\,+\, \,\big( \partial_{\pm}\,x(t, \sigma) \big) \, \big( \partial_{\pm}\,x(t, \sigma) \big)
    \\[.2cm]
    & = \,  -\,\big( \partial_{\pm}\,t \big) \, \big( \partial_{\pm}\,t \big)\,+\, \,\Big[ \partial_{\pm}\, \Big(x_0 \, +\, \frac{1}{2}\, \big( f(\sigma^+) \, - \,  f(\sigma^-)\big) \Big) \Big]^2 \, . \label{VirasoroConstraints_FlatSpace}
\end{aligned} &
\end{flalign}
\vspace{0.1cm}

The first line follows from Eq.(\ref{StaticGuageAnsatz2}), and Eq.(\ref{GenEoMFlatSpace2}) was used in the final line. Eq.(\ref{VirasoroConstraints_FlatSpace}) breaks up into two, $(+)$ and $(-)$, constraint equations. For the $(+)$ equation, Eq.(\ref{VirasoroConstraints_FlatSpace}) becomes

\begin{flalign}
& \begin{aligned}
    &  0 \, =\, -\,\bigg(\frac{1}{2}\bigg)\,\bigg(\frac{1}{2}\bigg)\, +\, \bigg( \frac{1}{2} \partial_+\,  f(\sigma^+) \bigg)^2\\[.2cm]
    & \Rightarrow \, F(\sigma^+)^2\,=\, 1 \, . \label{VirasoroConstraints_FlatSpace+}
\end{aligned} &
\end{flalign}
\vspace{0.1cm}

\begin{figure}[!ht]
\centering
\begin{subfigure}{.5\textwidth}
  \centering
  \includegraphics[width=0.95\linewidth]{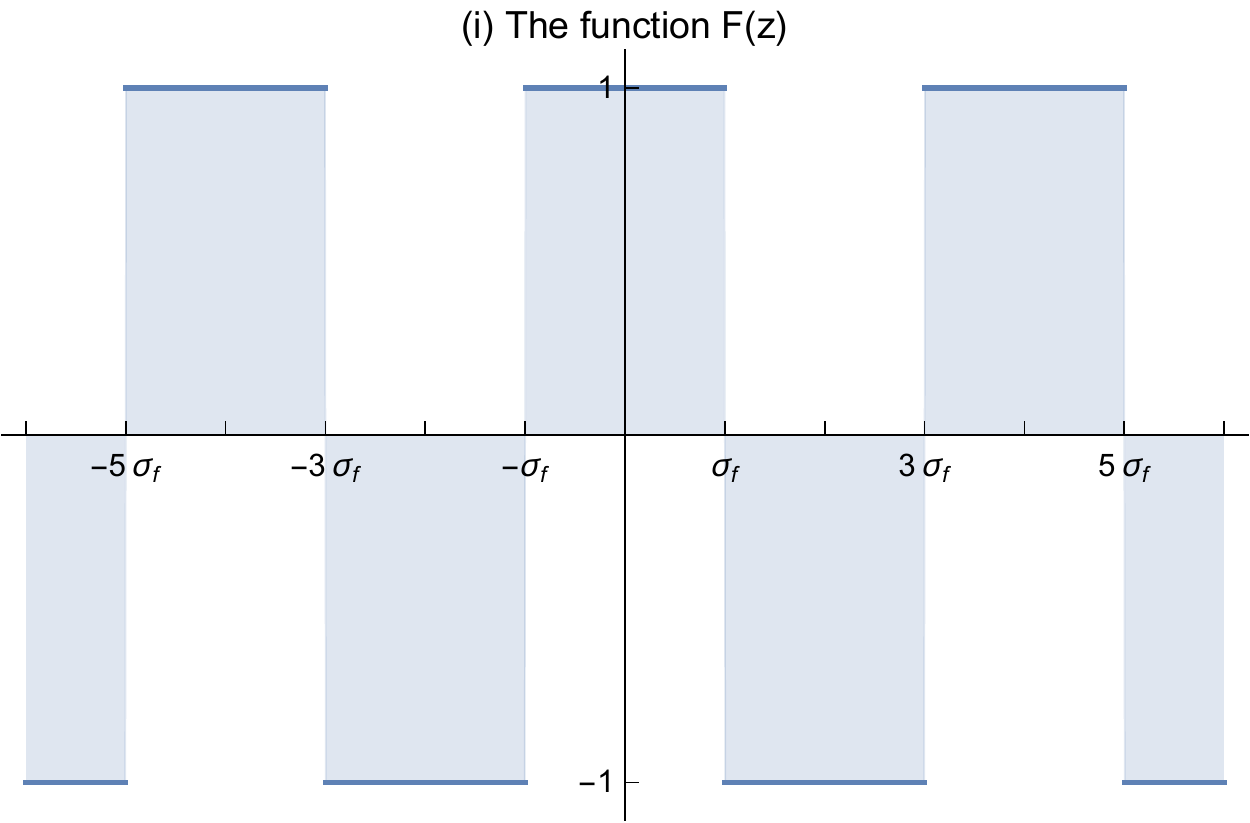}
\end{subfigure}%
\begin{subfigure}{.5\textwidth}
  \centering
  \includegraphics[width=0.95\linewidth]{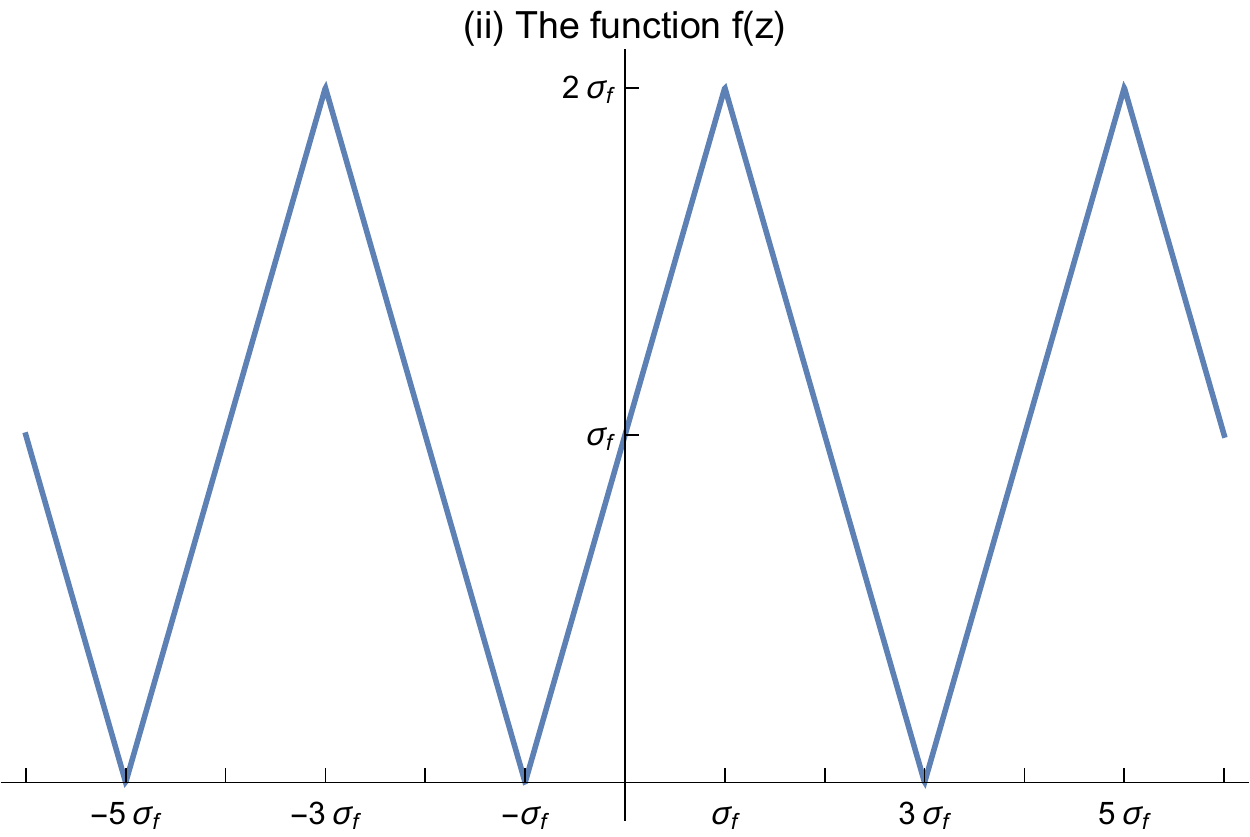}
\end{subfigure}
\caption[The functions (i) $F(z)$ and (ii) $f(z)$ over the range $(-6 \sigma_f, \, 6 \sigma_f)$]{The functions (i) $F(z)$ and (ii) $f(z)$ over the range $[-6 \sigma_f, \, 6 \sigma_f]$, given by Eq.(\ref{Fu_Solution}) and Eq.(\ref{fu_Solution}) respectively.
 \label{fig:functions_fu_Fu}}
\end{figure}

Similarly, for the $(-)$ equation $ F(\sigma^-)^2\,=\, 1$. Hence the Virasoro constraint equations become

\begin{equation}
F(\sigma^{\pm})^2\,=\, 1 \,, \label{VirasoroConstraints_FlatSpace_FINAL}
\end{equation}
\vspace{0.1cm}

which implies $ F(\sigma^-)^2 \rvert_{\sigma = \sigma_f}\,=\, 1$ or equivalently $F(z)^2\,=\, 1$. What remains to be determined is the sign of $F(z)$. This is given by the anti-periodicity condition Eq.(\ref{Neumann_BC_FlatSpace2}) which results in

\begin{equation}
F(z)\,=\, \big( -1 \big)^{\big\lfloor \frac{z + \sigma_f}{2 \sigma_f} \big\rfloor} \,, \label{Fu_Solution}
\end{equation}
\vspace{0.1cm}

where $\lfloor x \rfloor$ is the floor operation which returns the largest integer less than or equal to $x$. The function $F(z)$, plotted in figure (\ref{fig:functions_fu_Fu}), is a step function. The definition of $F(z)$ as the derivative of $f(z)$ with respect to $z$, implies that $f(z)$ can be found through integrating Eq.(\ref{Fu_Solution}). But $F(z)$ is a discrete function -- performing this integration involves integrating each segment of $F(z)$ separately.
Since it is already known that $F(z)$ is alternating $(+1)$ or $(-1)$ in each $2\, \sigma_f$ interval, and the integral of a constant is a straight monotonic function; all that remains to be determined is if the function $f(z)$ is monotonically increasing or decreasing on each interval. Hence $f(z)$ can be defined by

\begin{equation}
f(z)\,=\, \big( -1 \big)^{\big\lfloor \,\frac{z + \sigma_f}{2 \sigma_f}\, \big\rfloor} \, \left( \left[(z \,+ \, \sigma_f) \,\,
\text{mod}\,\, 2\sigma_f \right] \, - \, \sigma_f\right) \, + \, \sigma_f \,, \label{fu_Solution}
\end{equation}
\vspace{0.1cm}

where {\footnotesize $\big( -1 \big)^{\big\lfloor \,\frac{z + \sigma_f}{2 \sigma_f}\, \big\rfloor}$} defines the domain of each segment to be integrated,
and $\left( \left[(z \,+ \, \sigma_f) \,\,\text{mod}\,\, 2\sigma_f \right] \, - \, \sigma_f\right)$ yields $0$ if $z  \in  z_1\,\sigma_f$ (where $z_1$ is an even integer)
and $-\sigma_f$ if $z  \in  z_2\,\sigma_f$ (where $z_2$ is an odd integer).
The function $f(z)$ is a triangular wave function and is plotted in figure (\ref{fig:functions_fu_Fu}). \\

\vspace{0.1cm}

Now that a specific form of the function $f(z)$ has been obtained, $x(t, \sigma)$ (Eq.(\ref{GenEoMFlatSpace2})) is completely known.  For example, at $t \, = \, \sigma_f$, Eq.(\ref{GenEoMFlatSpace2}) becomes

\begin{equation}
x(\sigma_f, \sigma) \,  = \, x_0 \, +\, \frac{1}{2}\, \left( f(\sigma_f \,+\, \sigma) \, - \,  f(\sigma_f \,-\, \sigma)\right) = \, x_0 \, ,
\label{t=sigmaf_GenSoltuion_FlatSpace}
\end{equation}
\vspace{0.1cm}

where the second equality follows from using the definition Eq.(\ref{fu_Solution}). In studying the Limp Noodle, the primary interest is in the behaviour of the string as it shrinks from a length of $\ell_0$ to a single point $x_0$. From Eq.(\ref{t=sigmaf_GenSoltuion_FlatSpace}) this clearly is achieved in a $t= \sigma_f$ amount of time; hence the choice $t_f \, = \, \sigma_f$ is made.
The worldsheet parameter space is therefore given by $\mathcal{M} \, = \, [0, \sigma_f] \times [0, \sigma_f]$, i.e. the parameter space is square.\\

Given the insight of \cite{Bars_1994, Bars_1995, Ficnar_2014} the parameter space $\mathcal{M}$ can be minimally partitioned into two exclusive triangular regions

\begin{flalign}
& \begin{aligned}
    & \mathcal{M}_1 \, := \, \big\{ (t, \sigma)  \in  \mathcal{M} \rvert_\sigma \, \in \,[0, \sigma_f -t] \big\}\\[.2cm]
    & \mathcal{M}_2 \, := \, \big\{ (t, \sigma)  \in  \mathcal{M} \rvert_\sigma \, \in  \, (\sigma_f -t, \sigma_f] \big\}\, . \label{M1M2_ParameterSpace}
\end{aligned} &
\end{flalign}
\vspace{0.1cm}

In order to satisfy the boundary and initial conditions two classes of solutions exist, one in each of the two parameter space regions: the upper triangular region $(\mathcal{M}_2)$ and the lower triangular region $(\mathcal{M}_1)$. The embedding functions of the Limp Noodle in $\mathbb{R}^{1+1}$ is therefore given by

\begin{equation}
X^\mu_\text{Mink} (t, \sigma) \, = \, \left( t, \, x_0 \, + \, \begin{Bmatrix}
   \sigma, \,\,\,\,\,\,\,\,\,\,\,\,\,\,\,\,\,\,\,\,\,\,\, \text{if} \,\,\,\ (t, \sigma) \, \in \, \mathcal{M}_1 \\
   \sigma_f  - t, \,\,\,\,\,\,\,\,\,\,\, \text{if} \,\,\,\ (t, \sigma) \, \in \, \mathcal{M}_2
  \end{Bmatrix} \right)^\mu \,. \label{LN_FlatSpaceSolution}
\end{equation}
\vspace{0.1cm}

The Limp Noodle embedding functions satisfy the given boundary and initial conditions. The boundary conditions can be checked first: for $\sigma = 0$, $(t, \sigma)  \in  \mathcal{M}_1$ and the Dirichlet condition Eq.(\ref{Dirichlet_BC_R11}) is satisfied; while for $\sigma = \sigma_f $, $(t, \sigma)  \in  \mathcal{M}_2$ and the Neumann condition Eq.(\ref{Neumann_BC_R11}) is satisfied.
The initial conditions are also fulfilled. For $t =  0$, $(t, \sigma)  \in  \mathcal{M}_1$ and the initial condition Eq.(\ref{Static_t0_IC_R11}) is satisfied. Further the initial condition Eq.(\ref{Spatial_x0_IC_R11}) is satisfied, since $\sigma_f \, = \, \ell_0$ is a requirement of the Virasoro constraints in flat space. \\

\begin{figure}[!ht]
\centering
\begin{subfigure}{.5\textwidth}
  \centering
  \includegraphics[width=0.9\linewidth]{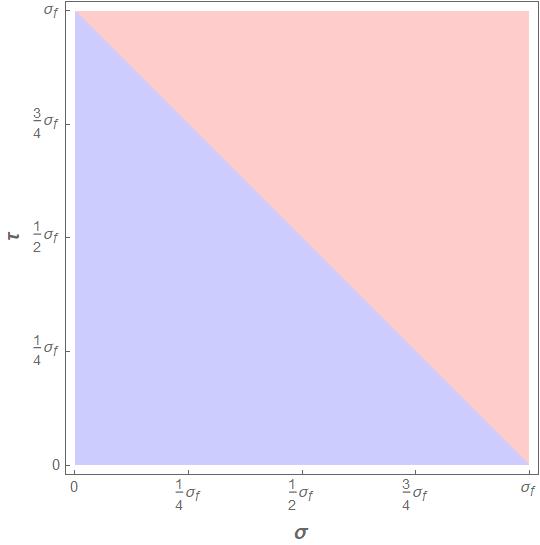}
\end{subfigure}%
\begin{subfigure}{.5\textwidth}
  \centering
  \includegraphics[width=0.93\linewidth]{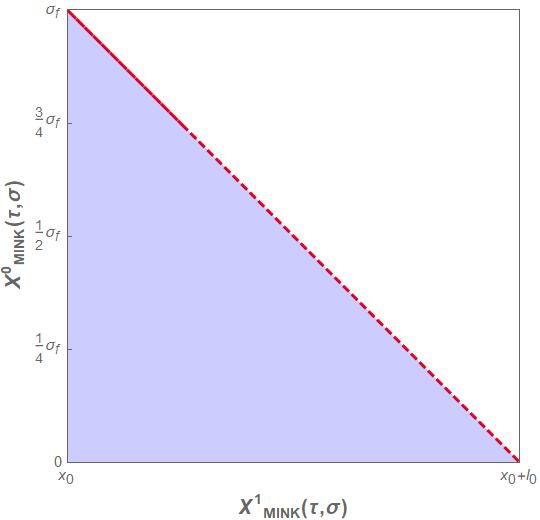}
\end{subfigure}
\caption[The worldsheet parameter space and embedding functions in $\mathbb{R}^{1+1}$]{Left: The worldsheet parameter space, where $\mathcal{M}_1$ is the blue region and $\mathcal{M}_2$ is the red region (Eq.(\ref{M1M2_ParameterSpace})).
Right: The embedding functions $X^\mu_\text{Mink} (t, \sigma)$, Eq.(\ref{LN_FlatSpaceSolution}), where $\mathcal{M}_1$ is mapped to a space-like region (blue region) and $\mathcal{M}_2$ is mapped to a null geodesic (red, dashed line) in the target spacetime. \label{fig:R11WorldSheet_Embedding}}
\end{figure}

\begin{figure}[!ht]
\centering
\begin{subfigure}{.5\textwidth}
  \centering
  \includegraphics[width=0.97\linewidth]{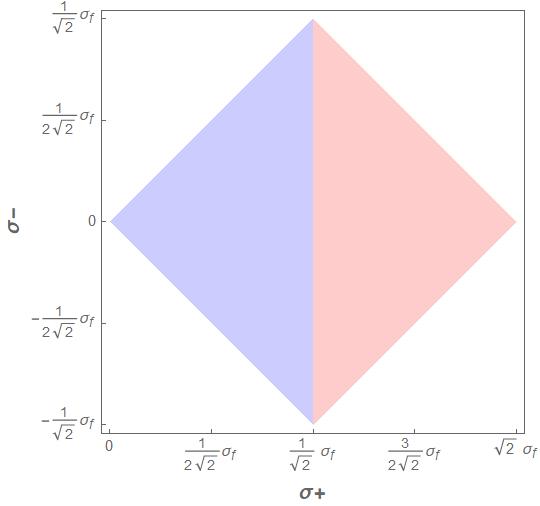}
\end{subfigure}%
\begin{subfigure}{.5\textwidth}
  \centering
  \includegraphics[width=0.9\linewidth]{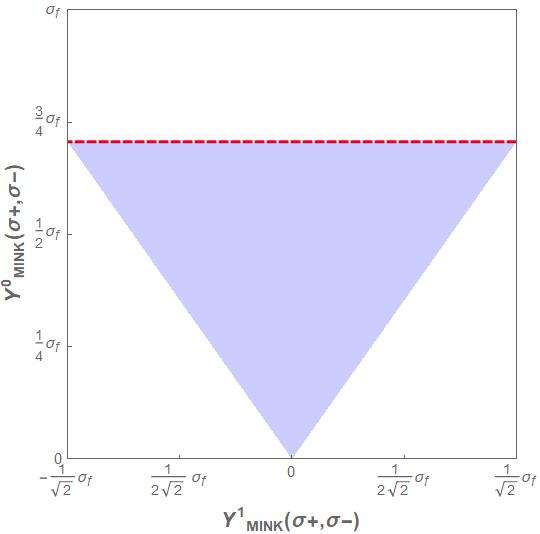}
\end{subfigure}
\caption[The worldsheet parameter space in light-cone coordinates and isothermal embedding functions in $\mathbb{R}^{1+1}$]{Left: The worldsheet parameter space in light-cone coordinates ($\sigma^+, \sigma^-$), where $\mathcal{M}_1$ is the blue region and $\mathcal{M}_2$ is the red region. The parameter space regions $\mathcal{M}_1$ and $\mathcal{M}_2$ are plotted from Eq.(\ref{M1M2_ParameterSpace}) which is converted to ($\sigma^+, \sigma^-$) coordinates using Eq.(\ref{LCDef}). Right: The isothermal embedding functions $Y^\mu_\text{Mink} (\sigma^+, \sigma^-)$, Eqs.(\ref{IsoEmbeddingR11_Y0}, \ref{IsoEmbeddingR11_Y1}, \ref{IsoEmbeddingR11_Y0_M2}, \ref{IsoEmbeddingR11_Y1_M2}), where $\mathcal{M}_1$ is mapped to a space-like region (blue region) and $\mathcal{M}_2$ is mapped to a null geodesic (red, dashed line) in the target spacetime. \label{fig:R11WorldSheet_EmbeddingLC}}
\end{figure}

From Eq.(\ref{LN_FlatSpaceSolution}), observe that the functions $X^\mu_\text{Mink} (t, \sigma)$ map the parameter space $\mathcal{M}_1$ region to an extended, space-like region in the target spacetime; while the parameter space $\mathcal{M}_2$ region is mapped to a null geodesic in the target spacetime. Null geodesics describe the trajectories of massless `point-like' objects, and as such the worldsheet is considered to be wrapped up to a point along this geodesic. Hence, the string solution on $\mathcal{M}_1$ describes the entire string sans the falling endpoint, while the string solution on $\mathcal{M}_2$ describes the falling endpoint.
This is evident from figure (\ref{fig:R11WorldSheet_Embedding})\footnote{Details regarding figures (\ref{fig:R11WorldSheet_Embedding})-(\ref{fig:AdS3WorldSheet_Embedding}) can be found in Mathematica Notebook [a] (\texttt{MappingWorldSheetToTarget.nb}) -- see appendix (\ref{appCode}) for access.}.\\

$X^\mu_\text{Mink}(\mathcal{M}_2)$ lies on the boundary of $X^\mu_\text{Mink}(\mathcal{M}_1)$, and momentum flows between these two regions in the target spacetime. In order to continue to satisfy the Neumann boundary condition Eq.(\ref{Neumann_BC_R11}), momentum flows out of the $\mathcal{M}_1$ region at $\sigma \, = \, \sigma_f -t$, but immediately stops at $\mathcal{M}_2$ since any momentum flowing out of the string endpoint would change the physics in the boundary theory\footnote{An infinite force is required to stop the flow of momentum abruptly at the $\mathcal{M}_1/\mathcal{M}_2$ divide, and yet the string has a finite energy density (since the string has a finite tension). This discrepancy stems from the fact that strings are fundamentally one-dimensional objects and are not made up of $d=0$ constituent objects.}.\\

If Eq.(\ref{LN_FlatSpaceSolution}) is converted into isothermal coordinates in terms of light-cone parameter space ($\sigma^+, \sigma^-$), $Y^\mu_\text{Mink} (\sigma^+, \sigma^-)$ defined on the region $\mathcal{M}_1$ is a Bars \textit{et al.} \cite{Bars_1994, Bars_1995} class $A$ solution\footnote{See subsection (\ref{subsubsecBarsMethod}) for details.} and the solution defined on the region $\mathcal{M}_2$ is a class $C$ solution, as expected. To see this, use the definition Eq.(\ref{IsoMapFunc}) to define the isothermal embedding functions on $\mathcal{M}_1$,

\begin{flalign}
& \begin{aligned}
    Y^{0'} & \, = \,\frac{1}{\sqrt{2}}\Big( t \,+\, \left( x_0 + \sigma \right) \Big) \\[.2cm]
    & \, = \,\frac{1}{\sqrt{2}}\left( \frac{1}{\sqrt{2}}\left( \sigma^+ \,+\, \sigma^- \right) \,+\, x_0 \,+\, \frac{1}{\sqrt{2}}\left( \sigma^+ \,- \, \sigma^- \right)  \right)   \\[.2cm]
    & \, = \sigma^+  \,+\, \frac{1}{\sqrt{2}} \,x_0 \, =: \, Y^{0'}\left(\sigma^+\right)   \, , \label{IsoEmbeddingR11_Y0}
\end{aligned} &
\end{flalign}
\vspace{0.1cm}

where the definition of the light-cone coordinates (Eq.(\ref{LCDef})) is used in the second line. Similarly, 

\begin{flalign}
& \begin{aligned}
    Y^{1'} \, = \, \sigma^- \,-\, \frac{1}{\sqrt{2}} \,x_0 \, =: Y^{1'}\left(\sigma^-\right)   \, . \label{IsoEmbeddingR11_Y1}
\end{aligned} &
\end{flalign}
\vspace{0.1cm}

Comparing Eqs.(\ref{IsoEmbeddingR11_Y0}, \ref{IsoEmbeddingR11_Y1}) against the four classes of Bars \textit{et al.} solutions (Eq.(\ref{ABCDsolutions})),
it is obvious that the isothermal embedding functions $Y^\mu_\text{Mink} (\sigma^+, \sigma^-)$ defined on the region $\mathcal{M}_1$ are a class $A$ solution. This is depicted in figure (\ref{fig:R11WorldSheet_EmbeddingLC}).\\

The isothermal embedding functions on $\mathcal{M}_2$ are given by

\begin{flalign}
& \begin{aligned}
    Y^{0'}  \, = \,\frac{1}{\sqrt{2}}\Big( t \,+\, \left( x_0 + \sigma_f - t \right) \Big)
     \, = \, \frac{1}{\sqrt{2}} \left( x_0 + \sigma_f \right)\, =: \, Y^{0'}_C  \, , \label{IsoEmbeddingR11_Y0_M2}
\end{aligned} &
\end{flalign}
\vspace{0.1cm}

and

\begin{flalign}
& \begin{aligned}
    Y^{1'}  \, = \,\frac{1}{\sqrt{2}}\Big( t \,-\, \left( x_0 + \sigma_f - t \right) \Big) 
     \, = \, -  \left( x_0 + \sigma_f - \left(\sigma^+ + \sigma^- \right)\right)\, =: \, Y^{1'}\left(\sigma^+, \sigma^-\right)   \, . \label{IsoEmbeddingR11_Y1_M2}
\end{aligned} &
\end{flalign}
\vspace{0.3cm}

Matching Eqs.(\ref{IsoEmbeddingR11_Y0_M2}, \ref{IsoEmbeddingR11_Y1_M2}) with four classes of Bars \textit{et al.} solutions (Eq.(\ref{ABCDsolutions}))
it is easily apparent that the embedding functions $Y^\mu_\text{Mink} (\sigma^+, \sigma^-)$ defined on the region $\mathcal{M}_2$ are a class $C$ solution. Figure (\ref{fig:R11WorldSheet_EmbeddingLC}) displays this graphically.\\

\vspace{0.1cm}

\subsubsection{Test Strings in \texorpdfstring{AdS$_3$-Schwarzschild}{AdS3-Schwarzschild}} \label{subsubsecLQinAdS3}

In the previous subsection, the general embedding functions for the Limp Noodle in $\mathbb{R}^{1+1}$, Eq.(\ref{LN_FlatSpaceSolution}), were discovered. From this the embedding functions for the Limp Noodle in AdS$_3$-Schwarzschild can be found using the same minimal partitioning of the parameter space and the determined pattern of solution classes on the patches of the parameter space.\\

In order to trivially write down the general solution for the Limp Noodle in AdS$_3$-Schwarzschild a set of isothermal coordinates is needed. As was shown in subsection (\ref{subsubsecHQinAdS3}), the tortoise coordinate $r_*$ together with the temporal coordinate $t$ forms such a set of coordinates in AdS$_d$-Schwarzschild. \\

\begin{figure}[!ht]
\centering
\begin{subfigure}{.5\textwidth}
  \centering
  \includegraphics[width=0.9\linewidth]{R11_WorldSheet.jpg}
\end{subfigure}%
\begin{subfigure}{.5\textwidth}
  \centering
  \includegraphics[width=0.93\linewidth]{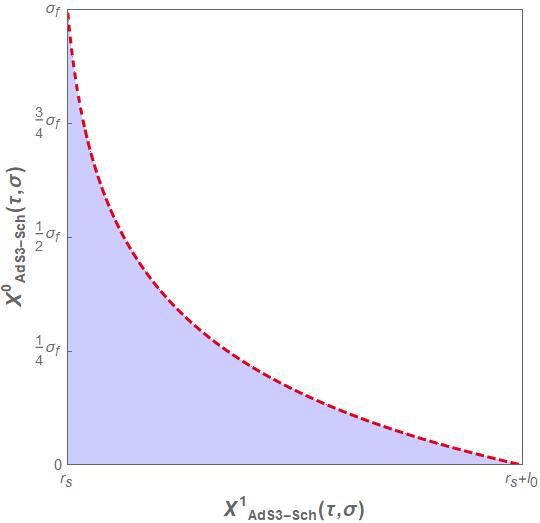}
\end{subfigure}
\caption[The worldsheet parameter space and embedding functions in AdS$_3$-Schwarzschild]{Left: The worldsheet parameter space, where $\mathcal{M}_1$ is the blue region and $\mathcal{M}_2$ is the red region (Eq.(\ref{M1M2_ParameterSpace})).
Right: The embedding functions $X^\mu_\text{AdS3-Sch} (t, \sigma)$, Eq.(\ref{LN_AdS3Sol}), where $\mathcal{M}_1$ is mapped to a space-like region (blue region) and $\mathcal{M}_2$ is mapped to a null geodesic (red, dashed line) in the target spacetime. \label{fig:AdS3WorldSheet_Embedding}}
\end{figure}

The Dirichlet and Neumann conditions, Eqs.(\ref{Dirichlet_BC_AdS3}, \ref{Neumann_BC_AdS3}), are chosen as boundary conditions for the Limp Noodle in AdS$_3$-Schwarzschild. The boundary conditions can be rewritten in terms of the isothermal $(t, r_*)$ coordinate set.
In subsection (\ref{subsubsecHQinAdS3}), it was noticed that these boundary conditions in the conformally flat description of AdS$_3$-Schwarzschild, Eqs.(\ref{Dirichlet_BC_AdS3_TortoiseCoord}, \ref{Neumann_BC_AdS3_TortoiseCoord}), are respectively analogous to the boundary conditions for the
Limp Noodle (and the heavy quark test string) in $\mathbb{R}^{1+1}$, Eqs.(\ref{Dirichlet_BC_R11}, \ref{Neumann_BC_R11}). Hence, the embedding functions of the Limp Noodle in the conformally flat description of AdS$_3$-Schwarzschild are the direct analogue of Eq.(\ref{LN_FlatSpaceSolution}). In $(t, r_*)$ coordinates, the embedding functions are

\begin{equation}
X^{\mu}_\text{\text{AdS$_3$-Sch}} (t, \sigma) \, = \, \left( t, \, r_{s*} \, + \, \begin{Bmatrix}
   \sigma, \,\,\,\,\,\,\,\,\,\,\,\,\,\,\,\,\,\,\,\,\,\,\, \text{if} \,\,\,\ (t, \sigma) \, \in \, \mathcal{M}_1 \\
   \sigma_f  - t, \,\,\,\,\,\,\,\,\,\,\, \text{if} \,\,\,\ (t, \sigma) \, \in \, \mathcal{M}_2
  \end{Bmatrix}\, , \, 0 \right)^{\mu} \,, \label{LN_AdS3Sol_TortoiseCoord}
\end{equation}
\vspace{0.1cm}

where the $\mathcal{M}_1$ and $\mathcal{M}_2$ parameter space regions are given by Eq.(\ref{M1M2_ParameterSpace}). Using the inverse tortoise transformation Eq.(\ref{d=3_TortoiseCoordINVERTED}),
the embedding functions for the AdS$_3$-Schwarzschild Limp Noodle (Eq.(\ref{LN_AdS3Sol_TortoiseCoord})) can be rewritten in $(t,r)$ coordinates

\begin{equation}
X^{\mu}_\text{\text{AdS$_3$-Sch}} (t, \sigma) \, = \, \left( t, \, \begin{Bmatrix}
   - r_H\, \coth \big(\frac{r_H}{l^2}\,\left(r_{s*} \,+\,\sigma\right) \big), \,\,\,\,\,\,\,\,\,\,\,\,\,\,\,\,\,\,\,\,\,\,\, \text{if} \,\,\,\ (t, \sigma) \, \in \, \mathcal{M}_1 \\
   - r_H\, \coth \big(\frac{r_H}{l^2}\, \left(r_{s*} \,+\,\sigma_f  - t \right) \big) , \,\,\,\,\,\,\,\,\,\,\, \text{if} \,\,\,\ (t, \sigma) \, \in \, \mathcal{M}_2
  \end{Bmatrix}\, , \, 0 \right)^{\mu} \,, \label{LN_AdS3Sol}
\end{equation}
\vspace{0.1cm}

where the position of the fixed string endpoint attached to the stretched horizon $r_{s*}$ is given by Eq.(\ref{StrHor_TortoiseCoord}) and the length of the string in tortoise coordinates is given by Eq.(\ref{simga_f_TortoiseCoor}).
Notice that the identification Eq.(\ref{InverseTortoiseCoord_Iden}) still holds. The embedding functions are plotted in figure (\ref{fig:AdS3WorldSheet_Embedding}).\\

\vspace{0.1cm}

\subsection{Fluctuating Test String Dynamics} \label{subsecLQ_TSDynamics}

As discussed previously, the boundary endpoint of the Limp Noodle falls at the local speed of light -- the same speed at which information propagates. Hence, any part of the string below the falling endpoint does not \textit{know} if the string is stretched like an on-mass-shell quark or its endpoint is falling like an off-mass-shell quark. Therefore, the transverse fluctuations of all parts of the Limp Noodle except the falling endpoint are described by the same transverse equations of motion as those of the on-mass-shell heavy quark's test string and its solution (discussed in subsections (\ref{subsubsecHQ_NambuTransverse}) and (\ref{subsubsecHQ_Displacement})). \\

The leading order solution to the string equations of motion -- renamed $X^\mu_0(t, \sigma)$ -- is given by Eq.(\ref{LN_AdS3Sol}) in AdS$_3$-Schwarzschild. Adding non-zero fluctuations in the transverse $x$-direction, the Limp Noodle solution becomes

\begin{equation}
X^{\mu}_\text{\text{AdS$_3$-Sch}} (t, \sigma) \Big\rvert_{\mathcal{M}_1} = \, X^\mu_0(t, \sigma) \, + \, \big(0, \,0,\, X(t, \sigma)\, \big)^\mu \,
= \,\left( t, \, - r_H \coth \left(\frac{r_H}{l^2}\,\left(r_{s*} +\sigma\right) \right) , \, X(t, \sigma) \right)^{\mu}\,. \label{LN_AdS3Sol_XI}
\end{equation}
\vspace{0.1cm}

The transverse string equations of motion are supplied in Eq.(\ref{StringEoM_XI_d=3_tsig}); and its solution is given by Eq.(\ref{X_operator}) in AdS$_3$-Schwarzschild. Continuity of the string solution in spacetime means that it is unnecessary to find the explicit string solution on the $\mathcal{M}_2$ parameter space region which maps onto the falling string endpoint.

\subsubsection{Calculating the Light Quark's Mean-Squared Displacement \texorpdfstring{$s^2(t)$}{s2t} in \texorpdfstring{AdS$_3$-Schwarzschild}{AdS3}} \label{subsubsecLQ_Displacement}

Unlike in the case of the heavy quark's test string where the boundary endpoint is fixed in the radial direction (subsection (\ref{subsubsecHQ_Displacement})); for the Limp Noodle the position of the boundary endpoint is falling at the local speed of light and is defined by the operator

\begin{equation}
    \hat{X}_{\text{end}}(t) \, := \, \hat{X}(t, \, \sigma_f -t) \, . \label{Free_Endpoint_Operator}
\end{equation}
\vspace{0.1cm}

As in subsection (\ref{subsubsecHQ_Displacement}), in order to determine $s^2(t)$ -- the falling string endpoint's mean-squared transverse displacement defined in Eq.(\ref{displacement_def}) -- begin by calculating the expectation value of the position of the free endpoint at two different times. Using Eq.(\ref{X_operator}),

\begin{flalign}
& \begin{aligned}
      \langle  \normord{\hat{X}_{\text{end}}(t_1) \hat{X}_{\text{end}}(t_2)}  \rangle
     &  =  \frac{\beta^2}{4  \pi   \sqrt{\lambda}}  \int_0^\infty \frac{d \omega \, d \omega^\prime}{(2 \pi)^2}  \frac{1}{\sqrt{\omega  \omega^\prime}} \Big[f_\omega (\sigma_f -t_1) f^{\,*}_{\omega^\prime} (\sigma_f -t_2)  e^{- i \omega t_1 + i {\omega^\prime} t_2} \langle \normord{ \hat{a}_\omega \hat{a}^{\,\dagger}_{\omega^\prime}} \rangle \, \\[0.2cm]
     & \,\,\,\,\,+  f^{\,*}_\omega (\sigma_f - t_1) f_{\omega^\prime} (\sigma_f -t_2) e^{i \omega t_1 - i {\omega^\prime} t_2} \langle \normord{ \hat{a}^{\,\dagger}_\omega \hat{a}_{\omega^\prime}} \rangle \Big]\\[0.2cm]
     &  =  \frac{\beta^2}{4  \pi   \sqrt{\lambda}}  \int_0^\infty \frac{d \omega \, d \omega^\prime}{2 \pi} \, \frac{1}{\sqrt{\omega \, \omega^\prime}} \bigg[\delta( \omega^\prime - \omega) f_\omega (\sigma_f -t_1) f^{\,*}_{\omega^\prime} (\sigma_f -t_2) \frac{e^{- i \omega t_1 + i {\omega^\prime} t_2}}{e^{\beta \omega^\prime} - 1}\, \\[0.2cm]
     & \,\,\,\,\, + \delta(\omega - \omega^\prime) f^{\,*}_\omega (\sigma_f - t_1) f_{\omega^\prime} (\sigma_f -t_2) \frac{ e^{i \omega t_1 - i {\omega^\prime} t_2}}{e^{\beta \omega} - 1} \bigg]\\[0.2cm]
     &  =  \frac{\beta^2}{4  \pi^2   \sqrt{\lambda}}  \int_0^\infty \frac{d \omega}{\omega}  \frac{1}{e^{\beta \omega} - 1} \, \text{Re}\left(f_\omega (\sigma_f -t_1) f^{\,*}_{\omega} (\sigma_f -t_2)  e^{- i \omega (t_1 - t_2)}\right) \, , \label{Xt1_Xt2_expectationLQ}
\end{aligned} &
\end{flalign}
\vspace{0.1cm}

where $\normord{ \hat{a}^{\,\dagger}_\omega \hat{a}_{\omega^\prime}} \, = \, \normord{ \hat{a}_{\omega^\prime} \hat{a}^{\,\dagger}_\omega} \,
= \, \hat{a}^{\,\dagger}_\omega \hat{a}_{\omega^\prime}$ is the normal ordering operator\footnote{Normal ordering of the annihilation and creation operators $\hat{a}_{\omega}, \hat{a}^{\,\dagger}_\omega$ is used to remove logarithmic UV divergences.}, and the definition of the Bose-Einstein distribution (Eq.(\ref{Bose_Einstein_Distribution})) is used in the third equality.
The final equality follows from the property of the Dirac delta function $\int_{0}^\infty dk \, e^{- k x}\, \delta(k - a) \, = \, e^{- a x}$; and the complex conjugate property $\text{Re}(z) \, =\, (z + \bar{z})/2$ has also been used. At specific times, the regularized correlator Eq.(\ref{Xt1_Xt2_expectationLQ}) becomes

\begin{equation}
\langle  \normord{\hat{X}_{\text{end}}(t) \hat{X}_{\text{end}}(0)}  \rangle  \,
= \,\frac{\beta^2}{4 \, \pi^2  \, \sqrt{\lambda}} \, \int_0^\infty \frac{d \omega}{\omega} \, \frac{1}{e^{\beta \omega} - 1} \,
\text{Re}\big(f_\omega (\sigma_f -t) f^{\,*}_{\omega} (\sigma_f)  e^{- i \omega t}\big) \, , \label{Xt_X0_expectationLQ}
\end{equation}
\vspace{0.1cm}

\begin{equation}
\langle  \normord{\hat{X}_{\text{end}}(0) \hat{X}_{\text{end}}(0)}  \rangle  \,
= \,\frac{\beta^2}{4 \, \pi^2  \, \sqrt{\lambda}} \, \int_0^\infty \frac{d \omega}{\omega} \, \frac{1}{e^{\beta \omega} - 1} \,
\text{Re}\big(f_\omega (\sigma_f) f^{\,*}_{\omega} (\sigma_f)\big) \, , \label{X0_X0_expectationLQ}
\end{equation}
\vspace{0.1cm}

\begin{equation}
\langle  \normord{\hat{X}_{\text{end}}(t) \hat{X}_{\text{end}}(t)}  \rangle  \,
= \,\frac{\beta^2}{4 \, \pi^2  \, \sqrt{\lambda}} \, \int_0^\infty \frac{d \omega}{\omega} \, \frac{1}{e^{\beta \omega} - 1} \,
\text{Re}\big(f_\omega (\sigma_f -t) f^{\,*}_{\omega} (\sigma_f-t)\big) \, . \label{Xt_Xt_expectationLQ}
\end{equation}
\vspace{0.1cm}

Inputting Eqs.(\ref{Xt_X0_expectationLQ}, \ref{X0_X0_expectationLQ}, \ref{Xt_Xt_expectationLQ}) into the definition Eq.(\ref{displacement_def}), the string falling endpoint's mean-squared transverse displacement can be calculated as

\begin{flalign}
& \begin{aligned}
     s^2(t) \, \, &  =  \frac{\beta^2}{4  \pi^2   \sqrt{\lambda}}  \int_0^\infty \frac{d \omega}{\omega} \frac{1}{e^{\beta \omega} - 1}  \text{Re}\left(f_\omega (\sigma_f -t) f^{\,*}_{\omega} (\sigma_f-t) + f_\omega (\sigma_f) f^{\,*}_{\omega} (\sigma_f) - 2 f_\omega (\sigma_f -t) f^{\,*}_{\omega} (\sigma_f)  e^{- i \omega t} \right) \\[0.2cm]
     &  =  \frac{\beta^2}{4 \pi^2   \sqrt{\lambda}} \int_0^\infty \frac{d \omega}{\omega} \frac{1}{e^{\beta \omega} - 1}  \big \rvert f_\omega (\sigma_f -t) -  f_{\omega} (\sigma_f)  e^{ i \omega t}\big \rvert^2 \, , \label{st_squaredLQ}
\end{aligned} &
\end{flalign}

where in the second line the complex conjugate property $\text{Re}(z) \, =\, (z + \bar{z})/2$ is again used, and in the final line another complex conjugate property $\rvert z \rvert^2 \, = \, z \, \bar{z}$ is used.

\subsubsection{The Limiting Cases of \texorpdfstring{$s^2(t)$}{s2t} } \label{subsubsecLQ_LimitCases}

The mass (or \textit{virtuality}) of the probe, off-mass-shell light quark\footnote{An off-mass-shell quark does not satisfy Einstein's energy and momentum relation $E^2 = (pc)^2 +(2mc)^2$. In the case of the light quark it starts at $t=0$ as an off-mass-shell particle (corresponding to the initial static string), radiates energy as it travels through the thermal medium (string contracts as the boundary endpoint falls at the local speed of light), and finally stops radiating as it becomes an on-mass-shell particle.} in the boundary theory determines the length of the test string $\ell_0$ in the bulk. Specifically\footnote{See table (\ref{tabletranslation}), page \pageref{tabletranslation}, for a comprehensive dictionary between quantities in the boundary theory and the bulk theory.},

\begin{equation}
\tilde{r}_0 r_H \, \equiv \, r_c \, := \, (r_s + \ell_0)  \, \, \iff \,\, 2 \,\pi\, \alpha^{\prime} (M_{\text{rest}}+ \Delta m) \, , \label{Mass_Length_Rel}
\end{equation}
\vspace{0.1cm}

where $M_{\text{rest}}(T)$ is the static thermal mass of external particle\footnote{In the AdS$_5$/$\mathcal{N}=4$ SYM case, this is the free energy of a quark at rest in $\mathcal{N}=4$ SYM plasma. In the limit of zero temperature it is equal to the QCD Lagrangian quark mass, $m_q$ \cite{tanabashi2018m}.} and $\Delta m(T)$ is the thermal rest mass shift. In the boundary theory this corresponds to the virtuality of the probe light quark being much less than the temperature of the thermal plasma, i.e. $Q^2 \ll T^2$. Equivalently in the dual gravitational picture,

\begin{equation}
Q^2 \, \ll \,  T^2 \,\,\,\, \implies \,\,\,\,  \frac{\lambda  \,\ell_0^2}{4 \, \pi^2 \,l^4} \, \ll \, \frac{1}{\beta^2}\,\,\,\, \implies \,\,\,\,  \frac{4 \lambda}{(d-1)^2}\,\frac{\ell_0^2}{r_H^2} \, \ll \, 1\,\,\,\, \implies \,\,\,\,   \ell_0 \, \ll \, r_H \, , \label{SV_Condition}
\end{equation}
\vspace{0.1cm}

for fixed $\lambda$. In the second equality, the Hawking temperature Eq.(\ref{HawkingTemp}) and the definition of the quark mass (or \textit{virtuality}) Eq.(\ref{FlatSpaceMass}) is used; and in the third equality, Eq.(\ref{HawkingTemp}) is used again. In the small virtuality limit, the minimum radius of the space filling D7-brane is small, i.e. $\tilde{r}_0 \rightarrow 1 + \epsilon$, where $0 < \epsilon \ll 1$ and $\tilde{r}_0$ is defined in Eq.(\ref{dim_r0}). In juxtaposition, the large virtuality limit refers $\tilde{r}_0 \rightarrow \infty$.\\

In the current subsection the string falling endpoint's mean-squared transverse displacement, $s^2(t)$, is examined in a number of different limiting cases. Videlicet, the small virtuality case and the arbitrary virtuality case. In both of these cases, the asymptotically early time and the asymptotically late time limit is explored. For the arbitrary virtuality case, a further limit of small or large virtuality can be considered. It is expected that the large virtuality, early time limit will correspond exactly with the early time limit of the on-mass-shell heavy quark -- since, at early times the light quark's test string is static (i.e. the boundary endpoint has not yet begun to fall).\\

\textit{Small Virtuality}\label{parSmallVirtuality}

In this subsection the behaviour of $s^2(t)$, Eq.(\ref{st_squaredLQ}), in the small virtuality limit is considered; and after obtaining a suitable expression for $s_{\text{small}}^2(t)$, the asymptotic early and late time dynamics are explored. The defining condition for the small virtuality limit is given in Eq.(\ref{SV_Condition}). This equation implies\footnote{ Explicitly, for the case $\sigma = \sigma_f$,

\begin{equation*}
    \coth^2 \left( \frac{r_H \, (r_{s*} + \sigma)}{l^2}\right) = \coth^2 \left( \frac{r_H \, (r_{s*} + \sigma_f)}{l^2}\right) =  \coth^2 \left(\coth^{-1} \left(- \frac{r_s\, + \, \ell_0}{r_H} \right) \right) \approx \, \coth^2 \left(\coth^{-1} \left(- \frac{r_s}{r_H} \right) \right) \stackrel{(\epsilon \rightarrow 0)}{\longrightarrow}\, 1
\end{equation*}
\vspace{0.05cm}

where the second equality follows from using the definition for $\sigma_f$, Eq.(\ref{simga_f_TortoiseCoor}); the approximation follows since $\ell_0 \, \ll \, r_H$; and in taking the limit the definition of the stretched horizon $r_s \, = \, (1+ \epsilon)\,r_H$ (where $0 < \epsilon \ll 1$) is used. The limit follows since, $\coth^{-1} (- (1+ \epsilon)) \rightarrow -\infty$ as $\epsilon \rightarrow 0$; and $\coth^2(x) \rightarrow 1$ as $x \rightarrow -\infty$. Further, for the case $\sigma = 0$,

\begin{equation*}
    \coth^2 \left( \frac{r_H \, (r_{s*} + \sigma)}{l^2}\right) =   \coth^2 \left( \frac{r_H \, r_{s*} }{l^2}\right) = \coth^2 \left(\coth^{-1} \left(- \frac{r_s}{r_H} \right) \right) \stackrel{(\epsilon \rightarrow 0)}{\longrightarrow}\, 1
\end{equation*}
\vspace{0.05cm}

where the first equality follows from the definition for $r_{s*}$, Eq.(\ref{StrHor_TortoiseCoord}).}\\

\begin{equation}
    \coth^2 \left( \frac{r_H \, (r_{s*} + \sigma)}{l^2}\right) \, \rightarrow \, 1, \,\,\,\,\,\, \text{since} \,\,\,\,\,\,  \ell_0 \, \ll \, r_H \, , \label{SV_Condition2}
\end{equation}
\vspace{0.1cm}

where $\sigma \, \in \, [0, \sigma_f]$. Using Eq.(\ref{SV_Condition2}) the equations of motion in AdS$_3$-Schwarzschild, Eq.(\ref{StringEoM_XI_d=3_tsig}), simplifies to the wave equation

\begin{equation}
    - \, \partial_t^2 \, X(t,\sigma)\,+\,  \partial_\sigma^2\,  X(t,\sigma) \, = \, 0 \, . \label{d=3_wave_equation}
\end{equation}
\vspace{0.1cm}

The plane wave solutions to Eq.(\ref{d=3_wave_equation}) are the near-horizon solutions described in Eq.(\ref{IC_horizon_fw_TC}). Hence in the small virtuality limit the in-falling $(+)$ and out-going $(-)$ modes of the general solution $f_\omega (\sigma)$ are given by,\\

\begin{equation}
    f^{(\pm)}_\omega (\sigma) \, =\, e^{\pm i \omega \, ( r_{s*}\, + \, \sigma)}  \, . \label{fModes_SV}
\end{equation}
\vspace{0.1cm}

The general solution $f_\omega (\sigma)$ is a linear combination of the in-falling and out-going modes (defined in Eq.(\ref{gen_solution_sum_modes})). The coefficient $B_\omega$ is fixed in the same way as in subsection (\ref{subsubsecHQ_Displacement}): by imposing a Neumann boundary condition at the falling string endpoint in the radial direction at $t=0$,

\begin{equation}
 \partial_\sigma \, f_\omega(\sigma)\big\rvert_{\sigma = \sigma_f} \, = \,  0 \, . \label{NeumannBC_fixBw_SV}
\end{equation}
\vspace{0.1cm}

Using Eqs.(\ref{gen_solution_sum_modes}, \ref{fModes_SV}), Eq.(\ref{NeumannBC_fixBw_SV}) can be easily solved in Mathematica\footnote{See Mathematica Notebook [b]: \texttt{BrownianMotion.nb}.} to yield

\begin{equation}
 B_\omega \,=\, e^{i 2 \omega \, ( r_{s*}\, + \, \sigma_f)}\, . \label{Bw_SV}
\end{equation}
\vspace{0.1cm}

Hence, the general solution $f_\omega (\sigma)$ is given by,

\begin{flalign}
& \begin{aligned}
 f_\omega(\sigma) \, & = \,  f_\omega^{(+)}(\sigma) \, + \, B_\omega \, f_\omega^{(-)}(\sigma) \\[0.2cm]
 & = \, 2 e^{i \omega \, ( r_{s*}\, + \, \sigma_f)} \, \cos \big(\omega \, (\sigma-\sigma_f) \big) \,, \label{SV_gen_solution}
\end{aligned} &
\end{flalign}
\vspace{0.1cm}

where Eqs.(\ref{fModes_SV}, \ref{Bw_SV}) and the trigonometric identity $\cos (x) \, = \,( e^{i x} + e^{-i x})/2$ are used in the second line. In order to calculate the falling string endpoint's mean-squared transverse displacement in the small virtuality limit, Eq.(\ref{SV_gen_solution}) is substituted into Eq.(\ref{st_squaredLQ}):

\begin{flalign}
& \begin{aligned}
     s^2_{\text{small}}(t) := s^2(t) \big\rvert_{Q^2 \ll T^2}\, &  =  \frac{\beta^2}{4 \pi^2  \sqrt{\lambda}}  \int_0^\infty \frac{d \omega}{\omega}  \frac{1}{e^{\beta \omega} - 1} \Big \rvert 2 e^{i \omega  ( r_{s*} + \sigma_f)}  \cos \big(\omega t \big)   -   2 e^{i \omega \, ( r_{s*} +  \sigma_f)} e^{ i \omega t}\Big \rvert^2 \\[0.2cm]
     & = \frac{\beta^2}{4 \pi^2  \sqrt{\lambda}}  \int_0^\infty \frac{d \omega}{\omega}  \frac{1}{e^{\beta \omega} - 1} \big( 2 - e^{-2 i \omega t} - e^{2 i \omega t} \big) \\[0.2cm]
     &  = \frac{\beta^2}{\pi^2  \sqrt{\lambda}}  \int_0^\infty \frac{d \omega}{\omega}  \frac{1}{e^{\beta \omega} - 1} \sin^2 (\omega t)  \, , \label{st_squaredLQ_SV}
\end{aligned} &
\end{flalign}
\vspace{0.1cm}

where the complex conjugate property $\rvert z \rvert^2 \, = \, z \, \bar{z}$ is used in the second line, and the trigonometric identities $\cos (x) \, = \,( e^{i x} + e^{-i x})/2$ and $\sin ^2(x) \, =\, (1- \cos (2x))/2$ are used in the final line. The integral in Eq.(\ref{st_squaredLQ_SV}) can be solved analytically. Introducing a change of variables

\begin{equation}
x \, := \, \beta \omega \,\,\,\,\,\,\,\,\, \text{and} \,\,\,\,\,\,\,\,\, k \, := \, \frac{t}{\beta} \, , \label{Var_x_k}
\end{equation}
\vspace{0.1cm}

Eq.(\ref{st_squaredLQ_SV}) becomes

\begin{flalign}
& \begin{aligned}
     s^2_{\text{small}}(t) 
     & =  \frac{\beta^2}{\pi^2  \sqrt{\lambda}} \,\sum_{n=1}^\infty \,\int_0^\infty \frac{d x}{x} \,\sin^2 (kx) \,  e^{-nx} \\[0.2cm]
     & = \frac{\beta^2}{4\, \pi^2  \sqrt{\lambda}}  \,\ln \left( \prod_{n=1}^\infty \left( 1+ \frac{4 k^2}{n^2} \right) \right) \\[0.2cm]
     &  = \frac{\beta^2}{4\, \pi^2  \sqrt{\lambda}}  \,\ln \left( \frac{\beta}{2 \, \pi \, t} \sinh \left( \frac{2 \,\pi \, t}{\beta} \right) \right) \, , \label{st_squaredLQ_SV2}
\end{aligned} &
\end{flalign}
\vspace{0.1cm}

where in the second line the geometric series $1/(1-x) \, = \, \sum_{n=0}^\infty x^n$ is used; and in the third line the integral was performed using Mathematica\footnote{See Mathematica Notebook [b]: \texttt{BrownianMotion.nb}.}. In the final line, Mathematica was again used, followed by a change of variables using Eq.(\ref{Var_x_k}). Notice from Eq.(\ref{st_squaredLQ_SV2}) that $\beta$ naturally defines a cross-over time scale between the dynamics of early and late times\footnote{See footnote \ref{footnoteCrossOverTime}, page \pageref{footnoteCrossOverTime}.}, and that the cross-over time is independent of the length of the string $\ell_0$.\\
\vspace{0.3cm}

\textbf{(I) Asymptotic Early Time Dynamics}\\

Since, in the early time limit $t \ll \beta$, Eq.(\ref{st_squaredLQ_SV2}) can be expanded in powers of $k = t/\beta$. For small $k$ a Taylor expansion (about $k=0$) can be performed. Explicitly, $\ln(\sinh(x)/x) = x^2/6 - \,x^4/180 + \, x^6/2835 + \, \mathcal{O}(x^8)$ (where $x =2 k \pi$). Therefore, Eq.(\ref{st_squaredLQ_SV2}) becomes

\begin{equation}
s^2_{\text{small}}(t) \,\stackrel{(t \ll \beta)}{\longrightarrow} \, \frac{\beta^2}{4 \, \pi^2  \sqrt{\lambda}}  \, \left[ \frac{2 \pi^2 \, t^2}{3\,\beta^2} \,+ \, \mathcal{O}\left(\frac{t}{\beta}\right)^4 \right] \, = \,  \frac{t^2}{6  \sqrt{\lambda}} \,+ \, \mathcal{O}\left(\frac{t}{\beta}\right)^4  \, . \label{st_squaredLQ_SV_Early}
\end{equation}
\vspace{0.1cm}

Since $ s^2_{\text{small}}(t) \sim t^2$, the early time dynamics exhibit ballistic behaviour\footnote{The early time limit of the on-mass-shell heavy quark also displays ballistic behaviour (see Eq.(\ref{st_squaredHQ_Early})).} (see Eq.(\ref{DisplacementSqLimit1})).\\

\textbf{(II) Asymptotic Late Time Dynamics}\\

At asymptotically late times ($t \gg \beta$) the integral in  Eq.(\ref{st_squaredLQ_SV2}) can be expanded in powers of $1/k = \beta/t$. Specifically,
\begin{flalign}
& \begin{aligned}
     s^2_{\text{small}}(t) &  \,\,\, = \,\,\,\,\,\,\,   \frac{\beta^2}{4 \, \pi^2  \sqrt{\lambda}}  \, \Big[ \ln \big( \sinh \big( 2 \,\pi \, k \big) \big) \, + \, \ln \Big(\frac{1}{2 \, \pi \, k}\Big) \Big] \\[0.2cm]
     & \,\,\, = \,\,\,\,\,\,\,  \frac{\beta^2}{4 \, \pi^2  \sqrt{\lambda}}  \, \bigg[\frac{2 \,\pi \, t}{\beta}  \,+\, \ln\big(1 \,-\, e^{-4 \,\pi \, k}\big) \, + \, \ln \bigg(\frac{\beta}{4 \, \pi \, t}\bigg) \bigg] \\[0.2cm]
     & \,\stackrel{(\beta \ll t)}{\longrightarrow}\,\, \frac{\beta \,t}{2 \, \pi  \sqrt{\lambda}}  \, + \, \frac{\beta^2}{4 \, \pi^2  \sqrt{\lambda}}  \ln \bigg(\frac{\beta}{4 \, \pi \, t}\bigg) \,+\, \mathcal{O}\Big(\beta^2 \, e^{\frac{-4 \,\pi \, t }{\beta}}\Big)\, , \label{st_squaredLQ_SV_Late}
\end{aligned} &
\end{flalign}
\vspace{0.1cm}

where the second line follows from trigonometric identity $\sinh(x)=(e^x - e^{-x})/2$; and, in the final line, the expansion $\ln{(1-x)} = -x -\,x^2/2 - \,x^3/3\, + \mathcal{O}(x^4)$ (where $x=e^{- 4 \,\pi \, k}$) is performed, since $e^{- 4 \,\pi \, k} \rightarrow 0$ in the late time regime ($k \gg 1$).\\

Since $s^2_{\text{small}}(t) \sim t$, the late time dynamics exhibit diffusive behaviour\footnote{The late time limit of the on-mass-shell heavy quark also displays diffusive behaviour (see Eq.(\ref{st_squaredHQ_Late})).}. From Eq.(\ref{DisplacementSqLimit2}) it is expected that $s^2_{\text{small}}(t) \, =\, 2\, D \, t$ at late times, where $D$ is the diffusion coefficient. Comparing this to Eq.(\ref{st_squaredLQ_SV_Late}), it is easy to see that the diffusion coefficient is given by

\begin{equation}
D_{\text{LQ}}^{\text{AdS$_3$}} \, = \, \frac{\beta}{4 \, \pi  \sqrt{\lambda}}\, . \label{DifCoeff_LQ}
\end{equation}
\vspace{0.1cm}

The diffusion coefficient for the on-mass-shell heavy quark is given by Eq.(\ref{DifCoeff_HQ}). By comparing Eq.(\ref{DifCoeff_LQ}) with Eq.(\ref{DifCoeff_HQ}), notice that an off-mass-shell small virtuality light quark which is initially at rest in a strongly-coupled thermal plasma has a diffusion coefficient related to that of a massive on-mass-shell heavy quark by

\begin{equation}
D_{\text{LQ}}^{\text{AdS$_3$}} \, = \, \frac{1}{2}\,D_{\text{HQ}}^{\text{AdS$_3$}}\, . \label{DifCoeff_comparison}
\end{equation}
\vspace{0.1cm}

The factor of $1/2$ may arise through the differences in partitioning the worldsheet for the heavy and light quark test strings. This is explored further in subsection (\ref{subsecTRAIL_GenAdSd}).\\

\textit{Arbitrary Virtuality}\\

\textbf{(I) Asymptotic Early Time Dynamics}\\

Now consider a probe light quark with arbitrary virtuality. The behaviour of $s^2(t)$, Eq.(\ref{st_squaredLQ}), at asymptotically early times is considered. Since $t \ll \beta$, Eq.(\ref{st_squaredLQ}) can be expanded in powers of $k = t/\beta$,

\begin{flalign}
& \begin{aligned}
      \big \rvert f_\omega (\sigma_f -t) -  f_{\omega} (\sigma_f)  e^{ i \omega t}\big \rvert^2  
       &\stackrel{(t \ll \beta)}{\longrightarrow}\,\left( f_\omega (\sigma_f) -  f_{\omega} (\sigma_f)\,  e^{ i \omega t}\right) \left(f_\omega^* (\sigma_f) -  f_{\omega}^* (\sigma_f) \, e^{- i \omega t} \right)\\[0.2cm]
     & = \big \rvert f_\omega (\sigma_f) \big \rvert^2 \left(2 \,-\, e^{- i \omega t} \,-\, e^{ i \omega t}  \right) \\[0.2cm]
     & =  \big \rvert f_\omega (\sigma_f) \big \rvert^2 \left(\omega^2 t^2   +  \mathcal{O}(\omega t)^4\right) \\[0.2cm]
     & = \,  4 \,\frac{1+\nu^2}{1+\tilde{r}_0\nu^2} \left(\omega^2 t^2   +  \mathcal{O}(\omega t)^4\right) \, , \label{LQ_Early_Times}
\end{aligned} &
\end{flalign}
\vspace{0.1cm}

where the first line follows since in the early time regime $t \ll \beta$ implies $f_\omega (\sigma_f -t) \approx f_{\omega} (\sigma_f)$; and in the third line $e^{i\omega t}$ and $e^{- i\omega t}$ are Taylor expanded. The final line follows\footnote{Mathematica is used to simplify the algebra. This computation (and others in this section) are explicitly shown in Mathematica Notebook [b] (\texttt{BrownianMotion.nb}). For access, see appendix (\ref{appCode}).} from
using Eqs.(\ref{TC_nu_fw_r_ODE_Solution}, \ref{gen_solution_sum_modes}, \ref{B_w_TC}) to calculate $\big \rvert f_\omega (\sigma_f) \big \rvert^2 =  f_\omega (\sigma_f) f_\omega^{\,*} (\sigma_f)$. Inputting Eq.(\ref{LQ_Early_Times}) into Eq.(\ref{st_squaredLQ}), yields


\begin{equation}
      s^2(t) \big \rvert_{t \ll \beta} \, =  \, \frac{4\, t^2}{\sqrt{\lambda}} \, \int_0^\infty \, d \nu \, \frac{\nu}{e^{2 \pi \nu} - 1}  \, \frac{1+\nu^2 }{ 1+\tilde{r}_0\nu^2}  \, , \label{st_LQ_Early_Times}
\end{equation}
\vspace{0.1cm}

where a change of variables was performed using  Eqs.(\ref{HawkingTemp}, \ref{dim_angular_f_def}). Eq.(\ref{st_LQ_Early_Times}) is evaluated by breaking the integral into two parts and recognising that these terms are the series expansion to $\mathcal{O}(k)^0$ (around $k=0$, where $k=t/\beta$) of the second derivatives with respect to $k$ of the integrals $I_1$ and $I_2$ given in Eq.(\ref{I1}) and Eq.(\ref{I2}) respectively.
To clarify, the second derivative of $I_1$ and $I_2$ are given by

\begin{flalign}
& \begin{aligned}
     & I_1   \, = \, 4 \int_0^{\infty} \, \frac{d x}{x (1 + a^2 x^2)} \, \frac{\sin^2 \big(\frac{kx}{2} \big)}{e^{x} -1}\\[0.2cm]
     & \Rightarrow  \frac{\partial^2 I_1}{\partial k^2} \bigg \rvert_{k=0}  = \, 4 \pi \int_0^{\infty} \, d \nu \, \frac{\nu}{e^{2 \pi \nu} -1} \, \frac{1}{1+\tilde{r}_0^2 \,\nu^2}\, , \label{derI_1}
\end{aligned} &
\end{flalign}
\vspace{0.1cm}

and,

\begin{flalign}
& \begin{aligned}
     & I_2   = \, 4 \int_0^{\infty} \, \frac{d x}{x } \, \frac{\sin^2 \big(\frac{kx}{2} \big)}{e^{x} -1}\\[0.2cm]
     & \Rightarrow  \frac{\partial^2 I_2}{\partial k^2} \bigg \rvert_{k=0}  = \, 4 \pi \int_0^{\infty} \, d \nu \, \frac{\nu}{e^{2 \pi \nu} -1}  \,  , \label{derI_2}
\end{aligned} &
\end{flalign}
\vspace{0.1cm}

where $x= 2 \pi \nu$, $a = \tilde{r}_0/(2\pi)$ and $k =t/ \beta$. Explicitly rewriting Eq.(\ref{st_LQ_Early_Times}) in terms of $\partial_k^2 I_1$ and $\partial_k^2 I_2$ yields

\begin{flalign}
& \begin{aligned}
      s^2(t) \big \rvert_{t \ll \beta} &  = \, \frac{4\, t^2}{\sqrt{\lambda}}  \int_0^\infty \, d \nu  \left(\frac{\tilde{r}_0^2 -1}{\tilde{r}_0^2}  \frac{ \nu}{e^{2 \pi \nu} - 1}  \, \frac{1}{ 1+\tilde{r}_0^2 \nu^2} \, + \, \frac{1}{\tilde{r}_0^2} \frac{ \nu}{e^{2 \pi \nu} - 1} \right)\\[0.2cm]
     & =  \, \frac{ t^2}{\pi \sqrt{\lambda}}  \left(\frac{\tilde{r}_0^2 -1}{\tilde{r}_0^2} \,\frac{\partial^2 I_1}{\partial k^2} \bigg \rvert_{k=0} + \, \frac{1}{\tilde{r}_0^2}\,\frac{\partial^2 I_2}{\partial k^2} \bigg \rvert_{k=0} \right)  \, . \label{st_LQ_Early_Times_I1I2}
\end{aligned} &
\end{flalign}
\vspace{0.1cm}
The solution to the integrals $I_1$ and $I_2$ are given by Eqs.(\ref{I1Sol}, \ref{I2Sol}) respectively.
Taking the second derivative with respect to $k$ followed by a series expansion to $\mathcal{O}(k)^0$ (around $k=0$) for each of these solutions, yields

\begin{equation}
\frac{\partial^2 I_1}{\partial k^2} \bigg \rvert_{k=0}  \, = \, \frac{2 \left(-\pi ^2 \ln \left(\frac{\tilde{r}_0}{2 \pi }\right)+\pi ^2 \ln \left(\frac{2 \pi }{\tilde{r}_0}\right)+\pi ^3 \cot \left(\frac{\pi }{\tilde{r}_0}\right)-\pi ^2 \psi\left(1+\frac{1}{\tilde{r}_0}\right)-\pi ^2 \psi\left(1-\frac{1}{\tilde{r}_0}\right)-2 \pi ^2 \ln (2 \pi )\right)}{\tilde{r}_0^2} \, , \label{derI1_sol}
\end{equation}
\vspace{0.1cm}

where it was recognised in the early time case\footnote{For the physically relevant solution $t \geq 0$, the early time limit $t \ll \beta$ implies $\beta > 0$. Hence, $k = t/\beta \geq 0$, and $\rvert k \rvert = k$.} that $\rvert k \rvert = k$, and $\psi(z)$ is the digamma function. In addition,

\begin{equation}
\frac{\partial^2 I_2}{\partial k^2} \bigg \rvert_{k=0}  \, = \, \frac{\pi ^2}{3}\, . \label{derI2_sol}
\end{equation}
\vspace{0.1cm}

Inputting Eq.(\ref{derI1_sol}) and Eq.(\ref{derI2_sol}) into Eq.(\ref{st_LQ_Early_Times_I1I2}), yields\footnote{Eq.(\ref{st_LQ_Early_Times_2}) agrees with Eq.(3.42) in Moerman \textit{et al.}'s \cite{moerman2016semi} if the the digamma function is rewritten in terms of the harmonic numbers. Specifically,  $\psi(n) = H_{(n-1)} - \gamma_{E}$, where $\gamma_{E}$ is the Euler-Mascheroni constant and $n$ is a positive integer. It is worth remarking that this form was avoid here since $\big(1+\frac{1}{\tilde{r}_0}\big)$ and $\big(1-\frac{1}{\tilde{r}_0}\big)$ are not necessarily positive integers.}

\begin{equation}
s^2(t) \big \rvert_{t \ll \beta}  \, = \, \frac{t^2}{6 \tilde{r}_0^4 \sqrt{\lambda}}\,  \left[\tilde{r}_0^2+6 \left(1-\tilde{r}_0^2\right) \left(2 \ln (\tilde{r}_0)-\pi  \cot \left(\frac{\pi }{\tilde{r}_0}\right)+\psi\left(1+\frac{1}{\tilde{r}_0}\right)+\psi\left(1-\frac{1}{\tilde{r}_0}\right)\right)\right]\, . \label{st_LQ_Early_Times_2}
\end{equation}
\vspace{0.1cm}

The limit of small or large virtuality can now be considered.

\begin{enumerate}[(i)]
    \item \textbf{Small Virtuality Limit ($\tilde{r}_0 \rightarrow 1 + \epsilon $)}\\
    Using Mathematica to take the limit of Eq.(\ref{st_LQ_Early_Times_2}) as $\tilde{r}_0 \rightarrow 1 + \epsilon $, yields

    \begin{equation}
    s^2(t) \big \rvert_{t \ll \beta}  \,  \stackrel{(\tilde{r}_0 \rightarrow 1+ \epsilon)}{\longrightarrow}  \, \frac{t^2}{6 \sqrt{\lambda}} \,+\, \mathcal{O}(\epsilon)  \, . \label{st_LQ_Early_Times_SV}
    \end{equation}
    \vspace{0.1cm}

    This agrees exactly with Eq.(\ref{st_squaredLQ_SV_Early}), i.e. taking the early time limit followed by the small virtuality limit is equivalent to taking small virtuality limit followed by the early time limit -- a necessary consistency check.

    \item \textbf{Large Virtuality Limit ($\tilde{r}_0 \rightarrow \infty $)}\\
    Since $\tilde{r}_0 \gg 0$, define a variable $y=1/\tilde{r}_0$ (where $y \ll 0$) so that a Taylor expansion in $y$ can be performed. Rewriting Eq.(\ref{st_LQ_Early_Times_2}) in terms of $y$ and performing a series expansion (around $y=0$) in Mathematica, yields

    \begin{equation}
    s^2(t) \big \rvert_{t \ll \beta}  \,  \stackrel{(\tilde{r}_0 \rightarrow \infty)}{\longrightarrow}  \, \frac{t^2}{\tilde{r}_0 \sqrt{\lambda}} \,+\, \mathcal{O}\left(\frac{1}{\tilde{r}_0 }\right)^2  \, . \label{st_LQ_Early_Times_LV}
    \end{equation}
    \vspace{0.1cm}

    As expected, the large virtuality early time behaviour corresponds exactly with the early time limit of the on-mass-shell heavy quark (i.e. the static string solution at early times, Eq.(\ref{st_squaredHQ_Late})).
\end{enumerate}
\vspace{0.1cm}

\textbf{(II) Asymptotic Late Time Dynamics}\\

The behaviour of $s^2(t)$, Eq.(\ref{st_squaredLQ}), at asymptotically late times is considered. Defining two dimensionless quantities

\begin{equation}
u \, := \, \frac{\beta}{t} \, = \, \frac{1}{k}\,\,\,\,\,\,\,\,\, \text{and} \,\,\,\,\,\,\,\,\, z \, := \, \omega t \, , \label{Var_x_z}
\end{equation}
\vspace{0.1cm}
the late time dynamics of $s^2(t)$ can be examined. At asymptotically late times $t\gg \beta$, or equivalently $u \ll 1$. The integral Eq.(\ref{st_squaredLQ}) can then be rewritten

\begin{equation}
s^2(t)   = \frac{\beta^2}{4 \, \pi^2   \sqrt{\lambda}}  \int_0^\infty \frac{d z}{z}  \frac{1}{e^{z u} - 1}  \big \rvert f_{z/t} \left(\sigma_f - \beta/u \right) -  f_{z/t} (\sigma_f)  e^{ i z}\big \rvert^2 \, . \label{st_squared_LQ_x_z}
\end{equation}
\vspace{0.1cm}

Using Eqs.(\ref{TC_nu_fw_r_ODE_Solution}, \ref{gen_solution_sum_modes}, \ref{B_w_TC}) to simplify the integrand, yields

 \begin{flalign}
& \begin{aligned}
     \big \rvert f_{z/t} \left(\sigma_f - \beta/u \right) -  f_{z/t} (\sigma_f)  e^{ i z}\big \rvert^2
    &  =  \left(f_{z/t} \left(\sigma_f - \beta/u \right) -  f_{z/t} (\sigma_f)  e^{ i z}\right) \, \left(f_{z/t}^{\,*} \left(\sigma_f - \beta/u \right) -  f_{z/t}^{\,*} (\sigma_f)  e^{ i z} \right) \\[0.2cm]
      &\stackrel{(u \ll 1)}{\longrightarrow}  \, 4 \sin^2(\omega t) \,+\, \mathcal{O}\left(\beta/t\right)\, , \label{LQ_Late_Times}
 \end{aligned} &
 \end{flalign}
 \vspace{0.1cm}

where the final line is written in terms of $\omega$, $\beta$ and $t$ using Eq.(\ref{Var_x_z}). Further, it has also been recognised that at asymptotically late times the test string's falling endpoint is in the near-horizon region, therefore $r \approx r_H$. Substituting Eq.(\ref{LQ_Late_Times}) into Eq.(\ref{st_squared_LQ_x_z}), yields

\begin{equation}
s^2(t) \big\rvert_{t \gg \beta}   = \frac{\beta^2}{\pi^2  \sqrt{\lambda}}  \int_0^\infty \frac{d \omega}{\omega}  \frac{1}{e^{\beta \omega} - 1} \sin^2 (\omega t) \, \equiv \, s^2_{\text{small}}(t) \, . \label{LQ_Late_Times_HQ}
\end{equation}
\vspace{0.1cm}

Therefore, the late time dynamics of an arbitrary virtuality light quark in a thermal plasma is diffusive $s^2_{\text{small}}(t) \sim t$, and the diffusion coefficient is given by Eq.(\ref{DifCoeff_LQ}). This statement can be made due to a key insight of Moerman \textit{et al.} \cite{moerman2016semi} -- that the integral in Eq.(\ref{LQ_Late_Times_HQ}) is identical to $s^2_{\text{small}}(t)$, the string falling endpoint’s mean-squared transverse displacement in the limit of small virtualities (see subsection (\ref{parSmallVirtuality})). This allows them to conclude that the behaviour of arbitrary virtuality quarks at asymptotically late times (i.e. in the near-horizon region) is encoded in the small virtuality case. This is intuitive, since the length of the Limp Noodle is arbitrarily short at asymptotically late times\footnote{Comparing Eq.(\ref{SV_Condition2}) with Eq.(\ref{near_horizon_limit}) the limit of small virtualities appears to be equivalent to the near-horizon limit.}. Since $s^2_{\text{small}}(t;d)$ can be solved for any dimensions $d \geq 3$ in AdS$_d$-Schwarzschild, the late time behaviour (and therefore the diffusion coefficient) of Limp Noodles with arbitrary length (equivalently, light quarks of arbitrary virtuality) in any number of transverse spatial dimensions are able to be determined. This generalisation to AdS$_d$-Schwarzschild is presented in the following subsection.

\subsection{Generalising to  \texorpdfstring{AdS$_d$-Schwarzschild}{AdSd-Schwarzschild}} \label{subsubsecLQ_GenAdSd}

To determine $s^2_{\text{small}}(t;d)$ the AdS$_d$-Schwarzschild metric (Eq.(\ref{AdSd_Metric})) is examined in the near-horizon limit\footnote{As discussed in subsection (\ref{secAdSSpaceTime}), `dropping the 1' from the function $H(r)$ in the AdS$_d \times S^d$ metric corresponds to being in the near-horizon region (or in the `deep-throat') of AdS$_d \times S^d$ spacetime -- see figure (\ref{fig:DeepThroat}), page \pageref{fig:DeepThroat}. Here, a further limit is considered -- letting $r = (1+ \tilde{\epsilon}) r_H$ and studying the near-horizon geometry of the black-brane metric -- which is the near-horizon limit refered to in this subsection.
}, which surmounts to letting $r = (1+ \tilde{\epsilon}) r_H$, where ($0 < \epsilon \leq \tilde{\epsilon} \ll 1 $) and series expanding each term individually around $\tilde{\epsilon} =0$.\\

Setting $r = (1+ \tilde{\epsilon}) r_H$, the AdS$_d$-Schwarzschild metric Eq.(\ref{AdSd_Metric}) becomes

\begin{equation}
ds_d^2 = \frac{r_H^2}{l^2} (1+\tilde{\epsilon})^2 \, d\Vec{X}_{I}^2 \, - \, \frac{r_H^2}{l^2} (1+\tilde{\epsilon})^2 \left(1- \Big(\frac{1}{1+\tilde{\epsilon}} \Big)^{d-1} \right)\, dt^2  \,+\, \frac{l^2}{(1+\tilde{\epsilon})^2}\frac{1}{1- \left(1/(1+\tilde{\epsilon}) \right)^{d-1} }\,d\tilde{\epsilon}^2 \, , \label{AdSd_Metric_1+epsilon_rH}
\end{equation}
\vspace{0.1cm}
where each term is now able to be expanded in $\tilde{\epsilon}$ (around $\tilde{\epsilon}=0$). Specifically\footnote{In Eq.(\ref{dt_expand}), all terms greater than $\mathcal{O}\left(\tilde{\epsilon}\right)^2$ disappear when $d=3$.},

\begin{equation}
\frac{r_H^2}{l^2} (1+\tilde{\epsilon})^2 \, d\Vec{X}_{I}^2  \, = \,\left( \frac{r_H^2}{l^2}\,+\,\frac{2 r_H^2 \tilde{\epsilon}  }{l^2}\,+\,\frac{r_H^2\tilde{\epsilon}^2 }{l^2} \right) d\Vec{X}_{I}^2\, , \label{dx_expand}
\end{equation}

\begin{equation}
\frac{r_H^2}{l^2} (1+\tilde{\epsilon})^2 \left(1- \Big(\frac{1}{1+\tilde{\epsilon}} \Big)^{d-1} \right)\, dt^2  \, \longrightarrow \,\left( -\frac{(d-1) r_H^2 \tilde{\epsilon}  }{l^2}\,+\,\frac{\left(d^2-5 d+4\right)r_H^2 \tilde{\epsilon}^2 }{2 l^2}\,+\,\mathcal{O}\left(\tilde{\epsilon}\right)^3 \right) dt^2  \, , \label{dt_expand}
\end{equation}

\begin{equation}
\frac{l^2}{(1+\tilde{\epsilon})^2}\frac{1}{1- \left(1/(1+\tilde{\epsilon}) \right)^{d-1} }\,d\tilde{\epsilon}^2 \, \longrightarrow \,\left( \frac{l^2}{(d-1) \tilde{\epsilon} }\,+\,\frac{(d-4) l^2}{2 (d-1)}\,+\,\frac{\left(d^2-14 d+36\right)  l^2\,\tilde{\epsilon}}{12 (d-1)}\,+\, \mathcal{O}\left(\tilde{\epsilon}\right)^2 \right) d\tilde{\epsilon}^2\, . \label{depsilon_expand}
\end{equation}
\vspace{0.1cm}

The question now arises as how to truncate the series expansions (Eqs.(\ref{dx_expand})-(\ref{depsilon_expand})), such that the $dt^2, d\Vec{X}_{I}^2$ and $d\tilde{\epsilon}^2$ terms in the metric are all to the same $\mathcal{O}(\tilde{\epsilon})$. To help clarify this task of consistent truncation, there are actually only two options available:
\begin{enumerate}[(i)]
    \item assume $dt^2 \sim \mathcal{O}(\tilde{\epsilon})^0$ and $d\Vec{X}_{I}^2 \sim \mathcal{O}(\tilde{\epsilon})^0$, such that the AdS$_d$-Schwarzschild metric becomes

    \begin{equation}
    ds_d^2 = \left( \frac{r_H^2}{l^2}\,+\,\frac{2 r_H^2 \tilde{\epsilon}}{l^2} \right) d\Vec{X}_{I}^2 \, -\, \frac{(d-1) r_H^2 \tilde{\epsilon}  }{l^2} \, dt^2  \,+\, \frac{l^2}{(d-1) \tilde{\epsilon} }\,d\tilde{\epsilon}^2 \,+\, \mathcal{O}\left(\tilde{\epsilon}\right)^2 \, , \label{AdSd_Metric_OptI}
    \end{equation}
    \vspace{0.1cm}

    \item or, assume $dt^2 \sim \mathcal{O}(\tilde{\epsilon})^0$ and $d\Vec{X}_{I}^2 \sim \mathcal{O}(\tilde{\epsilon})$, such that the AdS$_d$-Schwarzschild metric becomes

    \begin{equation}
    ds_d^2 =  \frac{r_H^2}{l^2} \, d\Vec{X}_{I}^2 \, -\, \frac{(d-1) r_H^2 \tilde{\epsilon}  }{l^2} \, dt^2  \,+\, \frac{l^2}{(d-1) \tilde{\epsilon} }\,d\tilde{\epsilon}^2 \,+\, \mathcal{O}\left(\tilde{\epsilon}\right)^2 \, . \label{AdSd_Metric_OptII}
    \end{equation}
    \vspace{0.1cm}
\end{enumerate}

These are the only options, since if higher order terms in the $dt^2$ or $d\tilde{\epsilon}^2$ series are included, the different signs in Eqs.(\ref{dt_expand}, \ref{depsilon_expand}) will result in a metric which is not conformally flat in ($t, r_*$) coordinates. The leading order metric includes only the first term in the $d\tilde{\epsilon}^2$ series expansion (Eq.(\ref{depsilon_expand})), therefore -- to be consistent -- all terms in the leading order metric must be of $\mathcal{O}(\tilde{\epsilon})$. One of the ramifications of this is that it can not be assumed that $d\Vec{X}_{I} \sim \mathcal{O}(\tilde{\epsilon})$ (or higher) at leading order, since the $d\Vec{X}_{I}^2$ term in the metric would then disappear.\\

To decide between options (i) or (ii), establish which option ensures that $t,\, \Vec{X}_{I}, \, \tilde{\epsilon}$ are \textit{on equal footing}, and that the $dt, \, d\Vec{X}_{I}, \, d\tilde{\epsilon}$ terms scale the same way as $\tilde{\epsilon} \rightarrow \lambda \,\tilde{\epsilon}$. This will ensure the resulting near-horizon AdS$_d$-Schwarzschild metric has directions which encode transverse fluctuations\footnote{It is clear that the near-horizon AdS$_d$-Schwarzschild metric needs to have transverse directions since these are present in the near-horizon AdS$_3$-Schwarzschild metric (which was found directly from inverting the tortoise coordinate in $d=3$, and calculating the transverse string equations of motion without series expanding the parameters -- see subsection (\ref{subsecLQ_TSDynamics})).}. For option (ii), considering the scaling $\tilde{\epsilon} \rightarrow \lambda \,\tilde{\epsilon}$, the $d \tilde{\epsilon}^2$ term in Eq.(\ref{AdSd_Metric_OptII}) scales as
\begin{equation*}
\frac{l^2}{(d-1) \tilde{\epsilon} }\,d\tilde{\epsilon}^2 \rightarrow \lambda \,\frac{l^2}{(d-1) \tilde{\epsilon} }\,d\tilde{\epsilon}^2  \,.
\end{equation*}
Under the same transformation, in order for the $d t^2$  term in Eq.(\ref{AdSd_Metric_OptII}) to scale as
\begin{equation*}
  \frac{- (d-1) r_H^2 \tilde{\epsilon}  }{l^2} \, dt^2 \rightarrow \lambda \, \frac{- (d-1) r_H^2 \tilde{\epsilon}  }{l^2} \, dt^2  \,,
\end{equation*}
the temporal variable needs to scale like $t \rightarrow t$. Similarly, in order for the $d \Vec{X}_{I}^2$  term in Eq.(\ref{AdSd_Metric_OptII}) to scale as
\begin{equation*}
 \frac{r_H^2}{l^2} \, d\Vec{X}_{I}^2 \rightarrow \lambda \, \frac{r_H^2}{l^2} \, d\Vec{X}_{I}^2  \,,
\end{equation*}
the transverse variable needs to scale like $\Vec{X}_{I} \rightarrow\sqrt{\lambda} \, \Vec{X}_{I}$\footnote{Accepting that $t$ and $\Vec{X}_{I}$ scale differently isn't far-fetched considering this is the near-horizon region of the black-brane.}. A consistent way to transform $t$ and $\Vec{X}_{I}$, such that the $dt, \, d\Vec{X}_{I}, \, d\tilde{\epsilon}$ terms in the AdS$_d$-Schwarzschild metric scale the same way under the transformation $\tilde{\epsilon} \rightarrow \lambda \,\tilde{\epsilon}$, can not be found for option (i). This strongly suggests that option (ii) is the correct way to truncate the near-horizon AdS$_d$-Schwarzschild metric.\\

Intuitive arguments aside, a rigorous mathematical approach to proving $dt^2 \sim \mathcal{O}(\tilde{\epsilon})^0$ and $d\Vec{X}_{I}^2 \sim \mathcal{O}(\tilde{\epsilon})$ can be applied in the $d=3$ case. Since in AdS$_3$-Schwarzschild the tortoise coordinate is invertible, the transverse string equations of motion and its solution can be exactly found. The equations of motion and its solution can be series expanded in $\tilde{\epsilon}$ in the near-horizon region. Order by order these should match with the equations of motion and its solution derived from the near-horizon AdS$_3$-Schwarzschild (found by expanding the metric in $\tilde{\epsilon}$ and assuming $d t^2 \sim \mathcal{O}(\tilde{\epsilon})^0$ and $d x^2 \sim \mathcal{O}(\tilde{\epsilon})$)\footnote{This is done in detail Mathematica Notebook [c] (\texttt{NearHorizonAdSd.nb}) -- see appendix (\ref{appCode}) for access.}. If this method of matching terms proves to be consistent, the order of $\tilde{\epsilon}$ of the temporal and transverse directions will be known, and this remains true for a general number of dimensions, $d$. The near-horizon AdS$_d$-Schwarzschild metric will, therefore, also be correctly known. The remainder of the subsection will follow the standard method of finding the equations of motion for the transverse fluctuations, its solution, the string falling endpoint’s mean-squared transverse displacement $s^2(t)$, and -- finally -- the diffusion coefficient for the light quark in AdS$_d$-Schwarzschild.

\subsubsection{Expanding the \texorpdfstring{AdS$_3$-Schwarzschild}{AdS3-Schwarzschild} Transverse Equations of Motion and its Solution in the Near-Horizon Region}\label{subsubsecExpandEoMAdS3}

The AdS$_3$-Schwarzschild metric, the resulting transverse equation of motion, and its solution are given by Eqs.(\ref{AdS3_Metric}, \ref{StringEoM_XI_d=3_tr}, \ref{ModeExpansion_X}) respectively\footnote{The full solution to the transverse equation of motion is given by Eq.(\ref{gen_solution_sum_modes}), where the constant $B_\omega$ is defined in Eq.(\ref{B_w_TC}) and the modes $f^{(\pm)}_{\omega}(r)$ are given in Eq.(\ref{nu_fw_r_ODE_Solution}). The linearly independent solutions $f^{(+)}_{\omega}(r)$ or $f^{(-)}_{\omega}(r)$ are sufficient to work with; since, if an ODE is solved by $f^{(+)}_{\omega}(r)$ or $f^{(-)}_{\omega}(r)$, it will also be solved by the superposition of those solutions, $f_{\omega}(r)$. Therefore, in order to simplify the algebra, take $f_{\omega}(r) = f^{(+)}_{\omega}(r)$ (equivalently $f_{\omega}(r) = f^{(-)}_{\omega}(r)$ could have been chosen).}. In the near-horizon region (which corresponds to setting $r = (1+ \tilde{\epsilon})r_H$), Eq.(\ref{AdS3_Metric}) becomes

\begin{equation}
ds_3^2 = -\frac{r_H^2}{l^2} \, \tilde{\epsilon}\,  (2 + \tilde{\epsilon}) \, dt^2 \, + \, \frac{r_H^2}{l^2} \, (1+\tilde{\epsilon}^2)\, dx^2 \,+\, \frac{l^2}{\tilde{\epsilon}\,  (2 + \tilde{\epsilon}) }\,d\tilde{\epsilon}^2  \, , \label{AdS3_Metric_NH_noexpand}
\end{equation}

which is equivalent to Eq.(\ref{AdSd_Metric_1+epsilon_rH}) when $d=3$;
Eq.(\ref{StringEoM_XI_d=3_tr}) becomes

 \begin{flalign}
& \begin{aligned}
     0\, &= \, - \, \partial_t^2 X(t,\tilde{\epsilon})\,+\, \frac{r_H^2}{l^4}\,  \frac{\tilde{\epsilon}\,  (2 + \tilde{\epsilon})}{(1+\tilde{\epsilon})^2}\,\partial_{\tilde{\epsilon}} \left( \tilde{\epsilon} \, (1+\tilde{\epsilon})^2  (2 + \tilde{\epsilon})\, {\partial_{\tilde{\epsilon}}} X(t,\tilde{\epsilon})\right)\\[0.2cm]
     & = -\, \partial_t^2  X_{\text{reg}}(t,\tilde{\epsilon}) +\frac{(\tilde{\epsilon} +2)}{4 l^4 (\tilde{\epsilon} +1)} \Big[-l^2 \omega  \left(l^2 \omega  (\tilde{\epsilon} +1) (\tilde{\epsilon} +2)-2 i r_H \tilde{\epsilon}  (3 \tilde{\epsilon} +5)\right) X_{\text{reg}}(t,\tilde{\epsilon}) \\[0.2cm]
     &\,\,\,\,\,\,\, + \, 4 r_H \tilde{\epsilon}   \left(i l^2 \omega  (\tilde{\epsilon} +1) (\tilde{\epsilon} +2)+4 r_H \tilde{\epsilon}  (\tilde{\epsilon} +2)+2 r_H\right)\partial_{\tilde{\epsilon}}  X_{\text{reg}}(t,\tilde{\epsilon})  \, + \, 4 r_H^2 \tilde{\epsilon}^2  (\tilde{\epsilon} +1) (\tilde{\epsilon} +2) \partial^2_{\tilde{\epsilon}}  X_{\text{reg}}(t,\tilde{\epsilon})\Big] \,, \label{StringEoM_XI_d=3_NH}
 \end{aligned} &
 \end{flalign}
 \vspace{0.1cm}

where $r=(1+\tilde{\epsilon})r_H$ and $\partial_{\tilde{\epsilon}} = r_H \partial_r$ are used in the first line; and Eq.(\ref{ModeExpansion_X}) becomes

\begin{equation}
X(t,\tilde{\epsilon}) \, = \, \left(\frac{\tilde{\epsilon} }{\tilde{\epsilon} +2}\right)^{\frac{i l^2 \omega }{2 rH}} \frac{i\,l^2 \omega +r_H \tilde{\epsilon} +r_H}{(\tilde{\epsilon} +1) \left(r_H+ i\,l^2 \omega \right)} \, e^{- i \omega t} \, =: \, \tilde{\epsilon}^{-\frac{i l^2 \omega }{2 rH}} \, X_{\text{reg}}(t,\tilde{\epsilon}) \,, \label{StringSol_XI_d=3_NH}
\end{equation}
\vspace{0.1cm}

where $f_{\omega}(r) = f^{(+)}_{\omega}(r)$ (which is defined in Eq.(\ref{nu_fw_r_ODE_Solution})), and the regular part of the solution is given by $X_{\text{reg}}(t,\tilde{\epsilon})$. The regular equation of motion (Eq.(\ref{StringEoM_XI_d=3_NH})) can now be series expanded in $\tilde{\epsilon}$ (around $\tilde{\epsilon} = 0$). However, in order to perform this expansion, an ansatz for the expansion of $X_{\text{reg}}(t,\tilde{\epsilon})$ is required. Up to $\mathcal{O}(\tilde{\epsilon})^2$, the simple ansatz $X_{\text{reg}}(t,\tilde{\epsilon}) = x_0(t) + \tilde{\epsilon} \,x_1(t) + \tilde{\epsilon}^2\, x_2(t)$ can be used. The regular equation of motion then becomes

 \begin{flalign}
& \begin{aligned}
     0\, &=    \left(-x_0''(t)-\omega ^2 x_0(t)\right) +\tilde{\epsilon} \left(\frac{\omega \left(-l^2 \omega +5 i r_H\right)}{l^2}x_0(t) + \frac{\left(2 r_H+i l^2 \omega \right)^2}{l^4}x_1(t) - x_1''(t)\right)\\[0.2cm]
      & +\tilde{\epsilon}^2 \left(-\frac{\omega \left(l^2 \omega -2 i r_H\right)}{4 l^2}x_0(t) +\frac{ \left( -l^4\omega ^2 +9 i l^2 r_H \omega +14 r_H^2\right)}{l^4}x_1(t)  +\frac{ \left(4 r_H+i l^2 \omega \right)^2}{l^4}x_2(t)  -x_2''(t)\right) + \mathcal{O}(\tilde{\epsilon})^3 , \label{StringEoM_XI_d=3_NH_Expand}
 \end{aligned} &
 \end{flalign}
 \vspace{0.1cm}

where $x''(t) = d^2 x/ dt^2$. Series expanding the regular part of the solution (Eq.(\ref{StringSol_XI_d=3_NH})) in $\tilde{\epsilon}$ (around $\tilde{\epsilon} = 0$), yields
 \begin{flalign}
& \begin{aligned}
     X_{\text{reg}}(t,\tilde{\epsilon})  & =  e^{-i t \omega } \, 2^{-\frac{i l^2 \omega }{2 rH}} \left( 1  - \frac{i l^2 \omega  \left(l^2 \omega -5 i r_H\right)}{4\, r_H \left(l^2 \omega -i r_H\right)} \tilde{\epsilon}  - \frac{l^2 \omega  \left(l^4 \omega ^2-11 i l^2 r_H \omega -34 r_H^2\right)}{32 \,r_H^2 \left(l^2 \omega -i r_H\right)}\tilde{\epsilon}^2 + \mathcal{O}(\tilde{\epsilon})^3\right)\\[0.2cm]
     & =: x_0(t) + \tilde{\epsilon} \,x_1(t) + \tilde{\epsilon}^2\, x_2(t) + \mathcal{O}(\tilde{\epsilon})^3\,. \label{StringSol_XI_d=3_NH_Expand}
 \end{aligned} &
 \end{flalign}
 \vspace{0.1cm}

As a consistency check, notice that the solution Eq.(\ref{StringSol_XI_d=3_NH_Expand}) solves the transverse equation of motion Eq.(\ref{StringEoM_XI_d=3_NH_Expand}) at each respective order of $\tilde{\epsilon}$.

\subsubsection{Expanding the \texorpdfstring{AdS$_3$-Schwarzschild}{AdS3-Schwarzschild} Metric in the Near-Horizon Region} \label{ExpandNHAdS3}

The transverse equation of motion and its solution which have been derived from the AdS$_3$-Schwarzschild metric and then series expanded to yield the near-horizon limit (Eqs.(\ref{StringEoM_XI_d=3_NH_Expand}, \ref{StringSol_XI_d=3_NH_Expand})) should match exactly with the transverse equation of motion and its solution derived from the series expanded, near-horizon AdS$_3$-Schwarzschild metric. Going on physical intuition, it seems like a good starting point would be to assume $d t^2 \sim \mathcal{O}(\tilde{\epsilon})^0$ and $d x^2 \sim \mathcal{O}(\tilde{\epsilon})$ and check if the transverse equation of motion and its solution derived from expanding this metric in the near-horizon limit match with those found in the previous subsection. To this end, assuming $d t^2 \sim \mathcal{O}(\tilde{\epsilon})^0$ and $d x^2 \sim \mathcal{O}(\tilde{\epsilon})$ results in an AdS$_3$-Schwarzschild metric

\begin{equation}
ds_3^2 = \frac{r_H^2}{l^2} \, d x^2 \, -\, \frac{2 r_H^2 \tilde{\epsilon}  }{l^2} \, dt^2  \,+\, \frac{l^2}{2 \tilde{\epsilon} }\,d\tilde{\epsilon}^2 \,+\, \mathcal{O}\left(\tilde{\epsilon}\right)^2 \, , \label{AdS3_Metric_NH}
\end{equation}
\vspace{0.1cm}

where Eq.(\ref{AdSd_Metric_OptII}) for $d=3$ has been used. The near-horizon tortoise coordinate, defined by setting $r = (1+ \tilde{\epsilon}) r_H$ in the definition of the tortoise coordinate Eq.(\ref{gen_d_TortoiseCoord}), can also be expanded and truncated around $\tilde{\epsilon}=0$. To leading order it is sufficient to take\footnote{It's proven that the leading order tortoise coordinate is given by Eq.(\ref{d_3_TortoiseCoord_NH}) in appendix (\ref{appTortoiseCoord}).}
\begin{equation}
\tilde{\epsilon}_* := \frac{l^2}{2\,r_H^2} \ln(\tilde{\epsilon}) \, , \label{d_3_TortoiseCoord_NH}
\end{equation}

where $r_* = r_H \tilde{\epsilon}_*$, and the number of dimensions $d=3$ has been set posthumously. Barring an integration constant of $\mathcal{O}(\tilde{\epsilon})^0$ (specifically $-l^2/(2r_H) \ln(2)$), Eq.(\ref{d_3_TortoiseCoord_NH}) agrees with expanding and truncating Eq.(\ref{d=3_TortoiseCoord}) after setting $r = (1+ \tilde{\epsilon}) r_H$\footnote{See Mathematica Notebook [c] (\texttt{NearHorizonAdSd.nb}) for details: access in appendix (\ref{appCode}).}. Inverting and calculating the differential yields

\begin{equation}
\tilde{\epsilon} = e^{ \frac{2\, r_H^2}{l^2}\, \tilde{\epsilon}_*}\,, \,\,\,\,\,\,\,\,\,\,\, \text{and} \,\,\,\,\,\,\,\,\,\,\,\, d\tilde{\epsilon} = \frac{2\, r_H^2}{l^2} \,e^{ \frac{2\, r_H^2}{l^2} \, \tilde{\epsilon}_*} \, d\tilde{\epsilon}_* \, . \label{d_3_TortoiseCoord_InverseDiff_NH}
\end{equation}
 \vspace{0.1cm}
 
 Eqs.(\ref{d_3_TortoiseCoord_NH}, \ref{d_3_TortoiseCoord_InverseDiff_NH}) are used to transform the expanded and truncated metric into ($t, \tilde{\epsilon}_*$) coordinates. The metric Eq.(\ref{AdS3_Metric_NH}) becomes
 
\begin{flalign}
 & \begin{aligned}
 ds_3^2 & =  2\,\frac{r_H^2}{l^2} \, \,e^{\frac{2\,r_H^2}{l^2} \, \tilde{\epsilon}_*} \left( -dt^2  \, + \,r_H^2\, d\tilde{\epsilon}_*^2 \right)  \, + \, \frac{r_H^2}{l^2} \,dx^2 \\[0.2cm]
 & =  2\, \frac{r_H^2}{l^2}  \,e^{\frac{2\, r_H \, r_*}{l^2} } \left( -dt^2  \, + \, dr_*^2 \right)  \, + \, \frac{r_H^2}{l^2} \,dx^2 \, , \label{AdS3_Metric_NH_t_epsilonstar}
 \end{aligned} &
 \end{flalign}
 \vspace{0.1cm}
where the final line is a conformally flat description of near-horizon AdS$_3$-Schwarzschild in $(t, r_*)$ coordinates. It follows from the near-horizon definition $r = (1+ \tilde{\epsilon})r_H$, and Eqs.(\ref{d_3_TortoiseCoord_NH}, \ref{d_3_TortoiseCoord_InverseDiff_NH}) that

\begin{equation}
r = r_H \, \left( 1 \, +\, e^{\frac{2\, r_H\, r_*}{l^2}} \right)\,, \,\,\,\,\,\,\,\,\,\,\, \text{and} \,\,\,\,\,\,\,\,\,\,\,\, r_* = \frac{l^2}{2\, r_H} \, \ln \left( \frac{r}{r_H} -  1 \right) \, . \label{d_3_TortoiseCoord_r_NH}
\end{equation}
\vspace{0.1cm}

As discussed in subsection (\ref{subsubsecLQinAdS3}), in the conformally flat $(t, r_*)$ coordinate system the boundary conditions for the Limp Noodle in AdS$_3$-Schwarzschild are given by  Eqs.(\ref{Dirichlet_BC_AdS3_TortoiseCoord}, \ref{Neumann_BC_AdS3_TortoiseCoord}), which are respectively analogous to the boundary conditions for the Limp Noodle in $\mathbb{R}^{1+1}$, Eqs.(\ref{Dirichlet_BC_R11}, \ref{Neumann_BC_R11}). Hence, the leading order solution for the Limp Noodle in AdS$_3$-Schwarzschild in conformal ($t, r_*$) coordinates can be written down (Eq.(\ref{LN_AdS3Sol_TortoiseCoord})). Converting then to $(t, r)$ coordinates, the string solution for the $d=3$ Limp Noodle in the near-horizon limit at leading order is given by

\begin{equation}
X^{\mu}_\text{\text{AdS$_3$-Sch}} (t,\sigma) \, = \, \left( t, \, \begin{Bmatrix}
 r_H \Big(1 \, + \, e^{\frac{2\, r_H}{l^2} \, \left( r_{s*} \,+\, \sigma \right)} \Big), \,\,\,\,\,\,\,\,\,\,\,\,\,\,\,\,\,\,\,\,\,\,\, \text{if} \,\,\,\ (t, \sigma) \, \in \, \mathcal{M}_1 \\
 r_H \Big(1 \,+ \,  e^{\frac{2\, r_H}{l^2} \, \left( r_{s*} \,+\, \sigma_f \, - \, t \right)} \Big), \,\,\,\,\,\,\,\,\,\,\,\, \text{if} \,\,\,\ (t, \sigma) \, \in \, \mathcal{M}_2
  \end{Bmatrix}\, , \,0 \right)^{\mu} \,, \label{LN_AdS3Solution_NH}
\end{equation}
\vspace{0.1cm}

where Eq.(\ref{d_3_TortoiseCoord_r_NH}) is used. The position of the fixed string endpoint attached to the stretched horizon is given by
\begin{equation}
r_{s*} \,= \, \frac{l^2}{2 \, r_H} \, \ln \Big(\frac{r_s}{r_H} \, - \, 1 \Big) \,\,\,\,\,\,\,\,\,\, \text{or, equivalently} \,\,\,\,\,\,\,\,\,\, \tilde{\epsilon}_{s*}\, = \, \frac{l^2}{2\, r_H^2} \, \ln(\epsilon) \,,  \label{TRUC_rs_TortoiseCoordinate_d=3}
\end{equation}
\vspace{0.1cm}

since the stretched horizon is defined as $r_s = (1+ \epsilon)r_H$, where $0 < \epsilon \leq \tilde{\epsilon} \ll 1$. The length of the string in tortoise coordinates is given by

\begin{equation}
\sigma_f \,= \, \frac{l^2}{2 \, r_H} \, \ln \Big(\frac{r_s\,+\, \ell_0}{r_H} \, - \, 1 \Big) \, - \, r_{s*} \,\,\,\,\,\,\,\,\,\, \text{or, equivalently} \,\,\,\,\,\,\,\,\,\, \sigma_f \,= \, \frac{l^2}{2\, r_H} \, \ln\left(\epsilon + \frac{\ell_0}{r_H}\right) \, - \,r_H \, \tilde{\epsilon}_{s*} \,,  \label{TRUC_simga_f_TortoiseCoordinate_d=3}
\end{equation}
\vspace{0.1cm}

and the $\mathcal{M}_1$ and $\mathcal{M}_2$ parameter space regions are still given by Eq.(\ref{M1M2_ParameterSpace}). Using the definition of the near-horizon limit ($r = (1+\tilde{\epsilon})r_H$) and $r_* = r_H \tilde{\epsilon}_*$, the leading order string solution Eq.(\ref{LN_AdS3Solution_NH}) can be converted into $(t, \tilde{\epsilon})$ coordinates. Specifically,

\begin{equation}
X^{\mu}_\text{\text{AdS$_3$-Sch}} (t, \sigma) \, = \, \left( t, \, \begin{Bmatrix}
 e^{\frac{2\,r_H}{l^2} \, \left( r_H \,\tilde{\epsilon}_{s*} \,+\, \sigma \right)}, \,\,\,\,\,\,\,\,\,\,\,\,\,\,\,\,\,\,\,\,\,\,\, \text{if} \,\,\,\ (t, \sigma) \, \in \, \mathcal{M}_1 \\
  e^{\frac{2\, r_H}{l^2}  \, \left( r_H \, \tilde{\epsilon}_{s*} \,+\, \sigma_f \, - \, t \right)} , \,\,\,\,\,\,\,\,\,\,\,\, \text{if} \,\,\,\ (t, \sigma) \, \in \, \mathcal{M}_2
  \end{Bmatrix}\, , \,0 \right)^{\mu} \,. \label{LN_AdS3Solution_NH_epsilon}
\end{equation}
\vspace{0.1cm}

Therefore, the identification $\tilde{\epsilon} \,= \,e^{\frac{2\,r_H}{l^2}  \, \left( r_H \, \tilde{\epsilon}_{s*} \,+\, \sigma \right)}$ is made. The equation of motion for the transverse fluctuations in the near-horizon limit of AdS$_3$-Schwarzschild can be derived from Eq.(\ref{StringEoM_XI_d=3}),

\begin{flalign}
& \begin{aligned}
0 \, &= \,  \partial_a \left(  - \frac{1}{2 \pi \alpha'} \, \left(\sqrt{-g}\, g^{ab} \, G_{IJ} \right) \big\rvert_{X^\mu_0} {\partial_b} X^J\right) \\[0.2cm]
 &= \, \partial_a  \left(  \left(\sqrt{-g}\, g^{ab} \, \frac{r_H^2}{l^2} \right) \bigg\rvert_{X^\mu_0} {\partial_b} X(t, \tilde{\epsilon})\right) \,, \label{StringEoM_XI_d=3_NH_trunc}
\end{aligned} &
\end{flalign}
\vspace{0.1cm}

where the spacetime metric in the near-horizon limit $G_{\mu \nu}$ is given explicitly by

\begin{equation}
G_{\mu \nu} \, =\,
  \begin{bmatrix}
  -\frac{2\,r_H^2\, \tilde{\epsilon}}{l^2} & 0 & 0  \\
    0 & \frac{l^2}{2\, \tilde{\epsilon}} & 0 \\
    0 & 0 & \frac{r_H^2}{l^2}
  \end{bmatrix} \, ,\label{AdS3_NH_TargetMatrix}
\end{equation}
\vspace{0.3cm}

where $\mu,\nu$ index over the directions $(t,\tilde{\epsilon},x)$. The explicit entries of the induced worldsheet metric are calculated from Eq.(\ref{InducedMetric})\footnote{This calculation is similar to Eq.(\ref{appIndMetric2}) where the induced metric is calculated explicitly in ($t,r$) coordinates (see appendix (\ref{appEnergyCurvedSpace})).}. In the near-horizon limit, for $d=3$, the leading order induced metric in ($t,\tilde{\epsilon}$) coordinates is given by 

\begin{equation}
g_{a b}\big\rvert_{X^\mu_0}\, := \,
  \begin{bmatrix}
  g_{tt} & g_{t \sigma} \\
  g_{\sigma t} & g_{\sigma \sigma}
  \end{bmatrix} \Bigg\rvert_{X^\mu_0} \, = \,
  \begin{bmatrix}
    G_{tt} & 0 \\
  0 &  \tilde{\epsilon}'^2 \,G_{\tilde{\epsilon} \tilde{\epsilon}}
  \end{bmatrix} \, = \,
  \begin{bmatrix}
     -\frac{2\,r_H^2\,\tilde{\epsilon}}{l^2}  & 0 \\
  0 & \tilde{\epsilon}'^2 \,\frac{l^2}{2\, \tilde{\epsilon}}
  \end{bmatrix} \,, \label{InducedMatrix_LO_d=3_NH}
\end{equation}
\vspace{0.1cm}

where Eq.(\ref{AdS3_NH_TargetMatrix}) is used, and $\tilde{\epsilon}' = \partial_\sigma \tilde{\epsilon} $. Hence, its inverse is given by

\begin{equation}
g^{a b}\big\rvert_{X^\mu_0}\, := \,
  \begin{bmatrix}
  g^{tt} & g^{t \sigma} \\
  g^{\sigma t} & g^{\sigma \sigma}
  \end{bmatrix} \Bigg\rvert_{X^\mu_0} \, = \,
  \frac{1}{\det \big(g_{a b}\rvert_{X^\mu_0} \big) }\,\begin{bmatrix}
   \tilde{\epsilon}'^2 \, G_{\tilde{\epsilon} \tilde{\epsilon}} & 0 \\
  0 & G_{tt}
  \end{bmatrix} \, = \,
  \begin{bmatrix}
    -\frac{l^2}{2\,r_H^2 \, \tilde{\epsilon}}  & 0 \\
  0 & \frac{2\, \tilde{\epsilon}}{l^2 \, \tilde{\epsilon}'^2 }
  \end{bmatrix} \,, \label{InducedMatrix_LO_d=3_INVERSE_NH}
\end{equation}
\vspace{0.3cm}

where $\det \big(g_{a b}\rvert_{X^\mu_0} \big)$ is calculated from Eq.(\ref{InducedMatrix_LO_d=3_NH}),

\begin{equation}
g\rvert_{X^\mu_0} \, \equiv \, \det \big(g_{a b}\rvert_{X^\mu_0} \big) \,= \, \tilde{\epsilon}'^2 \, G_{\tilde{\epsilon} \tilde{\epsilon}}\,G_{tt} \, = \,  \bigg(\tilde{\epsilon}'^2 \,\frac{l^2}{2\, \tilde{\epsilon}}  \bigg) \, \bigg( -\frac{2\,r_H^2}{l^2} \, \tilde{\epsilon} \bigg) \, = \, -r_H^2 \, \tilde{\epsilon}'^2 \,. \label{InducedMatrix_LO_d=3_det_NH}
\end{equation}
\vspace{0.1cm}

Expanding the indices of Eq.(\ref{StringEoM_XI_d=3_NH_trunc}) using Eqs.(\ref{InducedMatrix_LO_d=3_NH})-(\ref{InducedMatrix_LO_d=3_det_NH}), the transverse equation of motion become

\begin{flalign}
& \begin{aligned}
0 \, & = \, \partial_{\sigma} \left(  \bigg(\sqrt{-g}\, g^{\sigma \sigma} \, \frac{r_H^2}{l^2} \bigg) \bigg\rvert_{X^\mu_0}\,\, {\partial_\sigma} X(t,\sigma)\right) \,+\, \partial_t  \left(  \bigg(\sqrt{-g}\, g^{tt} \, \frac{r_H^2}{l^2} \bigg) \bigg\rvert_{X^\mu_0}\,\, {\partial_t} X(t,\sigma)\right)\\[0.2cm]
 &= \, {\partial_\sigma}  \left(r_H \tilde{\epsilon}' \left(\frac{2\, \tilde{\epsilon}}{l^2 \, \tilde{\epsilon}'^2 } \right) \, \frac{r_H^2}{l^2}\, {\partial_\sigma} X(t,\sigma)\right) \,+\, \partial_t \left( r_H \tilde{\epsilon}' \left(- \frac{l^2}{2\, r_H^2 \, \tilde{\epsilon}} \right) \, \frac{r_H^2}{l^2}\, {\partial_t} X(t,\sigma)\right)\, \\[0.2cm]
 & = \, - \, \partial_t^2 X(t,\sigma)\,+\, \frac{4\, r_H^2}{l^4}  \,  \tilde{\epsilon}\, \frac{1}{\partial_\sigma \tilde{\epsilon}}\,\partial_\sigma \left(\frac{1}{\partial_\sigma \tilde{\epsilon}}\, \tilde{\epsilon}\, {\partial_\sigma} X(t,\sigma) \right)\, \\[0.2cm]
 & = \, - \, \partial_t^2 X(t,\tilde{\epsilon})\,+\, \frac{4\, r_H^2}{l^4}  \,  \tilde{\epsilon}\,\partial_{\tilde{\epsilon}} \big( \tilde{\epsilon}\, {\partial_{\tilde{\epsilon}}} X(t,\tilde{\epsilon})\big) \,. \label{StringEoM_XI_d=3_NH_epsilon}
\end{aligned} &
\end{flalign}
\vspace{0.1cm}

where the last line is written completely in terms of near-horizon spacetime coordinates ($t, \tilde{\epsilon}$) by differentiating the identification $\tilde{\epsilon} \,= \,e^{\frac{2\,r_H}{l^2}  \, \left( r_H \, \tilde{\epsilon}_{s*} \,+\, \sigma \right)}$, 


\begin{equation}
\partial_\sigma \, =  \,2 \, \frac{r_H}{l^2}\,e^{2\frac{r_H}{l^2} \, \big( r_H \, \tilde{\epsilon}_{s*} \,+\, \sigma \big)} \, \partial_{\tilde{\epsilon}} \,=\, \,2 \, \frac{r_H}{l^2}\, \tilde{\epsilon} \, \partial_{\tilde{\epsilon}} \,. \label{d=3_NH_Iden_differential}
\end{equation}
\vspace{0.1cm}
    
In terms of $X_{\text{reg}}(t,\tilde{\epsilon})$ the equation of motion is given by

\begin{equation}
0 \, = \, - \partial_t^2 X_{\text{reg}}(t,\tilde{\epsilon}) + \frac{4 r_H \tilde{\epsilon}  \left(r_H \tilde{\epsilon}  \, \partial_{\tilde{\epsilon}}^2 X_{\text{reg}}(t,\tilde{\epsilon} )+\left(r_H+i l^2 \omega \right) \partial_{\tilde{\epsilon}} X_{\text{reg}}(t,\tilde{\epsilon} )\right)}{l^4}   -\omega ^2 X_{\text{reg}}(t,\tilde{\epsilon} )\,, \label{StringEoM_XREG_d=3_NH}
\end{equation}
\vspace{0.1cm}

where the definition of the regular solution in Eq.(\ref{StringSol_XI_d=3_NH}) has been used. In order to check whether, at leading order, the regular equation of motion agrees with the series expanded regular equation of motion (Eq.(\ref{StringEoM_XI_d=3_NH_Expand})), the leading order ansatz $X_{\text{reg}}(t,\tilde{\epsilon}) = x_0(t)$ should be used. The leading order contribution to Eq.(\ref{StringEoM_XREG_d=3_NH}) then becomes

 \begin{equation}
     0\, = \,  -x_0''(t)-\omega^2 x_0(t)  \, , \label{StringEoM_XREG_d=3_NH_Expand}
 \end{equation}
 \vspace{0.1cm}

which agrees exactly with the leading order contribution of Eq.(\ref{StringEoM_XI_d=3_NH_Expand}). This proves that the expanded near-horizon metric is equivalent to the near-horizon metric at leading order in AdS$_3$-Schwarzschild. The expanded near-horizon metric given in Eq.(\ref{AdS3_Metric_NH}) is therefore correct, and $d t^2 \sim \mathcal{O}(\tilde{\epsilon})^0$ and $d x^2 \sim \mathcal{O}(\tilde{\epsilon})$.
As a final check, the calculations in this subsection are repeated assuming $d t^2 \sim \mathcal{O}(\tilde{\epsilon})^0$ and $d x^2 \sim \mathcal{O}(\tilde{\epsilon})^0$. The equation of motion found is,

 \begin{flalign}
& \begin{aligned}
     0\, & \,\,\, = \,\,\, - \, \partial_t^2 \, X(t,\tilde{\epsilon})\,+\, \frac{4\, r_H^2}{l^4}  \,  \frac{\tilde{\epsilon}}{\left(1+\,2\,\tilde{\epsilon}\right)}\,\partial_{\tilde{\epsilon}} \big( \tilde{\epsilon}\,\left(1+\,2\,\tilde{\epsilon}\right)\, {\partial_{\tilde{\epsilon}}} X(t,\tilde{\epsilon})\big) \\[0.2cm]
     & \,\,\, = \,\,\, -\, \partial_t^2 X_{\text{reg}}(t,\tilde{\epsilon}) +  \frac{4 r_H^2}{l^4} \tilde{\epsilon} ^2 \partial^2_{\tilde{\epsilon}} X_{\text{reg}}(t,\tilde{\epsilon} )
     + \frac{4 r_H \tilde{\epsilon} \, \partial_{\tilde{\epsilon}} X_{\text{reg}}(t,\tilde{\epsilon} ) \left(i l^2 \omega  (2 \tilde{\epsilon} +1)+4 r_H \tilde{\epsilon} +r_H\right)}{l^4 (2 \tilde{\epsilon} +1)} \\[0.2cm]
     & \,\,\,\,\,\,\,\,\,\,\,\,\, - \, \frac{\omega  X_{\text{reg}}(t,\tilde{\epsilon} ) \left(l^2 \omega  (2 \tilde{\epsilon} +1)-4 i r_H \tilde{\epsilon} \right)}{l^2 (2 \tilde{\epsilon} +1)} \\[0.2cm]
     & \stackrel{LO}{\longrightarrow} \, - \, x_0''(t) + \left(-\omega +\frac{4 i r_H}{l^2 }\frac{ \tilde{\epsilon} }{(2 \tilde{\epsilon} +1)}\right) \omega  x_0(t) \,, \label{WrongOption}
 \end{aligned} &
 \end{flalign}
 \vspace{0.1cm}

which does not match the leading order contribution of Eq.(\ref{StringEoM_XI_d=3_NH_Expand}).

\subsubsection{The Diffusion Constant in \texorpdfstring{AdS$_d$-Schwarzschild}{AdSd-Schwarzschild}}

In the previous two subsections it was proven that in order for a near-horizon expansion of the metric to be consistent, the leading order metric is taken to $\mathcal{O}(\tilde{\epsilon})$ -- where $d t^2 \sim \mathcal{O}(\tilde{\epsilon})^0$ and $d x^2 \sim \mathcal{O}(\tilde{\epsilon})$ (equivalently $t \sim \mathcal{O}(\tilde{\epsilon})^0$ and $x \sim \mathcal{O}(\sqrt{\tilde{\epsilon}})$). Generalising to $d$ dimensions this holds true, where $X_I$ (instead of $x$) denotes the transverse directions in AdS$_d$-Schwarzschild. Hence the leading order, near-horizon AdS$_d$-Schwarzschild metric is given by Eq.(\ref{AdSd_Metric_OptII}).\\

From the near-horizon metric, which is conformal in $(t, r_*)$ coordinates, the leading order string solution followed by the transverse equations of motion can be found. A similar method to subsection (\ref{ExpandNHAdS3}) is followed. The near-horizon tortoise coordinate, defined by setting $r = (1+ \tilde{\epsilon}) r_H$ in the definition of the tortoise coordinate Eq.(\ref{gen_d_TortoiseCoord}), can  be expanded and truncated around $\tilde{\epsilon}=0$. To leading order it is sufficient to take\footnote{It's proven that the leading order tortoise coordinate is given by Eq.(\ref{d_TortoiseCoord_NH}) in appendix (\ref{appTortoiseCoord}).}

\begin{equation}
\tilde{\epsilon}_* :=  \frac{l^2}{r_H^2} \, \frac{\ln(\tilde{\epsilon})}{(d-1)} \, , \label{d_TortoiseCoord_NH}
\end{equation}
\vspace{0.1cm}

where $r_* = r_H \tilde{\epsilon}_*$. Inverting and calculating the differential yields

\begin{equation}
\tilde{\epsilon} = e^{\frac{r_H^2}{l^2} \, (d-1) \, \tilde{\epsilon}_*} \,, \,\,\,\,\,\,\,\,\,\,\, \text{and} \,\,\,\,\,\,\,\,\,\,\,\, d\tilde{\epsilon} = \frac{r_H^2}{l^2} \, (d-1) \,e^{\frac{r_H^2}{l^2} \, (d-1) \, \tilde{\epsilon}_*} \, d\tilde{\epsilon}_* \, . \label{d_TortoiseCoord_InverseDiff_NH}
\end{equation}
\vspace{0.1cm}

The definition of the near-horizon tortoise coordinate, its inverse and differential (Eqs.(\ref{d_TortoiseCoord_NH}, \ref{d_TortoiseCoord_InverseDiff_NH})) can be used to convert the metric Eq.(\ref{AdSd_Metric_OptII}) into $(t, \tilde{\epsilon}_*)$ coordinates

\begin{flalign}
 & \begin{aligned}
 ds_d^2 & =  \frac{r_H^2}{l^2} \, (d-1) \,e^{\frac{r_H^2}{l^2} \, (d-1) \, \tilde{\epsilon}_*} \left( -dt^2  \, + \,r_H^2\, d\tilde{\epsilon}_*^2 \right)  \, + \, \frac{r_H^2}{l^2} \,d\Vec{X}_{I}^2 \\[0.2cm]
 & =  \frac{r_H^2}{l^2} \, (d-1) \,e^{\frac{r_H \, r_*}{l^2} \, (d-1)} \left( -dt^2  \, + \, dr_*^2 \right)  \, + \, \frac{r_H^2}{l^2} \,d\Vec{X}_{I}^2 \, . \label{AdS_Metric_NH_t_epsilonstar}
 \end{aligned} &
 \end{flalign}
 \vspace{0.1cm}
From the final line in Eq.(\ref{AdS_Metric_NH_t_epsilonstar}) notice that the metric -- like the near-horizon AdS$_3$-Schwarzschild metric (Eq.(\ref{AdS3_Metric_NH_t_epsilonstar})) -- is conformal in $(t, r_*)$ coordinates. It follows from the near-horizon definition $r = (1+ \tilde{\epsilon})r_H$, and Eqs.(\ref{d_TortoiseCoord_NH}, \ref{d_TortoiseCoord_InverseDiff_NH}) that

\begin{equation}
r = r_H \, \left( 1 \, +\, e^{\frac{r_H\, r_*}{l^2} \, (d-1)} \right)\,, \,\,\,\,\,\,\,\,\,\,\, \text{and} \,\,\,\,\,\,\,\,\,\,\,\, r_* = \frac{l^2}{(d-1)\, r_H} \, \ln \left( \frac{r}{r_H} \,- \, 1 \right) \, , \label{d_TortoiseCoord_r_NH}
\end{equation}
\vspace{0.1cm}

which reduces down to Eq.(\ref{d_3_TortoiseCoord_r_NH}) for $d=3$. From the metric Eq.(\ref{AdS_Metric_NH_t_epsilonstar}) the method laid out in subsection (\ref{subsubsecLQinAdS3}) can be followed in order to write down the leading order solution for the Limp Noodle in AdS$_d$-Schwarzschild in conformal ($t, r_*$) coordinates, which is given by Eq.(\ref{LN_AdS3Sol_TortoiseCoord})\footnote{This equation -- derived from the worldsheet parameter space partitioning of the test string in $\mathbb{R}^{1,1}$, Eq.(\ref{LN_FlatSpaceSolution}) -- is independent of $d$, the number of spacetime dimensions.}. Converting then to $(t, r)$ coordinates, the string solution for the Limp Noodle in the near-horizon limit at leading order is given by

\begin{equation}
X^{\mu}_\text{\text{AdS$_d$-Sch}} (t, \sigma) \, = \, \left( t, \, \begin{Bmatrix}
 r_H \Big(1 \, + \, e^{\frac{ r_H}{l^2} \, (d-1) \,  \left( r_{s*} \,+\, \sigma \right)} \Big), \,\,\,\,\,\,\,\,\,\,\,\,\,\,\,\,\,\,\,\,\,\,\, \text{if} \,\,\,\ (t, \sigma) \, \in \, \mathcal{M}_1 \\
 r_H \Big(1 \,+ \,  e^{\frac{r_H}{l^2} \, (d-1) \,  \left( r_{s*} \,+\, \sigma_f \, - \, t \right)} \Big), \,\,\,\,\,\,\,\,\,\,\,\, \text{if} \,\,\,\ (t, \sigma) \, \in \, \mathcal{M}_2
  \end{Bmatrix}\, , \,0 \right)^{\mu} \,, \label{LN_AdSSolution_NH}
\end{equation}
\vspace{0.1cm}

where Eq.(\ref{d_TortoiseCoord_r_NH}) is used\footnote{It will be much harder to find next-to-leading order corrections to the string solution in the near-horizon limit, since if the near-horizon metric includes higher order terms ($\mathcal{O}(\tilde{\epsilon})^2$ terms) it will no longer be conformal in ($t, r_*$) coordinates.}. The position of the fixed string endpoint attached to the stretched horizon is given by

\begin{equation}
r_{s*} \,= \, \frac{l^2}{(d-1) \, r_H} \, \ln \Big(\frac{r_s}{r_H} \, - \, 1 \Big) \,\,\,\,\,\,\,\,\,\, \text{or, equivalently} \,\,\,\,\,\,\,\,\,\, \tilde{\epsilon}_{s*}\, = \, \frac{l^2}{(d-1) \, r_H} \, \ln(\epsilon) \,,  \label{rs_TortoiseCoordinate_d}
\end{equation}
\vspace{0.1cm}

since the stretched horizon is defined as $r_s = (1+ \epsilon)r_H$, where $0 < \epsilon \leq \tilde{\epsilon} \ll 1$. The length of the string in tortoise coordinates is given by

\begin{equation}
\sigma_f \,= \, \frac{l^2}{(d-1) \, r_H} \, \ln \Big(\frac{r_s\,+\, \ell_0}{r_H} \, - \, 1 \Big) \, - \, r_{s*} \,\,\,\,\,\,\,\,\text{or, equivalently} \,\,\,\,\,\,\,\, \sigma_f \,= \, \frac{l^2}{(d-1) \, r_H} \, \ln\left(\epsilon + \frac{\ell_0}{r_H}\right) \, - \,r_H \, \tilde{\epsilon}_{s*} \,,  \label{sigma_f_TortoiseCoordinate_d}
\end{equation}
\vspace{0.1cm}

and the $\mathcal{M}_1$ and $\mathcal{M}_2$ parameter space regions are still given by Eq.(\ref{M1M2_ParameterSpace}). Using the definition of the near-horizon limit ($r = (1+\tilde{\epsilon})r_H$) and $r_* = r_H \tilde{\epsilon}_*$, the leading order string solution Eq.(\ref{LN_AdSSolution_NH}) can be converted into $(t, \tilde{\epsilon})$ coordinates. Specifically,

\begin{equation}
X^{\mu}_\text{\text{AdS$_d$-Sch}} (t, \sigma) \, = \, \left( t, \, \begin{Bmatrix}
 e^{\frac{r_H}{l^2} \, (d-1) \, \left( r_H \,\tilde{\epsilon}_{s*} \,+\, \sigma \right)}, \,\,\,\,\,\,\,\,\,\,\,\,\,\,\,\,\,\,\,\,\,\,\, \text{if} \,\,\,\ (t, \sigma) \, \in \, \mathcal{M}_1 \\
  e^{\frac{ r_H}{l^2}  \, (d-1) \, \left( r_H \, \tilde{\epsilon}_{s*} \,+\, \sigma_f \, - \, t \right)} , \,\,\,\,\,\,\,\,\,\,\,\, \text{if} \,\,\,\ (t, \sigma) \, \in \, \mathcal{M}_2
  \end{Bmatrix}\, , \,0 \right)^{\mu} \,, \label{LN_AdSSolution_NH_epsilon}
\end{equation}
\vspace{0.1cm}

where $\tilde{\epsilon}_{s*}$ and $\sigma_f$ are defined in Eqs.(\ref{rs_TortoiseCoordinate_d}, \ref{sigma_f_TortoiseCoordinate_d}). Therefore, the identification $\tilde{\epsilon} \,= \,e^{\frac{r_H}{l^2}   \, (d-1) \, \left( r_H \, \tilde{\epsilon}_{s*} \,+\, \sigma \right)}$ is made.  The equations of motion for the transverse fluctuations in the near-horizon limit of AdS$_d$-Schwarzschild can be derived from Eq.(\ref{StringEoM_XI}), and are equivalent to Eq.(\ref{StringEoM_XI_d=3_NH_trunc}) with the generalised spacetime metric in the near-horizon limit ($G_{\mu \nu}$) given by

\begin{equation}
G_{\mu \nu} \, =\,
  \begin{bmatrix}
   -\frac{r_H^2}{l^2}\,(d-1) \, \tilde{\epsilon}  & 0 & 0 & 0 & 0 & 0 \\
    0 & \frac{l^2}{(d-1)}\,\frac{1}{ \tilde{\epsilon}}  & 0 & 0 & 0 & 0 \\
    0 & 0 & \frac{r_H^2}{l^2} & 0 & 0 & 0 \\
    0 & 0 & 0 & \frac{r_H^2}{l^2}\ & \ddots & 0 \\
    0 & 0 & 0 & \ddots & \ddots & 0 \\
    0 & 0 & 0 & 0 & 0 &\frac{r_H^2}{l^2}
  \end{bmatrix} \, ,\label{AdS_NH_TargetMatrix}
\end{equation}
\vspace{0.1cm}
where $\mu,\nu$ index over $\left(t,\, r,\, 2,\, ...\, ,\, d-1\right)$. The explicit entries of the induced worldsheet metric are calculated from Eq.(\ref{InducedMetric}). In the near-horizon limit, the leading order induced metric in ($t,\tilde{\epsilon}$) coordinates is given by

\begin{equation}
g_{a b}\big\rvert_{X^\mu_0}\, := \,
  \begin{bmatrix}
  g_{tt} & g_{t \sigma} \\
  g_{\sigma t} & g_{\sigma \sigma}
  \end{bmatrix} \Bigg\rvert_{X^\mu_0} \, = \,
  \begin{bmatrix}
    G_{tt} & 0 \\
  0 &  \tilde{\epsilon}'^2 \, G_{\tilde{\epsilon} \tilde{\epsilon}}
  \end{bmatrix} \, = \,
  \begin{bmatrix}
     -\frac{r_H^2}{l^2}\,(d-1) \, \tilde{\epsilon}  & 0 \\
  0 & \tilde{\epsilon}'^2 \, \frac{l^2}{(d-1)}\,\frac{1}{\tilde{\epsilon}}
  \end{bmatrix} \,, \label{InducedMatrix_LO_d_NH}
\end{equation}
\vspace{0.1cm}

where Eq.(\ref{AdS_NH_TargetMatrix}) is used, and $\tilde{\epsilon}' = \partial_\sigma \tilde{\epsilon}$. Hence, its inverse is given by

\begin{equation}
g^{a b}\big\rvert_{X^\mu_0}\, := \,
  \begin{bmatrix}
  g^{tt} & g^{t \sigma} \\
  g^{\sigma t} & g^{\sigma \sigma}
  \end{bmatrix} \Bigg\rvert_{X^\mu_0} \, = \,
  \frac{1}{\det \big(g_{a b}\rvert_{X^\mu_0} \big) }\,\begin{bmatrix}
    \tilde{\epsilon}'^2 \, G_{\tilde{\epsilon} \tilde{\epsilon}} & 0 \\
  0 & G_{tt}
  \end{bmatrix} \, = \,
  \begin{bmatrix}
    - \frac{l^2}{(d-1)r_H^2 \,\tilde{\epsilon}}  & 0 \\
  0 & \frac{(d-1) \, \tilde{\epsilon}}{l^2 \, \tilde{\epsilon}'^2}
  \end{bmatrix} \,, \label{InducedMatrix_LO_d_INVERSE_NH}
\end{equation}
\vspace{0.2cm}

where $\det \big(g_{a b}\rvert_{X^\mu_0} \big)$ is calculated from Eq.(\ref{InducedMatrix_LO_d_NH}),

\begin{equation}
g\rvert_{X^\mu_0} \, \equiv \, \det \big(g_{a b}\rvert_{X^\mu_0} \big) \,= \, \tilde{\epsilon}'^2 \, G_{\tilde{\epsilon} \tilde{\epsilon}}\,G_{tt} \, = \,  \bigg(\tilde{\epsilon}'^2 \frac{l^2}{(d-1)}\,\frac{1}{\tilde{\epsilon}}  \bigg) \, \bigg( -\frac{r_H^2}{l^2}\,(d-1) \, \tilde{\epsilon} \bigg) \, = \, -r_H^2\,\tilde{\epsilon}'^2  \,. \label{InducedMatrix_LO_d_det_NH}
\end{equation}
\vspace{0.1cm}

Expanding the indices of Eq.(\ref{StringEoM_XI_d=3_NH_trunc}) using Eqs.(\ref{InducedMatrix_LO_d_NH})-(\ref{InducedMatrix_LO_d_det_NH}), the transverse equation of motion become

\begin{flalign}
& \begin{aligned}
0 \, &  = \, \partial_{\sigma} \left(  \bigg(\sqrt{-g}\, g^{\sigma \sigma} \, \frac{r_H^2}{l^2} \bigg) \bigg\rvert_{X^\mu_0}\,\, {\partial_\sigma} X^I(t,\sigma)\right) \,+\, \partial_t  \left(  \bigg(\sqrt{-g}\, g^{tt} \, \frac{r_H^2}{l^2} \bigg) \bigg\rvert_{X^\mu_0}\,\, {\partial_t} X^I(t,\sigma)\right)\\[0.2cm]
 &= \, \partial_{\sigma}  \left(r_H \tilde{\epsilon}' \left(\frac{(d-1) \, \tilde{\epsilon}}{l^2 \, \tilde{\epsilon}'^2} \right) \, \frac{r_H^2}{l^2}\, {\partial_\sigma} X^I(t,\sigma)\right) \,+\, \partial_t \left( r_H\tilde{\epsilon}' \left(- \frac{l^2}{(d-1)r_H^2}\,\frac{1}{\tilde{\epsilon}} \right) \, \frac{r_H^2}{l^2}\, {\partial_t} X^I(t,\sigma)\right)\, \\[0.2cm]
 & = \, - \, \partial_t^2 X^I(t,\tilde{\epsilon})\,+\, \frac{r_H^2}{l^4}  \, (d-1)^2 \,  \tilde{\epsilon}\,\partial_{\tilde{\epsilon}} \big( \tilde{\epsilon}\, {\partial_{\tilde{\epsilon}}} X^I(t,\tilde{\epsilon})\big) \,, \label{StringEoM_XI_d_NH_epsilon}
\end{aligned} &
\end{flalign}
\vspace{0.1cm}

where $X^I$ denotes the transverse direction ($I \in (2,\, 3,\, ...\,,\, d-1)$); and the last line is written completely in terms of near-horizon spacetime coordinates ($t, \tilde{\epsilon}$) by differentiating the identification $\tilde{\epsilon} \,= \,e^{\frac{r_H}{l^2}   \, (d-1) \, \left( r_H \, \tilde{\epsilon}_{s*} \,+\, \sigma \right)}$,

    \begin{flalign}
    & \begin{aligned}
    & d \tilde{\epsilon} \,= \,(d-1) \, \frac{r_H}{l^2}\,e^{\frac{r_H}{l^2} \, (d-1) \, \big( r_H \, \tilde{\epsilon}_{s*} \,+\, \sigma \big)}  \, d \sigma \\[0.2cm]
     &\Rightarrow \, \partial_\sigma \, =  \,(d-1) \, \frac{r_H}{l^2}\,e^{\frac{r_H}{l^2} \, (d-1) \, \big( r_H \, \tilde{\epsilon}_{s*} \,+\, \sigma \big)} \, \partial_{\tilde{\epsilon}} \\[0.2cm]
     &\Rightarrow \, \partial_\sigma \,=\, \,(d-1) \, \frac{r_H}{l^2}\, \tilde{\epsilon} \, \partial_{\tilde{\epsilon}} \,. \label{d_NH_Iden_differential}
    \end{aligned} &
    \end{flalign}
    \vspace{0.1cm}

Eq.(\ref{d_NH_Iden_differential}) can also be used to convert the equations of motion Eq.(\ref{StringEoM_XI_d_NH_epsilon}) to parameter space coordinates $(t,\sigma)$,

\begin{flalign}
& \begin{aligned}
0 \,&=  \, - \, \partial_t^2 \, X^I(t,\sigma)\,+\, \frac{r_H^2}{l^4} \, (d-1)^2\,  \tilde{\epsilon}\, \frac{l^2}{(d-1)\, r_H} \, \frac{1}{\tilde{\epsilon}}\, \partial_\sigma \left( \tilde{\epsilon}\, \frac{l^2}{(d-1)\, r_H} \, \frac{1}{\tilde{\epsilon}}\, \partial_\sigma X^I(t,\sigma)\right)  \\[0.2cm]
 &=\, - \, \partial_t^2 \, X^I(t,\sigma)\,+\,  \partial_\sigma^2 \,  X^I(t,\sigma)  \,. \label{StringEoM_XI_d_NH_sigma}
\end{aligned} &
\end{flalign}
\vspace{0.1cm}

The near-horizon string equations of motion for the transverse fluctuations on the Limp Noodle is given by wave equation. Remember that the small virtuality limit of the $d=3$ Limp Noodle also yielded the wave equation as the transverse equation of motion (Eq.(\ref{d=3_wave_equation}))\footnote{The solution to Eq.(\ref{StringEoM_XI_d_NH_sigma}) is therefore given by Eq.(\ref{fModes_SV}), or alternatively by $f_\omega^{(\pm)}(\tilde{\epsilon}) = \tilde{\epsilon}^{\pm \frac{1}{(d-1)} \frac{i l^2 \omega }{r_H}}$, where the identification $\tilde{\epsilon} \,= \,e^{\frac{r_H}{l^2}   \, (d-1) \, \left( r_H \, \tilde{\epsilon}_{s*} \,+\, \sigma \right)}$ is used to convert from parameter space coordinates ($t, \sigma$) to near-horizon spacetime coordinates ($t, \tilde{\epsilon}$). Notice when $f_\omega^{(+)}(\tilde{\epsilon})$ is used as the solution (as in subsection (\ref{subsubsecExpandEoMAdS3})), $f_\omega^{(-)}(\tilde{\epsilon})$ is used in the definition of $X_{\text{reg}}(t,\tilde{\epsilon})$, Eq.(\ref{StringSol_XI_d=3_NH}).}.
Since this is true, repeating the calculations laid out in subsection (\ref{parSmallVirtuality}) is sufficient to calculate $s^2(t;d)$. The only change is that $\beta$ scales with the number of spacetime dimensions, so Eq.(\ref{st_squaredLQ_SV}) needs to be adapted by adding in the appropriate dimensionally dependent factor. In order to calculate this, notice that $\beta$ enters the calculation of $s^2_{\text{small}}(t)$ in subsection (\ref{parSmallVirtuality}) through the use of the normalization constant $A_\omega$. For $d \geq 3$, Eq.(\ref{Aw_normalization}) is generalised

\begin{equation}
    A_\omega(d) \, := \, \frac{l}{r_H} \, \sqrt{\frac{\pi \alpha'}{\omega}} \,
    = \,  \frac{(d-1) \beta}{4 \, \sqrt{\pi \omega} \, \lambda^{1/4}} \, \equiv \, \frac{(d-1)}{2} \, A_\omega \, , \label{Aw_normalization_d}
\end{equation}
\vspace{0.1cm}

where the second equality follows from using the definition of the AdS radius of curvature $l$ (Eq.(\ref{alpha_def})) and the Hawking temperature (Eq.(\ref{HawkingTemp})). Using the generalised normalization constant $A_\omega(d)$, the expectation value of the position of the boundary endpoint at two different times (Eq.(\ref{Xt1_Xt2_expectationLQ})) becomes\footnote{Note that the Bose-Einstein distribution Eq.(\ref{Bose_Einstein_Distribution}) does not carry any dimensional dependence.}
\begin{flalign}
& \begin{aligned}
     \langle  \normord{\hat{X}_{\text{end}}(t_1;d) \hat{X}_{\text{end}}(t_2;d)}  \rangle  & = \frac{(d-1)^2 \beta^2}{16 \pi^2 \sqrt{\lambda}} \int_0^\infty \frac{d \omega}{\omega}  \frac{1}{e^{\beta \omega} - 1}  \text{Re}\left(f_\omega (\sigma_f-t_1) f^{\,*}_{\omega} (\sigma_f-t_2)  e^{- i \omega (t_1 - t_2)}\right) \\[0.2cm]
     & \equiv  \frac{(d-1)^2}{4} \langle  \normord{\hat{X}_{\text{end}}(t_1) \hat{X}_{\text{end}}(t_2)}  \rangle
      \, , \label{Xt1_Xt2_expectation_d}
\end{aligned} &
\end{flalign}

while the string falling endpoint's mean-squared transverse displacement, Eq.(\ref{st_squaredLQ}), becomes

\begin{flalign}
& \begin{aligned}
    s^2(t;d) & =  \frac{(d-1)^2 \beta^2}{16 \pi^2 \sqrt{\lambda}} \int_0^\infty \frac{d \omega}{\omega} \frac{1}{e^{\beta \omega} - 1}  \big \rvert f_\omega (\sigma_f -t) -  f_{\omega} (\sigma_f)  e^{ i \omega t}\big \rvert^2 \\[0.2cm]
     & \equiv  \frac{(d-1)^2}{4} s^2 (t)  \, , \label{st_squaredLQ_d}
\end{aligned} &
\end{flalign}
\vspace{0.1cm}

due to the factors of $A_\omega A_{\omega'}$ appearing in the calculations of both  $\langle  \normord{\hat{X}_{\text{end}}(t_1;d) \hat{X}_{\text{end}}(t_2;d)}  \rangle $ and $s^2(t;d)$.\\

In subsection (\ref{parSmallVirtuality}), the general solution to the wave equation, $f_\omega (\sigma)$, is given by Eqs.(\ref{fModes_SV})-(\ref{SV_gen_solution}). In order to calculate the falling string endpoint's mean-squared transverse displacement in the small virtuality limit, Eq.(\ref{SV_gen_solution}) is substituted into Eq.(\ref{st_squaredLQ_d}). Specifically,

\begin{flalign}
& \begin{aligned}
     s^2(t;d)\, &  =   \frac{(d-1)^2 \beta^2}{16 \pi^2 \sqrt{\lambda}}  \int_0^\infty \frac{d \omega}{\omega}  \frac{1}{e^{\beta \omega} - 1} \Big \rvert 2 e^{i \omega  ( r_{s*} + \sigma_f)}  \cos \big(\omega t \big)   -   2 e^{i \omega \, ( r_{s*} +  \sigma_f)} e^{ i \omega t}\Big \rvert^2 \\[0.2cm]
     &  = \frac{(d-1)^2 \beta^2}{4 \pi^2 \sqrt{\lambda}}  \int_0^\infty \frac{d \omega}{\omega}  \frac{1}{e^{\beta \omega} - 1} \sin^2 (\omega t)\\[0.2cm]
     &\equiv \frac{(d-1)^2}{4} s^2_{\text{small}}(t)\\[0.2cm]
     &  = \frac{(d-1)^2 \beta^2}{16 \pi^2 \sqrt{\lambda}}  \,\ln \left( \frac{\beta}{2 \pi t} \sinh \left( \frac{2\pi t}{\beta} \right) \right) \, , \label{st_squaredLQ_SV_d}
\end{aligned} &
\end{flalign}
\vspace{0.1cm}

where Eq.(\ref{st_squaredLQ_SV}) is used in the third line, and Eq.(\ref{st_squaredLQ_SV2}) in the final line. In order to calculate the diffusion coefficient, the late time dynamics of Eq.(\ref{st_squaredLQ_SV_d}) must be explored. Expanding in powers of $1/k = \beta/ t$ yields

\begin{flalign}
& \begin{aligned}
     s^2(t;d) &  \,\,\, = \,\,\,\,\,\,\,   \frac{(d-1)^2 \beta^2}{16 \pi^2 \sqrt{\lambda}}  \, \Big[ \ln \big( \sinh \big( 2 \pi k \big) \big) \, + \, \ln \Big(\frac{1}{2 \pi  k}\Big) \Big] \\[0.2cm]
      & \,\stackrel{(\beta \ll t)}{\longrightarrow}\,\, \frac{(d-1)^2 \beta t }{8 \pi \sqrt{\lambda}}  \, + \, \frac{(d-1)^2 \beta^2}{16 \pi^2 \sqrt{\lambda}}  \ln \bigg(\frac{\beta}{4 \, \pi \, t}\bigg) \,+\, \mathcal{O}\Big(\beta^2 \, e^{\frac{-4 \pi t }{\beta}}\Big)\, , \label{st_squaredLQ_SV_d_Late}
\end{aligned} &
\end{flalign}
\vspace{0.1cm}

where the calculation steps in Eq.(\ref{st_squaredLQ_SV_Late}) have been followed closely. Since $s^2(t;d) \sim t$, the late time dynamics exhibit diffusive behaviour. From Eq.(\ref{DisplacementSqLimit2}) it is expected that $s^2_{\text{small}}(t) \, =\, 2\, D \, t$ at late times, where $D$ is the diffusion coefficient. Comparing this to Eq.(\ref{st_squaredLQ_SV_d_Late}) it is easy to see that the diffusion coefficient is given by

\begin{equation}
D_{\text{LQ}}^{\text{AdS$d$}}(d) \, = \, \frac{(d-1)^2 \beta}{16 \pi \sqrt{\lambda}}\, . \label{DifCoeff_LQ_d}
\end{equation}

For $d=3$, this reduces down to Eq.(\ref{DifCoeff_LQ}) as expected.\\

Moerman \textit{et al.} in \cite{moerman2016semi} present an expression for the light quark diffusion coefficient in $d$ dimensions which agrees with Eq.(\ref{DifCoeff_LQ_d}). However, their method of expanding the near-horizon AdS$_d$-Schwarzschild metric (described in subsection (3.3) of \cite{moerman2016semi}) is inconsistent. The near-horizon AdS$_d$-Schwarzschild metric given in Eq.(3.48) of \cite{moerman2016semi} is dimensionally incorrect\footnote{Since $r_H^2$, $l^2$ $dx_{d-2}^2$ and $dt^2$ all go like $Energy^{-1}$, and $\tilde{\epsilon}$ (as well as $d\tilde{\epsilon}^2$) are defined as dimensionless quantities; the first term in the metric Eq.(3.48) of \cite{moerman2016semi} goes like $Energy^{-2}$, the second term also goes like $Energy^{-2}$, but the third term is dimensionless. In juxtaposition, all terms in the correct, leading order near-horizon AdS$_d$-Schwarzschild metric given in this subsection (Eq.(\ref{AdSd_Metric_OptII})) consistently go like $Energy^{-2}$.}\footnote{\label{footnoteTerminology}A note on terminology: $\epsilon$ in \cite{moerman2016semi} corresponds to $\tilde{\epsilon}$ in this article.}. Using the near-horizon tortoise coordinate Eq.(\ref{d_TortoiseCoord_NH}) (no attempt is made to explain how the order of  truncation is consistent with the leading order near-horizon metric\footnote{The paper \cite{moerman2016semi} simply states "Expanding (and truncating)
each term in the AdS$_d$ metric, Eq.(2.32), to lowest non-vanishing order in $\epsilon$". There does not appear to have been an attempt to examine if each term is consistently taken to the same order of $\epsilon$ ($\tilde{\epsilon}$ -- see footnote \ref{footnoteTerminology}).}), the metric in Eq.(3.48) of \cite{moerman2016semi} is converted to ($t, \tilde{\epsilon}_*$) coordinates, and found to be conformal in this system\footnote{The correct near-horizon AdS$_d$-Schwarzschild metric (Eq.(\ref{AdSd_Metric_OptII})) is actually conformal in ($t, r_*$) coordinates (as shown in Eq.(\ref{AdS_Metric_NH_t_epsilonstar})).}. Hence, the leading order string solution can be written down in ($t, \tilde{\epsilon}_*$) coordinates, from which Moerman \textit{et al.} proceeded to finding the transverse equations of motion -- this, coincidentally, being the wave equation.
Therefore, although there is agreement between the final results, Eq.(\ref{DifCoeff_LQ_d}) and the diffusion coefficient in $d$ dimensions derived by Moerman \textit{et al.}, the authors conclude that the calculations in this subsection arriving at Eq.(\ref{DifCoeff_LQ_d}), present the first complete, consistent derivation of the light quark diffusion coefficient in $d$ dimensions.

\section{Drag Force in AdS/CFT}\label{secDragForce}

\begin{figure}[!htb]
\centering
\includegraphics[width=0.75\textwidth]{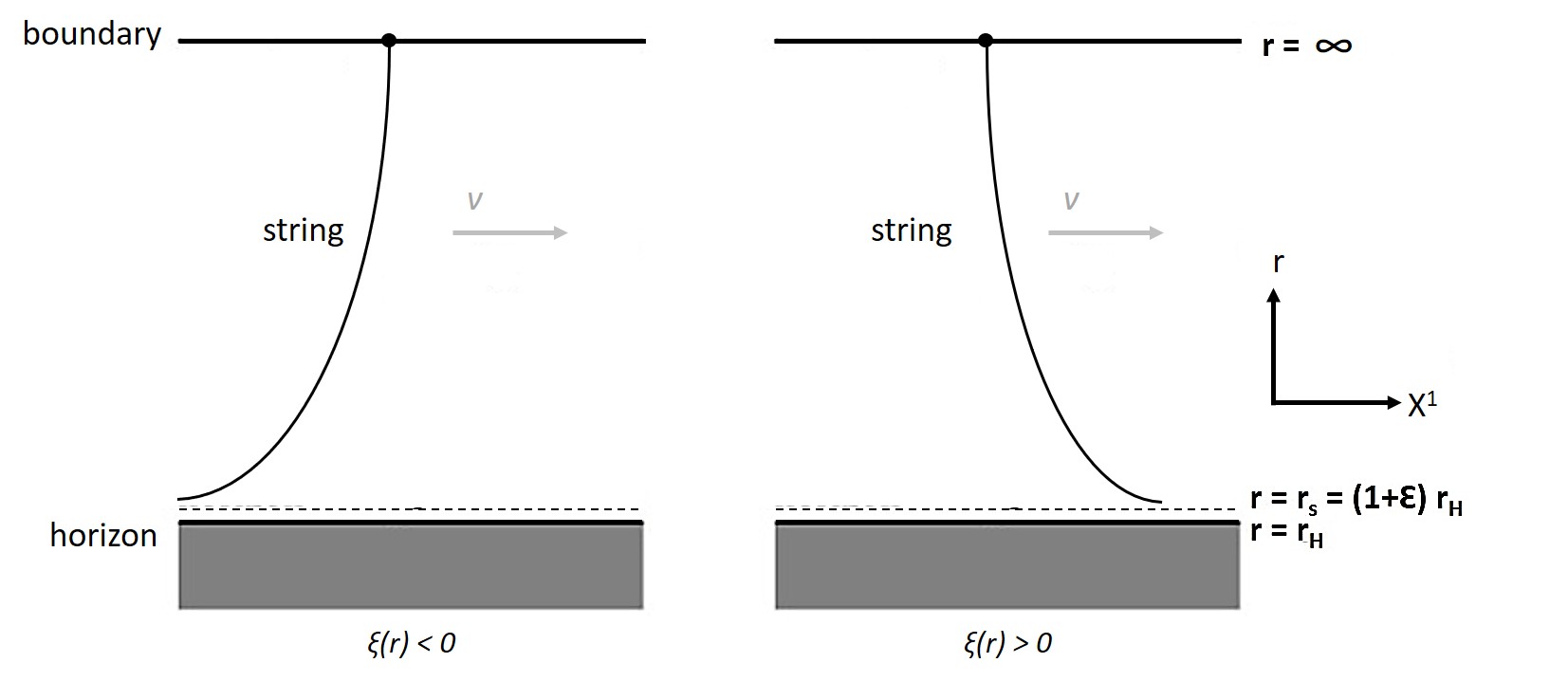}
\caption[A fundamental open string used as a probe in an AdS black hole background to model an infinitely massive quark moving through a thermal plasma at a constant velocity $v$ on the boundary]{\label{fig:stringdiagramTRAILING} A fundamental open string used as a probe in an AdS black hole background to model an infinitely massive quark moving through a thermal plasma at a constant velocity $v$ on the boundary.
One endpoint of the test string is attached to the boundary of anti-de Sitter spacetime at infinity ($r= \infty$), from which the string hangs down to a \textit{stretched horizon} ($r_s \, = \, (1+ \epsilon)\,r_H$ where $0 < \epsilon \ll 1$) placed just above the Schwarzschild black hole horizon.
Left: A string which trails behind its boundary endpoint moving with velocity $v$ in the $X^2$ direction. This is the physical solution. Energy flows from the boundary, down the string, and towards the horizon. Right: A string which trails in front of its boundary endpoint moving with velocity $v$ in the $X^2$ direction. This solution is unphysical. Energy would need to be flowing away from the horizon, up the string, and towards the boundary.}
\end{figure}

Besides the test string set-ups depicted in figures (\ref{fig:stringdiagram}) and (\ref{fig:stringdiagramLQ}) (see section (\ref{secHeavyQuark}), page \pageref{fig:stringdiagram}, and section (\ref{secLightQuark}), page \pageref{fig:stringdiagramLQ}, respectively), there are many other string set-ups -- illuminating various aspects or properties of the thermal plasma in the boundary theory -- that could be considered. One such set-up is a trailing string in AdS$_5$-Schwarzschild which models an infinitely massive probe quark\footnote{Due to its infinite mass such a quark would not undergo Brownian motion in thermal plasma.} moving with a constant velocity $v$ in a $\mathcal{N}=4$ SYM thermal plasma\footnote{The constant velocity is measured with respect to the rest frame of the plasma.}.
This was considered by Gubser \cite{Gubser_2006} and Herzog \textit{et al.} \cite{herzog2006energy} independently in 2006, who aimed to approximately explain jet-quenching -- the phenomenon whereby energy loss is experienced by high-energy quarks travelling through the quark-gluon plasma\footnote{Around the same time, studies examining trailing strings in the AdS spacetime in order to investigate the dissipative and diffusive behaviour of a massive quark moving through a field theory plasma were common (see \cite{Liu_2006, Herzog_2006, Casalderrey_Solana_2006, Liu_2007, Casalderrey_Solana_2007, Gubser_2008}).}.
The succeeding subsection follows the calculations of \cite{Gubser_2006,herzog2006energy} in using the AdS/CFT correspondence to calculate the drag force experienced by a massive probe quark in the $\mathcal{N}=4$ SYM thermal medium. The string set-up used is depicted in figure (\ref{fig:stringdiagramTRAILING}). Further, in subsection (\ref{subsecTRAIL_GenAdSd}), the set-up is generalised to AdS$_d$-Schwarzschild.

\subsection{Test Strings in \texorpdfstring{AdS$_5$-Schwarzschild}{AdS5-Schwarzschild}} \label{subsec_TRAILinAdS5}

Consider a static open string in $AdS_5$-Schwarzschild, with one endpoint attached to the boundary at $r= \infty$ and the other allowed to hang down towards the horizon. Since the string is infinitely long (which models an infinitely massive quark in the boundary theory), the transverse fluctuations that the string experiences due to the presence of the black hole horizon never reach the boundary, i.e. the infinitely massive quark does not experience Brownian motion in the thermal medium. Regardless of its length, this test string is still a classical relativistic string and, as such, is described by the standard Polyakov action (Eq.(\ref{PolyakovAct})) or Nambu-Goto action (Eq.(\ref{NGAction})). It is convenient to use the Nambu-Goto action here\footnote{The Nambu-Goto action given in Eq.(\ref{NGAction}) is defined in the String Frame. In the Einstein Frame, the Nambu-Goto action is constructed with an added dilaton factor $e^{\phi/2}$,
\begin{equation*}
S_{NG}^{E} \, := \, - \frac{1}{2 \pi \alpha'} \, \int_{\mathcal{M}} d^2 \sigma \, e^{\phi/2} \, \sqrt{-g} \, ,
\end{equation*}
where $g \, := \,\text{det}\,( g_{a b})$ and the induced worldsheet metric $g_{a b}$ is given by Eq.(\ref{InducedMetric}). In \cite{Gubser_2006} the Einstein Frame is chosen for the proceeding calculations. The phenomenologically relevant case of AdS$_5$-Schwarzschild is considered. In this spacetime the dilaton factor can be ignored, since $e^{\phi/2} = H^{(5-d)/4}$ (where $H$ is the warping factor and $e^{\phi/2}$ disappears when $d=5$).
It is worth a footnote to remark, however, that the Einstein Frame is related to the String Frame by a conformal rescaling of the metric, and that the frames are considered interchangeable when describing the
physics of massless modes on the string \cite{Alvarez_2002}. The calculations in this article are in the String Frame.}.\\ 

From Eq.(\ref{AdSd_Metric}), the AdS$_5$-Schwarzschild metric is given by

\begin{equation}
ds_5^2 = \frac{r^2}{l^2} \, \left(-h(r) \, dt^2 \, + \, d\Vec{X}_{I}^2 \,\right) \,+\, \frac{l^2}{r^2}\frac{dr^2}{h(r)} \, , \,\,\,\,\,\,\,\, \text{where} \,\,\,\,\,\,\,\,  h(r) \,=\, 1 - \left( \frac{r_H}{r} \right)^{4} \, \in [0,1]  \label{AdS5_Metric}
\end{equation}
\vspace{0.1cm}

is the blackening factor of the Schwarzschild black hole situated at the horizon; $t \in [0, \infty)$ is the temporal coordinate; $r \in [0, \infty)$ is the radial coordinate; and the transverse spatial directions along which the D3-brane is extended are denoted by $\Vec{X}_{I} = \left(X^2, X^3, X^4 \right) \in \mathbb{R}^{3}$. Further, $l \in \mathbb{R}^+$ is the curvature radius of AdS$_d$ and $S^d$.\\

Following \cite{Gubser_2006, herzog2006energy}, a gauge choice is made where $\tau = t$ and $\sigma = r$ are the worldsheet coordinates. The target spacetime metric $G_{\mu \nu}$ -- where $\mu,\nu$ index over the spacetime directions $\left(t,\, r,\, 2,\, 3,\, 4 \right)$ -- is given explicitly in Eq.(\ref{appTargetMatrix}); and the induced worldsheet metric $g_{a b}$, defined in Eq.(\ref{InducedMetric}), takes the form Eq.(\ref{appIndMetric}). The explicit entries of the metric $g_{a b}$, its inverse $g^{a b}$, and its determinant $g := \text{det} (g_{a b })$ are given by Eqs.(\ref{appIndMetric2}, \ref{appInverseIndMetric}, \ref{appIndMetricDet}) respectively. Using the target spacetime metric $G_{\mu \nu}$ and remembering the gauge choice $\tau = t$ and $\sigma = r$, the determinant of the induced metric becomes
\begin{equation}
   g =  - \left( 1 \, -\, \frac{1}{h(r)}\Dot{X}_{I}^2 + \, \frac{r^4}{l^4} h(r) \,X_{I}'^2 \right)
   \, , \label{IndMetricDet_AdS5}
\end{equation}
\vspace{0.1cm}

where $\Dot{X} \equiv \partial_\tau X = \partial_t X$ and $X' \equiv \partial_\sigma X = \partial_r X$ (in both definitions the second equality follows from the worldsheet gauge choice). Inputting Eq.(\ref{IndMetricDet_AdS5}), into the definition of the Nambu-Goto action Eq.(\ref{NGAction}) yields

\begin{equation}
    S_{NG}  =  - \frac{1}{2 \pi \alpha'}  \int_{\mathcal{M}} d^2 \sigma  \, \sqrt{1 \, -\, \frac{1}{h}\Dot{X}_{I}^2 + \, \frac{r^4}{l^4} h \,X_{I}'^2}\, , \label{NG_Gubser}
\end{equation}
\vspace{0.1cm}

where the notational adjustment $h(r) \equiv h$ is used for brevity. The Lagrangian density, also defined in Eq.(\ref{NGAction}), is therefore given by

\begin{equation}
    \mathcal{L}  =  - \,\sqrt{1 \, -\, \frac{1}{h}\Dot{X}_{I}^2 + \, \frac{r^4}{l^4} h \,X_{I}'^2}  \, . \label{LDensity_Gubser}
\end{equation}
\vspace{0.1cm}

To proceed, consider movement of the test string with a constant velocity $v$ in one of the transverse directions, the $X^2$ direction\footnote{Due to a symmetry of anti-de Sitter spacetime all transverse directions are identical. Therefore any of the transverse directions can be chosen -- considering motion in either the $X^2$, $X^3$ or $X^4$ directions would be equivalent.}. Following \cite{Gubser_2006}, an ansatz to describe the late-time behaviour of the string is made. Specifically\footnote{To clarify the notation: the mapping function for the second transverse direction will be denoted $X^2$, while squaring this mapping function (if necessary) will be denoted $(X^2)^2$.},
\begin{equation}
    X^2(t, r)\, = \, v t \,+ \, \xi(r) \, + \, o(t)  \, , \label{LateTime_Gubser}
\end{equation}
\vspace{0.1cm}

where, at late times, all other motions are damped out and the $o(t)$ term disappears. The ansatz Eq.(\ref{LateTime_Gubser}) relies on the assumption that steady state behaviour of the string's motion is achieved at late times. Substituting this ansatz into the Lagrangian density Eq.(\ref{LDensity_Gubser}) -- remembering that $X^2$ is the only direction in which the string is considered to move -- yields


\begin{equation}
    \mathcal{L}\, = \, -\,\sqrt{1 - \frac{v^2}{h} + \frac{r^4}{l^4} h \, \xi'^2}  \, ,\label{LDensity2_Gubser}
\end{equation}
\vspace{0.1cm}

where $\xi(r) \equiv \xi$ is implied, and $\xi' = \partial_r \xi$. In order to determine $\xi(r)$, the Euler-Lagrange equations are derived from the Lagrangian density (Eq.(\ref{LDensity2_Gubser})), and are given by

\begin{equation}
    \frac{{\partial}}{{\partial t}} \,\frac{{\partial \mathcal{L}}}{\partial \xi'}  - \frac{{\partial \mathcal{L}}}{{\partial \,\xi}} \, = \, 0 \, , \label{EulerLagrange_Guber}
\end{equation}
\vspace{0.1cm}

where the quantity $\pi_\xi$ can be defined as

\begin{equation}
    \pi_\xi\, := \, \frac{{\partial \mathcal{L}}}{{\partial \big({\partial_r}\,\xi \big)}} \, . \label{MomDensity_Gubser}
\end{equation}
\vspace{0.1cm}

Since the Lagrangian density (Eq.(\ref{LDensity2_Gubser})) does not depend on $\xi$ (i.e. ${\partial \mathcal{L}}/{\partial \,\xi} = 0$), the Euler-Lagrange equations
become

\begin{flalign}
& \begin{aligned}
     \pi_\xi \, &:= \, \frac{{\partial \mathcal{L}}}{{\partial\, \xi'}} \, =\, \mathcal{C}\\[.2cm]
  & = \, - \, \frac{r^4}{l^4} h \,\xi' \left(1 - \frac{v^2}{h} + \frac{r^4}{l^4} h \,\xi'^2 \right)^{-\frac{1}{2}}  \, ,\label{EoM_Gubser}
\end{aligned} &
\end{flalign}
\vspace{0.1cm}

where Eq.(\ref{LDensity2_Gubser}) is used in the second line. Hence, the equation of motion of the string is $\pi_\xi$ being a constant of motion. Solving Eq.(\ref{EoM_Gubser}) yields a defining relation for $\xi'(r)$. Mathematica's $\tt{Solve}$ function is used\footnote{Mathematica is used to help simplify the algebra. See Mathematica Notebook [d] (\texttt{DragForce.nb}) for details on the calculations presented in section (\ref{secDragForce}): access in appendix (\ref{appCode}).}, to find

\begin{equation}
    \xi'(r) \,=\, \pm  \, \pi_\xi \, \frac{l^4}{r^4 \, h} \, \sqrt{\frac{h \, - v^2}{h \, - \, \frac{l^4}{r^4} \pi_\xi^2}} \, . \label{Define_XiPrime}
\end{equation}
\vspace{0.1cm}

Finding $\xi(r)$ will involve integrating Eq.(\ref{Define_XiPrime}) with respect to $r$. However, $\xi(r)$ is required to be real everywhere, and the square root in Eq.(\ref{Define_XiPrime}) is not necessarily always real. The only variable which can be adjusted is $\pi_\xi$. Defining $\pi_\xi$ such that $(h \, - v^2)/(h \, - \, \frac{l^4}{r^4} \pi_\xi^2)$ is always positive (thereby ensuring the square root is always real) is equivalent to setting the appropriate boundary conditions. Since $h(r) \in [0,1]$, at some intermediate radius $h \, = \, h^*$ the numerator $(h \,-\,v^2)$ changes sign. In order for the square root to be real everywhere the denominator must change sign at the same point, i.e. at $h\,=\,h^*$,

\begin{equation}
    h \, - \, \frac{l^4}{r^4} \pi_\xi^2 \, = \, h \,-\,v^2 \, = \, 0 \, . \label{FindingBC_Gubser}
\end{equation}
\vspace{0.1cm}

Rearranging the first equality leaves

\begin{equation}
    \pi_\xi^2 \, = \,\frac{r^4}{l^4} \, v^2\, . \label{FindingBC2_Gubser}
\end{equation}
\vspace{0.1cm}

While the second equality in Eq.(\ref{FindingBC_Gubser}) is rearranged to

\begin{equation}
    r^4 \, = \, \frac{r_H^4}{1 - v^2} \, . \label{FindingBC3_Gubser}
\end{equation}
\vspace{0.1cm}

where the definition of $h(r)$ (Eq.(\ref{AdS5_Metric})) is used. Inputting Eq.(\ref{FindingBC3_Gubser}) into Eq.(\ref{FindingBC2_Gubser}) results in an appropriate expression for $\pi_\xi$,

\begin{equation}
    \pi_\xi \, = \, \pm \, \frac{v}{\sqrt{1 - v^2}}\,\frac{r_H^2}{l^2} \, . \label{FindingBC4_Gubser}
\end{equation}
\vspace{0.1cm}

Plugging this defining relation for $\pi_\xi$ into Eq.(\ref{Define_XiPrime}) guarantees that the square root in Eq.(\ref{Define_XiPrime}) is always positive, and $\xi(r)$ is everywhere real. The $(\pm)$ in Eq.(\ref{FindingBC4_Gubser}) determines if the quark is moving in the $(+)$ or $(-)$ $X^2$ direction. Consider the case where $\pi_\xi > 0$ for the remainder of the calculation. Eq.(\ref{Define_XiPrime}) becomes,

\begin{equation}
    \xi'(r) \, = \,\pm \, v \, \frac{r_H^2\, l^2}{r^4 \, - \, r_H^4}\, . \label{XiPrime_Gubser}
\end{equation}
\vspace{0.1cm}

An integrable expression for $\xi'$ has finally been found. Integrating Eq.(\ref{XiPrime_Gubser}) with respect to $r$ yields\footnote{Mathematica is used to easily integrate (see Mathematica Notebook [d]: \texttt{DragForce.nb}).}


\begin{equation}
    \xi(r) \, = \,  \mp\, \frac{l^2\,v}{2 \, r_H} \, \left(\tan ^{-1}\left(\frac{r}{r_H}\right)\, + \,\ln \sqrt{ \frac{r\,+\,r_H}{r \,-\, r_H}} \right) \,, \label{XiPrime2_Gubser}
\end{equation}
\vspace{0.1cm}

where $\ln (-x) \, = \, \ln (x) \, + \, i \, \pi$ for $x > 0$ is used and it is recognised that $i \, \pi$ can be absorbed into the integration constant, which (if the calculation was entirely rigorous) would be present in Eq.(\ref{XiPrime2_Gubser}). The function $\xi(r)$ can take on two values: (i) for $( \xi'(r) > 0, \, \xi(r) < 0)$ the string trails behind its boundary endpoint which is moving with a velocity $v$ in the $(+)$ $X^2$ direction, and (ii) for $(\xi'(r) < 0,\, \xi(r) > 0)$ the string trails in front of its boundary endpoint which is moving with a velocity $v$ in the $(+)$ $X^2$ direction. The `tail wagging the dog' scenario depicted in (ii) is an unphysical solution \cite{herzog2006energy}; however both of these cases are depicted in figure (\ref{fig:stringdiagramTRAILING}), page \pageref{fig:stringdiagramTRAILING}.\\

Using the expression for $\xi'(r)$ given in Eq.(\ref{XiPrime_Gubser}) the Lagrangian density (Eq.(\ref{LDensity2_Gubser})) becomes

\begin{equation}
    \mathcal{L}\, = \,- \, \sqrt{1-v^2}\, . \label{LDensity3_Gubser}
\end{equation}
\vspace{0.1cm}

Since the definition of the Nambu-Goto action (Eq.(\ref{NGAction})) implies $\mathcal{L}= -\sqrt{-g}$, the determinant of the induced metric becomes 
\begin{equation}
    g\, = \,- \, \left(1-v^2\right)\, . \label{IndMetricDet_Gubser}
\end{equation}
\vspace{0.1cm}

To calculate the flow of momentum down the string $dp_{2}/dt$, the canonical momentum densities $\Pi^a_\mu (t, \sigma)$ (Eq.(\ref{MomDensity})),
and the string equations of motion (Eq.(\ref{StringEoM})) are needed. The conserved charges associated with the momentum densities $\Pi^a_\mu$ are defined \cite{carroll2004spacetime, Morad_2014} for a general curve $\gamma$ on the worldsheet, as

\begin{equation}
    p^\gamma_\mu \, := \, \int_\gamma \, d\sigma^b \, \Tilde{\epsilon}_{a b} \, \Pi^a_\mu\, , \label{ConservedCharge_Gubser}
\end{equation}
\vspace{0.1cm}

where $\Tilde{\epsilon}_{a b}$ is the Levi-Civita symbol\footnote{The Levi-Civita symbol is defined in two dimensions as
\begin{equation*}
\Tilde{\epsilon}_{a b} \, = \, \begin{Bmatrix}
   +1 \, , \,\,\,\,\,\,\,\,\,  \text{if} \,\,\,\ (a, b) =  (1,2) \\
   -1 \, , \,\,\,\,\,\,\,\,\,  \text{if} \,\,\,\ (a, b) =  (2,1)\\
   0 \, , \,\,\,\,\,\,\,\,\,\,\,\,\,\,\,\,\,\,\,\,\,\,\,\,\,\,\, \,\,\,\,\,\,  \text{if} \,\,\,\ a = b
  \end{Bmatrix} \,.
\end{equation*}}, and $g_{a b}$ is the induced metric on the worldsheet in curved spacetime (Eq.(\ref{InducedMetric})). Since the string moves only in the $X^2$ direction by construction, the string's transverse momentum densities are only non-zero in this direction. Using Eq.(\ref{ConservedCharge_Gubser}) the $X^2$ component of the spacetime momentum flowing over some general time interval $\mathcal{I}$ of length $\Delta t$ is

\begin{equation}
    \frac{d p_{2}}{d  t} \, \Delta t \, = \, p^\mathcal{I}_{2} \, = \, \int_\mathcal{I} dt  \, \Pi^{r}_{2} \, , \label{ConservedcCharge2_Gubser}
\end{equation}
\vspace{0.1cm}

where the gauge choice $\tau = t$ and $\sigma = r$ is applied, and $\Tilde{\epsilon}_{r t} =  +1$ from the definition of the Levi-Civita symbol. The radius at which the integral in Eq.(\ref{ConservedcCharge2_Gubser}) is evaluated does not matter, since the momentum densities  $\Pi^a_\mu (t, \sigma)$ are conserved (Eq.(\ref{StringEoM})). For the physical solution where the string trails behind its boundary endpoint $(\xi'(r) > 0, \, \xi(r) < 0)$, the quantity $d p_{2}/d t$ is identified as the drag force and the orientation of the integral is chosen such that the drag force is negative (i.e. acts to oppose the motion of the string). Therefore, $d p_{2}/d t   =  \Pi^{r}_{2}$. According to the definition of the momentum densities Eq.(\ref{MomDensity}),

\begin{flalign}
& \begin{aligned}
     \frac{d p_{2}}{d  t} \, := \, \Pi^{r}_{2} \, & = \,- \frac{1}{2 \pi \alpha'} \, \sqrt{-\gamma} \, \gamma^{r b}\, G_{2  \,\nu}\, {\partial_b} X^\nu\\[0.2cm]
     & =  \, - \frac{1}{2 \pi \alpha'} \, \sqrt{1-v^2} \, g^{r b}\, G_{2 \,\nu}\, {\partial_b} X^\nu\, , \label{MomDensity2_Gubser}
     \end{aligned} &
\end{flalign}
\vspace{0.1cm}

where the constraint equation, Eq.(\ref{Constraint2}), and Eq.(\ref{IndMetricDet_Gubser}) is used in the second line. Expanding the implied summation over repeated indices in Eq.(\ref{MomDensity2_Gubser}) yields
\begin{flalign}
& \begin{aligned}
       \frac{d p_{2}}{d  t}   \, & = \, - \frac{1}{2 \pi \alpha'} \sqrt{1- v^2} \, G_{2 \,\nu}\, \left(g^{r r}\, \partial_r X^\nu  \, + \, g^{r t}\, \partial_t X^\nu  \right)\\[0.2cm]
       & = \, \frac{1}{2 \pi \alpha'} \frac{1}{\sqrt{1- v^2}} \, G_{2 \,2}\, \left(\left(G_{tt}\,+ \, G_{2 \,2}\, \partial_t X^2 \partial_t X^2 \right)\, \partial_r X^2 \, + \, \left(- G_{2 \,2}\,\partial_r X^2 \,\partial_t X^2\right)\, \partial_t X^2  \right)\\[0.2cm]
       & = \, \frac{1}{2 \pi \alpha'} \frac{1}{\sqrt{1- v^2}} \, G_{2 \,2}\, G_{tt} \, \partial_r \xi\\[0.2cm]
       & = \, - \frac{1}{2 \pi \alpha'} \frac{r_H^2}{l^2} \frac{v}{\sqrt{1- v^2}} \, , \label{DragForce_Gubser}
\end{aligned} &
\end{flalign}
\vspace{0.1cm}

where, in the second line, it was recognised that $G_{2 \,\nu} = G_{2 \,2}$  is the only non-zero possibility since $G_{\mu \nu}$ is a diagonal matrix, and Eq.(\ref{appInverseIndMetric}) was used to find the explicit entries of the inverse induced metric $g^{ab}$ (remembering that the gauge choice $\tau = t$ and $\sigma = r$ must be made). Further, Eq.(\ref{IndMetricDet_Gubser}) was used and it was noted that, by construction, the only non-zero transverse direction was the $X^2$ direction. In the third line the commutative property of partial derivatives and the ansatz Eq.(\ref{LateTime_Gubser}) was used. The final line follows from inputting the target spacetime metric entries (Eq.(\ref{appTargetMatrix})) and the defining relation for $\xi'(r)$, Eq.(\ref{XiPrime_Gubser}) (where $\xi'(r)>0$).\\

Comparing Eq.(\ref{FindingBC4_Gubser}) (where $\pi_\xi < 0$) and Eq.(\ref{DragForce_Gubser}), notice that $\pi_\xi$ is understood to be a momentum density, since

\begin{equation}
     \Pi_{2}^r \, = \, \frac{1}{2 \pi \alpha'} \pi_\xi \, . \label{MomDensity3_Gubser}
\end{equation}
\vspace{0.1cm}

Eq.(\ref{MomDensity3_Gubser}) follows from Eq.(\ref{MomDensity}) which gives $ \Pi_{2}^\sigma  :=  \frac{1}{2 \pi \alpha'} \partial_{({\partial_\sigma} X^2)} \mathcal{L}$; and from the worldsheet gauge choice $\sigma=r$, from which the momentum density becomes $ \Pi_{2}^r \,=\, \frac{1}{2 \pi \alpha'} \partial_{({\partial_r}\,\xi )} \mathcal{L}  \,= \,\frac{1}{2 \pi \alpha'} \pi_\xi$ (where the ansatz Eq.(\ref{LateTime_Gubser}) is used in the first equality and Eq.(\ref{MomDensity_Gubser}) in the second equality).\\

For the phenomenologically relevant AdS$_5$/$\mathcal{N}=4$ SYM case, the `t Hooft coupling is $\lambda \, \equiv \, g_{\text{YM}}^2\,N_c$, where $N_c$ is the number of colours and the Yang-Mills coupling is related to the string coupling by: $g_{\text{YM}}\, = \, 2 \sqrt{\pi \,g_s}$. The fundamental string length scale Eq.(\ref{alpha_def}) becomes

\begin{equation}
     \alpha' \, = \, \frac{l^2}{\sqrt{g_{\text{YM}}^2\,N_c}}\, . \label{AlphaDef_Gubser}
\end{equation}
\vspace{0.1cm}

Using Eq.(\ref{AlphaDef_Gubser}) and the definition of Hawking temperature (Eq.(\ref{HawkingTemp})) with $d=5$, the drag force (Eq.(\ref{DragForce_Gubser})) can be expressed in terms of phenomenologically relevant gauge theory variables,

\begin{equation}
     \frac{d p_{2}}{d  t} \, = \, - \frac{\pi\, \sqrt{g_{YM}^2  N_c}}{2}\, T^2\, \frac{v}{\sqrt{1  -  v^2}}\, . \label{DragForce2_Gubser}
\end{equation}
\vspace{0.1cm}

Following the formal frameworki of Gubser \cite{Gubser_2006}, in the gauge theory the momentum $p_2$ of the probe quark can be theoretically\footnote{Both the momentum $p_2$ and the mass $m$ of the external quark are, in fact, infinite.} related to its mass $m$ through
\begin{equation}
     p_2(t) \ = \, \frac{v}{\sqrt{1 \, - \, v^2}} \, m \, , \label{Mandp2_Gubser}
\end{equation}
\vspace{0.1cm}

where the $1/ \sqrt{1-v^2} \equiv \gamma$ is identified as the Lorentz factor from special relativity. Eq.(\ref{Mandp2_Gubser}) holds regardless of the dimension of the spacetime. Inserting Eq.(\ref{Mandp2_Gubser}) into Eq.(\ref{DragForce2_Gubser}) yields

\begin{equation}
\frac{dp_2}{d  t} \, = \, - \frac{\pi\, T^2}{2}\, \sqrt{g_{YM}^2  N_c}\, \frac{p_2(t)}{m}\, , \label{Mandp2_DragForce_Gubser}
\end{equation}
\vspace{0.1cm}
which is a linearly separable differential equation and easily integrable to find $p_2(t)$. Specifically,


\begin{equation}
p_2(t) \, = \, p_2(0) \, e^{-\, \frac{t}{t_0}}\,\, , \,\,\,\,\,\,\,\,\,\,\,\,\,\,\,\, \text{where} \,\,\,\,\,\,\,\,\, t_0 \, = \, \frac{2}{\pi \,\sqrt{g_{YM}^2  N_c}}\, \frac{m}{T^2}\, . \label{Mandp2_Integrate_Gubser}
\end{equation}
\vspace{0.1cm}

The expression for the drag force experienced by a probe quark in a thermal plasma found in Eq.(\ref{Mandp2_DragForce_Gubser}) agrees with the central result of Gubser \cite{Gubser_2006}, and Herzog \textit{et al.} \cite{herzog2006energy}. Agreement with the later rests on the identification $\xi' = x_{\text{Herzog}}'/l^2$ and $g = -g_{\text{Herzog}}/l^4$.\\

Recall the Langevin Model discussed in section (\ref{secBrownMot}) describes a non-relativistic particle of mass $m$, undergoing Brownian motion in one spatial dimension. The non-retarded Langevin equation Eq.(\ref{LangEq}) is applicable if -- as in the case presented in this section -- the Brownian particle is
taken to have infinite mass with respect to the constituent fluid particles. Since the infinitely massive probe quark travels at a constant velocity $v$ under the influence of the external force $K(t)$, Newton's First Law of Motion ensures that the random force $F(t)$ present in the system must be of equal and opposite magnitude ($K(t) = - F(t)$). The Langevin equation is adjusted accordingly and becomes $\Dot{p}(t) = - \gamma_0 \, p(t)$, where $\gamma_0$ is the friction coefficient. It is therefore apparent, from Eq.(\ref{Mandp2_DragForce_Gubser}), that in the non-relativistic limit $v \ll 1$ the friction coefficient in AdS$_5$-Schwarzschild is given by

\begin{equation}
\gamma_0^{\text{AdS$_5$}} \, = \,  \frac{\pi\, T^2}{2 m}\, \sqrt{g_{YM}^2  N_c}\, . \label{FrictionCoeff_Gubser}
\end{equation}
\vspace{0.1cm}

Using the Einstein-Sutherland relation Eq.(\ref{DiffusionC}) an expression for the diffusion coefficient can be obtained. Specifically,
\begin{equation}
D_{\text{HQ}}^{\text{AdS$_5$}} \, = \,  \frac{2}{\pi\, T}\, \frac{1}{\sqrt{g_{YM}^2  N_c}} \, = \,\frac{2 \, \beta}{\pi \sqrt{\lambda}} \, , \label{DiffusionC_Gubser}
\end{equation}
\vspace{0.1cm}

where the second equality follows from the `t Hooft coupling $\lambda \, \equiv \, g_{\text{YM}}^2\,N_c$, and the definition of Hawking temperature Eq.(\ref{HawkingTemp}).

\subsection{Generalising to  \texorpdfstring{AdS$_d$-Schwarzschild}{AdSd-Schwarzschild}} \label{subsecTRAIL_GenAdSd}

The result for the drag force experienced by a probe quark in a thermal plasma found in Eq.(\ref{Mandp2_DragForce_Gubser}) can be generalised by repeating the calculation presented in subsection (\ref{subsec_TRAILinAdS5}) in AdS$_d$-Schwarzschild. The AdS-Schwarzschild metric in $d$ dimensions is given by Eq.(\ref{AdSd_Metric}), where the blackening factor of the Schwarzschild black hole situated at the horizon, $h(r;d)$, is defined in Eq.(\ref{BlackFactors}). Explicitly, the target spacetime metric $G_{\mu \nu}$ is given by Eq.(\ref{appTargetMatrix}), where $\mu,\nu$ index over the spacetime directions $\left(t,\, r,\, 2,\, ...,\, d-1 \right)$. As in subsection (\ref{subsec_TRAILinAdS5}), the reparameterization offered by the gauge choice separates the temporal and radial coordinates such that the worldsheet parameter space coordinates become $\tau = t$ and $\sigma = r$, and the mapping functions between the worldsheet and the target spacetime are specified by $X^{\mu}(t,r)$.\\

The Hawking temperature of the black-brane in AdS-Schwarzschild scales with the number of dimensions of the spacetime and -- in AdS$_d$-Schwarzschild -- is given by Eq.(\ref{HawkingTemp}). Unchanged as the calculation is generalised to AdS$_d$, the dynamics of the string are still described by the Nambu-Goto action (Eq.(\ref{NGAction})); and the induced metric $g_{a b}$, its inverse $g^{a b}$, and its determinant $g := \text{det} (g_{a b })$ are still given by Eqs.(\ref{appIndMetric2}, \ref{appInverseIndMetric}, \ref{appIndMetricDet}) respectively. Because of this, Eqs.(\ref{IndMetricDet_AdS5})-(\ref{Define_XiPrime}) all remain true in AdS$_d$-Schwarzschild.\\

Consider movement of the test string in AdS$_d$-Schwarzschild, again with a constant velocity $v$ in one of the transverse directions, the $X^2$ direction. Following the same reasoning as for the AdS$_5$ case, Eq.(\ref{Define_XiPrime}) can be solved to yield an expression for $\pi_\xi$. Since the blackening factor is now given by Eq.(\ref{BlackFactors}), $\pi_\xi$ will depend on the number of dimensions of the spacetime. Rearranging the first equality in Eq.(\ref{FindingBC_Gubser}) leaves

\begin{equation}
    \pi_\xi^2 \, = \,\frac{r^4}{l^4} \, v^2\, . \label{FindingBC2_AdSd_Gubser}
\end{equation}
\vspace{0.1cm}

While, for AdS$_d$-Schwarzschild, the second equality in Eq.(\ref{FindingBC_Gubser}) is rearranged to

\begin{equation}
    r \, = \, \frac{r_H}{(1 \, - \, v^2)^{1/(d-1)}} \, , \label{FindingBC3_AdSd_Gubser}
\end{equation}
\vspace{0.1cm}

where $h=h(r;d)$ (Eq.(\ref{BlackFactors})) is used. Inputting Eq.(\ref{FindingBC3_AdSd_Gubser}) into Eq.(\ref{FindingBC2_AdSd_Gubser}) results in an appropriate expression for $\pi_\xi$,
\begin{equation}
    \pi_\xi \, = \, \pm \, \frac{v}{(1\, - \, v^2)^{2/(d-1)}}\,\frac{r_H^2}{l^2} \, . \label{FindingBC4_AdSd_Gubser}
\end{equation}
\vspace{0.1cm}

Plugging this generalised relation for $\pi_\xi$ into Eq.(\ref{Define_XiPrime}) guarantees that the square root in Eq.(\ref{Define_XiPrime}) is always positive, and $\xi(r)$ is everywhere real. The $(\pm)$ in Eq.(\ref{FindingBC4_AdSd_Gubser}) determines if the quark is moving in the $(+)$ or $(-)$ $X^2$ direction. Again consider the case where $\pi_\xi > 0$. In AdS$_d$-Schwarzschild, Eq.(\ref{Define_XiPrime}) becomes\footnote{Mathematica is used to help simplify the algebra (see Mathematica Notebook [d]: \texttt{DragForce.nb}).}

\begin{equation}
    \xi'(r) \, = \, \pm \, \frac{l^2 \,r_H^3 \,r^{d-4}\,v \, \left(1-v^2\right)^{-\frac{2}{d-1}} \,\sqrt{\frac{r^4 \,\left(r_H\, \left(v^2-1\right)\, r^d \, +\, r\, r_H^d \right)}{-r_H \,r^{d+4}\, + \, r_H^5 v^2 \, r^d\, \left(1-v^2\right)^{-\frac{4}{d-1}}+r^5 r_H^d}}}{r_H \,r^d\, -\,r \,r_H^d}\, , \label{XiPrime_AdSd_Gubser}
\end{equation}
\vspace{0.1cm}

which reduces down to the familiar Eq.(\ref{XiPrime_Gubser}) for $d = 5$. The function $\xi'(r)$ can take on two values: (i) for $( \xi'(r) > 0, \, \xi(r) < 0)$  the string trails behind its boundary endpoint which is moving with a velocity $v$ in the $(+)$ $X^2$ direction, and (ii) for $(\xi'(r) < 0,\, \xi(r) > 0)$ the string trails in front of its boundary endpoint which is moving with a velocity $v$ in the $(+)$ $X^2$ direction. Using the expression for $\xi'(r)$ in Eq.(\ref{XiPrime_AdSd_Gubser}) and the definition for $h(r;d)$ in Eq.(\ref{BlackFactors}), the Lagrangian density (Eq.(\ref{LDensity2_Gubser})) becomes

\begin{equation}
    \mathcal{L}\, := \, -\sqrt{-g} \, = \,- \, \sqrt{\frac{r^4 \left(r_H \left(v^2-1\right) r^d \,+\, r \,r_H^d \right)}{-  r_H \, r^{d+4}\, +\, r_H^5\, v^2\, r^d\, \left(1-v^2 \right)^{-\frac{4}{d-1}}\, +\, r^5\, r_H^d}}\, . \label{LDensity3_AdSd_Gubser}
\end{equation}
\vspace{0.1cm}



To proceed with calculating the drag force $dp_2/dt$ note that the canonical momentum densities $\Pi^a_\mu (t, \sigma)$ (Eq.(\ref{MomDensity})), the string equations of motion (Eq.(\ref{StringEoM})), and the definition of the conserved charges associated with the momentum densities Eq.(\ref{ConservedCharge_Gubser}) all remain the same in AdS$_d$-Schwarzschild. As in the $d=5$ case, the string moves only in the $X^2$ direction by construction, and its transverse momentum densities are only non-zero in this direction. The $X^2$ component of the spacetime momentum flowing over some general time interval $\mathcal{I}$ of length $\Delta t$, is
still given by Eq.(\ref{ConservedcCharge2_Gubser}) in AdS$_d$-Schwarzschild. For the physical solution where the string trails behind its boundary endpoint $(\xi'(r) > 0, \, \xi(r) < 0)$, $d p_{2}/d t   =  \Pi^{r}_{2}$. According to the definition of the momentum densities Eq.(\ref{MomDensity}),

\begin{flalign}
& \begin{aligned}
     \frac{d p_{2}}{d  t} \, := \, \Pi^{r}_{2} \, 
     & =  \,- \frac{1}{2 \pi \alpha'} \, \sqrt{-g} \, g^{r b}\, G_{2 \,\nu}\, {\partial_b} X^\nu\\[0.2cm]
     & = \, \frac{1}{2 \pi \alpha'} \left(\sqrt{\frac{r^4 \left(r_H \left(v^2-1\right) r^d \,+\, r \,r_H^d \right)}{-  r_H \, r^{d+4}\, +\, r_H^5\, v^2\, r^d\, \left(1-v^2 \right)^{-\frac{4}{d-1}}\, +\, r^5\, r_H^d}}\right)^{-\frac{1}{2}}G_{2 \,2}\, G_{tt} \, \partial_r \xi \\[0.2cm]
     & = \, - \frac{1}{2 \, \pi \, \alpha'}\,\frac{r_H^2}{l^2}\, \frac{v}{ \left(1-v^2\right)^{\frac{2}{d-1}}} \, , \label{MomDensity2_AdSd_Gubser}
     \end{aligned} &
\end{flalign}
\vspace{0.1cm}

where the constraint equation Eq.(\ref{Constraint2}) is used in the first line; while in the second line, the summation over repeated indices is expanded and Eq.(\ref{LDensity3_AdSd_Gubser}) is used. Following a similar calculation to Eq.(\ref{DragForce_Gubser}), recognize that $G_{2 \,\nu} = G_{2 \,2}$  is the only non-zero possibility, Eq.(\ref{appInverseIndMetric}) can be used to find the explicit entries of the inverse induced metric $g^{ab}$ and the ansatz Eq.(\ref{LateTime_Gubser}) can be used to simplify the derivative terms. The final line follows from inputting the target spacetime metric entries (Eq.(\ref{appTargetMatrix})) and the defining relation for $\xi'(r)$, Eq.(\ref{XiPrime_AdSd_Gubser}) (where $\xi'(r) >0$).\\

The drag force given in Eq.(\ref{MomDensity2_AdSd_Gubser}) reduces down to Eq.(\ref{DragForce_Gubser}) when $d=5$ -- a necessary consistency check. Using the definition of Hawking temperature (Eq.(\ref{HawkingTemp})) to insert a unit factor $\left(T^2 \times 1/T^2\right)$, the drag force can be rewritten
\begin{equation}
     \frac{d p_{2}}{d  t} \, = \,  - \, \frac{8 \,\pi \,l^2 \, T^2}{(d-1)^2\, \alpha'} \, \frac{v}{ \left(1-v^2\right)^{\frac{2}{d-1}}}  \, , \label{DragForce_AdSd_Gubser}
\end{equation}
\vspace{0.1cm}

where the momentum $p_2(t)$ is given by Eq.(\ref{Mandp2_Gubser}), which holds regardless of the dimension of the spacetime. From Eq.(\ref{DragForce_AdSd_Gubser}), in the non-relativistic limit $v \ll 1$, the friction coefficient in AdS$_d$-Schwarzschild is given by
\begin{equation}
\gamma_0^{\text{AdS$_d$}} \, = \,  \frac{8 \,\pi \,l^2 \, T^2}{(d-1)^2\, \alpha'\, m}\, . \label{FrictionCoeff_AdSd_Gubser}
\end{equation}
\vspace{0.1cm}

Using the Einstein-Sutherland relation Eq.(\ref{DiffusionC}) an expression for the diffusion coefficient in general $d$ dimensions can be obtained. Specifically,

\begin{equation}
D_{\text{HQ}}^{\text{AdS$_d$}}(d) \, = \,  \frac{(d-1)^2\, \alpha'}{8 \,\pi \,l^2 \, T}\, = \, \frac{(d-1)^2 \, \beta}{8 \pi \sqrt{\lambda}} \, , \label{DiffusionC_AdSd_Gubser}
\end{equation}
\vspace{0.1cm}

where the second equality follows from the fundamental string length scale Eq.(\ref{alpha_def}), and the definition of Hawking temperature Eq.(\ref{HawkingTemp}). For $d=3$, complete agreement is found between this equation and the diffusion coefficient found in the case of a finite mass heavy quark undergoing Brownian motion in the thermal plasma (subsection (\ref{subsubsecHQ_GenAdSd}), Eq.(\ref{DifCoeff_HQ})) -- a comforting consistency check for this work\footnote{Eq.(\ref{DiffusionC_AdSd_Gubser}) also agrees with the result given in Eq.(3.10) of \cite{Boer_2009}; the results from subsection (3.3) in \cite{herzog2006energy}; the results from section (3) in \cite{Gubser_2006}; and the result -- for $d=5$ -- from section (V) in \cite{Casalderrey_Solana_2006}.
The rigorous reader might also be interested in \cite{Liu_2006}, a topical study which computed an ultra-relativistic quark's transverse momentum diffusion.}.\\

The heavy quark diffusion coefficient in general $d$ dimensions (Eq.(\ref{DiffusionC_AdSd_Gubser})) can be compared to the light quark diffusion coefficient in general $d$ dimensions (Eq.(\ref{DifCoeff_LQ_d})), to find

\begin{equation}
D_{\text{LQ}}^{\text{AdS$_d$}}(d) \, = \, \frac{1}{2}\,D_{\text{HQ}}^{\text{AdS$_d$}}(d)\, , \label{DifCoeff_comparison_d}
\end{equation}
\vspace{0.1cm}

which agrees with Eq.(\ref{DifCoeff_comparison}) for $d=3$. The factor of $1/2$ in Eq.(\ref{DifCoeff_comparison_d}) may arise through differences in partitioning the worldsheet for the heavy and light quark test strings. Moerman \textit{et al.} postulated in \cite{moerman2016semi} that by introducing a factor $a$ which determines the fraction of the local speed of light the boundary endpoint of the test string falls at, a general expression can be found for the mean-squared displacement of said endpoint (and therefore the diffusion coefficient) which interpolates between the heavy and light quark results\footnote{The heavy and light quark set-ups to study Brownian motion are considered in sections (\ref{secHeavyQuark}) and (\ref{secLightQuark}), respectively.}. Explicitly, the string falling endpoint’s mean-squared transverse displacement is given by

\begin{equation}
s^2_{\text{small}} (t;a;d) \, = \, \frac{1}{\sqrt{\lambda}} \left(\frac{(d-1) \beta}{4 \pi}\right)^2  \ln \left(\frac{2 a \beta^3}{\pi^3 (a^2-1)^2 t^3} \sinh^2\left(\frac{\pi (a+1) t}{\beta}\right) \sinh^2 \left( \frac{\pi (a-1) t}{\beta}\right)  \csch \left(\frac{2 \pi a t}{\beta}\right) \right) \, , \label{displacement_a}
\end{equation}
\vspace{0.1cm}

which, at late times, becomes\footnote{See Mathematica notebook [b] (\texttt{BrownianMotion.nb}) for a derivation of Eqs.(\ref{displacement_a})-(\ref{DiffusionCoeff_a}) -- access in appendix (\ref{appCode}).}
\begin{equation}
s^2_{\text{small}} (t;a;d) \,\stackrel{(\beta \ll t)}{\longrightarrow}\,\, \frac{(d-1)^2 \beta t }{4 \pi \sqrt{\lambda}} \left[1 - \frac{a}{2} \right] \, + \, \frac{(d-1)^2 \beta^2}{16 \pi^2 \sqrt{\lambda}}  \begin{Bmatrix}
   4 \ln \left(\frac{\beta}{2 \pi t} \right) \, , \,\,\,\,\,\,\,\,\,\,\,\,\,\,\,\,\,\,\,\,\,\,\,\,\,\,\,\,\,\,\,\,\,\,\,\,\,  \text{if} \,\,\,\ a = 0  \\
   \ln \left(\frac{a \beta^3}{4 \pi^3 (a^2-1)^2 t^3} \right) \, , \,\,\,\,\,\,\,\,\,  \text{if} \,\,\,\ 0 <a < 1\\
   \ln \left(\frac{\beta}{4 \pi t} \right) \, , \,\,\,\,\,\,\,\,\,\,\,\,\,\,\,\,\,\,\,\,\,\,\,\,\,\,\, \,\,\,\,\,\,\,\,\,\,\,\,\,\,  \text{if} \,\,\,\ a = 1
  \end{Bmatrix} \,+\, \mathcal{O}\left(\frac{\beta}{t}\right)^0\, .  \label{displacement_a_Late}
\end{equation}
\vspace{0.1cm}

The diffusion coefficient for a quark in a ($d-1$)-dimensional thermal plasma can be extracted from Eq.(\ref{displacement_a_Late}). Specifically,

\begin{equation}
D(a;d) \, = \, \left[1- \frac{a}{2}\,  \right] \, \frac{(d-1)^2 \, \beta}{8 \pi \sqrt{\lambda}} \,  . \label{DiffusionCoeff_a}
\end{equation}
\vspace{0.1cm}

When $a=0$, Eq.(\ref{DiffusionCoeff_a}) reduces to the heavy quark diffusion coefficient, Eq.(\ref{DiffusionC_AdSd_Gubser}); and when $a=1$, Eq.(\ref{DiffusionCoeff_a}) becomes the light quark diffusion coefficient, Eq.(\ref{DifCoeff_LQ_d}). Hence Eq.(\ref{displacement_a}) provides a natural interpolation between a test string set-up where the boundary endpoint is held stationary, and a test string set-up where the boundary endpoint is allowed to fall at the local speed of light. However, values of $a \notin \{0,1\}$ correspond to a test string whose boundary endpoint is falling at velocity smaller than the local speed of light. An external force opposing the motion of the falling string must be present in order for this to be the case\footnote{The solution where $a \notin \{0,1\}$ is not a physically obtainable string solution in the current set-up since the Virasoro constraints (Eq.(\ref{Virasoro})) are violated and, consequently, the late time next-to-leading order behaviour in Eq.(\ref{displacement_a_Late}) does not smoothly interpolate between the light and heavy quark results. This can be rectified by adding in an external field which will alter the energy-momentum tensor (Eq.(\ref{EMdef})) and, by construction, cause $T_{ab}$ to vanish (which ensures the Virasoro constraints are satisfied).}. This might be realised by the introduction of a flavour D7-brane with a world-volume electric field on it at the boundary\footnote{This corresponds to an additional force acting on the external Brownian particle in the boundary theory whose motion can now be characterised by the generalised Langevin model (described in subsection (\ref{subsecGenLangevin})).}\footnote{For the case of the heavy quark, adding forced motion has been studied by de Boer \textit{et al.} \cite{Boer_2009}, subsection (3.2).}.
Considering an external force acting to retard the motion of the falling string's endpoint is a natural and interesting extension of this work -- a possible starting point for future research.

\section{Conclusions and Future Outlook}\label{secConclusion}

In this article we present a pedagogic account of applications of the AdS/CFT correspondence for understanding the dynamical behaviour of probe heavy and light quarks immersed in a thermal plasma, such as the quark-gluon plasma created in heavy-ion experiments. As a central result, we give the first self-consistent derivation of the light quark diffusion coefficient in an arbitrary number of dimensions, correcting previous errors and giving some of the vaguer statements found in the literature a rigorous foundation.\\

To summarise, the gauge/string duality is briefly introduced in section (\ref{secGaugeString}). Particular attention is paid to the \textit{throat} construction of anti-de Sitter spacetime as a limit of D3-brane geometry, and the justification of the AdS/CFT conjecture. Section (\ref{secBrownMot}) explored the basic theory of Brownian motion and Langevin dynamics in the boundary theory. Specifically, particles undergoing Brownian motion in the absence of an external force can be described by the non-retarded Langevin equation which is parametrized by two constants: (i) the friction coefficient $\gamma_0$, and (ii) the magnitude of the random force $\kappa_0$. The second fluctuation-dissipation theorem relates these two constants. It is possible to calculate the mean-squared displacement (Eq.(\ref{LangDisplacementSquaredAv3})), from which the time dependence of $s^2(t)$ is apparent. At early times, the Brownian particle’s behaviour is proportional to time and the motion is expected to be ballistic $s(t) \sim t$; while, at late times, the Brownian particle motion is diffusive, $s(t) \sim \sqrt{t}$. The cross-over time sets the scale for early and late time behaviour and is given by $t_{\text{relax}}$, which represents the time it takes for a Brownian particle which had some initial velocity at $t=0$ to thermalize in the medium. The section concludes by describing the generalised Langevin model which adapts the friction term to depend on the past trajectory of the Brownian particle and takes an additional external force acting on the system into account. In section (\ref{secDragForce}) it was seen that the generalised Langevin model can be used in a number of applications: to model heavy quark trailing strings or light quark strings whose endpoint falls with a velocity that is less than the local speed of light.\\

Sections (\ref{secHeavyQuark}), (\ref{secLightQuark}) and (\ref{secDragForce}) turn towards the bulk theory. In the bulk theory, the dual description of these probe quarks are realised as test strings in an asymptotically anti-de Sitter-Schwarzschild background. Calculations are computed in AdS-Schwarzschild, and then related to quantities in the boundary theory using the AdS/CFT dictionary. In section (\ref{secHeavyQuark}) an on-mass-shell external heavy quark is modelled as a fundamental open string of length $\ell_0$ attached at the boundary of anti-de Sitter spacetime and hanging towards the stretched horizon.
At the semiclassical level, the Hawking radiation due to the Schwarzschild black hole environment excites the modes on the string -- resulting in the string's boundary endpoint enduring irregular motion. This motion can be related to the Brownian motion of the external heavy quark in the boundary gauge theory. The main results of this section include the derivation of the leading order, static string solution in AdS$_3$-Schwarzschild (Eq.(\ref{HQ_Sol_AdS3})); the transverse equations of motion found by expanding the Nambu-Goto action up to quadratic order and varying this action with respect to the transverse string worldsheet coordinates $X^I$ (Eq.(\ref{StringEoM_XI})); and from there the mean-squared displacement of the test string’s boundary endpoint, $s^2(t)$, in AdS$_3$-Schwarzschild (Eq.(\ref{st_squared})). The cross-over time from early to late time dynamics is found to be independent of the initial length of the string $\ell_0$ and solely dependant on the Hawking temperature $\beta = 1/T$. As expected, in the early time limit ($t \ll \beta$) the motion is ballistic (Eq.(\ref{st_squaredHQ_Early})); while in the late time limit ($t \gg \beta$) the motion is diffusive, and the diffusion coefficient can be extracted from $s^2(t)\rvert_{t\gg \beta}$ (Eq.(\ref{DifCoeff_HQ})). We further showed that the diffusion coefficient computed in this way and the friction coefficient computed in a completely independent manner are exactly related by the fluctuation dissipation theorem as required for thermalizing motion.\\

An off-mass-shell external light quark is modelled in section (\ref{secLightQuark}) as an open string, initially stretched between the AdS boundary and just above the horizon, whose AdS boundary endpoint is released to fall at the local speed of light. The set-up is termed the \textit{Limp Noodle} configuration \cite{moerman2016semi}. This initial set-up is held for an asymptotically long amount of time, allowing for a full population of thermal modes on the initially stretched string. Using the Bars \textit{et al.} method \cite{Bars_1994, Bars_1995}, the worldsheet of the Limp Noodle is partitioned into two regions and the leading order string solution for the string once the endpoint near the boundary is released is found in AdS$_3$-Schwarzschild (Eq.(\ref{LN_AdS3Sol})). Again, the transverse fluctuations on the string excited by the Schwarzschild black hole are considered. From the equations of motion of these fluctuations, the string falling endpoint’s mean-squared transverse displacement $s^2(t)$ is calculated (Eq.(\ref{st_squaredLQ})).
The limiting cases of $s^2(t)$ are then examined. For the small virtuality case (small string lengths $\ell_0$ compared to the radial position of the black-brane horizon $r_H$), $s_{\text{small}}^2(t)$ is analytically evaluated (Eq.(\ref{st_squaredLQ_SV2})) and the early and late time behaviour found to be ballistic and diffusive respectively. For the arbitrary virtuality case, $s^2(t)$ can be analytically found in the early time limit (Eq.(\ref{st_LQ_Early_Times_2})); while the behaviour of arbitrary virtuality quarks at asymptotically late times (i.e. in the near-horizon region) is found to be encoded in the small virtuality case, i.e. $s^2(t)\rvert_{t\gg \beta} = s_{\text{small}}^2(t)$ (Eq.(\ref{LQ_Late_Times_HQ})). Since $s_{\text{small}}^2(t;d)$ can be solved in any $d \geq 3$ dimensions, this universality of late times motivates the generalisation of the mean-squared displacement $s^2(t)\rvert_{t\gg \beta}$ to AdS$_d$-Schwarzschild. In terms of original advancement, subsection (\ref{subsubsecLQ_GenAdSd}) presents a very important part of this article: correcting Moerman \textit{et al.} \cite{moerman2016semi} by presenting the proper method in which to generalise to $s^2(t;d)$. Finally, the diffusion coefficient in AdS$_d$-Schwarzschild is extricated from $s^2(t;d)\rvert_{t\gg \beta}$ and given in Eq.(\ref{DifCoeff_LQ_d}) by
\begin{equation*}
D_{\text{LQ}}^{\text{AdS$d$}}(d) \, = \, \frac{(d-1)^2 \beta}{16 \pi \sqrt{\lambda}}\, ,
\end{equation*}

which is -- to the authors' knowledge -- the first complete, consistent derivation of the light quark diffusion coefficient in $d$ dimensions.\\

In section (\ref{secDragForce}), a trailing string in AdS$_5$-Schwarzschild which models an infinitely massive probe quark moving with a constant velocity $v$ in a $\mathcal{N}=4$ SYM thermal plasma is considered. By making use of the late-time behaviour ansatz (Eq.(\ref{LateTime_Gubser})), the drag force on the test string is calculated in the bulk and rewritten -- via the AdS/CFT correspondence -- in terms of relevant quantities in the gauge theory (Eq.(\ref{DragForce2_Gubser})). Because the external quark is moving with a constant velocity in the boundary theory, it can be modelled by the Langevin equation adapted such that the external force $K(t)$ acting on the system is of equal and opposite magnitude to the random force ($K(t) = −F(t)$). Hence the friction coefficient $\gamma_0$ can be read off from Eq.(\ref{Mandp2_DragForce_Gubser}), and related -- using the Einstein-Sutherland relation -- to the diffusion coefficient in $d=5$ dimensions. In the second subsection, the drag force calculation is generalised to AdS$_d$-Schwarzschild and the diffusion coefficient is found to be given in Eq.(\ref{DiffusionC_AdSd_Gubser}) by

\begin{equation*}
D_{\text{HQ}}^{\text{AdS$_d$}}(d)\, = \,\frac{(d-1)^2 \, \beta}{8 \pi \sqrt{\lambda}} \, ,
\end{equation*}

which agrees, for $d=3$, with the diffusion coefficient found by studying the transverse fluctuations on the heavy quark's test string in section (\ref{secHeavyQuark}) (Eq.(\ref{DifCoeff_HQ})).\\

The heavy and light quark's diffusion constants in general $d$ dimensions are related by a factor of a $1/2$ (Eq.(\ref{DifCoeff_comparison_d})). This disparity may arise through the differences in partitioning the worldsheet for the heavy and light quark test strings. As was briefly discussed towards the end of section (\ref{secDragForce}), in order to understand the physical origin of the $1/2$ factor an interpolation between $D_{\text{LQ}}^{\text{AdS$d$}}(d)$ and $D_{\text{HQ}}^{\text{AdS$_d$}}(d)$ can be sought. This leads to the realisation that considering test strings with boundary endpoints falling slower than the local speed of light will only yield physically valid solutions if an external force is introduced. Considering an external force acting to retard the motion of the falling string's endpoint is a natural and interesting extension of this work, and a possible starting point for future research. The publications \cite{Boer_2009, Matsuo_2006, Kim_2011, cederwall1996dirac} might provide a good commencement for this exercise.\\

The main contribution of this article to the field of AdS/CFT calculations is presenting a definitive, consolidated theoretical derivation of the light and heavy quark diffusion constants in general $d$ dimensions. The work acts as a \textit{springboard} from which students might pursue further research avenues. Some of these that the authors have considered -- besides the addition of an external electric field -- are (i) repeating these calculations using a numerical framework in order to confirm the analytic results; (ii) examining the fluctuation-dissipation theorem in the bulk; (iii) considering next-to-leading order transverse fluctuations on the leading order solution (this would surely prove difficult as the expanded, near-horizon metric in AdS$_d$-Schwarzschild would no longer be conformal in ($t, r_*$) coordinates); and (iv) considering different test string configurations to illuminate other aspects of the thermal plasma. With regards to the latter, one such example is studying the quantum fluctuations in the non-transverse directions of a trailing string where the boundary endpoint is allowed to fall. Another excellent example might be the consideration of a light quark string setup that more closely models the phenomenologically relevant situation in which a light quark--anti-quark pair is created highly off-shell and without a pre-population of thermal modes along the string. Determining the fluctuations on top of the leading order classical motion of this string as it evolves in time would require knowledge of the thermal spectrum of metric fluctuations due to the presence of the black brane.

\section*{Acknowledgements}
The authors thank Steven Gubser for the idea which inspired the Limp Noodle construction. The news of his untimely passing is tragic; he left an indelible impact on the community. May he rest in peace.
\vspace{0.2cm}

\noindent \textbf{Funding information}  A.K.M. appreciates the support of the Mandela Rhodes Foundation, and W.A.H. gratefully acknowledges support from the South African National Research Foundation and the South Africa-CERN Collaboration.

\newpage

\appendix
\section*{Appendix} \label{secAppendix}
\vspace{0.2cm}

\section{Polyakov String Equations of Motion} \label{appPolyakov}


Working in the static gauge, the string equations of motion are derived in this appendix by calculating the functional derivative of the Polyakov Action with respect to the string worldsheet coordinates and setting this variation to zero. The calculation of the string equations of motion presented here follows the layout of appendix (A) in Moerman \textit{et al.}'s exposition on light quark Brownian motion \cite{moerman2016semi}.\\

Using the definition Eq.(\ref{MomDensity}), the Polyakov Action Eq.(\ref{PolyakovAct}) is rewritten in terms of the canonical momentum densities

\begin{equation}
S_P \, = \, \frac{1}{2}\, \int_{\mathcal{M}} \, d^2 \sigma \, \Pi_\mu^a(t,\sigma)\, \partial_a\,X^\mu(t,\sigma) \, . \label{appPolyakovMomDensities}
\end{equation}
\vspace{0.1cm}

Determining the functional derivative of $S_P$ with respect to $X^\mu$ and setting this variation to vanish, yields

\begin{flalign}
& \begin{aligned}
    0 \, = \, \delta_X  S_P\, & = \, \frac{1}{2}\, \int_{\mathcal{M}} \, d^2 \sigma \,\Big[ \big( \delta \Pi_\mu^a \big)\, \partial_a X^\mu \,+\, \Pi_\mu^a\, \partial_a \big(\delta X^\mu \big) \Big] \\[.2cm]
    & = \, \frac{1}{2}\, \int_{\mathcal{M}} \, d^2 \sigma \,\Big[ \big( \delta  \Pi_\mu^a \big)\, \partial_a X^\mu \,+\,  \partial_a \big(\Pi_\mu^a\,\delta X^\mu \big) \, - \, \delta X^\mu\, \partial_a \big(\Pi_\mu^a \big)\Big]  \, , \label{appVaryS_P}
\end{aligned} &
\end{flalign}
\vspace{0.1cm}

where the product rule is used in the second line.\\

The functional derivative of the momentum densities with respect to $X^\mu$ gives two terms

\begin{equation}
 \delta_X \Pi_\mu^a \, = \, -\frac{1}{2\pi \alpha'}\,\sqrt{-\gamma} \, \gamma^{ab} \Big[ G_{\mu \nu}\, {\partial_b}\big(\delta X^\nu \big) \,
 +\, \partial_\rho \, G_{\mu \nu}\,\delta X^\rho {\partial_b} X^\nu\,\Big] \, , \label{appVaryMomDen}
\end{equation}
\vspace{0.1cm}

since, in a curved background, the worldsheet coordinates $X^\mu$ as well as the spacetime metric $G_{\mu \nu}$ (which also depends on $X^\mu$) are varied. Hence the first term in Eq.(\ref{appVaryS_P}), $\big( \delta \,\Pi_\mu^a \big)\, \partial_a X^\mu$, becomes

\begin{flalign}
& \begin{aligned}
    \big( \delta_X \Pi_\mu^a \big)\, \partial_a\,X^\mu\, & = \, \frac{-1}{2\pi \alpha'}\,\sqrt{-\gamma} \, \gamma^{ab} \, G_{\mu \nu}\, {\partial_b}\big(\delta X^\nu \big)\, \partial_a\,X^\mu \, +\, \partial_\rho \, G_{\mu \nu}\,\delta X^\rho \, \Big(\frac{-1}{2\pi \alpha'}\,\sqrt{-\gamma} \, \gamma^{ab} \, {\partial_b} X^\nu\, \Big)\, \partial_a\,X^\mu  \\[.2cm]
    & = \, \frac{-1}{2\pi \alpha'}\,\sqrt{-\gamma} \, \gamma^{ab} \, G_{\mu \nu}\, {\partial_b} X^\nu \, \partial_a \big(\delta X^\mu \big) \, +\, \partial_\rho \, G_{\mu \nu}\,\delta X^\rho \, \Big(\frac{-1}{2\pi \alpha'}\,\sqrt{-\gamma} \, \gamma^{ab} \, G_{\gamma \nu} \, {\partial_b} X^\nu\, \Big)\, G^{\gamma \nu} \, \partial_a\,X^\mu \\[.2cm]
    &  = \, \Pi_\mu^a \, \partial_a \big(\delta X^\mu \big) \, +\, \partial_\rho \, G_{\mu \nu}\,\delta X^\rho \, \Pi_\gamma^a\, G^{\gamma \nu} \, \partial_a\,X^\mu \\[.2cm]
    &= \, \partial_a \big(\Pi_\mu^a \,\delta X^\mu \big)\, - \,\delta \,X^\mu\, \partial_a \big(\Pi_\mu^a \big) \, +\, 2\,\delta X^\rho \, \Big(\frac{1}{2}\,G^{\gamma \nu} \,\partial_\rho \, G_{\mu \nu}\Big)\, \Pi_\gamma^a \, \partial_a\,X^\mu\\[.2cm]
    &= \, \partial_a \big(\Pi_\mu^a \,\delta X^\mu \big)\, - \,\delta \,X^\mu\, \partial_a \big(\Pi_\mu^a \big) \, +\, 2\,\delta X^\rho \, \Big(\frac{1}{2}\,G^{\gamma \nu} \big(\partial_\rho \, G_{\mu \nu}  + \partial_\mu\,G_{\rho \nu}  - \partial_\nu\,G_{\rho \mu}\big)\Big)\, \Pi_\gamma^a \, \partial_a\,X^\mu\, , \label{appVaryS_P1Term}
\end{aligned} &
\end{flalign}
\vspace{0.1cm}

where the definition of the momentum densities Eq.(\ref{MomDensity}) is used in the third line, and the product rule is used in the fourth line. In the final line two extra terms are added. This can be done since

\begin{equation}
\Pi_\gamma^a \, \partial_a\,X^\mu\, \big(\partial_\mu\,G_{\rho \nu}  - \partial_\nu\,G_{\rho \mu}\big) \, = - \frac{1}{2 \pi \alpha'} \, \sqrt{-\eta} \,G_{\mu \nu} \,
\Big[ \eta^{ab} \, \partial_b \, X^\nu\,\partial_a \, X^\mu  \big(\partial_\mu\,G_{\rho \nu}  - \partial_\nu\,G_{\rho \mu}\big)\,\Big]\,
= \, 0  \,, \label{appConformalSimp}
\end{equation}
\vspace{0.1cm}

where the first equality holds in the conformal gauge ($\gamma^{ab} = \eta^{ab}$) -- which, as discussed in subsection (\ref{subsecPolyakovEoM}), is chosen to eliminate the Weyl invariance in the Polyakov action. To prove Eq.(\ref{appConformalSimp}), notice that the Minkowski metric can be expanded

\begin{equation*}
\eta^{ab} \, \partial_a \, X^\mu \, \partial_b \, X^\nu \big(\partial_\mu\,G_{\rho \nu}  - \partial_\nu\,G_{\rho \mu}\big)\,
= \, \partial_\tau \, X^\mu \, \partial_\tau \, X^\nu \big(\partial_\mu\,G_{\rho \nu}  - \partial_\nu\,G_{\rho \mu}\big)\,+\, \partial_\sigma \, X^\mu \,
\partial_\sigma \, X^\nu \big(\partial_\mu\,G_{\rho \nu}  - \partial_\nu\,G_{\rho \mu}\big)
\end{equation*}
\vspace{0.1cm}

and that for each set of reflective cases ($\mu = 1, \, \nu = 2$ and $\mu=2, \, \nu=1$), the terms generated by the first case will cancel those generated by the second. For the identical case (which doesn't have a reflective partner case), i.e. $\mu = \nu = 1$ or  $\mu = \nu = 2$, the bracket $\big(\partial_\mu\,G_{\rho \nu}  - \partial_\nu\,G_{\rho \mu}\big)$ vanishes.\\

The Christoffel symbols are now able to be defined

\begin{equation}
\Gamma^\alpha_{\mu \nu}\,
:= \, \frac{1}{2}\, G^{\alpha \gamma} \big(\partial_\mu \, G_{\nu \gamma}  + \partial_\nu\,G_{\mu \gamma}
- \partial_\gamma\,G_{\mu \nu}\big)  \, . \label{appChristoffelSymbol}
\end{equation}
\vspace{0.1cm}

These symbols are a set of numbers which characterise a metric connection. A metric connection defines precisely how distances are measured on a surface, by invoking the notion of consistently transporting data in a parallel manner along a family of curves \cite{carroll2004spacetime}.\\

With this definition, Eq.(\ref{appVaryS_P1Term}) becomes

\begin{equation}
 \big( \delta_X \Pi_\mu^a \big)\, \partial_a\,X^\mu\, = \, \partial_a \big(\Pi_\mu^a \,\delta X^\mu \big)\,
 - \,\delta X^\mu\, \partial_a \big(\Pi_\mu^a \big) \, +\, 2\,\delta X^\rho \, \Gamma^\gamma_{\rho \mu}\, \Pi_\gamma^a \, \partial_a X^\mu  \, . \label{appVaryS_P1Term2}
\end{equation}
\vspace{0.1cm}

Finally, inputting Eq.(\ref{appVaryS_P1Term2}) into Eq.(\ref{appVaryS_P}) returns


\begin{flalign}
& \begin{aligned}
    0 \, = \, \delta_X  S_P\, & = \, \frac{1}{2}\, \int_{\mathcal{M}} \, d^2 \sigma \,\Big[- 2\, \delta X^\mu\, \partial_a \big(\Pi_\mu^a \big) \, + 2\,\delta X^\rho \, \Gamma^\gamma_{\rho \mu}\, \Pi_\gamma^a \, \partial_a\,X^\mu \,+\,  2\,\partial_a \big(\Pi_\mu^a\,\delta X^\mu \big) \Big]  \\[.2cm]
    & = - \int_{\mathcal{M}} \, d^2 \sigma \,\delta X^\mu\,\Big[\partial_a \,\Pi_\mu^a \,-\, \Gamma^\gamma_{\mu \nu}\, \Pi_\gamma^a \, \partial_a X^\nu \Big] \, + \, \int_{\mathcal{M}} \, d^2 \sigma \,\partial_a \big(\Pi_\mu^a\,\delta X^\mu \big)
    \\[.2cm]
    & = - \int_{\mathcal{M}} \, d^2 \sigma \,\delta X^\mu\,\Big[\partial_a \,\Pi_\mu^a \,-\, \Gamma^\gamma_{\mu \nu}\, \Pi_\gamma^a \, \partial_a X^\nu \Big] \, + \, \int_{\partial \mathcal{M}} \, d \sigma^b\, \epsilon_{ba}\, \big(\Pi_\mu^a\,\delta X^\mu \big)\\[.2cm]
    & = - \int_{\mathcal{M}} \, d^2 \sigma \,\delta X^\mu\,\Big[\partial_a \,\Pi_\mu^a \,-\, \Gamma^\gamma_{\mu \nu}\, \Pi_\gamma^a \, \partial_a X^\nu \Big] \, + \, \int_{0}^{\tau_f} \, d \tau \,\Big[\delta X^\mu\, \Pi_\mu^a \big\rvert^{\sigma = \sigma_f}_{\sigma = 0} \Big]\, , \label{appVaryS_P2}
\end{aligned} &
\end{flalign}
\vspace{0.1cm}

where Stokes' Theorem in d-dimensions \cite{carroll2004spacetime} is used in the third line. In the final line the coordinates on the worldsheet parameter space are chosen to be $\sigma^a = (t, \sigma)^a$ where $\sigma^a \in \mathcal{M} = [0,t_f] \times [0, \sigma_f]$, and one of the integrals of the last term is simplified by using $\delta X^\mu \rvert _{t \in \{ 0, t_f\}} = 0$.\\

Choosing the boundary conditions

\begin{equation}
 \delta X^\mu\, \Pi_\mu^a \big\rvert^{\sigma = \sigma_f}_{\sigma = 0} \, = \, 0 \, , \label{appBC}
\end{equation}
\vspace{0.1cm}

the last term in Eq.(\ref{appVaryS_P2}) disappears. Hence, the string equations of motion are given by

\begin{equation}
0 \, = \, \partial_a \,\Pi_\mu^a \, - \, \Gamma^\alpha_{\mu \nu}\, \partial_a \,X^\nu\,\Pi^a_\alpha \, =:\, \nabla_a\,\Pi^a_\mu\, , \label{appStringEoM}
\end{equation}
\vspace{0.1cm}

which is the string analogue of the geodesic equation for a point particle \cite{carroll2004spacetime}.

\section{The Virasoro Constraints and String Equations of Motion in Isothermal Coordinates} \label{appVirasoroEoMIso}

Since the metric of any $(1+1)$-dimensional subspace can be transformed into a conformally flat metric \cite{Bars_1994, Bars_1995} a new, isothermal coordinate system is introduced. These isothermal coordinates $y^{\mu'}(\sigma^+, \sigma^-)$ are given by Eq.(\ref{IsothemalCo-ord}), while the corresponding string mapping functions $Y^{\mu'}(\sigma^+, \sigma^-)$ are defined in Eq.(\ref{IsoMapFunc}). Notice that light-cone coordinates have been chosen for the parameter space.
In this appendix, the Virasoro constraint equations and the string equations of motion, Eqs.(\ref{Virasoro}, \ref{StringEoM}), are rewritten in terms of $Y^{\mu'}(\sigma^+, \sigma^-)$.\\

In the conformal gauge, the Virasoro constraint equations (Eqs.(\ref{Virasoro})) become

\begin{flalign}
& \begin{aligned}
  0 \, & = \, G \, \eta_{\mu \nu}\,\partial_{\pm} X^\mu \, \partial_{\pm} X^\nu \\[.2cm]
   & = \, - G  \,\partial_{\pm} X^0 \, \partial_{\pm} X^0 \,+\, G  \,\partial_{\pm} X^1 \, \partial_{\pm} X^1\\[.2cm]
    & = \, - \frac{G}{2}  \, \left[\partial_{\pm} \left( Y^{0'}+Y^{1'}\right) \, \partial_{\pm} \left( Y^{0'}+Y^{1'}\right) \right]\,+\, \frac{G}{2}\, \left[\partial_{\pm} \left( Y^{0'}-Y^{1'}\right) \, \partial_{\pm} \left( Y^{0'}-Y^{1'}\right) \right]\\[.2cm]
    & = \, - 2\, G \partial_{\pm} Y^{0'}  \, \partial_{\pm} Y^{1'} \\[.2cm]
    & = \, \partial_{\pm} Y^{0}\, \partial_{\pm} Y^{1} \, , \label{appVirasoroLC}
\end{aligned} &
\end{flalign}
\vspace{0.1cm}

where $\mu, \nu  \in \{0, 1 \}$ are spacetime coordinates, $\partial_{\pm} := \frac{\partial}{\partial \sigma^{\pm}}$, and the definition of the new mapping functions Eq.(\ref{IsoMapFunc}) is used in the third line. In the final line, recognise that $G$ is defined as a non-zero scalar function and that the indices have been renamed ($\mu' = \mu$). This is precisely the Virasoro constraint equations given in subsection (\ref{subsubsecBarsMethod}), Eq.(\ref{Virasoro_EoM_Iso}). \\

The string equations of motion, Eq.(\ref{StringEoM}), become

\begin{equation}
\partial_\pm \,\Pi_\mu^\pm \, - \, \Gamma^\alpha_{\mu \nu}\, \partial_\pm \,X^\nu\,\Pi^\pm_\alpha \,=\, 0 \,, \label{StringEoMLC}
\end{equation}
\vspace{0.1cm}

in the light-cone coordinate frame. This breaks into two equations of motion, since $\mu = (0, 1)$. For $\mu  =  0$, the string equations of motion are

\begin{equation}
\big(\partial_+ \Pi_0^+ \, - \, \Gamma^\alpha_{0 \nu}\, \partial_+ X^\nu\,\Pi^+_\alpha\big) \,+\, \big(\partial_- \Pi_0^- \,
 - \, \Gamma^\alpha_{0 \nu}\, \partial_- X^\nu\,\Pi^-_\alpha \big) \,=\, 0 \,. \label{StringEoMLC_mu=0}
\end{equation}
\vspace{0.1cm}

The first bracket in Eq.(\ref{StringEoMLC_mu=0}) can be simplified

\begin{flalign}
& \begin{aligned}
  & \partial_+ \Pi_0^+ \, - \, \Gamma^\alpha_{0 \nu}\, \partial_+ X^\nu\,\Pi^+_\alpha \\[.2cm]
  & = \, \partial_+ \left( - \frac{1}{2 \pi \alpha'} \, G\, {\partial_-} X^0 \right) \, - \, \Gamma^\alpha_{0 \nu}\, \partial_+ X^\nu \left(\frac{1}{2 \pi \alpha'} \, G \, \eta_{\alpha \alpha}\, {\partial_-} X^{\alpha}\right)\\[.2cm]
    & = \, \frac{1}{2 \pi \alpha'} \left[ - \partial_+ \left(G\, {\partial_-} X^0 \right) \, + \, G \eta_{\alpha \alpha} \left( - \Gamma^\alpha_{0 0}\, \partial_+ X^0\, {\partial_-} X^{\alpha}  \, - \, \Gamma^\alpha_{0 1}\, \partial_+ X^1 \, {\partial_-} X^{\alpha} \right)  \right] \\[.2cm]
    & =  \, \frac{1}{2 \pi \alpha'} \left[ - \partial_+ \left(G\, {\partial_-} X^0 \right)  \,+  G \left( \Gamma^0_{0 0}\, \partial_+ X^0\, {\partial_-} X^0   -  \Gamma^1_{0 0}\, \partial_+ X^0\, {\partial_-} X^1 + \Gamma^0_{0 1}\, \partial_+ X^1 \, {\partial_-} X^0 - \Gamma^1_{0 1}\, \partial_+ X^1 \, {\partial_-} X^1\right)  \right]\, , \label{StringEoMLC_mu=0_T1}
\end{aligned} &
\end{flalign}
\vspace{0.1cm}

where, in the first line, the canonical momentum densities Eq.(\ref{MomDensityLC}) are used and the spacetime metric simplifies due to the conformal gauge choice (i.e. $G_{\mu \nu}(x) = G \,\eta_{\mu \nu}$, $G^{\mu \nu}(x) = 1/G \,\eta^{\mu \nu}$). Further, the Christoffel symbols, defined in Eq.(\ref{ChristoffelDef}), can be calculated

\begin{flalign}
& \begin{aligned}
   &\Gamma^0_{0 0}\, = \, \frac{1}{2 \, G} \,  \eta^{0 \gamma} \left(\eta_{0 \gamma}\, \partial_0 G \,+\, \eta_{0 \gamma}\,\partial_0 G  \,-\, \eta_{0 0}\,\partial_\gamma G \right) \,= \, \frac{1}{2 \, G} \, \eta^{0 0}\, \left(-\, \partial_0 G \,-\, \partial_0 G \,+\, \partial_0 G \right) \, = \, \frac{1}{2 \, G}  \, \left(\partial_0 G \right) \\[.2cm]
   & \Gamma^1_{0 0}\, = \,  \frac{1}{2 \, G} \,  \eta^{1 \gamma} \left(\eta_{0 \gamma}\, \partial_0 G \,+\, \eta_{0 \gamma}\,\partial_0 G  \,-\, \eta_{0 0}\,\partial_\gamma G \right) \,= \, \frac{1}{2 \, G} \, \eta^{1 1}\, \left(\partial_1 G \right) \, = \, \frac{1}{2 \, G}  \, \left(\partial_1  G \right) \\[.2cm]
   & \Gamma^0_{0 1}\, = \,  \frac{1}{2 \, G} \,  \eta^{0 \gamma} \left(\eta_{1 \gamma}\, \partial_0 G \,+\, \eta_{0 \gamma}\,\partial_1 G  \,-\, \eta_{0 1}\,\partial_\gamma G \right) \,= \, \frac{1}{2 \, G} \, \eta^{0 0}\, \left(\partial_1 G \right) \, = \, \frac{1}{2 \, G}  \, \left(\partial_1  G \right)  \\[.2cm]
   &\Gamma^1_{0 1}\, = \,  \frac{1}{2 \, G} \,  \eta^{1 \gamma} \left(\eta_{1 \gamma}\, \partial_0 G \,+\, \eta_{0 \gamma}\,\partial_1 G  \,-\, \eta_{0 1}\,\partial_\gamma G \right) \,= \, \frac{1}{2 \, G} \, \eta^{1 1}\, \left(\partial_0 G \right) \, = \, \frac{1}{2 \, G}  \, \left(\partial_0  G \right)    \, . \label{ChristoffelComponents}
\end{aligned} &
\end{flalign}
\vspace{0.1cm}

Hence Eq.(\ref{StringEoMLC_mu=0_T1}) becomes

\begin{equation}
   \, \frac{1}{2 \pi \alpha'} \left[ - \partial_+ \left(G\, {\partial_-} X^0 \right) +  \frac{1}{2} \left( \partial_0  G\, \partial_+ X^0\, {\partial_-} X^0   -  \partial_1  G\, \partial_+ X^0\, {\partial_-} X^1 + \partial_1  G\, \partial_+ X^1 \, {\partial_-} X^0 - \partial_0  G\, \partial_+ X^1 \, {\partial_-} X^1\right)  \right] \, , \label{StringEoMLC_mu=0_T1_2}
\end{equation}
\vspace{0.1cm}

where $\partial_{\mu} := \frac{\partial}{\partial X^{\mu}}$ for $\mu = (0,1)$. The second bracket in Eq.(\ref{StringEoMLC_mu=0}) can be similarly simplified

\begin{flalign}
& \begin{aligned}
  & \partial_- \Pi_0^- \, - \, \Gamma^\alpha_{0 \nu}\, \partial_- X^\nu\,\Pi^-_\alpha \\[.2cm]
  & =  \partial_- \left( - \frac{1}{2 \pi \alpha'} \, G {\partial_+} X^0 \right) \, - \, \Gamma^\alpha_{0 \nu}\, \partial_- X^\nu \left(\frac{1}{2 \pi \alpha'} \, G  \eta_{\alpha \alpha}\, {\partial_+} X^{\alpha}\right)\\[.2cm]
    & =  \frac{1}{2 \pi \alpha'} \left[ - \partial_- \left(G {\partial_+} X^0 \right)  +  G \eta_{\alpha \alpha} \left( - \Gamma^\alpha_{0 0}\, \partial_- X^0\, {\partial_+} X^{\alpha}  \, - \, \Gamma^\alpha_{0 1}\, \partial_- X^1 \, {\partial_+} X^{\alpha} \right)  \right] \\[.2cm]
    & =   \frac{1}{2 \pi \alpha'} \left[ - \partial_- \left(G {\partial_+} X^0 \right) +  G \left( \Gamma^0_{0 0}\, \partial_- X^0\, {\partial_+} X^0   -  \Gamma^1_{0 0}\, \partial_- X^0\, {\partial_+} X^1 + \Gamma^0_{0 1}\, \partial_- X^1 \, {\partial_+} X^0 - \Gamma^1_{0 1}\, \partial_- X^1 \, {\partial_+} X^1\right)  \right]\\[.2cm]
    & =  \frac{1}{2 \pi \alpha'} \left[ - \partial_- \left(G {\partial_+} X^0 \right) +  \frac{1}{2} \left( \partial_0  G\, \partial_- X^0\, {\partial_+} X^0   -  \partial_1  G\, \partial_- X^0\, {\partial_+} X^1 +\partial_1  G\, \partial_- X^1 \, {\partial_+} X^0 - \partial_0  G\, \partial_- X^1 \, {\partial_+} X^1\right)  \right] , \label{StringEoMLC_mu=0_T2}
\end{aligned} &
\end{flalign}
\vspace{0.1cm}

where the calculated Christoffel symbols (Eq.(\ref{ChristoffelComponents})) are used in the final line. Using the simplifications for the first and second brackets (Eqs.(\ref{StringEoMLC_mu=0_T1_2}, \ref{StringEoMLC_mu=0_T2})), the first string equation of motion Eq.(\ref{StringEoMLC_mu=0}) becomes

\begin{equation}
 - \partial_+ \left(G\, {\partial_-} X^0 \right) \,- \,\partial_- \left(G\, {\partial_+} X^0 \right) \,+\,   \partial_0  G\, \partial_+ X^0\, {\partial_-} X^0  \,
 -\, \partial_0  G\, \partial_+ X^1 \, {\partial_-} X^1 \, = \, 0\,. \label{StringEoM_mu=0_final}
\end{equation}
\vspace{0.1cm}

Following a similar calculation as before, the second string equation of motion (where $\mu \, = \, 1$) is

\begin{equation}
\big(\partial_+ \Pi_1^+ \, - \, \Gamma^\alpha_{1 \nu}\, \partial_+ X^\nu\,\Pi^+_\alpha\big) \,+\, \big(\partial_- \Pi_1^- \,
- \, \Gamma^\alpha_{1 \nu}\, \partial_- X^\nu\,\Pi^-_\alpha \big) \,=\, 0 \,. \label{StringEoMLC_mu=1}
\end{equation}
\vspace{0.1cm}

Again, the first bracket in Eq.(\ref{StringEoMLC_mu=1}) can be simplified

\begin{flalign}
& \begin{aligned}
  & \partial_+ \Pi_1^+ \, - \, \Gamma^\alpha_{1 \nu}\, \partial_+ X^\nu\,\Pi^+_\alpha \\[.2cm]
  & = \, \partial_+ \left( \frac{1}{2 \pi \alpha'} \, G\, {\partial_-} X^1 \right) \, - \, \Gamma^\alpha_{1 \nu}\, \partial_+ X^\nu \left(\frac{1}{2 \pi \alpha'} \, G \, \eta_{\alpha \alpha}\, {\partial_-} X^{\alpha}\right)\\[.2cm]
    & = \, \frac{1}{2 \pi \alpha'} \left[ \partial_+ \left(G\, {\partial_-} X^1 \right) \, + \, G \eta_{\alpha \alpha} \left( - \Gamma^\alpha_{1 0}\, \partial_+ X^0\, {\partial_-} X^{\alpha}  \, - \, \Gamma^\alpha_{1 1}\, \partial_+ X^1 \, {\partial_-} X^{\alpha} \right)  \right] \\[.2cm]
    & =  \, \frac{1}{2 \pi \alpha'} \left[\partial_+ \left(G\, {\partial_-} X^1 \right) \, + \, G \left( \Gamma^0_{1 0}\, \partial_+ X^0\, {\partial_-} X^0   - \Gamma^1_{1 0}\, \partial_+ X^0\, {\partial_-} X^1 + \Gamma^0_{1 1}\, \partial_+ X^1 \, {\partial_-} X^0 - \Gamma^1_{1 1}\, \partial_+ X^1 \, {\partial_-} X^1\right)  \right]\, , \label{StringEoMLC_mu=1_T1}
\end{aligned} &
\end{flalign}
\vspace{0.1cm}
where, in the first line, the canonical momentum densities Eq.(\ref{MomDensityLC}) are used and the spacetime metric simplifies due to the conformal gauge choice.
The Christoffel symbols now need to be calculated. However first note that the Christoffel symbols are symmetric on their lower indexes \cite{carroll2004spacetime}, i.e. $\Gamma^\gamma_{\alpha \beta} \, = \, \Gamma^\gamma_{\beta \alpha}$.
Therefore $\Gamma^0_{1 0} \, = \, \Gamma^0_{0 1}$ and $\Gamma^1_{1 0} \, = \, \Gamma^1_{0 1}$ (the latter of which have both been calculated in Eq.(\ref{ChristoffelComponents})). The remaining two Christoffel symbols can be calculated

\begin{flalign}
& \begin{aligned}
   &\Gamma^0_{1 1}\, = \, \frac{1}{2 \, G} \,  \eta^{0 \gamma} \left(\eta_{1 \gamma}\, \partial_1 G \,+\, \eta_{1 \gamma}\,\partial_1 G  \,-\, \eta_{1 1}\,\partial_\gamma G \right) \,= \, \frac{1}{2 \, G} \, \eta^{0 0}\, \left(-\, \partial_0 G \right) \, = \, \frac{1}{2 \, G}  \, \left(\partial_0 G \right) \\[.2cm]
    &\Gamma^1_{1 1}\, = \,  \frac{1}{2 \, G} \,  \eta^{1 \gamma} \left(\eta_{1 \gamma}\, \partial_1 G \,+\, \eta_{1 \gamma}\,\partial_1 G  \,-\, \eta_{1 1}\,\partial_\gamma G \right) \,= \, \frac{1}{2 \, G} \, \eta^{1 1}\, \left(\partial_1 G\,+\, \partial_1 G \,- \, \partial_1 G \right) \, = \, \frac{1}{2 \, G}  \, \left(\partial_1  G \right)    \, . \label{ChristoffelComponents_2}
\end{aligned} &
\end{flalign}
\vspace{0.1cm}

Hence Eq.(\ref{StringEoMLC_mu=1_T1}) becomes

\begin{equation}
   \, \frac{1}{2 \pi \alpha'} \left[ \partial_+ \left(G\, {\partial_-} X^1 \right) \, + \, \frac{1}{2} \left( \partial_1  G\, \partial_+ X^0\, {\partial_-} X^0  
    -  \partial_0  G\, \partial_+ X^0\, {\partial_-} X^1 + \partial_0  G\, \partial_+ X^1 \, {\partial_-} X^0 - \partial_1  G\,
    \partial_+ X^1 \, {\partial_-} X^1\right)  \right] \, . \label{StringEoMLC_mu=1_T1_2}
\end{equation}
\vspace{0.1cm}

The second bracket in Eq.(\ref{StringEoMLC_mu=1}) can be similarly simplified

\begin{flalign}
& \begin{aligned}
  & \partial_- \Pi_1^- \, - \, \Gamma^\alpha_{1 \nu}\, \partial_- X^\nu\,\Pi^-_\alpha \\[.2cm]
  & = \, \partial_- \left(\frac{1}{2 \pi \alpha'} \, G\, {\partial_+} X^1 \right) \, - \, \Gamma^\alpha_{1 \nu}\, \partial_- X^\nu \left(\frac{1}{2 \pi \alpha'} \, G \, \eta_{\alpha \alpha}\, {\partial_+} X^{\alpha}\right)\\[.2cm]
    & = \, \frac{1}{2 \pi \alpha'} \left[\partial_- \left(G\, {\partial_+} X^1 \right) \, + \, G \eta_{\alpha \alpha} \left( - \Gamma^\alpha_{1 0}\, \partial_- X^0\, {\partial_+} X^{\alpha}  \, - \, \Gamma^\alpha_{1 1}\, \partial_- X^1 \, {\partial_+} X^{\alpha} \right)  \right] \\[.2cm]
    & =  \, \frac{1}{2 \pi \alpha'} \left[ \partial_- \left(G\, {\partial_+} X^1 \right) \, + \, G \left( \Gamma^0_{1 0}\, \partial_- X^0\, {\partial_+} X^0   - \Gamma^1_{1 0}\, \partial_- X^0\, {\partial_+} X^1 + \Gamma^0_{1 1}\, \partial_- X^1 \, {\partial_+} X^0 - \Gamma^1_{1 1}\, \partial_- X^1 \, {\partial_+} X^1\right)  \right]\\[.2cm]
    & = \, \frac{1}{2 \pi \alpha'} \left[ \partial_- \left(G\, {\partial_+} X^1 \right)  +  \frac{1}{2} \left( \partial_1  G\, \partial_- X^0\, {\partial_+} X^0   -  \partial_0  G\, \partial_- X^0\, {\partial_+} X^1 \,+\, \partial_0  G\, \partial_- X^1 \, {\partial_+} X^0 - \partial_1  G\, \partial_- X^1 \, {\partial_+} X^1\right)  \right] , \label{StringEoMLC_mu=1_T2}
\end{aligned} &
\end{flalign}
\vspace{0.1cm}

where the calculated Christoffel symbols (Eqs.(\ref{ChristoffelComponents}, \ref{ChristoffelComponents_2})) are used in the final line. Using the simplifications for the first and second brackets (Eqs.(\ref{StringEoMLC_mu=1_T1_2}, \ref{StringEoMLC_mu=1_T2})), the second string equation of motion Eq.(\ref{StringEoMLC_mu=1}) becomes

\begin{equation}
 \partial_+ \left(G\, {\partial_-} X^1 \right) \,+ \,\partial_- \left(G\, {\partial_+} X^1 \right) \,
 +\,   \partial_1  G\, \partial_+ X^0\, {\partial_-} X^0  \,-\, \partial_1  G\, \partial_+ X^1 \, {\partial_-} X^1 \, = \, 0\,. \label{StringEoM_mu=1_final}
\end{equation}
\vspace{0.1cm}

Now, the two string equations of motion (Eqs.(\ref{StringEoM_mu=0_final}, \ref{StringEoM_mu=1_final})) can be rewritten in terms of the new string embedding functions $Y^{\mu'}(\sigma^+, \sigma^-)$. Combining Eq.(\ref{StringEoM_mu=0_final}) and Eq.(\ref{StringEoM_mu=1_final}) yields

\begin{equation}
 - \partial_+ \left(G\, {\partial_-} \left(X^0-X^1\right) \right) \,- \,\partial_- \left(G\, {\partial_+} \left(X^0-X^1\right) \right) \,
 +\, \left(\partial_0  G  + \partial_1  G\right) \left[   \partial_+ X^0\, {\partial_-} X^0  \,-\, \partial_+ X^1 \, {\partial_-} X^1 \right]\, = \, 0\,, \label{StringEoM_2}
\end{equation}
\vspace{0.1cm}

where the derivatives are defined as $\partial_{\mu} := \frac{\partial}{\partial X^{\mu}}$ for $\mu = ( 0,1)$. By the chain rule

\begin{equation}
\partial_0  G \,  = \, \left(\partial_{0'}  G + \partial_{1'}  G \right) \,\,\,\, \text{and} \,\,\,\, \partial_1  G \,  = \, \left(\partial_{0'}  G - \partial_{1'}  G \right)  \,, \label{ChainRule}
\end{equation}
\vspace{0.1cm}

where $\partial_{\mu'} := \frac{\partial}{\partial Y^{\mu'}}$ for $\mu' = \left( 0' , 1'\right)$. Using Eq.(\ref{ChainRule}) and the definition of $Y^{\mu'}(\sigma^+, \sigma^-)$ (Eq.(\ref{IsoMapFunc})),
the equation of motion Eq.(\ref{StringEoM_2}) becomes

\begin{flalign}
& \begin{aligned}
  0  & =  - \sqrt{2}\left[ \partial_+ \left(G\, {\partial_-} Y^{1'} \right) + \partial_- \left(G\, {\partial_+} Y^{1'}) \right) \right] \,+\, \sqrt{2}\, \partial_{0'}  G  \bigg[ \,\frac{1}{2}\partial_+ \left(Y^{0'}+Y^{1'} \right) {\partial_-} \left(Y^{0'}+Y^{1'} \right)\\[.2cm]
  & \,\,\,\,\,\,\, - \frac{1}{2} \partial_+\left(Y^{0'}-Y^{1'} \right)  {\partial_-} \left(Y^{0'}-Y^{1'} \right) \bigg] \\[.2cm]
   & = \, \partial_+ \left(G\, {\partial_-} Y^{1'} \right) + \partial_- \left(G\, {\partial_+} Y^{1'} \right) \,+\, \frac{1}{2}\, \partial_{0'}  G  \bigg[ \, \partial_+Y^{0'}\partial_-Y^{0'} +  \partial_+Y^{0'}\partial_-Y^{1'} +  \partial_+Y^{1'}\partial_-Y^{0'} +  \partial_+Y^{1'}\partial_-Y^{1'} \\[.2cm]
    & \,\,\,\,\,\,\,- \partial_+Y^{0'}\partial_-Y^{0'} + \partial_+Y^{0'}\partial_-Y^{1'}  + \partial_+Y^{1'}\partial_-Y^{0'} - \partial_+Y^{1'}\partial_-Y^{1'} \bigg] \\[.2cm]
    & = \, \partial_+ \left(G \partial_- Y^1\right) \,+\, \partial_- \left(G \partial_+ Y^1\right) \, - \, \left(\partial_0  G \right) \left[ \left(\partial_+ Y^0 \right)\left(\partial_- Y^1 \right) \,+\, \left(\partial_+ Y^1 \right)\left(\partial_- Y^0 \right)\right]  \, , \label{appStringEoMLC1_final}
\end{aligned} &
\end{flalign}

where the indices have been renamed ($\mu' = \mu$) in the last line. This is precisely one of the string equations of motion given in subsection (\ref{subsubsecBarsMethod}), Eq.(\ref{Virasoro_EoM_Iso}).
In order to find the other string equation of motion, subtract Eq.(\ref{StringEoM_mu=0_final}) from Eq.(\ref{StringEoM_mu=1_final}) to yield

\begin{flalign}
& \begin{aligned}
  0  & = - \partial_+ \left(G\, {\partial_-} \left(X^0+X^1\right) \right) \,- \,\partial_- \left(G\, {\partial_+} \left(X^0+X^1\right) \right) \,+\, \left(\partial_0  G  - \partial_1  G\right) \left[   \partial_+ X^0\, {\partial_-} X^0  \,-\, \partial_+ X^1 \, {\partial_-} X^1 \right]\\[.2cm]
  & - \sqrt{2}\left[ \partial_+ \left(G\, {\partial_-} Y^{0'} \right) + \partial_- \left(G\, {\partial_+} Y^{0'}) \right) \right] \,+\, \sqrt{2}\, \partial_{1'}  G  \bigg[ \,\frac{1}{2}\partial_+ \left(Y^{0'}+Y^{1'} \right) {\partial_-} \left(Y^{0'}+Y^{1'} \right)\\[.2cm]
  & \,\,\,\,\,\,\, - \frac{1}{2} \partial_+\left(Y^{0'}-Y^{1'} \right)  {\partial_-} \left(Y^{0'}-Y^{1'} \right) \bigg] \\[.2cm]
    & = \, \partial_+ \left(G \partial_- Y^0\right) \,+\, \partial_- \left(G \partial_+ Y^0\right) \, - \, \left(\partial_1  G \right)\left[ \left(\partial_+ Y^0 \right)\left(\partial_- Y^1 \right) \,+\, \left(\partial_+ Y^1 \right)\left(\partial_- Y^0 \right)\right]  \, . \label{appStringEoMLC2_final}
\end{aligned} &
\end{flalign}
\vspace{0.1cm}

The Virasoro constraints Eq.(\ref{appVirasoroLC}) and the string equations of motion Eqs.(\ref{appStringEoMLC1_final}, \ref{appStringEoMLC2_final}) agree with the equations given by
Bars \textit{et al.} in \cite{Bars_1994, Bars_1995} where $u = Y^0$ and $v = Y^1$. Further, using the product and chain rules, Eq.(\ref{appStringEoMLC1_final}) becomes

\begin{flalign}
& \begin{aligned}
  0  & =  \left(\partial_+ G \right)\left(\partial_- Y^1\right) \,+\, G \partial_+\left(\partial_- Y^1\right) \,+\, \left(\partial_- G \right)\left(\partial_+ Y^1\right)\,+\, G \partial_-\left(\partial_+ Y^1\right)\\[.2cm]
  &\,\,\,\,\,\,\, - \, \left(\partial_0  G \right) \left[ \left(\partial_+ Y^0 \right)\left(\partial_- Y^1 \right) \,+\, \left(\partial_+ Y^1 \right)\left(\partial_- Y^0 \right)\right] \\[.2cm]
  & = \left[\left(\partial_{0} G\right)\left(\partial_+ Y^0\right) + \left(\partial_{1} G\right)\left(\partial_+ Y^1\right) \right]\left(\partial_- Y^1\right) \,+\, \left[\left(\partial_{0} G\right)\left(\partial_- Y^0\right) + \left(\partial_{1} G\right)\left(\partial_- Y^1\right) \right]\left(\partial_+ Y^1\right)  \,+\, 2 \, G \partial_+\left(\partial_- Y^1\right)\\[.2cm]
  &\,\,\,\,\,\,\, - \, \left(\partial_0  G \right) \left[ \left(\partial_+ Y^0 \right)\left(\partial_- Y^1 \right) \,+\, \left(\partial_+ Y^1 \right)\left(\partial_- Y^0 \right)\right] \\[.2cm]
    & = \, G \partial_+ \partial_- Y^1 \, + \, \left(\partial_{1} G\right)\left(\partial_+ Y^1\right)\left(\partial_- Y^1\right) \, , \label{appStringEoM_Moerman1}
\end{aligned} &
\end{flalign}

and Eq.(\ref{appStringEoMLC2_final}) becomes

\begin{flalign}
& \begin{aligned}
  0  & =  \left(\partial_+ G \right)\left(\partial_- Y^0\right) \,+\, G \partial_+\left(\partial_- Y^0\right) \,+\, \left(\partial_- G \right)\left(\partial_+ Y^0\right)\,+\, G \partial_-\left(\partial_+ Y^0\right)\\[.2cm]
  &\,\,\,\,\,\,\, - \, \left(\partial_1  G \right) \left[ \left(\partial_+ Y^0 \right)\left(\partial_- Y^1 \right) \,+\, \left(\partial_+ Y^1 \right)\left(\partial_- Y^0 \right)\right] \\[.2cm]
  & = \left[\left(\partial_{0} G\right)\left(\partial_+ Y^0\right) + \left(\partial_{1} G\right)\left(\partial_+ Y^1\right) \right]\left(\partial_- Y^0\right) \,+\, \left[\left(\partial_{0} G\right)\left(\partial_- Y^0\right) + \left(\partial_{1} G\right)\left(\partial_- Y^1\right) \right]\left(\partial_+ Y^0\right)  \,+\, 2 \, G \partial_+\left(\partial_- Y^0\right)\\[.2cm]
  &\,\,\,\,\,\,\, - \, \left(\partial_1  G \right) \left[ \left(\partial_+ Y^0 \right)\left(\partial_- Y^1 \right) \,+\, \left(\partial_+ Y^1 \right)\left(\partial_- Y^0 \right)\right] \\[.2cm]
    & = \, G \partial_+ \partial_- Y^0 \, + \, \left(\partial_{0} G\right)\left(\partial_+ Y^0\right)\left(\partial_- Y^0\right) \, , \label{appStringEoM_Moerman2}
\end{aligned} &
\end{flalign}

which agrees exactly with the string equations of motion given by Moerman \textit{et al.} in \cite{moerman2016semi}.

\section{Energy of a Test String in an AdS-Schwarzschild Background} \label{appEnergyCurvedSpace}

At $t=0$ the test string set-up for the heavy quark agrees with the test string set-up for the light quark, since the string is initially static in both cases  -- i.e. $X^I  =  0$ (where $I \in (2,\,3, \,  ...\, ,\, d-1)$) at time $t=0$, with embedding functions  $X^\mu (t, \sigma) \, = \, \left(t, r(t, \sigma), 0 \right)^\mu$. The on-mass-shell heavy quark's mass or the off-mass-shell light quark's mass\footnote{Particles are \textit{on the mass shell}, or simply \textit{on-mass-shell}, if their behaviour satisfies Einstein's energy and momentum relation $E^2 = (pc)^2 +(2mc)^2$. Particles whose behaviour violates this relation are known as \textit{off-mass-shell}. In the case of the light quark it starts at $t=0$ as an off-mass-shell particle (corresponding to the initial static string), radiates energy as it travels through the thermal medium (string contracts as the boundary endpoint falls at the local speed of light), and finally stops radiating as it becomes an on-mass-shell particle.
This appendix is focused on calculating the mass (or \textit{virtuality}) of the light quark as an initially off-mass-shell particle; and the mass of an on-mass-shell heavy quark.} can be calculated by finding the total energy of the relevant test string in AdS$_d$-Schwarzschild. In order to calculate the string's total energy, the configuration of the string is chosen to be stretched between the radial positions $r=0$ and $r=\ell_0$. This choice will simplify the calculation. \\

The AdS$_d$-Schwarzschild metric in $d$ dimensions is given by

\begin{flalign}
& \begin{aligned}
ds_d^2 & := G_{\mu \nu}\, dx^\mu \, dx^\nu \\[0.25cm]
& = G_{tt}\, dt^2 \, + \, G_{rr}\, dr^2 \, + \, G_{I I}\, d\Vec{X}_{I}^2 \\[0.25cm]
& = \frac{r^2}{l^2} \, \Big(-h(r;d) \, dt^2 \, + \, d\Vec{X}_{I}^2 \,\Big) \,+\, \frac{l^2}{r^2}\frac{dr^2}{h(r;d)} \, , \label{appAdSd_Metric}
\end{aligned} &
\end{flalign}
\vspace{0.1cm}

where $t \in [0, \infty)$ is the temporal coordinate, $r \in [0, \infty)$ is the radial coordinate, and the transverse spatial directions are denoted by $\Vec{X}_{I} = \left(X^2, X^3, ...., X^{(d-1)}\right) \in \mathbb{R}^{d-2}$. Further, $l \in \mathbb{R}^+$ is the curvature radius of AdS$_d$ and $S^d$; and the blackening factor of the Schwarzschild black hole situated at the horizon, $h(r;d)$, is given by Eq.(\ref{BlackFactors}).
Explicitly, the target spacetime metric $G_{\mu \nu}$ is given by

\begin{equation}
G_{\mu \nu} \, =\,
  \begin{bmatrix}
   \Big( -\frac{r^2}{l^2}\,h(r;d) \Big) & 0 & 0 & 0 & 0 & 0 \\
    0 & \Big(\frac{l^2}{r^2}\,\frac{1}{h(r;d)} \Big) & 0 & 0 & 0 & 0 \\
    0 & 0 & \Big(\frac{r^2}{l^2}\Big) & 0 & 0 & 0 \\
    0 & 0 & 0 & \Big(\frac{r^2}{l^2}\Big) & \ddots & 0 \\
    0 & 0 & 0 & \ddots & \ddots & 0 \\
    0 & 0 & 0 & 0 & 0 &\Big(\frac{r^2}{l^2}\Big)
  \end{bmatrix} \, ,\label{appTargetMatrix}
\end{equation}

\vspace{0.3cm}

where $\mu,\nu$ index over $\left(t,\, r,\, 2,\, ... \,,\, d-1\right)$. The induced worldsheet metric $g_{a b}$, defined in Eq.(\ref{InducedMetric}), takes the explicit form

\begin{equation}
g_{a b}=
  \begin{bmatrix}
   g_{tt} & g_{t \sigma} \\
   g_{\sigma t} & g_{\sigma \sigma}
  \end{bmatrix} \, , \label{appIndMetric}
\end{equation}
\vspace{0.1cm}

since $(t, \sigma)$ are the two parameter space coordinates in the static gauge ($\tau = t$). In order to determine the explicit entries of the induced metric $g_{a b}$, the definition Eq.(\ref{InducedMetric}) is used to calculate each component. Specifically,

\begin{equation}
g_{a b}=
  \begin{bmatrix}
    G_{tt}\,+ \, G_{I I}\, \Dot{X}_{I}^2 & G_{I I}\,\Dot{X}_{I}\,X_{I}' \\
   G_{I I}\,X_{I}'\,\Dot{X}_{I} & r'^2\, G_{rr} \, + \,G_{I I} \, X_{I}'^{\,2}
  \end{bmatrix} \, , \label{appIndMetric2}
\end{equation}
\vspace{0.1cm}

where $\Dot{X} \equiv \partial_t X$ and $X' \equiv \partial_\sigma X$; $I$ indexes over the transverse directions $X_I=\left(X^2,X^3, ...., X^{(d-1)}\right)$; and the fact that  $G_{\mu \nu}$ (Eq.(\ref{appTargetMatrix})) is a diagonal matrix is used.\\

For later use, the inverse metric $g^{a b}$ (where $g^{a b} \, \equiv \, (g_{a b})^{-1}$) is also written out explicitly

\begin{equation}
g^{a b}= \, \frac{1}{\text{det}(g_{a b})}
  \begin{bmatrix}
    r'^2\,G_{rr} \, + \,G_{I I} \, X_{I}'^{\,2} & - G_{I I}\,\Dot{X}_{I}\,X_{I}'  \\
   - G_{I I}\,X_{I}'\,\Dot{X}_{I} & G_{tt}\,+ \, G_{I I}\, \Dot{X}_{I}^2
  \end{bmatrix} \, . \label{appInverseIndMetric}
\end{equation}
\vspace{0.5cm}

The determinant of the induced metric is

\begin{flalign}
& \begin{aligned}
  g \, := \, \text{det}\, (g_{a b }) \, &= \,  \big(r'^{2}\, G_{rr} \, + \,G_{I I} \, X_{I}'^{\,2} \big)\, \big( G_{tt}\,+ \, G_{I I}\, \Dot{X}_{I}^{\,2} \big) \, - \, \big( G^2_{I I}\,\Dot{X}_{I}^{\,2}\,X_{I}'^{\,2} \big)\\[.2cm]
  &= \,  r'^2\,G_{rr}\,G_{tt} \, + \, r'^2\,G_{rr}\,G_{I I}\,\Dot{X}_{I}^{\,2}\, +\, G_{tt}\,G_{I I}\, X_{I}'^{\,2} \, + \, G_{I I}^{\,2} \, X_{I}'^{\,2}\, \Dot{X}_{I}^{\,2} \, -  \, G_{I I}^{\,2}\,\Dot{X}_{I}^{\,2} \, X_{I}'^{\,2}
  \\[.2cm]
  &= \,  G_{rr}\,G_{tt}\, \Big( r'^2 \, + \, \frac{r'^2}{G_{tt}}\,G_{I I}\,\Dot{X}_{I}^{\,2} \, +\, \frac{1}{G_{rr}}\,G_{I I}\, X_{I}'^{\,2} \Big) \,. \label{appIndMetricDet}
\end{aligned} &
\end{flalign}
\vspace{0.1cm}

Now that the ground work has been laid, the total energy of the static string can be calculated
\begin{flalign}
& \begin{aligned}
    \text{E}\, & = \, \int_0^{\ell_0} \, d\sigma\, \Pi^{\tau}_{\,t} \\[.2cm]
    & = \, - \frac{1}{2 \pi \alpha'}\, \int_0^{\ell_0}\, d \sigma \, \sqrt{-\gamma} \, \gamma^{\tau b} \, G_{t \nu} \, \partial_b \, X^\nu \\[.2cm]
    & = \, - \frac{1}{2 \pi \alpha'}\, \int_0^{\ell_0}\, d \sigma \, \sqrt{-g} \, g^{t b} \, G_{t t} \, \partial_b \, X^t\\[.2cm]
    & = \,-  \frac{1}{2 \pi \alpha'}\, \int_0^{\ell_0}\, d \sigma \, \sqrt{-g} \, \big( g^{t t}  \, \partial_t \, X^t \, + \, g^{t \sigma}  \, \partial_\sigma \, X^t\big) \, \bigg(- \frac{r^2}{l^2} \, h(r;d) \bigg)
    \\[.2cm]
    & = \, \frac{1}{2 \pi \alpha'}\, \int_0^{\ell_0}\, d \sigma \, \sqrt{-g} \, g^{t t} \, \bigg( \frac{r^2}{l^2} \, h(r;d) \bigg)\\[.2cm]
    & = \, \frac{1}{2 \pi \alpha'}\, \int_0^{\ell_0}\, d \sigma \, \sqrt{-g} \, \big( r'^2\,G_{rr} \, + \,G_{I I} \, X_{I}'^{\,2} \big) \, \bigg(\frac{r^2}{l^2} \, h(r;d) \bigg)\\[.2cm]
    & = \, \frac{1}{2 \pi \alpha'}\, \int_0^{\ell_0}\, d \sigma \, \sqrt{-g} \,r'^2\, \bigg( \frac{l^2}{r^2}\, \frac{1}{h(r;d)} \bigg) \, \bigg( \frac{r^2}{l^2} \, h(r;d) \bigg)\, , \label{appCurveEnergy}
\end{aligned} &
\end{flalign}
\vspace{0.1cm}

where the constraint equation Eq.(\ref{Constraint2}) and the static gauge choice ($\tau =t$) are used in the third line, and Eq.(\ref{appTargetMatrix}) is used in the fourth line. The fifth line follows from remembering the string is static ($X_I  =0$), and Eq.(\ref{appInverseIndMetric}) is used in the sixth line.\\

Using Eq.(\ref{appIndMetricDet}), the total energy of the static string (Eq.(\ref{appCurveEnergy})) becomes

\begin{flalign}
& \begin{aligned}
    \text{E}\, & = \, \frac{1}{2 \pi \alpha'}\, \int_0^{\ell_0}\, d \sigma \, r'^2\,\sqrt{- G_{rr}\,G_{tt}\, \Big( r'^2 \, + \, \frac{r'^2}{G_{tt}}\,G_{I I}\,\Dot{X}_{I}^{\,2} \, +\, \frac{1}{G_{rr}}\,G_{I I}\, X_{I}'^{\,2} \Big)} \\[.2cm]
    & = \, \frac{1}{2 \pi \alpha'}\, \int_0^{\ell_0}\, d \sigma \, r'^2\,\sqrt{- r'^2\,G_{rr}\,G_{tt}} \\[.2cm]
    & = \, \frac{1}{2 \pi \alpha'}\,\int_0^{\ell_0}\, d \sigma \, \partial_\sigma^3 r \, , \label{appCurveEnergy2}
\end{aligned} &
\end{flalign}
\vspace{0.1cm}
where Eq.(\ref{appIndMetricDet}) is used in the first line, the second line follows again from $X_I  =0$, and Eq.(\ref{appTargetMatrix}) is used the final line.\\ 

From Eq.(\ref{appCurveEnergy2}) it is apparent that the energy of the string is contingent on how the radial coordinate $r$ depends on the worldsheet parameter space coordinate $\sigma$. In the static gauge (which is employed in sections (\ref{secHeavyQuark}) and (\ref{secLightQuark})) the energy of the test string is easily found in AdS$_3$-Schwarzschild by using the identification Eq.(\ref{InverseTortoiseCoord_Iden}) and solving the integral in Eq.(\ref{appCurveEnergy2}). Since the tortoise coordinate in anti-de Sitter-Schwarzschild spacetime can only be inverted for $d = 3$, an equivalent identification to Eq.(\ref{InverseTortoiseCoord_Iden}) in AdS$_d$-Schwarzschild can not be found. However, a different gauge choice can be made. Taking, for example, the gauge choice\footnote{This gauge choice is also made in \cite{Gubser_2006, herzog2006energy}.} in section (\ref{secDragForce}) where $\tau = t$ and $\sigma = r$ are the worldsheet coordinates, Eq.(\ref{appCurveEnergy2}) becomes 
\begin{equation}
\text{E}\, =\, \frac{1}{2 \pi \alpha'}\, \big[ \ell_0\, - \, 0 \big]  \,, \label{appCurveVirtual_i}
\end{equation}

in AdS$_d$-Schwarzschild. Therefore
\begin{equation}
\text{E}^2\, \equiv \, m_q^2 \, = \, \frac{\ell_0^2}{4 \pi^2 \alpha'^2} \,, \label{appCurveVirtual}
\end{equation}
\vspace{0.1cm}

where the first equivalence follows from the AdS/CFT correspondence: the energy of a test string in the bulk theory is equivalent to the mass of the given probe quark in the boundary theory. Notice that Eq.(\ref{appCurveVirtual}) agrees with Eq.(\ref{FlatSpaceMass}), which is a reflection of the fact that the spacetime in the bulk theory is curved precisely along the additional radial direction $r$, and not along the temporal or transverse directions. Since the boundary theory does not include this radial direction, from the perspective of the probe quark its mass will be the same whether the string is in Minkowski or AdS-Schwarzschild spacetime\footnote{In an AdS/CFT context, this implies that the energy of the string is equivalent in Minkowski or AdS-Schwarzschild spacetime.}. Considering that 

\begin{equation}
\sqrt{\alpha'} \, \equiv \, \frac{l}{\lambda^{1/4}}  \,, \label{apptHooftcoupling_and_alpha_def}
\end{equation}
\vspace{0.1cm}

where $l \in \mathbb{R}^+$ is the radius of curvature of AdS$_d$ and $\lambda = g_{YM}^2\,N_c$ is the 't Hooft coupling (see table (\ref{tabletranslation}), page \pageref{tabletranslation}); the quark's mass\footnote{Remembering this is either the initial, off-mass-shell mass for the light quark or the on-mass-shell mass for the heavy quark.} becomes

\begin{equation}
m_q^2 \, = \, \frac{\lambda  \,\ell_0^2}{4 \, \pi^2 \,l^4} \,. \label{appCurveVirtualFINAL}
\end{equation}
\vspace{0.1cm}

Alternatively, Eq.(\ref{appCurveVirtualFINAL}) can be rewritten as

\begin{equation}
m_q \, = \, \frac{\sqrt{\lambda}  \,\ell_0}{\beta \,r_H}\, =\, \frac{\sqrt{\lambda}}{\beta}\, \frac{(r_s + \ell_0 - r_s)}{r_H} \, = \, \frac{\sqrt{\lambda}}{\beta}\, \left(\Tilde{r}_0 - \frac{r_s}{r_H}\right) \,\approx \, \frac{\sqrt{\lambda}}{\beta}\, \Tilde{r}_0\,, \label{appCurveVirtualFINAL2}
\end{equation}
\vspace{0.1cm}

where the first equality follows from using the definition of the Hawking temperature Eq.(\ref{HawkingTemp}); the second equality from Eq.(\ref{dim_r0}); and the approximation from the definition of the stretched horizon $r_s = (1+ \epsilon)r_H$ (where $0< \epsilon \ll 1$). Using the definition of the Hawking temperature yields $m_q = \sqrt{\lambda}\, \Tilde{r}_0 \, T$. Eq.(\ref{appCurveVirtualFINAL2}) might be a convenient form of the quark mass $m_q$ in some cases.

\section{Nambu-Goto String Equations of Motion for Transverse Fluctuations} \label{appNambuGoto}

The calculation of the string equations of motion for the transverse fluctuations presented in this appendix, follows the layout of appendix (B) in Moerman \textit{et al.}'s exposition on light quark Brownian motion \cite{moerman2016semi}. Working in the static gauge, the transverse string equations of motion are derived by varying the effective action for the transverse fluctuations with respect to the transverse string worldsheet coordinates, and setting this functional variation to zero.\\

Using the definition Eq.(\ref{MomDensity}), the effective transverse action Eq.(\ref{NG_Quadratic2}) can be written in terms of the transverse canonical momentum densities

\begin{equation}
 S_{NG}^{(2)} \, = \, \frac{1}{2}\, \int_{\mathcal{M}} \, d^2 \sigma \, \Pi_{\,I}^a(t, \sigma)\, \partial_a X^I(t, \sigma) \, . \label{NG_XI_MomDen}
\end{equation}
\vspace{0.1cm}

Determining the functional derivative of $S_{NG}^{(2)}$ with respect to $X^I$ and setting this variation to vanish, results in
\begin{flalign}
& \begin{aligned}
    0 \, = \,  \delta_X  S_{NG}^{(2)}\, & = \, \frac{1}{2}\, \int_{\mathcal{M}} \, d^2 \sigma \,\Big[ \big( \delta \Pi_{\,I}^a \big)\, \partial_a\,X^I \,+\, \Pi_{\,I}^a\, \partial_a \big(\delta X^I\big) \Big] \\[.2cm]
    & = \, \frac{1}{2}\, \int_{\mathcal{M}} \, d^2 \sigma \,\Big[ \big( \delta  \Pi_{\,I}^a \big)\, \partial_a\,X^I \,+\,  \partial_a \big(\Pi_{\,I}^a\,\delta X^I \big) \, - \, \partial_a \big(\Pi_{\,I}^a \big)\, \delta X^I \Big]  \, , \label{VaryNG_XI}
\end{aligned} &
\end{flalign}
\vspace{0.1cm}

where the product rule is used in the second line.\\

The functional derivative of the momentum densities with respect to $X^I$ is given by

\begin{equation}
 \delta_X \Pi_{\,I}^a \, = \, -\frac{1}{2\pi \alpha'}\, \big(\sqrt{-g}\, g^{ab} \,
  G_{IJ} \big) \big\rvert_{X^\mu_0} \, \, {\partial_b}\big(\delta X^J \big) \, , \label{Vary_MomDen_XI}
\end{equation}
\vspace{0.1cm}

where the spacetime metric $G_{\mu \nu}$ does not need to be varied since it is independent of the transverse directions $X^I$ (Eq.(\ref{appTargetMatrix})).
Hence the first term in Eq.(\ref{VaryNG_XI}), $\big( \delta \Pi_{\,I}^a \big) \partial_a X^I$, becomes

\begin{flalign}
& \begin{aligned}
  \big( \delta_X \Pi_{\,I}^a \big)\, \partial_a\,X^I \, & =  \left( -\frac{1}{2\pi \alpha'}\, \big(\sqrt{-g}\, g^{ab} \, G_{IJ} \big) \big\rvert_{X^\mu_0} \,  {\partial_b}\big(\delta X^J \big)\right)  \partial_a X^\mu  \\[.2cm]
    & = \left( -\frac{1}{2\pi \alpha'}\, \big(\sqrt{-g}\, g^{ab} \, G_{IJ} \big) \big\rvert_{X^\mu_0} \,  \partial_a X^\mu \right) {\partial_b}\big(\delta X^J \big) \\[.2cm]
    & = \, \Pi_{\,J}^b \, {\partial_b}\big(\delta X^J \big)  \\[.2cm]
    &  = \, \partial_b \big(\Pi_{\,J}^b \delta X^J \big) - \big(\partial_b \Pi_{\,J}^b \big) \delta X^J \, , \label{VaryNG_XI_1Term}
\end{aligned} &
\end{flalign}
\vspace{0.1cm}

where the definition of the momentum densities Eq.(\ref{MomDen_XI}) are used in the third line, and the product rule is used in the final line. Inputting Eq.(\ref{VaryNG_XI_1Term}) into Eq.(\ref{VaryNG_XI}), yields

\begin{flalign}
& \begin{aligned}
    0 \, = \, \delta_X  S_{NG}^{(2)}\, & = \, \frac{1}{2}\, \int_{\mathcal{M}} \, d^2 \sigma \,\Big[ \partial_b \big(\Pi_{\,J}^b \,\delta X^J \big)\, - \,\big(\partial_b \,\Pi_{\,J}^b \big)\,\delta X^J \,+\,  \partial_a \big(\Pi_{\,I}^a\,\delta X^I \big) \, - \, \partial_a \big(\Pi_{\,I}^a \big)\, \delta X^I \Big] \\[0.2cm]
    &  = \, -  \int_{\mathcal{M}} \, d^2 \sigma  \, \delta X^I \big(\partial_a  \Pi_{\,I}^a \big)  \, +\, \int_{\mathcal{M}} \, d^2 \sigma \, \partial_a \big(\Pi_{\,I}^a\,\delta X^I \big)  \\[.2cm]
    &  = \, -  \int_{\mathcal{M}} \, d^2 \sigma  \, \delta X^I \big(\partial_a  \Pi_{\,I}^a \big)  \, +\, \int_{\partial \mathcal{M}} \, d \sigma^b\, \epsilon_{ba} \, \big(\Pi_{\,I}^a\,\delta X^I \big)\\[.2cm]
    &  = \, -  \int_{\mathcal{M}} \, d^2 \sigma  \, \delta X^I \big(\partial_a  \Pi_{\,I}^a \big)  \, +\, \, \int_{0}^{\tau_f} \, d \tau \,\Big[\Pi_{\,I}^a\,\delta X^I \big\rvert^{\sigma = \sigma_f}_{\sigma = 0} \Big]  \, , \label{Vary_NG_XI2}
\end{aligned} &
\end{flalign}
\vspace{0.1cm}

where, in the second line, the indices have been renamed and the terms grouped; and Stokes's Theorem in d-dimensions \cite{carroll2004spacetime} has been used in the third line. In the final line the coordinates on the worldsheet parameter space are chosen to be $\sigma^a = (t, \sigma)^a$ where $\sigma^a \in \mathcal{M} = [0, t_f] \times [0, \sigma_f]$,
and one of the integrals of the last term is simplified by using $\delta X^I \rvert _{t \in \{ 0, t_f\}} = 0$.\\

Choosing the boundary conditions

\begin{equation}
 \Pi_{\, I}^a \, \delta X^I\,  \big\rvert^{\sigma = \sigma_f}_{\sigma = 0} \, = \, 0 \, , \label{appendixboundaryconditions_XI}
\end{equation}
\vspace{0.1cm}

the last term in Eq.(\ref{Vary_NG_XI2}) disappears. Hence, the string equations of motion for the transverse fluctuations are given by

\begin{equation}
 0 \, = \, \partial_a \,\Pi_{\, I}^a \, =\, \nabla_a\,\Pi^a_{\,I}\, , \label{appStringEoM_XI}
\end{equation}
\vspace{0.1cm}

which agrees with Eq.(\ref{StringEoM_XI}). The second equality follows because $\nabla_a \Pi^a_{\,\mu} := \partial_a \Pi^a_{\,\mu} - \Gamma^\alpha_{\mu \nu}\, \partial_a X^\nu \Pi^a_{\,\mu}$ in curved spacetime, but the second term vanishes since the Christoffel symbols are identically zero ($\Gamma^\alpha_{I J} = 0$) due to symmetry between each of the transverse $X^I$ directions.

\section{Canonical Commutation Relations and Normalised Basis} \label{appCommutationRels}
\vspace{0.2cm}

This appendix aims to fix the normalization constant $A_\omega$, thereby completely determining the general solution for the transverse equations of motion (Eq.(\ref{gen_solution_linear_sup_X})) discussed in subsection (\ref{subsubsecHQ_Displacement}).
In order to determine $A_\omega$ the theory is quantized: the scalar field $X(t, \sigma)$ and its canonically conjugate momentum $P^t(t,\sigma)$ are promoted to operators (Eqs.(\ref{X_operator}, \ref{P_operator}) respectively), and suitable commutation relations are imposed

\begin{flalign}
& \begin{aligned}
    &  \left[\hat{X}(t,\sigma), \, n_t \,\hat{P}^t(t,\sigma^\prime)\right]_\Sigma \, = \, i \, \delta(\sigma, \sigma^\prime) \, = \, i  \frac{\delta(\sigma - \sigma^\prime)}{\sqrt{\Tilde{g}}\rvert_\Sigma} \, ,  \\[0.2cm]
    & \left[\hat{X}(t,\sigma), \hat{X}(t,\sigma^\prime) \right]_\Sigma \, = \, \left[n_t \,\hat{P}^t(t,\sigma), \,  n_t \,\hat{P}^t(t,\sigma^\prime) \right]_\Sigma \,= \, 0  \, , \label{app_commutation_relations}
\end{aligned} &
\end{flalign}
\vspace{0.1cm}

where $\Sigma$ is a Cauchy hypersurface in the $x^\mu = (t,r)^\mu$ part of spacetime that is chosen to be a constant time surface\footnote{Giving initial conditions on this hypersurface determines the future (and past) evolution uniquely.},
$\Tilde{g}$ is the induced metric on $\Sigma$, and $n_\mu$ is the future pointing normal to $\Sigma$ (where $n_\mu = \delta_{\mu t}/ \sqrt{-\Tilde{g}_{tt}}$). The commutation relations described in Eq.(\ref{app_commutation_relations}) are known as equal time commutation relations --
they encode the expectation that simultaneous measurements at different points on the string do not interfere with each other \cite{zwiebach2004first}.\\

Additionally, canonical creation and annihilation commutation relations on the Fourier coefficient operators ($\hat{a}_\omega, \hat{a}^{\dagger}_\omega$) are enforced:

\begin{flalign}
& \begin{aligned}
    &  \Big[\hat{a}_\omega, \, \hat{a}_{\omega^\prime}^{\dagger} \Big]_\Sigma \, = \, 2 \pi \delta(\omega - \omega^\prime) \, ,  \\[0.2cm]
    & \Big[\hat{a}_\omega, \, \hat{a}_{\omega^\prime} \Big]_\Sigma \, = \, \Big[\hat{a}_\omega^{\dagger} , \, \hat{a}_{\omega^\prime}^{\dagger} \Big]_\Sigma \, = \, 0 \, . \label{app_FC_commutation_relations}
\end{aligned} &
\end{flalign}
\vspace{0.1cm}

Consistency between the commutation relations Eq.(\ref{app_commutation_relations}) and Eq.(\ref{app_FC_commutation_relations}) is required\footnote{In order to ensure consistent quantization of the theory.}.
To this end, the commutator bracket $\left[\hat{X}(t,\sigma), \, n_t \,\hat{P}^t(t,\sigma^\prime)\right]_\Sigma$ is calculated from the
definitions Eqs.(\ref{X_operator}, \ref{P_operator}), and the result is equated to the relevant right hand side of Eq.(\ref{app_commutation_relations}). Specifically,

\begin{flalign}
& \begin{aligned}
     \left[\hat{X}(t,\sigma), \, n_t \hat{P}^t(t,\sigma^\prime)\right]_\Sigma  & =  \hat{X}(t,\sigma)\rvert_\Sigma \,\, n_t \hat{P}^t(t,\sigma^\prime)\rvert_\Sigma \, - \, n_t  \hat{P}^t(t,\sigma^\prime)\rvert_\Sigma \,\, \hat{X}(t,\sigma)\rvert_\Sigma \, ,  \\[0.2cm]
    &  = - \frac{i}{2 \pi \alpha'}  \, \frac{r^2}{l^2} \, \frac{\delta_{tt}}{\sqrt{\Tilde{g}}\rvert_\Sigma }\int_0^\infty \frac{d \omega \, d \omega^\prime}{(2 \pi)^2} \, \omega \, A_\omega \, A_{\omega^\prime} \,\,\,\, \times \, \\[0.2cm]
    & \,\,\,\,\,\,\,\, \Big( \big[f_\omega (\sigma) e^{- i \omega t} \hat{a}_\omega  +  f^{\,*}_\omega (\sigma)  e^{i \omega t}  \hat{a}^{\,\dagger}_\omega \big]\big[f_\omega (\sigma^\prime) e^{- i {\omega^\prime} t} \hat{a}_{\omega^\prime}  -  f^{\,*}_\omega (\sigma^\prime)  e^{i {\omega^\prime} t}  \hat{a}^{\,\dagger}_{\omega^\prime} \big]\\[0.2cm]
    & \,\,\,\,\,\,\,\,\,\,\,\,\,- \big[f_\omega (\sigma^\prime) e^{- i {\omega^\prime} t} \hat{a}_{\omega^\prime}  -  f^{\,*}_\omega (\sigma^\prime)  e^{i {\omega^\prime} t}  \hat{a}^{\,\dagger}_{\omega^\prime} \big] \big[f_\omega (\sigma) e^{- i \omega t} \hat{a}_\omega  +  f^{\,*}_\omega (\sigma)  e^{i \omega t}  \hat{a}^{\,\dagger}_\omega \big] \Big) \\[0.2cm]
    &  = - \frac{i}{2 \pi \alpha'}  \, \frac{r^2}{l^2} \, \frac{1}{\sqrt{\Tilde{g}}\rvert_\Sigma }\int_0^\infty \frac{d \omega \, d \omega^\prime}{2 \pi} \, \omega \, A_\omega \, A_{\omega^\prime} \,\,\,\, \times \, \\[0.2cm]
    & \,\,\,\,\,\,\,\, \Big( - f_\omega (\sigma) f^{\,*}_\omega (\sigma^\prime)  \,e^{i (\omega^\prime -\omega) t}\,\delta(\omega - \omega^\prime)  -  f_\omega (\sigma^\prime)f^{\,*}_\omega (\sigma) \,e^{ i (\omega - \omega^\prime) t} \, \delta(\omega^\prime - \omega ) \Big) \, , \label{XP_commutator_1}
\end{aligned} &
\end{flalign}
\vspace{0.1cm}

where the third line follows because the cross terms $(\hat{a}_{\omega}\hat{a}_{\omega^\prime} - \hat{a}_{\omega^\prime}\hat{a}_{\omega})$ and
 $(\hat{a}_{\omega}^{\dagger}\hat{a}_{\omega^\prime}^{\dagger} - \hat{a}_{\omega^\prime}^{\dagger}\hat{a}_{\omega}^{\dagger})$ vanish
 (using the second commutation relation in Eq.(\ref{app_FC_commutation_relations})). The first commutation relation in Eq.(\ref{app_FC_commutation_relations}) is also used in simplification. Collecting terms, and converting to $(t,r_*)$ coordinates using Eq.(\ref{InverseTortoiseCoord_Iden}), yields

\begin{flalign}
& \begin{aligned}
    \therefore \left[\hat{X}(t,\sigma), \, n_t \hat{P}^t(t,\sigma^\prime)\right]_\Sigma &  = - \frac{i}{2 \pi \alpha'}   \frac{r^2}{l^2}  \frac{1}{\sqrt{\Tilde{g}}\rvert_\Sigma }\int_0^\infty \frac{d \omega}{2 \pi}  \omega  A_\omega^2  \left( - f_\omega (\sigma) f^{\,*}_\omega (\sigma^\prime)    -  f_\omega (\sigma^\prime)f^{\,*}_\omega (\sigma) \right)\\[0.2cm]
    &  = \frac{i}{2 \pi \alpha'}   \frac{r_H^2}{l^2} \coth^2 \left( \frac{r_H  (r_{s*} + \sigma^\prime)}{l^2}\right) \frac{1}{\sqrt{\Tilde{g}}\rvert_\Sigma }\int_0^\infty \frac{d \omega}{2 \pi}  \omega \, A_\omega^2  \left(f_\omega (\sigma) f^{\,*}_\omega (\sigma^\prime)    +  f_\omega (\sigma^\prime)f^{\,*}_\omega (\sigma) \right)
     \, . \label{XP_commutator}
\end{aligned} &
\end{flalign}
\vspace{0.1cm}

Inputting Eqs.(\ref{TC_nu_fw_r_ODE_Solution}, \ref{gen_solution_sum_modes}, \ref{B_w_TC}), the commutator becomes\footnote{Mathematica is used to be spared from the tedious multiplication (see Mathematica Notebook [b]: \texttt{BrownianMotion.nb}).}

\begin{equation}
     \left[\hat{X}(t,\sigma), \, n_t \hat{P}^t(t,\sigma^\prime)\right]_\Sigma  =  \frac{1}{\sqrt{\Tilde{g}}\rvert_\Sigma } \frac{i}{ \pi \alpha'}
     \frac{r_H^2}{l^2} \int_0^\infty \frac{d \omega}{2 \pi}  \omega  A_\omega^2  \Big(e^{-i (\sigma + \sigma^\prime) \omega}+ e^{i (\sigma - \sigma^\prime) \omega} +e^{i (\sigma + \sigma^\prime) \omega} + e^{i ( \sigma^\prime-\sigma ) \omega} \Big)
     \, , \label{XP_commutator_2}
\end{equation}
\vspace{0.1cm}

where the near-horizon limit ($r \rightarrow r_H \, \equiv \, r_{s*}  \rightarrow  - \infty \, \equiv \, \tilde{r}_0 \rightarrow 1$) has been taken after multiplication\footnote{If the commutation relations hold in the near-horizon limit (i.e. for a specific value of $(t, \sigma)$) then they hold $\forall \, (t, \sigma)$.
In \cite{Boer_2009} de Boer \textit{et al.} determine the normalization constant $A$ by demanding the normalization of the modes through the Klein-Gordon inner product. The near-horizon limit is taken during this calculation.
In a later work \cite{Atmaja_2014}, Appendix A,  a similar collaboration Atmaja \textit{et al.} argue that the inner product contribution from the near-horizon region can be thought of as the overall inner product.
There is a contribution to the inner product from regions away from the horizon; however, the near-horizon region is semi-infinite in the tortoise coordinate $r_*$ ($r \rightarrow r_H \, \equiv \, r_{s*}  \rightarrow  - \infty$); and as such the normalization can be completely fixed by the near-horizon regime. Therefore, the near-horizon limit can be taken here without concern.}. Note that in the near-horizon limit, 

\begin{equation}
    \coth^2 \left( \frac{r_H \, (r_{s*} + \sigma)}{l^2}\right) \, \rightarrow \, 1, \,\,\,\,\,\,
    \text{as} \,\,\,\,\,\,  r_{s*} \, \rightarrow \, - \infty \, . \label{near_horizon_limit}
\end{equation}
\vspace{0.1cm}

The commutator Eq.(\ref{XP_commutator_2}) can be further simplified

\begin{flalign}
& \begin{aligned}
     &\left[\hat{X}(t,\sigma), \, n_t \hat{P}^t(t,\sigma^\prime)\right]_\Sigma \\[0.2cm]
     &  =  \frac{1}{\sqrt{\Tilde{g}}\rvert_\Sigma } \frac{i}{ \pi \alpha'}   \frac{r_H^2}{l^2} \left( \int_0^\infty \frac{d \omega}{2 \pi}  \omega  A_\omega^2  \left[ e^{-i (\sigma + \sigma^\prime) \omega}  +  e^{- i ( \sigma - \sigma^\prime ) \omega} \right] + \int_0^\infty \frac{d \omega}{2 \pi}  \omega  A_\omega^2  \left[ e^{i (\sigma + \sigma^\prime) \omega}  +  e^{i ( \sigma - \sigma^\prime ) \omega} \right] \right)\\[0.2cm]
    &  =  \frac{1}{\sqrt{\Tilde{g}}\rvert_\Sigma } \frac{i}{ \pi \alpha'}   \frac{r_H^2}{l^2} \left( - \int_\infty^0 \frac{d \omega}{2 \pi}  \omega  A_\omega^2  \left[ e^{-i (\sigma + \sigma^\prime) \omega} +  e^{- i ( \sigma - \sigma^\prime ) \omega} \right] + \int_0^\infty \frac{d \omega}{2 \pi}  \omega  A_\omega^2  \left[ e^{i (\sigma + \sigma^\prime) \omega} +  e^{i ( \sigma - \sigma^\prime ) \omega} \right] \right)\\[0.2cm]
    &  =  \frac{1}{\sqrt{\Tilde{g}}\rvert_\Sigma } \frac{i}{ \pi \alpha'}   \frac{r_H^2}{l^2} \left( - \int_{-\infty}^0 \frac{(-d \omega)}{2 \pi}  (-\omega)  (-A_\omega^2)  \left[ e^{i (\sigma + \sigma^\prime) \omega}  +  e^{i ( \sigma - \sigma^\prime ) \omega} \right] + \int_0^\infty \frac{d \omega}{2 \pi}  \omega  A_\omega^2  \left[ e^{i (\sigma + \sigma^\prime) \omega}  +  e^{i ( \sigma - \sigma^\prime ) \omega} \right] \right)\\[0.2cm]
    &  =  \frac{1}{\sqrt{\Tilde{g}}\rvert_\Sigma } \frac{i}{ \pi \alpha'}   \frac{r_H^2}{l^2} \int_{-\infty}^\infty \frac{d \omega}{2 \pi}  \omega  A_\omega^2  \left[ e^{i (\sigma + \sigma^\prime) \omega}  +  e^{i ( \sigma - \sigma^\prime ) \omega} \right] \, . \label{XP_commutator_3}
\end{aligned} &
\end{flalign}
\vspace{0.1cm}

In order for Eq.(\ref{XP_commutator_3}) to be consistent with the definition of the commutation relation Eq.(\ref{app_commutation_relations}), the normalization constant needs to be defined as

\begin{equation}
    A_\omega \, := \, \frac{l}{r_H} \, \sqrt{\frac{\pi \alpha'}{\omega}} \,
    = \, \frac{\beta}{2 \, \sqrt{\pi \omega} \, \lambda^{1/4}} \, , \label{app_Aw_normalization}
\end{equation}
\vspace{0.1cm}

where the second equality follows from using the definition of the AdS radius of curvature $l$ (Eq.(\ref{alpha_def})), and the relation $l  = \beta r_H^2 / 2 \pi$ from Eq.(\ref{HawkingTemp}). \\

To prove that the commutation relations Eq.(\ref{app_commutation_relations}) now hold, input the newly defined $A_\omega$ into Eq.(\ref{XP_commutator_3}). Explicitly,

\begin{flalign}
& \begin{aligned}
     \left[\hat{X}(t,\sigma), \, n_t \,\hat{P}^t(t,\sigma^\prime)\right]_\Sigma \, &  =  \frac{1}{\sqrt{\Tilde{g}}\rvert_\Sigma }\, \frac{i}{ \pi \alpha'}  \, \frac{r_H^2}{l^2}\,  \int_{-\infty}^\infty \frac{d \omega}{2 \pi}  \omega  \bigg(\frac{l^2}{r_H^2} \, \frac{\pi \alpha'}{\omega} \bigg) \left[ e^{i (\sigma + \sigma^\prime) \omega} \, + \, e^{i ( \sigma - \sigma^\prime ) \omega} \right] \, \\[0.2cm]
    & =  \frac{i}{\sqrt{\Tilde{g}}\rvert_\Sigma }  \,  \int_{-\infty}^\infty \frac{d \omega}{2 \pi} \, \omega \,  \, \left[ e^{i (\sigma + \sigma^\prime) \omega} \, + \, e^{i ( \sigma - \sigma^\prime ) \omega} \right]\\[0.2cm]
    &  = \frac{i}{\sqrt{\Tilde{g}}\rvert_\Sigma }  \, \left[ \delta(\sigma + \sigma^\prime)  +  \delta(\sigma - \sigma^\prime) \right]\\[0.2cm]
    & =  \frac{i}{\sqrt{\Tilde{g}}\rvert_\Sigma }  \, \delta(\sigma - \sigma^\prime) \, , \label{XP_commutator_4}
\end{aligned} &
\end{flalign}
\vspace{0.1cm}

where the third line follows from the Dirac delta function property $\delta(y-x) = \frac{1}{2 \pi}\, \int_{-\infty}^\infty dk \, e^{-ik (y-x)}$; and, in the final line, the first Dirac delta function has disappeared since $\sigma, \, \sigma^\prime  \in  [0, \sigma_f]$.
Hence, the scalar field and conjugate momentum commutation relations (Eq.(\ref{app_commutation_relations})) and the Fourier coefficient creation and annihilation commutation relations (Eq.(\ref{XP_commutator_3})) are consistent; proving that the normalization constant $A_\omega$ is correctly given by Eq.(\ref{app_Aw_normalization}).

\section{Leading Order Contributions of the Near-Horizon Tortoise Coordinate} \label{appTortoiseCoord}
\vspace{0.2cm}

The tortoise coordinate $r_*$ is defined in Eq.(\ref{gen_d_TortoiseCoord}). In the near-horizon limit ($r = (1+ \tilde{\epsilon}) r_H$) the tortoise coordinate can be expanded and truncated around $\tilde{\epsilon}=0$, in order to yield the near-horizon tortoise coordinate, $\tilde{\epsilon}_*$. This coordinate is given -- for general $d$ dimensions -- by

\begin{flalign}
& \begin{aligned}
    \tilde{\epsilon}_* &:= \frac{l^2}{r_H^2 (d-1)}\ln(\tilde{\epsilon}) + \frac{ (d-4) l^2}{2 r_H^2 (d-1)}\tilde{\epsilon}  + \frac{\left(d^2-14 d+36\right) l^2}{24 r_H^2 (d-1)}\tilde{\epsilon}^2 - \frac{ \left(5 d^2-46 d+96\right) l^2}{72 r_H^2 (d-1)}\tilde{\epsilon}^3 + \mathcal{O}\left(\tilde{\epsilon}\right)^4\\[0.2cm]
    & \, = a_0 \ln(\tilde{\epsilon}) + a_1 \tilde{\epsilon} + a_2 \tilde{\epsilon}^2 + a_3 \tilde{\epsilon}^3 + \mathcal{O}\left(\tilde{\epsilon}\right)^4 \, , \label{app_gen_TortoiseCoord_NH}
\end{aligned} &
\end{flalign}
\vspace{0.1cm}

where $r_* = r_H \tilde{\epsilon}_*$, and $a_0$, $a_1$, $a_2$, $a_3$ are constants. To leading order, the near-horizon AdS$_d$-Schwarzschild metric is given by

\begin{equation}
    ds_d^2 =  \frac{r_H^2}{l^2} \, d\Vec{X}_{I}^2 \, -\, \frac{(d-1) r_H^2 \tilde{\epsilon}  }{l^2} \, dt^2  \,+\, \frac{l^2}{(d-1) \tilde{\epsilon} }\,d\tilde{\epsilon}^2 \, , \label{appAdSd_Metric_OptII}
\end{equation}
\vspace{0.1cm}

where all terms are $\mathcal{O}(\tilde{\epsilon})$. To be consistent, at leading order $\tilde{\epsilon}_*$ must be considered to $\mathcal{O}(\tilde{\epsilon})$ -- hence, it has the form $ \tilde{\epsilon}_* = a_0 \ln(\tilde{\epsilon}) + a_1 \tilde{\epsilon} + \mathcal{O}\left(\tilde{\epsilon}\right)^2$. \\

The near-horizon tortoise coordinate is used to convert the near-horizon metric into $(t, \tilde{\epsilon}_*)$ coordinates, thereby arriving at a conformally flat description of near-horizon AdS$_d$-Schwarzschild, from which the leading order string solution and the transverse equations of motion can be found. Considering Eq.(\ref{AdS_Metric_NH_t_epsilonstar}) it is clear that the metric conversion factor between ($t, \tilde{\epsilon}$) and ($t, \tilde{\epsilon}_*$) coordinates is

\begin{flalign}
& \begin{aligned}
     e^{\frac{r_H^2(d-1) \tilde{\epsilon}_*}{l^2}} &= e^{\frac{r_H^2(d-1)}{l^2} \left(a_0 \ln(\tilde{\epsilon}) + a_1 \tilde{\epsilon} + \mathcal{O}\left(\tilde{\epsilon}\right)^2\right)}\\[0.2cm]
     &= e^{c\, a_0 \ln(\tilde{\epsilon})}\, e^{c\, a_1 \tilde{\epsilon}}\, e^{ \mathcal{O}\left(\tilde{\epsilon}\right)^2}\\[0.2cm]
     &= \tilde{\epsilon}^{c \, a_0} \left(1 + c\, a_1 \tilde{\epsilon} + \mathcal{O}\left(\tilde{\epsilon}\right)^2 \right)  \left(1 + \mathcal{O}\left(\tilde{\epsilon}\right)^2 \right)\\[0.2cm]
     &= \tilde{\epsilon} \left(1 +  \frac{(d-4)}{2} \tilde{\epsilon} + \mathcal{O}\left(\tilde{\epsilon}\right)^2 \right)  \left(1 + \mathcal{O}\left(\tilde{\epsilon}\right)^2 \right)\\[0.2cm]
     &= \tilde{\epsilon} \, +\, \mathcal{O}\left(\tilde{\epsilon}\right)^2 \, , \label{appConversionFactor}
\end{aligned} &
\end{flalign}
\vspace{0.1cm}

where the relevant leading order form of $\tilde{\epsilon}_*$ is used in the first equality, the constant $c = r_H^2(d-1)/l^2$ is defined in the second line, and in the third line the terms $e^{c\, a_1 \tilde{\epsilon}}$ and $e^{ \mathcal{O}\left(\tilde{\epsilon}\right)^2}$ are series expanded around $\tilde{\epsilon} =0$. While inputting the values for $c$, $a_0$, and $a_1$ (from Eq.(\ref{app_gen_TortoiseCoord_NH})) into the fourth line, notice that $c \,a_0 = 1$ by construction.  All terms of $\mathcal{O}\left(\tilde{\epsilon}\right)^2$ are disregarded, in order to be consistent with the leading order near-horizon AdS$_d$-Schwarzschild metric. Hence, from Eq.(\ref{appConversionFactor}), the inverse near-horizon tortoise coordinate is given by

\begin{equation}
\tilde{\epsilon} := e^{\frac{r_H^2(d-1) \tilde{\epsilon}_*}{l^2}} \, , \label{app_InverseTortoiseCoord_NH_LO}
\end{equation}
\vspace{0.1cm}

at leading order; while the near-horizon tortoise coordinate is given by

\begin{equation}
\tilde{\epsilon}_{*} := \frac{l^2}{r_H^2 (d-1)}\ln(\tilde{\epsilon}) \, , \label{app_TortoiseCoord_NH_LO}
\end{equation}
\vspace{0.1cm}

at leading order. Claims that Eq.(\ref{app_TortoiseCoord_NH_LO}) is the leading order contribution to the near-horizon tortoise coordinate have been made in \cite{moerman2016semi}. This is, however, the first time to the authors' knowledge a proof to this effect (even a heuristic one) has been provided.

\newpage

\section{Accessing the Mathematica Code} \label{appCode}
\vspace{0.2cm}

The Mathematica code used to generate all the analytic results and plots in this article is provided as supplementary material.\\

The code is organised into four comprehensively annotated Mathematica notebooks containing: (a) the mapping of the heavy and light quark test string solutions between the worldsheet parameter space and target spacetime (\texttt{MappingWorldSheetToTarget.nb}); (b) calculations pertaining to the analysis of heavy and light quark test strings undergoing Brownian motion (\texttt{BrownianMotion.nb}); (c) the expansion of the AdS$_d$-Schwarzschild metric in the near-horizon region for the light quark's test string configuration (\texttt{NearHorizonAdSd.nb}); and (d) drag force calculations in AdS/CFT (\texttt{DragForce.nb}). \\

For readers please always check for the latest versions of the notebooks in the GitHub repository:

\url{https://github.com/AlexesMes/brownian-motion-of-quarks}.\\

\begin{figure}[htb]
\centering
\includegraphics[width=\textwidth]{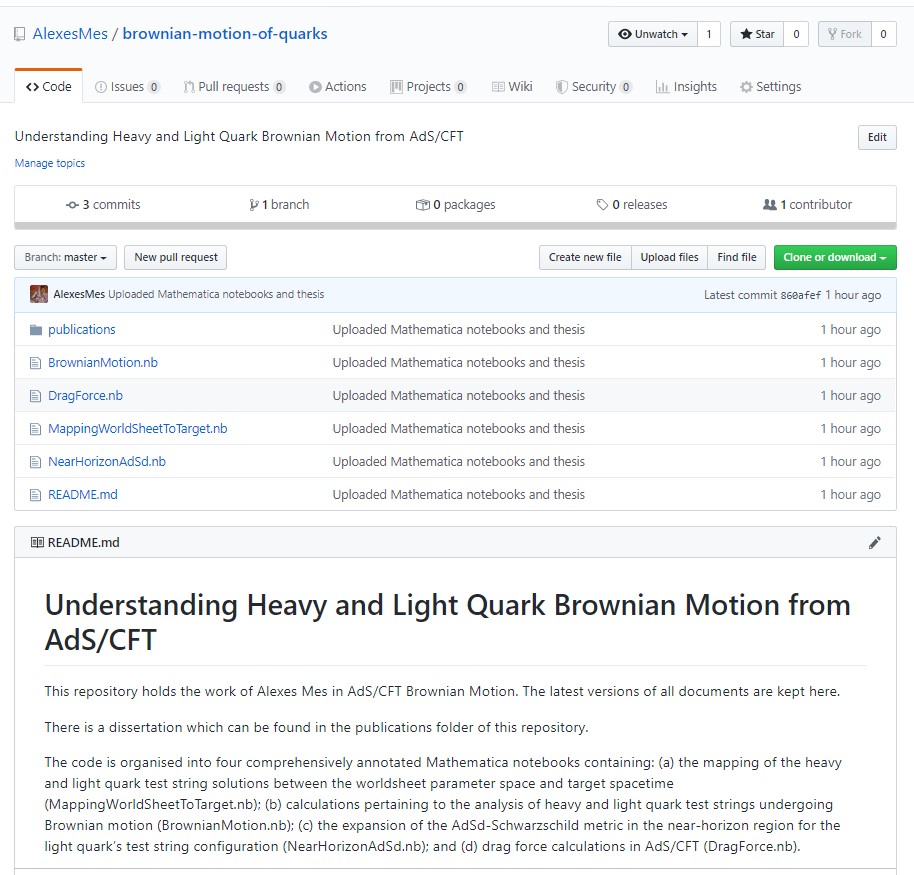}
\end{figure}

\newpage


\begin{thebibliography}{112}
\providecommand{\natexlab}[1]{#1}
\providecommand{\url}[1]{\texttt{#1}}
\expandafter\ifx\csname urlstyle\endcsname\relax
  \providecommand{\doi}[1]{doi: #1}\else
  \providecommand{\doi}{doi: \begingroup \urlstyle{rm}\Url}\fi

\bibitem[Aronson and Ludlam((2005))]{aronson2005hunting}
A.~Aronson and T.~Ludlam.
\newblock {Hunting the Quark Gluon Plasma: Results from the First 3 Years at
  the Relativistic Heavy Ion Collider (RHIC)}.
\newblock \emph{Upton, New York: Brookhaven National Laboratory. Formal Report:
  BNL-73847}, (2005).
\newblock {Available at \tt{http://www.bnl.gov/npp/docs/Hunting the QGP.pdf}}.

\bibitem[Heinz and Kolb((2002))]{Heinz_2002}
U.~Heinz and P.~Kolb.
\newblock Early {T}hermalization at {RHIC}.
\newblock \emph{Nuclear Physics A}, 702\penalty0 (1-4), (2002).
\newblock \doi{10.1016/s0375-9474(02)00714-5}.
\newblock {\tt{arXiv:hep-ph/0111075}}.

\bibitem[Gyulassy((2004))]{gyulassy2004qgp}
M.~Gyulassy.
\newblock The {QGP} discovered at {RHIC}.
\newblock In \emph{Structure and Dynamics of Elementary Matter (NATO Advanced
  Study Institute)}, pages 159--182. Springer, (2004).
\newblock {\tt{arXiv:nucl-th/0403032}}.

\bibitem[Heinz and Snellings((2013))]{Heinz_2013}
U.~Heinz and R.~Snellings.
\newblock Collective {F}low and {V}iscosity in {R}elativistic {H}eavy-{I}on
  {C}ollisions.
\newblock \emph{Annual Review of Nuclear and Particle Science}, 63\penalty0
  (1), (2013).
\newblock \doi{10.1146/annurev-nucl-102212-170540}.
\newblock {\tt{arXiv:1301.2826 [nucl-th]}}.

\bibitem[Gale et~al.((2013){\natexlab{a}})Gale, Jeon, and Schenke]{GALE_2013}
C.~Gale, S.~Jeon, and B.~Schenke.
\newblock Hydrodynamic {M}odeling of {H}eavy-{I}on {C}ollisions.
\newblock \emph{International Journal of Modern Physics A}, 28\penalty0 (11),
  (2013){\natexlab{a}}.
\newblock \doi{10.1142/s0217751x13400113}.
\newblock {\tt{arXiv:1301.5893 [nucl-th]}}.

\bibitem[Majumder and van Leeuwen((2011))]{Majumder_2011}
A.~Majumder and M.~van Leeuwen.
\newblock The {T}heory and {P}henomenology of {P}erturbative {QCD} based {J}et
  {Q}uenching.
\newblock \emph{Progress in Particle and Nuclear Physics}, 66\penalty0 (1),
  (2011).
\newblock \doi{10.1016/j.ppnp.2010.09.001}.
\newblock {\tt{arXiv:1002.2206 [hep-ph]}}.

\bibitem[Horowitz((2013))]{Horowitz_2013}
W.~Horowitz.
\newblock Heavy {Q}uark {P}roduction and {E}nergy {L}oss.
\newblock \emph{Nuclear Physics A}, 904-905, (2013).
\newblock \doi{10.1016/j.nuclphysa.2013.01.061}.
\newblock {\tt{arXiv:1210.8330 [nucl-th]}}.

\bibitem[Djordjevic et~al.((2014))Djordjevic, Djordjevic, and
  Blagojevic]{Djordjevic_2014}
M.~Djordjevic, M.~Djordjevic, and B.~Blagojevic.
\newblock {RHIC and LHC Jet Suppression in Non-central Collisions}.
\newblock \emph{Physics Letters B}, 737, (2014).
\newblock {\tt{arXiv:1405.4250 [nucl-th]}}.

\bibitem[Hirano et~al.((2011))Hirano, Huovinen, and Nara]{Hirano_2011}
T.~Hirano, P.~Huovinen, and Y.~Nara.
\newblock Elliptic {F}low in {Pb+Pb} {C}ollisions at {$\sqrt{s_{NN}}$ = 2.76
  TeV}: {H}ybrid {M}odel {A}ssessment of the {F}irst {D}ata.
\newblock \emph{Physical Review C}, 84\penalty0 (1), (2011).
\newblock \doi{10.1103/physrevc.84.011901}.
\newblock {\tt{arXiv:1012.3955 [nucl-th]}}.

\bibitem[Shen et~al.((2011))Shen, Heinz, Huovinen, and Song]{Shen_2011}
C.~Shen, U.~Heinz, P.~Huovinen, and H.~Song.
\newblock Radial and {E}lliptic {F}low in {Pb+Pb} {C}ollisions at {E}nergies
  {A}vailable at the {CERN} {L}arge {H}adron {C}ollider from {V}iscous
  {H}ydrodynamics.
\newblock \emph{Physical Review C}, 84\penalty0 (4), (2011).
\newblock \doi{10.1103/physrevc.84.044903}.
\newblock {\tt{arXiv:1105.3226 [nucl-th]}}.

\bibitem[Qiu et~al.((2012))Qiu, Shen, and Heinz]{Qiu_2012}
Z.~Qiu, C.~Shen, and U.~Heinz.
\newblock Hydrodynamic {E}lliptic and {T}riangular flow in {Pb$-$Pb}
  {C}ollisions at {$\sqrt{s}$ = 2.76ATeV}.
\newblock \emph{Physics Letters B}, 707\penalty0 (1), (2012).
\newblock \doi{10.1016/j.physletb.2011.12.041}.
\newblock {\tt{arXiv:1110.3033 [nucl-th]}}.

\bibitem[Gale et~al.((2013){\natexlab{b}})Gale, Jeon, Schenke, Tribedy, and
  Venugopalan]{Gale_event_2013}
C.~Gale, S.~Jeon, B.~Schenke, P.~Tribedy, and R.~Venugopalan.
\newblock Event-by-event {A}nisotropic {F}low in {H}eavy-{I}on {C}ollisions
  from {C}ombined {Y}ang-{M}ills and {V}iscous {F}luid {D}ynamics.
\newblock \emph{Physical Review Letters}, 110\penalty0 (1),
  (2013){\natexlab{b}}.
\newblock \doi{10.1103/physrevlett.110.012302}.
\newblock {\tt{arXiv:1209.6330 [nucl-th]}}.

\bibitem[Gubser et~al.((1996))Gubser, Klebanov, and Peet]{Gubser_1996}
S.~Gubser, I.~Klebanov, and A.~Peet.
\newblock {Entropy and Temperature of Black 3-Branes}.
\newblock \emph{Physical Review D}, 54\penalty0 (6), (1996).
\newblock \doi{10.1103/physrevd.54.3915}.
\newblock {\tt{arXiv:hep-th/9602135}}.

\bibitem[Casalderrey-Solana et~al.((2014){\natexlab{a}})Casalderrey-Solana,
  Liu, Mateos, Rajagopal, and Wiedemann]{casalderrey2014gauge}
J.~Casalderrey-Solana, H.~Liu, D.~Mateos, K.~Rajagopal, and U.~Wiedemann.
\newblock \emph{Gauge/{S}tring {D}uality, {H}ot {QCD} and {H}eavy {I}on
  {C}ollisions}.
\newblock Cambridge University Press, (2014){\natexlab{a}}.
\newblock {\tt{arXiv:1101.0618 [hep-th]}}.

\bibitem[Morad and Horowitz((2014))]{Morad_2014}
R.~Morad and W.~Horowitz.
\newblock Strong-coupling {J}et {E}nergy {L}oss from {AdS/CFT}.
\newblock \emph{Journal of High Energy Physics}, 2014\penalty0 (11), (2014).
\newblock \doi{10.1007/jhep11(2014)017}.
\newblock {\tt{arXiv:1409.75455 [hep-th]}}.

\bibitem[Casalderrey-Solana et~al.((2014){\natexlab{b}})Casalderrey-Solana,
  Gulhan, Milhano, Pablos, and Rajagopal]{Casalderrey_Solana_hybrid_2014}
J.~Casalderrey-Solana, D.~Gulhan, J.~Milhano, D.~Pablos, and K.~Rajagopal.
\newblock {A Hybrid Strong/Weak Coupling Approach to Jet Quenching}.
\newblock \emph{Journal of High Energy Physics}, 2014\penalty0 (10),
  (2014){\natexlab{b}}.
\newblock \doi{10.1007/jhep10(2014)019}.
\newblock {\tt{arXiv:1405.3864 [hep-ph]}}.

\bibitem[Horowitz((2015))]{Horowitz_2015}
W.~Horowitz.
\newblock {Fluctuating Heavy Quark Energy Loss in a Strongly Coupled
  Quark-Gluon Plasma}.
\newblock \emph{Physical Review D}, 91\penalty0 (8), (2015).
\newblock \doi{10.1103/physrevd.91.085019}.
\newblock {\tt{arXiv:1501.04693 [hep-ph]}}.

\bibitem['t~Hooft((1974))]{Hooft_1974}
G.~'t~Hooft.
\newblock A {P}lanar {D}iagram {T}heory for {S}trong {I}nteractions.
\newblock \emph{Nuclear Physics. B}, 72\penalty0 (3):\penalty0 461--473,
  (1974).

\bibitem[Klebanov((2001))]{Klebanov_2001}
I.~Klebanov.
\newblock {TASI} {L}ectures: {I}ntroduction to the {AdS/CFT} {C}orrespondence.
\newblock \emph{Strings, Branes and Gravity}, (2001).
\newblock \doi{10.1142/9789812799630_0007}.
\newblock {\tt{arXiv:hep-th/0009139}}.

\bibitem[Natsuume((2015))]{natsuume2015ads}
M.~Natsuume.
\newblock \emph{{AdS/CFT} {D}uality {U}ser {G}uide}, volume 903.
\newblock Springer, (2015).
\newblock {\tt{arXiv:1409.3575 [hep-th]}}.

\bibitem[Maldacena((1999))]{maldacena1999large}
J.~Maldacena.
\newblock The large-{N} {L}imit of {S}uperconformal {F}ield {T}heories and
  {S}upergravity.
\newblock \emph{International Journal of Theoretical Physics}, 38\penalty0
  (4):\penalty0 1113–1133, (1999).
\newblock \doi{10.1023/a:1026654312961}.
\newblock {\tt{arXiv:hep-th/9711200}}.

\bibitem[Gubser((2006))]{Gubser_2006}
S.~Gubser.
\newblock Drag force in {AdS/CFT}.
\newblock \emph{Physical Review D}, 74\penalty0 (12), (2006).
\newblock \doi{10.1103/physrevd.74.126005}.
\newblock {\tt{arXiv:hep-th/0605182}}.

\bibitem[Bak et~al.((2007))Bak, Karch, and Yaffe]{Bak_2007}
D.~Bak, A.~Karch, and L.~Yaffe.
\newblock {Debye Screening in Strongly Coupled $\mathcal{N}=4$ Supersymmetric
  Yang-Mills Plasma}.
\newblock \emph{Journal of High Energy Physics}, 2007\penalty0 (08), (2007).
\newblock \doi{10.1088/1126-6708/2007/08/049}.
\newblock {\tt{arXiv:0705.0994 [hep-th]}}.

\bibitem[{Witten}((1998))]{witten1998anti}
E.~{Witten}.
\newblock {Anti-de Sitter Space and Holography}.
\newblock \emph{Advances in Theoretical and Mathematical Physics}, 2:\penalty0
  253--291, (1998).
\newblock {\tt{arXiv:hep-th/9802150}}.

\bibitem[Gubser et~al.((1998))Gubser, Klebanov, and
  Polyakov]{GubserPolyakov_1998}
S.~Gubser, I.~Klebanov, and A.~Polyakov.
\newblock Gauge {T}heory {C}orrelators from {N}on-critical {S}tring {T}heory.
\newblock \emph{Physics Letters B}, 428\penalty0 (1-2):\penalty0 105–114,
  (1998).
\newblock \doi{10.1016/s0370-2693(98)00377-3}.
\newblock {\tt{arXiv:hep-th/9802109}}.

\bibitem[Aharony and Witten((1998))]{Aharony_1998}
O.~Aharony and E.~Witten.
\newblock Anti-de {S}itter {S}pace and the {C}enter of the {G}auge {G}roup.
\newblock \emph{Journal of High Energy Physics}, 1998\penalty0 (11):\penalty0
  018–018, (1998).
\newblock \doi{10.1088/1126-6708/1998/11/018}.
\newblock {\tt{arXiv:hep-th/9807205}}.

\bibitem[Aharony et~al.((2000))Aharony, Gubser, Maldacena, Ooguri, and
  Oz]{Aharony_2000}
O.~Aharony, S.~Gubser, J.~Maldacena, H.~Ooguri, and Y.~Oz.
\newblock Large {N} {F}ield {T}heories, {S}tring {T}heory and {G}ravity.
\newblock \emph{Physics Reports}, 323\penalty0 (3-4):\penalty0 183–386,
  (2000).
\newblock \doi{10.1016/s0370-1573(99)00083-6}.
\newblock {\tt{arXiv:hep-th/9905111}}.

\bibitem[Policastro et~al.((2001))Policastro, Son, and
  Starinets]{Policastro_2001}
G.~Policastro, D.~Son, and A.~Starinets.
\newblock Shear {V}iscosity of {S}trongly {C}oupled {$\mathcal{N}=4$}
  {S}upersymmetric {Y}ang-{M}ills {P}lasma.
\newblock \emph{Physical Review Letters}, 87\penalty0 (8), (2001).
\newblock \doi{10.1103/physrevlett.87.081601}.
\newblock {\tt{arXiv:hep-th/0104066}}.

\bibitem[Collaboration et~al.((2003))Collaboration, Adler, and
  et~al.]{Adler_2003}
{\textbf{PHENIX}}~Collaboration, S.~Adler, and et~al.
\newblock {Elliptic Flow of Identified Hadrons in Au$+$Au Collisions at
  {$\sqrt{s_{NN}}$ = 200 GeV}}.
\newblock \emph{Physical Review Letters}, 91\penalty0 (18), (2003).
\newblock \doi{10.1103/physrevlett.91.182301}.
\newblock {\tt{arXiv:nucl-ex/0305013}}.

\bibitem[Collaboration et~al.((2005))Collaboration, Adams, and
  et~al.]{Adams_2005}
{\textbf{STAR}}~Collaboration, J.~Adams, and et~al.
\newblock {Azimuthal Anisotropy in Au$+$Au collisions at {$\sqrt{s_{NN}}$ = 200
  GeV}}.
\newblock \emph{Physical Review C}, 72\penalty0 (1), (2005).
\newblock \doi{10.1103/physrevc.72.014904}.
\newblock {\tt{arXiv:nucl-ex/0409033}}.

\bibitem[Kovtun et~al.((2003))Kovtun, Son, and Starinets]{Kovtun_2003}
P.~Kovtun, D.~Son, and A.~Starinets.
\newblock Holography and {H}ydrodynamics: {D}iffusion on {S}tretched
  {H}orizons.
\newblock \emph{Journal of High Energy Physics}, 2003\penalty0 (10):\penalty0
  064–064, (2003).
\newblock \doi{10.1088/1126-6708/2003/10/064}.
\newblock {\tt{arXiv:hep-th/0309213}}.

\bibitem[Kovtun et~al.((2005))Kovtun, Son, and Starinets]{Kovtun_2005}
P.~Kovtun, D.~Son, and A.~Starinets.
\newblock Viscosity in {S}trongly {I}nteracting {Q}uantum {F}ield {T}heories
  from {B}lack {H}ole {P}hysics.
\newblock \emph{Physical Review Letters}, 94\penalty0 (11), (2005).
\newblock \doi{10.1103/physrevlett.94.111601}.
\newblock {\tt{arXiv:hep-th/0405231}}.

\bibitem[Buchel and Liu((2004))]{Buchel_2004}
A.~Buchel and J.~Liu.
\newblock Universality of the {S}hear {V}iscosity from {S}upergravity {D}uals.
\newblock \emph{Physical Review Letters}, 93\penalty0 (9), (2004).
\newblock \doi{10.1103/physrevlett.93.090602}.
\newblock {\tt{arXiv:hep-th/0311175}}.

\bibitem[Buchel((2005))]{Buchel_2005}
A.~Buchel.
\newblock On {U}niversality of {S}tress-{E}nergy {T}ensor {C}orrelation
  {F}unctions in {S}upergravity.
\newblock \emph{Physics Letters B}, 609\penalty0 (3-4):\penalty0 392–401,
  (2005).
\newblock \doi{10.1016/j.physletb.2005.01.052}.
\newblock {\tt{arXiv:hep-th/0408095}}.

\bibitem[Herzog et~al.((2006))Herzog, Karch, Kovtun, Kozcaz, and
  Yaffe]{herzog2006energy}
C.~Herzog, A.~Karch, P.~Kovtun, C.~Kozcaz, and L.~Yaffe.
\newblock Energy {L}oss of a {H}eavy {Q}uark {M}oving through $\mathcal{N}=4$
  {S}upersymmetric {Y}ang-{M}ills {P}lasma.
\newblock \emph{Journal of High Energy Physics}, 2006\penalty0 (07):\penalty0
  013, (2006).
\newblock \doi{10.1088/1126-6708/2006/07/013}.
\newblock {\tt{arXiv:hep-th/0605158}}.

\bibitem[Liu et~al.((2006))Liu, Rajagopal, and Wiedemann]{Liu_2006}
H.~Liu, K.~Rajagopal, and U.~Wiedemann.
\newblock Calculating the {J}et {Q}uenching {P}arameter.
\newblock \emph{Physical Review Letters}, 97\penalty0 (18), (2006).
\newblock \doi{10.1103/physrevlett.97.182301}.
\newblock {\tt{arXiv:hep-ph/0605178}}.

\bibitem[Herzog((2006))]{Herzog_2006}
C.~Herzog.
\newblock Energy {L}oss of a {H}eavy {Q}uark from {A}symptotically {AdS}
  {G}eometries.
\newblock \emph{Journal of High Energy Physics}, 2006\penalty0 (09):\penalty0
  032–032, (2006).
\newblock \doi{10.1088/1126-6708/2006/09/032}.
\newblock {\tt{arXiv:hep-th/0605191}}.

\bibitem[Casalderrey-Solana and Teaney((2006))]{Casalderrey_Solana_2006}
J.~Casalderrey-Solana and D.~Teaney.
\newblock Heavy {Q}uark {D}iffusion in {S}trongly {C}oupled $\mathcal{N}=4$
  {Y}ang {M}ills.
\newblock \emph{Physical Review D}, 74\penalty0 (8), (2006).
\newblock \doi{10.1103/physrevd.74.085012}.
\newblock {\tt{arXiv:hep-ph/0605199}}.

\bibitem[Liu et~al.((2007))Liu, Rajagopal, and Wiedemann]{Liu_2007}
H.~Liu, K.~Rajagopal, and U.~Wiedemann.
\newblock Wilson {L}oops in {H}eavy {I}on {C}ollisions and their {C}alculation
  in {AdS/CFT}.
\newblock \emph{Journal of High Energy Physics}, 2007\penalty0 (03):\penalty0
  066–066, (2007).
\newblock \doi{10.1088/1126-6708/2007/03/066}.
\newblock {\tt{arXiv:hep-ph/0612168}}.

\bibitem[Casalderrey-Solana and Teaney((2007))]{Casalderrey_Solana_2007}
J.~Casalderrey-Solana and D.~Teaney.
\newblock Transverse {M}omentum {B}roadening of a {F}ast {Q}uark in a
  $\mathcal{N}=4$ {Y}ang-{M}ills {P}lasma.
\newblock \emph{Journal of High Energy Physics}, 2007\penalty0 (04):\penalty0
  039–039, (2007).
\newblock \doi{10.1088/1126-6708/2007/04/039}.
\newblock {\tt{arXiv:hep-th/0701123}}.

\bibitem[Gubser((2008){\natexlab{a}})]{Gubser_2008}
S.~Gubser.
\newblock Momentum {F}luctuations of {H}eavy {Q}uarks in the {G}auge-{S}tring
  {D}uality.
\newblock \emph{Nuclear Physics B}, 790\penalty0 (1-2):\penalty0 175–199,
  (2008){\natexlab{a}}.
\newblock \doi{10.1016/j.nuclphysb.2007.09.017}.
\newblock {\tt{arXiv:hep-th/061214}}.

\bibitem[de~Boer et~al.((2009))de~Boer, Hubeny, Rangamani, and
  Shigemori]{Boer_2009}
J.~de~Boer, V.~Hubeny, M.~Rangamani, and M.~Shigemori.
\newblock Brownian {M}otion in {AdS/CFT}.
\newblock \emph{Journal of High Energy Physics}, 2009\penalty0 (07):\penalty0
  094–094, (2009).
\newblock \doi{10.1088/1126-6708/2009/07/094}.
\newblock {\tt{arXiv:0812.5112 [hep-th]}}.

\bibitem[Son and Teaney((2009))]{Son_2009}
D.~Son and D.~Teaney.
\newblock {Thermal Noise and Stochastic Strings in AdS/CFT}.
\newblock \emph{Journal of High Energy Physics}, 2009\penalty0 (07), (2009).
\newblock \doi{10.1088/1126-6708/2009/07/021}.
\newblock {\tt{arXiv:0901.2338 [hep-th]}}.

\bibitem[Fischler et~al.((2012))Fischler, Pedraza, and
  Tangarife~Garcia]{Fischler_2012}
W.~Fischler, J.~Pedraza, and W.~Tangarife~Garcia.
\newblock {Holographic Brownian Motion in Magnetic Environments}.
\newblock \emph{Journal of High Energy Physics}, 2012\penalty0 (12), (2012).
\newblock \doi{10.1007/jhep12(2012)002}.
\newblock {\tt{arXiv:1209.1044 [hep-th]}}.

\bibitem[Atmaja((2013))]{Atmaja_2013}
A.~Atmaja.
\newblock {Holographic Brownian Motion in Two Dimensional Rotating Fluid}.
\newblock \emph{Journal of High Energy Physics}, 2013\penalty0 (4), (2013).
\newblock \doi{10.1007/jhep04(2013)021}.
\newblock {\tt{arXiv:1212.5319 [hep-th]}}.

\bibitem[Atmaja et~al.((2014))Atmaja, de~Boer, and Shigemori]{Atmaja_2014}
A.~Atmaja, J.~de~Boer, and M.~Shigemori.
\newblock Holographic {B}rownian {M}otion and {T}ime {S}cales in {S}trongly
  {C}oupled {P}lasmas.
\newblock \emph{Nuclear Physics B}, 880:\penalty0 23–75, (2014).
\newblock \doi{10.1016/j.nuclphysb.2013.12.018}.
\newblock {\tt{arXiv:1002.2429 [hep-th]}}.

\bibitem[Banerjee and Sathiapalan((2014))]{Banerjee_2014}
P.~Banerjee and B.~Sathiapalan.
\newblock {Holographic Brownian Motion in $1+1$ Dimensions}.
\newblock \emph{Nuclear Physics B}, 884, (2014).
\newblock \doi{10.1016/j.nuclphysb.2014.04.016}.
\newblock {\tt{arXiv:1308.3352 [hep-th]}}.

\bibitem[Chakrabortty et~al.((2014))Chakrabortty, Chakraborty, and
  Haque]{Chakrabortty_2014}
S.~Chakrabortty, S.~Chakraborty, and N.~Haque.
\newblock Brownian {M}otion in {S}trongly {C}oupled, {A}nisotropic
  {Y}ang-{M}ills {P}lasma: {A} {H}olographic {A}pproach.
\newblock \emph{Physical Review D}, 89\penalty0 (6), (2014).
\newblock \doi{10.1103/physrevd.89.066013}.
\newblock {\tt{arXiv:1311.5023 [hep-th]}}.

\bibitem[Sadeghi et~al.((2014){\natexlab{a}})Sadeghi, Pourasadollah, and
  Vaez]{Sadeghi_2014_godel}
J.~Sadeghi, F.~Pourasadollah, and H.~Vaez.
\newblock {Holographic Brownian Motion in Three-Dimensional G{\"o}del Black
  Hole}.
\newblock \emph{Advances in High Energy Physics}, 2014, (2014){\natexlab{a}}.
\newblock \doi{10.1155/2014/762151}.
\newblock {\tt{arXiv:1308.2483 [hep-th]}}.

\bibitem[Sadeghi et~al.((2014){\natexlab{b}})Sadeghi, Pourhassan, and
  Pourasadollah]{Sadeghi_2014}
J.~Sadeghi, B.~Pourhassan, and F.~Pourasadollah.
\newblock {Holographic Brownian motion in $2+1$ Dimensional Hairy Black Holes}.
\newblock \emph{The European Physical Journal C}, 74\penalty0 (3),
  (2014){\natexlab{b}}.
\newblock \doi{10.1140/epjc/s10052-014-2793-7}.
\newblock {\tt{arXiv:1312.4906 [hep-th]}}.

\bibitem[Fischler et~al.((2014))Fischler, Nguyen, Pedraza, and
  Tangarife]{Fischler_2014}
W.~Fischler, P.~Nguyen, J.~Pedraza, and W.~Tangarife.
\newblock {Fluctuation and Dissipation in de Sitter Space}.
\newblock \emph{Journal of High Energy Physics}, 2014\penalty0 (8), (2014).
\newblock \doi{10.1007/jhep08(2014)028}.
\newblock {\tt{arXiv:1404.0347 [hep-th]}}.

\bibitem[Moerman and Horowitz((2016))]{moerman2016semi}
R.~Moerman and W.~Horowitz.
\newblock A {S}emi-classical {R}ecipe for {W}obbly {L}imp {N}oodles in
  {P}artonic {S}oup.
\newblock \emph{arXiv preprint}, (2016).
\newblock {\tt{arXiv:1605.09285 [hep-th]}}.

\bibitem[Brown((1828))]{brown1828}
R.~Brown.
\newblock A {B}rief {A}ccount of {M}icroscopical {O}bservations made in the
  {M}onths of {J}une, {J}uly and {A}ugust 1827, on the {P}articles contained in
  the {P}ollen of {P}lants; and on the {G}eneral {E}xistence of {A}ctive
  {M}olecules in {O}rganic and {I}norganic {B}odies.
\newblock \emph{The Philosophical Magazine}, 4\penalty0 (21):\penalty0
  161--173, (1828).

\bibitem[Karch and Katz((2002))]{Karch_2002}
A.~Karch and E.~Katz.
\newblock {Adding Flavor to AdS/CFT}.
\newblock \emph{Journal of High Energy Physics}, 2002\penalty0 (06), (2002).
\newblock \doi{10.1088/1126-6708/2002/06/043}.
\newblock {\tt{arXiv:hep-th/0205236}}.

\bibitem[Polyakov((1998))]{Polyakov_1998}
A.~Polyakov.
\newblock String {T}heory and {Q}uark {C}onfinement.
\newblock \emph{Nuclear Physics B}, 68\penalty0 (1-3):\penalty0 1–8, (1998).
\newblock \doi{10.1016/s0920-5632(98)00135-2}.
\newblock {\tt{arXiv:hep-th/9711002}}.

\bibitem['t~Hooft((1993))]{hooft1993dimensional}
G.~'t~Hooft.
\newblock Dimensional {R}eduction in {Q}uantum {G}ravity.
\newblock \emph{Salamfestschrift: a collection of talks}, 4\penalty0
  (A):\penalty0 1, (1993).
\newblock {\tt{arXiv:gr-qc/9310026}}.

\bibitem[Susskind((1995))]{Susskind_1995}
L.~Susskind.
\newblock The {W}orld as a {H}ologram.
\newblock \emph{Journal of Mathematical Physics}, 36\penalty0 (11):\penalty0
  6377–6396, (1995).
\newblock \doi{10.1063/1.531249}.
\newblock {\tt{arXiv:hep-th/9409089}}.

\bibitem[Thorn((1992))]{thorn1992reformulating}
C.~Thorn.
\newblock Reformulating {S}tring {T}heory with the {$1/N$} {E}xpansion.
\newblock In \emph{Proceedings of the first international Sakharov conference
  on physics}, (1992).
\newblock {\tt{arXiv:hep-th/9405069}}.

\bibitem[Gibbons and Maeda((1988))]{gibbons1988black}
G.~Gibbons and K.~Maeda.
\newblock Black {H}oles and {M}embranes in {H}igher-{D}imensional {T}heories
  with {D}ilaton {F}ields.
\newblock \emph{Nuclear Physics B}, 298\penalty0 (4):\penalty0 741--775,
  (1988).

\bibitem[Horowitz and Strominger((1991))]{horowitz1991black}
G.~Horowitz and A.~Strominger.
\newblock Black {S}trings and {P}-branes.
\newblock \emph{Nuclear Physics B}, 360\penalty0 (1):\penalty0 197--209,
  (1991).

\bibitem[Garfinkle et~al.((1991))Garfinkle, Horowitz, and
  Strominger]{garfinkle1991charged}
D.~Garfinkle, G.~Horowitz, and A.~Strominger.
\newblock Charged {B}lack {H}oles in {S}tring {T}heory.
\newblock \emph{Physical Review D}, 43\penalty0 (10):\penalty0 3140, (1991).

\bibitem[Ball{\'o}n~Bayona and Braga((2007))]{Bayona_2007}
C.~Ball{\'o}n~Bayona and N.~Braga.
\newblock Anti-de {S}itter {B}oundary in {P}oincar{\'e} {C}oordinates.
\newblock \emph{General Relativity and Gravitation}, 39\penalty0 (9), (2007).
\newblock \doi{10.1007/s10714-007-0446-y}.
\newblock {\tt{arXiv:hep-th/0512182}}.

\bibitem[Mateos((2007))]{Mateos_2007}
D.~Mateos.
\newblock String {T}heory and {Q}uantum {C}hromodynamics.
\newblock \emph{Classical and Quantum Gravity}, 24\penalty0 (21), (2007).
\newblock \doi{10.1088/0264-9381/24/21/s01}.
\newblock {\tt{arXiv:0709.1523 [hep-th]}}.

\bibitem[Witten((1998))]{witten1998anti_gauge}
E.~Witten.
\newblock Anti-de {S}itter {S}pace, {T}hermal {P}hase {T}ransition, and
  {C}onfinement in {G}auge {T}heories.
\newblock \emph{Adv. Theor. Math. Phys.}, 2\penalty0
  (IASSNS-HEP-98-21):\penalty0 89--117, (1998).
\newblock {\tt{arXiv:hep-th/9803131}}.

\bibitem[Douglas and Taylor((1998))]{douglas1998branes}
M.~Douglas and D.~Taylor.
\newblock Branes in the {B}ulk of {A}nti-de {S}itter {S}pace, (1998).
\newblock {\tt{arXiv:hep-th/9807225}}.

\bibitem[Das((1999))]{Das_1999}
S.~Das.
\newblock Holograms of {B}ranes in the {B}ulk and {A}cceleration terms in {SYM}
  {E}ffective {A}ction.
\newblock \emph{Journal of High Energy Physics}, 1999\penalty0 (06), (1999).
\newblock \doi{10.1088/1126-6708/1999/06/029}.
\newblock {\tt{arXiv:hep-th/9905037}}.

\bibitem[Gonzalez-Rey et~al.((1999))Gonzalez-Rey, Kulik, Park, and
  Ro{\v{c}}ek]{Gonzalez_Rey_1999}
F.~Gonzalez-Rey, B.~Kulik, I.~Park, and M.~Ro{\v{c}}ek.
\newblock Self-dual {E}ffective {A}ction of {$\mathcal{N}=4$}
  {super-{Y}ang-{M}ills}.
\newblock \emph{Nuclear Physics B}, 544\penalty0 (1-2), (1999).
\newblock \doi{10.1016/s0550-3213(99)00046-2}.
\newblock {\tt{arXiv:hep-th/9810152}}.

\bibitem[Tanabashi et~al.((2018))]{tanabashi2018m}
M.~Tanabashi et~al.
\newblock Particle {D}ata {G}roup {R}eview.
\newblock \emph{Physical Review D}, 98:\penalty0 030001, (2018).

\bibitem[Alday and Maldacena((2007))]{alday2007gluon}
L.~Alday and J.~Maldacena.
\newblock Gluon {S}cattering {A}mplitudes at {S}trong {C}oupling.
\newblock \emph{Journal of High Energy Physics}, 2007\penalty0 (06):\penalty0
  064, (2007).
\newblock {\tt{arXiv:0705.0303 [hep-th]}}.

\bibitem[Hartnoll et~al.((2008))Hartnoll, Herzog, and Horowitz]{Hartnoll_2008}
S.~Hartnoll, C.~Herzog, and G.~Horowitz.
\newblock Building a {H}olographic {S}uperconductor.
\newblock \emph{Physical Review Letters}, 101\penalty0 (3), (2008).
\newblock \doi{10.1103/physrevlett.101.031601}.
\newblock {\tt{arXiv:0803.3295 [hep-th]}}.

\bibitem[Gubser((2008){\natexlab{b}})]{Gubser_Breaking_2008}
S.~Gubser.
\newblock Breaking an {A}belian {G}auge {S}ymmetry {N}ear a {B}lack {H}ole
  {H}orizon.
\newblock \emph{Physical Review D}, 78\penalty0 (6), (2008){\natexlab{b}}.
\newblock \doi{10.1103/physrevd.78.065034}.
\newblock {\tt{arXiv:0801.2977 [hep-th]}}.

\bibitem[Maeda et~al.((2009))Maeda, Natsuume, and Okamura]{Maeda_2009}
K.~Maeda, M.~Natsuume, and T.~Okamura.
\newblock Universality {C}lass of {H}olographic {S}uperconductors.
\newblock \emph{Physical Review D}, 79\penalty0 (12), (2009).
\newblock \doi{10.1103/physrevd.79.126004}.
\newblock {\tt{arXiv:0904.1914 [hep-th]}}.

\bibitem[Hartnoll((2009))]{hartnoll2009lectures}
S.~Hartnoll.
\newblock Lectures on {H}olographic {M}ethods for {C}ondensed {M}atter
  {P}hysics.
\newblock \emph{Classical and Quantum Gravity}, 26\penalty0 (22):\penalty0
  224002, (2009).
\newblock {\tt{arXiv:0903.3246 [hep-th]}}.

\bibitem[Einstein((1905))]{einstein1905motion}
A.~Einstein.
\newblock {On the Motion of Small Particles Suspended in Liquids at Rest
  Required by the Molecular-Kinetic Theory of Heat}.
\newblock \emph{Annalen der physik}, 17\penalty0 (549-560):\penalty0 208,
  (1905).

\bibitem[Smoluchowski((1906))]{smoluchowski1906kinetic}
M.~Smoluchowski.
\newblock {On the Kinetic Theory of the Brownian Molecular Motion and of
  Suspensions}.
\newblock \emph{Annalen der Physik}, 21:\penalty0 756--780, (1906).

\bibitem[Brush((1968))]{brush1968history}
S.~Brush.
\newblock A {H}istory of {R}andom {P}rocesses: {I.} {B}rownian {M}ovement from
  {B}rown to {P}errin.
\newblock \emph{Archive for History of Exact Sciences}, 5\penalty0
  (1):\penalty0 1--36, (1968).

\bibitem[Bachelier((1900))]{bachelier1900theory}
L.~Bachelier.
\newblock Theory of {S}peculation.
\newblock \emph{Dimson, E. and M. Mussavian (1998), A brief history of market
  efficiency, European Financial Management}, 4\penalty0 (1):\penalty0 91--193,
  (1900).

\bibitem[Wiener((1923))]{wiener1923differential}
N.~Wiener.
\newblock Differential-{S}pace.
\newblock \emph{Journal of Mathematics and Physics}, 2\penalty0 (1-4):\penalty0
  131--174, (1923).

\bibitem[Langevin((1997))]{langevin1908}
P.~Langevin.
\newblock “{O}n the theory of {B}rownian {M}otion” (“{S}ur la
  {T}h{\'e}orie du {M}ouvement {B}rownien,” cr acad. sci.(paris) 146,
  530--533 (1908)).
\newblock \emph{American Journal of Physics}, 65\penalty0 (11):\penalty0
  1079--1081, (1997).

\bibitem[Uhlenbeck and Ornstein((1930))]{uhlenbeck1930theory}
G.~Uhlenbeck and L.~Ornstein.
\newblock On the {T}heory of the {B}rownian {M}otion.
\newblock \emph{Physical Review}, 36\penalty0 (5):\penalty0 823, (1930).

\bibitem[Pottier((2010))]{pottier2010nonequilibrium}
N.~Pottier.
\newblock \emph{Nonequilibrium {S}tatistical {P}hysics: {L}inear {I}rreversible
  {P}rocesses}.
\newblock Oxford University Press, (2010).

\bibitem[Kubo((1966))]{kubo1966fluctuation}
R.~Kubo.
\newblock The {F}luctuation-{D}issipation {T}heorem.
\newblock \emph{Reports on progress in physics}, 29\penalty0 (1):\penalty0 255,
  (1966).

\bibitem[Ornstein((1917))]{ornstein1917brownian}
L.~Ornstein.
\newblock On the {B}rownian {M}otion.
\newblock \emph{Acad. Amst}, 26:\penalty0 1005, (1917).

\bibitem[Mori((1965))]{mori1965transport}
H.~Mori.
\newblock Transport, {C}ollective {M}otion, and {B}rownian motion.
\newblock \emph{Progress of theoretical physics}, 33\penalty0 (3):\penalty0
  423--455, (1965).

\bibitem[Wiener((1930))]{wiener1930generalised}
N.~Wiener.
\newblock Generalised {H}armonic {A}nalysis.
\newblock \emph{Acta mathematica}, 55:\penalty0 117--258, (1930).

\bibitem[Wang and Uhlenbeck((1945))]{wang1945theory}
M.~Wang and G.~Uhlenbeck.
\newblock On the {T}heory of the {B}rownian {M}otion {II}.
\newblock \emph{Reviews of modern physics}, 17\penalty0 (2-3):\penalty0 323,
  (1945).

\bibitem[Dunkel and H{\"a}nggi((2009))]{Dunkel_2009}
J.~Dunkel and P.~H{\"a}nggi.
\newblock Relativistic {B}rownian {M}otion.
\newblock \emph{Physics Reports}, 471\penalty0 (1):\penalty0 1–73, (2009).
\newblock \doi{10.1016/j.physrep.2008.12.001}.
\newblock {\tt{arXiv:0812.1996 [hep-th]}}.

\bibitem[Kubo et~al.((2012))Kubo, Toda, and Hashitsume]{kubo2012statistical}
R.~Kubo, M.~Toda, and N.~Hashitsume.
\newblock \emph{Statistical {P}hysics {II}: {N}onequilibrium {S}tatistical
  {M}echanics}, volume~31.
\newblock Springer Science \& Business Media, (2012).

\bibitem[Brink et~al.((1976))Brink, Di~Vecchia, and Howe]{brink1976locally}
L.~Brink, P.~Di~Vecchia, and P.~Howe.
\newblock A {L}ocally {S}upersymmetric and {R}eparametrization {I}nvariant
  {A}ction for the {S}pinning {S}tring.
\newblock \emph{Physics Letters B}, 65\penalty0 (5):\penalty0 471--474, (1976).

\bibitem[Deser and Zumino((1976))]{deser1976complete}
S.~Deser and B.~Zumino.
\newblock A {C}omplete {A}ction for the {S}pinning {S}tring.
\newblock \emph{Physics Letters B}, 65\penalty0 (4):\penalty0 369--373, (1976).

\bibitem[Polyakov((1989))]{polyakov1989quantum}
A.~Polyakov.
\newblock Quantum {G}eometry of {B}osonic {S}trings.
\newblock In \emph{Supergravities in Diverse Dimensions: Commentary and
  Reprints (In 2 Volumes)}, pages 1197--1200. World Scientific, (1989).

\bibitem[Zwiebach((2004))]{zwiebach2004first}
B.~Zwiebach.
\newblock \emph{A {F}irst {C}ourse in {S}tring {T}heory}.
\newblock Cambridge University Press, (2004).

\bibitem[Polchinski((1998))]{polchinski1998string}
J.~Polchinski.
\newblock \emph{String {T}heory{:} {V}olume {1,} {A}n {I}ntroduction {T}o {T}he
  {B}osonic {S}tring}.
\newblock Cambridge University Press, (1998).

\bibitem[Nambu((1995))]{nambu1995duality}
Y.~Nambu.
\newblock Duality and {H}adrodynamics, {L}ectures at the {C}openhagen {H}igh
  {E}nergy {S}ymposium, 1970.
\newblock In \emph{Broken Symmetry: Selected Papers of Y Nambu}, pages
  280--301. World Scientific, (1995).

\bibitem[Got{\=o}((1971))]{goto1971relativistic}
T.~Got{\=o}.
\newblock Relativistic {Q}uantum {M}echanics of {O}ne-dimensional {M}echanical
  {C}ontinuum and {S}ubsidiary {C}ondition of {D}ual {R}esonance {M}odel.
\newblock \emph{Progress of Theoretical Physics}, 46\penalty0 (5):\penalty0
  1560--1569, (1971).

\bibitem[Carroll((2004))]{carroll2004spacetime}
S.~Carroll.
\newblock \emph{Spacetime and {G}eometry: {A}n {I}ntroduction to {G}eneral
  {R}elativity}.
\newblock Addison Wesley, (2004).

\bibitem[{W}olfram~{R}esearch{,} Inc.((2018))]{Mathematica}
{W}olfram~{R}esearch{,} Inc.
\newblock Mathematica, {V}ersion 11.3, (2018).
\newblock Champaign, IL.

\bibitem[Rich et~al.((2018))Rich, Scheibe, and Abbasi]{Rich2018}
A.~Rich, P.~Scheibe, and N.~Abbasi.
\newblock {R}ule-based {I}ntegration: {A}n {E}xtensive {S}ystem of {S}ymbolic
  {I}ntegration {R}ules.
\newblock \emph{Journal of Open Source Software}, 3\penalty0 (32):\penalty0
  1073, (2018).
\newblock \doi{10.21105/joss.01073}.
\newblock {Available at \tt{https://doi.org/10.21105/joss.01073}}.

\bibitem[Bañados et~al.((1992))Bañados, Teitelboim, and
  Zanelli]{Banados_1992}
M.~Bañados, C.~Teitelboim, and J.~Zanelli.
\newblock Black {H}ole in {T}hree-{D}imensional {S}pacetime.
\newblock \emph{Physical Review Letters}, 69\penalty0 (13):\penalty0
  1849–1851, (1992).
\newblock \doi{10.1103/physrevlett.69.1849}.
\newblock {\tt{arXiv:hep-th/9204099}}.

\bibitem[Bardeen et~al.((1976))Bardeen, Bars, Hanson, and
  Peccei]{bardeen1976study}
W.~Bardeen, I.~Bars, A.~Hanson, and R.~Peccei.
\newblock Study of the {L}ongitudinal {K}ink {M}odes of the {S}tring.
\newblock \emph{Physical Review D}, 13\penalty0 (8):\penalty0 2364, (1976).

\bibitem[Gubser et~al.((2008))Gubser, Gulotta, Pufu, and Rocha]{Gubser2_2008}
S.~Gubser, D.~Gulotta, S.~Pufu, and F.~Rocha.
\newblock Gluon {E}nergy {L}oss in the {G}auge-{S}tring {D}uality.
\newblock \emph{Journal of High Energy Physics}, 2008\penalty0 (10):\penalty0
  052–052, (2008).
\newblock \doi{10.1088/1126-6708/2008/10/052}.
\newblock {\tt{arXiv:0803.1470 [hep-th]}}.

\bibitem[Chesler et~al.((2009){\natexlab{a}})Chesler, Jensen, and
  Karch]{Chesler1_2009}
P.~Chesler, K.~Jensen, and A.~Karch.
\newblock {J}ets in {S}trongly {C}oupled $\mathcal{N}=4$ {Y}ang-{M}ills
  {T}heory.
\newblock \emph{Physical Review D}, 79\penalty0 (2), (2009){\natexlab{a}}.
\newblock \doi{10.1103/physrevd.79.025021}.
\newblock {\tt{arXiv:0804.3110 [hep-th]}}.

\bibitem[Chesler et~al.((2009){\natexlab{b}})Chesler, Jensen, Karch, and
  Yaffe]{Chesler_2009}
P.~Chesler, K.~Jensen, A.~Karch, and L.~Yaffe.
\newblock Light {Q}uark {E}nergy {L}oss in {S}trongly {C}oupled $\mathcal{N}=4$
  {S}upersymmetric {Y}ang-{M}ills {P}lasma.
\newblock \emph{Physical Review D}, 79\penalty0 (12), (2009){\natexlab{b}}.
\newblock \doi{10.1103/physrevd.79.125015}.
\newblock {\tt{arXiv:0810.1985 [hep-th]}}.

\bibitem[Ficnar((2012))]{Ficnar_2012}
A.~Ficnar.
\newblock {AdS/CFT} {E}nergy {L}oss in {T}ime-{D}ependent {S}tring
  {C}onfigurations.
\newblock \emph{Physical Review D}, 86\penalty0 (4), (2012).
\newblock \doi{10.1103/physrevd.86.046010}.
\newblock {\tt{arXiv:1201.1780 [hep-th]}}.

\bibitem[Ficnar et~al.((2013))Ficnar, Noronha, and Gyulassy]{Ficnar_2013}
A.~Ficnar, J.~Noronha, and M.~Gyulassy.
\newblock Falling {S}trings and {L}ight {Q}uark {J}et {Q}uenching at {LHC}.
\newblock \emph{Nuclear Physics A}, 910-911:\penalty0 252–255, (2013).
\newblock \doi{10.1016/j.nuclphysa.2012.12.030}.
\newblock {\tt{arXiv:1208.0305 [hep-th]}}.

\bibitem[Ficnar and Gubser((2014))]{Ficnar_2014}
A.~Ficnar and S.~Gubser.
\newblock Finite {M}omentum at {S}tring {E}ndpoints.
\newblock \emph{Physical Review D}, 89\penalty0 (2), (2014).
\newblock \doi{10.1103/physrevd.89.026002}.
\newblock {\tt{arXiv:1306.6648 [hep-th]}}.

\bibitem[Bars((1994))]{Bars_1994}
I.~Bars.
\newblock Classical {S}olutions of {2D} {S}tring {T}heory in any {C}urved
  {S}pacetime.
\newblock In \emph{Second Paris Cosmology Colloquium}, page 399. World
  Scientific, (1994).
\newblock {\tt{arXiv:hep-th/9411217}}.

\bibitem[Bars and Schulze((1995))]{Bars_1995}
I.~Bars and J.~Schulze.
\newblock Folded {S}trings {F}alling into a {B}lack {H}ole.
\newblock \emph{Physical Review D}, 51\penalty0 (4):\penalty0 1854–1868,
  (1995).
\newblock \doi{10.1103/physrevd.51.1854}.
\newblock {\tt{arXiv:hep-th/9405156}}.

\bibitem[Álvarez and Conde((2002))]{Alvarez_2002}
E.~Álvarez and J.~Conde.
\newblock Are the {S}ring and {E}instein {F}rames {E}quivalent{?}
\newblock \emph{Modern Physics Letters A}, 17\penalty0 (07):\penalty0
  413–420, (2002).
\newblock \doi{10.1142/s0217732302006606}.
\newblock {\tt{arXiv:gr-qc/0111031}}.

\bibitem[Matsuo et~al.((2006))Matsuo, Tomino, and Wen]{Matsuo_2006}
T.~Matsuo, D.~Tomino, and W.~Wen.
\newblock Drag {F}orce in {SYM} {P}lasma with {B} {F}ield from {AdS/CFT}.
\newblock \emph{Journal of High Energy Physics}, 2006\penalty0 (10), (2006).
\newblock \doi{10.1088/1126-6708/2006/10/055}.
\newblock {\tt{arXiv:hep-th/0607178}}.

\bibitem[Kim et~al.((2011))Kim, Shock, and Tarrío]{Kim_2011}
K.~Kim, J.~Shock, and J.~Tarrío.
\newblock The {O}pen {S}tring {M}embrane {P}aradigm with {E}xternal
  {E}lectromagnetic {F}ields.
\newblock \emph{Journal of High Energy Physics}, 2011\penalty0 (6), (2011).
\newblock \doi{10.1007/jhep06(2011)017}.
\newblock {\tt{arXiv:1103.4581 [hep-th]}}.

\bibitem[Cederwall et~al.((1997))Cederwall, von Gussich, Mikovic, Nilsson, and
  Westerberg]{cederwall1996dirac}
M.~Cederwall, A.~von Gussich, A.~Mikovic, B.~Nilsson, and A.~Westerberg.
\newblock On the {Dirac-Born-Infeld} {A}ction for {D}-branes.
\newblock \emph{Physics Letters B}, 390\penalty0 (1-4), (1997).
\newblock \doi{10.1016/s0370-2693(96)01367-6}.
\newblock {\tt{arXiv:hep-th/9606173}}.

\end{thebibliography}

\end{document}